\documentclass[iop,apjl,revtex4]{emulateapj}
\usepackage{graphicx,lscape}
\usepackage{appendix}
\usepackage{natbib,color}
\usepackage[yyyymmdd]{datetime}
\usepackage[breaklinks,colorlinks,citecolor=blue, backref]{hyperref}
\usepackage{breakurl}
\usepackage{enumitem}
\usepackage{url}
\usepackage{comment}
\usepackage{afterpage}

\slugcomment{Version \today~ \currenttime}

\shorttitle{NASA/NOAA MOU Annex Final Report}
\shortauthors{Mays et al.}

\begin{document}

\title{NASA/NOAA MOU Annex Final Report:\\ Evaluating Model Advancements for Predicting CME Arrival Time}

\author{{\bf CCMC:} M. L. Mays\altaffilmark{1}}
\author{P. J. MacNeice\altaffilmark{1}}
\author{A. Taktakishvili\altaffilmark{2,1}}
\author{C. P. Wiegand\altaffilmark{1}}
\author{J. Merka\altaffilmark{3,1}}
\author{\\{\bf SWPC:} E. T. Adamson\altaffilmark{4,5}}
\author{V. J. Pizzo\altaffilmark{5}}
\author{D. A. Biesecker\altaffilmark{5}}
\author{A. R. Marble\altaffilmark{4,5}}
\author{\\{\bf Model developers:} D. Odstrcil\altaffilmark{6,1}}
\author{C. J. Henney\altaffilmark{7}}
\author{C. N. Arge\altaffilmark{1}}
\author{S. I. Jones\altaffilmark{2,1}}
\author{S. Wallace\altaffilmark{8}}

\altaffiltext{1}{NASA Goddard Space Flight Center, Greenbelt, MD, USA}
\altaffiltext{2}{Catholic University of America, Washington, DC, USA}
\altaffiltext{3}{University of Maryland, Baltimore County, USA}
\altaffiltext{4}{Cooperative Institute for Research in Environmental Sciences, University of Colorado at Boulder, Boulder, CO}
\altaffiltext{5}{NOAA Space Weather Prediction Center, Boulder, CO}
\altaffiltext{6}{George Mason University, Fairfax, Virginia, USA}
\altaffiltext{7}{Air Force Research Laboratory, Kirtland Air Force Base, New Mexico, USA}
\altaffiltext{8}{Department of Physics and Astronomy, University of New Mexico}

\begin{abstract}
In 2017 NOAA SWPC and CCMC started a new project under an annex to a memorandum of understanding between NASA and NOAA. The purpose of this project was to assess improvements in CME arrival time forecasts at Earth using the Air Force Data Assimilative Photospheric Flux Transport (ADAPT) model driven by data from the Global Oscillation Network Group (GONG) ground observatories. These outputs are then fed into the coupled Wang-Sheeley-Arge (WSA) - ENLIL model and compared to an operational version of WSA-ENLIL (without ADAPT). The project was performed in close collaboration with the model developers.  SWPC selected a set of 38 historical events over the period of five years from 2012--2014 (33 events) and 2017--2019 (5 events). The overall three-year project consisted of multiple simulation validation studies for the entire event set (1292 simulations): (a) benchmark single map (operational version prior to May 2019) (b) time-dependent sequence of GONG maps driving WSA-ENLIL with 4 different model settings (c) single test simulation of a time-dependent sequence of GONG maps driving ADAPT-WSA-ENLIL (d) single GONG map driving ADAPT-WSA-ENLIL (e) time-dependent sequence of GONG maps driving ADAPT-WSA-ENLIL. We report that for all 38 events, within each model version/settings combination, the CME arrival time error decreased by 0.2 to 0.9 hours when using a sequence of time-dependent zero-point corrected magnetograms compared to using single magnetogram input. Overall, for all events, when using the older uncorrected magnetograms, the CME arrival time error increased for all new model versions/settings combination compared to the benchmark.  Notably for the 5 events in the period 2017--2019 when more reliable zero-point corrected magnetograms were available, the ADAPT-WSA-ENLIL (median arrival realization) CME arrival time error decreased by 3.1${\pm 4.0}$ hours for single map driven ADAPT compared to the single map driven benchmark, by 4.4${\pm 7.2}$ hours for time-dependent ADAPT compared to the time-dependent driven benchmark, and by 5.8$^{+8.6}_{-7.6}$ hours for time-dependent ADAPT compared to the single map driven benchmark. In this report we also discuss replicating the operational model, challenges in detecting CME arrival in simulations, and comparing zero-point corrected and uncorrected magnetogram inputs.
\end{abstract}

\section{Project Overview}\label{overview}
In May 2017, National Oceanic and Atmospheric Administration (NOAA) Space Weather Prediction Center (SWPC) and Community Coordinated Modeling Center CCMC started a new project under an annex to a memorandum of understanding between NASA and NOAA.   The purpose of this project was to assess potential improvements in CME arrival time forecasts at Earth afforded by incorporation of time-dependence within the inner boundary conditions of ENLIL and by coupling to the Air Force Data Assimilative Photospheric Flux Transport (ADAPT) model driven by data from the Global Oscillation Network Group {\it Global Oscillation Network Group} \citep[GONG:][]{harvey1996} ground observatories. These outputs are then fed into the coupled Wang-Sheeley-Arge (WSA) - ENLIL model and then compared to the operational version of WSA-ENLIL (without ADAPT). We also investigated how the CME arrival time changes when using a sequence of different model versions, settings, and magnetogram inputs. All of the simulations in this project were compared to a replicated operational benchmark.

Currently CME arrival time errors at Earth range between 9.8$\pm$2 hours for any model \citep{Vourlidas2019}.  With a sample size of 273 arrivals at 1 AU between 2010 and 2016, \citet{Wold2018} report an error of 10.4$\pm$0.9 hours. \citet{Riley2018} find an arrival time error of $\approx$13 hours for the CME arrival predictions submitted to the CME Scoreboard between 2013--2017  (\href{http://kauai.ccmc.gsfc.nasa.gov/CMEscoreboard/}{{\sf kauai.ccmc.gsfc.nasa.gov/CMEscoreboard/}}). There are multiple possible sources for the CME arrival time uncertainties. These include: the accuracy of the background solar wind through which the CME propagates, the uncertainties in the solar magnetograms used to drive global models, uncertainties associated the CME input parameters to the models, the manner in which the CME is specified in the model, and finally, forecaster error/bias such as level of experience or adapting model inputs to account for model limitations.  This project addresses the uncertainties in the background solar wind and input magnetograms by analyzing the change in CME arrival time error in a series of simulations using different model versions, settings and input magnetograms. Uncertainties related to CME input parameters are not addressed by this project, however our conclusions are isolated from these uncertainties, because the CME input parameters are kept fixed throughout all simulations for each event and error {\it differences} are studied.

\section{Models}\label{models}
The global 3D MHD heliospheric WSA-ENLIL model \citep{arge2004,odstrcil2004} uses synoptic solar magnetic field maps derived from magnetograms to provide a time-dependent description of the background solar-wind plasma and interplanetary magnetic field. The ENLIL inner boundary at 21.5 solar radii (0.1 AU) can be created from a single WSA outer boundary synoptic map computed from a photospheric magnetic field map, or a time-dependent boundary can be created from WSA maps computed from a sequence of photospheric magnetic field maps.

In general, the WSA model can be driven by synoptic magnetograms from any observatory. In this project we used magnetogram synoptic maps from GONG, and an ensemble of ADAPT maps \citep{arge2010,henney2012} computed from GONG magnetograms (see Section \ref{magnetograms}). ADAPT produces an ensemble of 12 model realizations based on varying different model parameters within the range of their uncertainties.

The current WSA-ENLIL construct, deployed in operations at SWPC, relies on a single, daily-updated zero-point uncorrected GONG map, referred to as ``GONGb'' hereafter (further discussed in Section \ref{magnetograms}) . This has proven to be a very reliable datastream, particularly considering the GONG observatory was never intended to serve as an operational platform. However, one notable drawback of terrestrial observatories is their inability to simultaneously view the entirety of the solar surface. Thus, these GONG synoptic maps comprise a sequence of individual photospheric magnetograms, merged, according to a spatial weighting, into a full-surface map, but representing approximately a month's history of observations. Thus, for a given GONG synoptic map, while observations across the solar meridian are current, the regions representing the far-side of the sun are relatively stale. While it is possible to drive the WSA-ENLIL model with a sequence of such synoptic maps, accounting for evolution of surface features across the central meridian, the issue of stale far-side observations persists. As the solar corona is a global structure, the convolution of such spatial and temporal features can lead to substantial errors, particularly in cases where an active region emerges on the far-side.

The ADAPT model aims to address this issue by accounting for photospheric flux transport processes driving the large-scale evolution of the solar photosphere including meridional flow, differential rotation, supergranular diffusion, random flux emergence, and constraining model results through a formal data assimilation paradigm. In this way, ADAPT strives to produce a substantially improved physical representation of the instantaneous, global solar photospheric flux distribution, affording the  potential for advancements in heliospheric modeling efforts by improving upon the nominal GONG synoptic map inputs to WSA-ENLIL.

The project was performed in close collaboration with model developers Carl Henney (ADAPT), Nick Arge (WSA), and Dusan Odstrcil (ENLIL).  The community is encouraged to follow the SWPC/CCMC project website which contains up to date information on the status of the project (\href{https://ccmc.gsfc.nasa.gov/annex/}{{\sf ccmc.gsfc.nasa.gov/annex/}}).  All simulations performed in support of this project are available for download from the project website.

\section{Methodology}\label{method}
At the start of this project SWPC operational forecasts used WSA version 2.2 and ENLIL version 2.6.2, driven by a single daily-updated zero-point uncorrected GONGb map. For the purposes of this project, we refer to these model versions as the SWPC operational benchmark, however we note that in May 2019 SWPC upgraded to ENLIL version 2.9e while still keeping WSA version 2.2, and still driven by a single daily-updated zero-point uncorrected GONGb map. We use WSA versions 2.2 and 4.5, ENLIL version 2.9e, and ENLIL version 2.6.2 as replicated by ENLIL version 2.9e with GONGb and GONGz (zero-point corrected) magnetograms for this project.

SWPC has selected a set of 38 historical events (Section \ref{events}) to test improvements in CME arrival time prediction. The overall validation project consists of multiple simulation experiments for the entire event set:\\

\noindent GONGb WSA 2.2 ENLIL 2.6.2 a3b2 (operational prior to May 2019):\\
\vspace{-6mm}
\begin{enumerate}[label=\alph*)]
\item[(a)] Benchmark: replicating single GONGb map driven WSA version 2.2 and ENLIL version 2.9e replicating  ENLIL version 2.6.2.
\item[(b)] Time-dependent sequence of GONGb maps driving WSA version 2.2 and ENLIL version 2.9e replicating  ENLIL version 2.6.2.
\item[(c)] For a single event, test simulation of: time-dependent sequence of GONG maps driving ADAPT, WSA version 4.5 and ENLIL version 2.9e.
\end{enumerate}
\noindent GONGz ADAPT WSA 4.5 ENLIL 2.9e a8b1:\\
\vspace{-6mm}
\begin{enumerate}[label=\alph*)]
\item[(d)] Single GONG map driving ADAPT, WSA version 4.5 and ENLIL version 2.9e.
\item[(e)] Time-dependent sequence of GONG maps driving ADAPT, WSA version 4.5 and ENLIL version 2.9e.
\end{enumerate}

Items (a)-(c) were selected for year 1 of the project and were presented in the interim year 1 report. To achieve (a), of the 38 events, a subset of 7 events were chosen to fully test that CCMC could replicate the operational ENLIL version 2.6.2 using ENLIL version 2.9e (Section \ref{rep}).  ENLIL ambient settings a3b2 and a8b1 were tested and this is further discussed in Section \ref{rep} with the full settings listed in see Table \ref{tbl:rep}.

WSA version 2.2 was used for (a) and (b) simulations using uncorrected GONGb observations {\sf{corobs=gongb}}. Note that, compared to version 2.2, WSA version 4.5 uses different coefficients in the velocity equation which are more appropriate for the more recent zero-point corrected GONGz observations ({\sf{corobs=gongz}}). A detailed discussion of the GONG corrections is provided in Section \ref{magnetograms}. SWPC desired to check the effect of updating models versions and settings such that only one setting is changed at a time, so we split up the work into sub-stages, resulting in a total of 1292 simulations.   Therefore, four more variations of (a - single map driven) and (b - time-dependent sequence of maps driven) were selected for year 2:\\
\begin{enumerate}
\item[1(a-b).] GONGb WSA 2.2 ENLIL 2.9e a8b1:\\
 WSA version 2.2, ENLIL version 2.9e with {\sf{amb=a8b1}} settings, {\sf{corobs=gongb}} (operational after May 2019).
\item[2(a-b).] GONGz WSA 4.5 ENLIL 2.6.2 a3b2:\\
  WSA version 4.5, ENLIL version 2.9e with {\sf{amb=a3b2}} replication settings, and zero point corrected magnetogram inputs.
\item[3(a-b).] GONGz WSA 4.5 ENLIL 2.9e a8b1:\\ 
  WSA version 4.5, ENLIL version 2.9e with {\sf{amb=a8b1}} settings, and zero point corrected magnetogram inputs. ({\sf{corobs=gongz}}).
\end{enumerate}

For each stage, ENLIL settings (Section \ref{settings}) were kept constant. After each stage, the performance of the new simulation results are compared to the benchmark single map driven stage (a) and time-dependent stage (b) and to other previous stages. Specific metrics for this validation study have been selected (Section \ref{metrics}).  

In Section \ref{adapttest} we perform a resolution test for ADAPT-WSA-ENLIL simulations for a single event (stage c) and present the results driven by all 12 realizations of ADAPT for all 38 events (stages d-e). The ADAPT maps for this study are downloaded directly from \href{ftp://gong2.nso.edu/adapt/maps/gong/}{{\sf ftp://gong2.nso.edu/adapt/maps/gong/}} which is the current operational source.

\section{Events}\label{events}
Initially SWPC selected a set of 33 historical events and 36 CME input parameters were been provided (3 events contain 2 CMEs) for the period of 2012--2014. Later in September 2019, SWPC added 5 more events in 2017--2019 in order to reflect the more reliable zero point corrections available during that period (see Section \ref{magnetograms}).  Recently in  January 2020, SWPC discovered that for some events in which multiple CMEs were simulated together, the CME input parameters for the additional CMEs that do not arrive at Earth were accidentally not provided.  SWPC identified cases in which the additional CMEs could have an influence on the CME arrival time of the main CME and provided the parameters for these additional CMEs.  At this time SWPC also decided to exclude some CME arrivals from the analysis and some of these exclusions are for the additional CMEs. Due to these last minute changes CCMC reran 224 simulations for 7 events and adjusted the validation algorithm to handle arrival exclusions. 

The final 38 chosen events are listed in Table \ref{tbl:events} with the event number, CME start time, input magnetogram time, CME parameters of latitude, longitude, half-width ($w$/2), and speed ($v$) in Heliocentric Earth Equatorial (HEEQ) coordinates, and the observed CME arrival time from the {\it Advanced Composition Explorer} (ACE: \citet{ace}) spacecraft.  CMEs that were included in the simulation but excluded from arrival time statistics are indicated by the asterisk next to the event number.   There are 42 CME arrivals included in the statistics for the 38 events, because there are 4 events in which two CME arrivals are included (event numbers 6, 8 19, 29). At least two of the events (2012-05-17 01:48 UT and  2013-05-22 08:48 UT), have ``glancing blow"/``flank" arrivals.  The other events are generally Earth-directed.

\begin{table*}[!p]
\tabletypesize{\scriptsize}
\caption{Summary of operational CME input parameters and observed arrival times for the 38 selected events to be validated, as described in Section \ref{events}. \label{tbl:events}}
\begin{center}
\begin{tabular}{lcccrrrrc}
\tableline
\tableline
Event & CME start time & CME time at 21.5 R$_{\odot}$ & Magnetogram Time$\dagger$ & \multicolumn{1}{c}{Lat.} & \multicolumn{1}{c}{Lon.} & \multicolumn{1}{c}{$w/2$} & \multicolumn{1}{c}{v} & Arrival Time$^{**}$\\
\# & {\scriptsize[UT]} & {\scriptsize[UT]} & {\scriptsize[UT]} &  {\scriptsize[$^{\circ}$]} &  \multicolumn{1}{c}{{\scriptsize[$^{\circ}$]}} &  \multicolumn{1}{c}{{\scriptsize[$^{\circ}$]}} & \multicolumn{1}{c}{\scriptsize{[km\,s$^{-1}$]}} & {\scriptsize[UT]}\\
\tableline
1	&	2012-01-23T04:00Z	&	2012-01-23T05:50Z	&	2012-01-23T06:00Z	&	29	&	17	&	52	&	1796	&	2012-01-24T14:31Z	\\
2$^*$	&	2012-02-09T21:17Z	&	2012-02-10T04:20Z	&	2012-02-11T00:00Z	&	15	&	-53	&	43	&	532	&	2012-02-13T18:57Z	\\
2	&	2012-02-10T20:00Z	&	2012-02-11T02:56Z	&	2012-02-11T00:00Z	&	16	&	-21	&	40	&	518	&	2012-02-14T07:03Z	\\
3	&	2012-02-24T03:46Z	&	2012-02-24T08:00Z	&	2012-02-24T18:00Z	&	23	&	-8	&	37	&	821	&	2012-02-26T20:58Z	\\
4	&	2012-03-07T01:25Z	&	2012-03-07T01:59Z	&	2012-03-07T06:00Z	&	24	&	-29	&	50	&	2040	&	2012-03-08T10:45Z	\\
4$^*$	&	2012-03-07T00:48Z	&	2012-03-07T03:09Z	&	2012-03-07T06:00Z	&	-10	&	-12	&	41	&	1650	&	---	\\
5	&	2012-04-02T02:12Z	&	2012-04-02T10:02Z	&	2012-04-02T16:00Z	&	26	&	-14	&	31	&	532	&	2012-04-05T20:03Z	\\
6	&	2012-04-18T17:24Z	&	2012-04-18T22:25Z	&	2012-04-20T04:00Z	&	-22	&	27	&	36	&	626	&	2012-04-21T09:25Z	\\
6$^*$	&	2012-04-19T07:24Z	&	2012-04-19T14:55Z	&	2012-04-20T04:00Z	&	-26	&	-6	&	26	&	313	&	---	\\
6	&	2012-04-19T00:00Z	&	2012-04-19T21:39Z	&	2012-04-20T04:00Z	&	-32	&	-35	&	39	&	607	&	2012-04-23T02:33Z	\\
7	&	2012-05-17T01:48Z	&	2012-05-17T04:34Z	&	2012-05-17T22:00Z	&	-17	&	71	&	54	&	1263	&	2012-05-20T01:36Z	\\
8	&	2012-06-13T13:25Z	&	2012-06-13T20:00Z	&	2012-06-14T18:00Z	&	-36	&	-33	&	49	&	597	&	2012-06-16T09:01Z	\\
8	&	2012-06-14T14:12Z	&	2012-06-14T16:57Z	&	2012-06-14T18:00Z	&	-27	&	-6	&	47	&	1177	&	2012-06-16T19:31Z	\\
9	&	2012-07-12T16:48Z	&	2012-07-12T19:13Z	&	2012-07-13T04:00Z	&	-14	&	-1	&	55	&	1453	&	2012-07-14T17:28Z	\\
10	&	2012-07-28T21:12Z	&	2012-07-29T05:25Z	&	2012-07-29T04:00Z	&	-21	&	-37	&	43	&	382	&	2012-08-02T09:22Z	\\
11	&	2012-08-31T20:00Z	&	2012-08-31T22:46Z	&	2012-09-01T06:00Z	&	6	&	-30	&	33	&	1010	&	2012-09-03T11:23Z	\\
12	&	2012-09-28T00:12Z	&	2012-09-28T03:49Z	&	2012-09-28T18:00Z	&	10	&	20	&	55	&	872	&	2012-09-30T22:13Z	\\
13	&	2012-10-05T02:48Z	&	2012-10-05T08:47Z	&	2012-10-05T18:00Z	&	-18	&	7	&	42	&	698	&	2012-10-08T04:31Z	\\
14	&	2012-10-27T12:48Z	&	2012-10-28T00:28Z	&	2012-10-28T22:00Z	&	6	&	11	&	30	&	375	&	2012-10-31T14:40Z	\\
14$^*$	&	2012-10-27T17:36Z	&	2012-10-28T03:59Z	&	2012-10-28T22:00Z	&	15	&	9	&	43	&	321	&	---	\\
15	&	2012-11-20T12:00Z	&	2012-11-20T17:40Z	&	2012-11-21T00:00Z	&	19	&	24	&	47	&	664	&	2012-11-23T21:12Z	\\
16	&	2013-01-13T06:00Z	&	2013-01-13T19:02Z	&	2013-01-13T20:00Z	&	1	&	-3	&	36	&	463	&	2013-01-16T23:00Z	\\
17$^*$	&	2013-01-13T08:30Z	&	2013-01-13T19:02Z	&	2013-01-15T20:00Z	&	1	&	-3	&	36	&	463	&	2013-01-16T23:00Z	\\
17	&	2013-01-15T08:24Z	&	2013-01-15T16:26Z	&	2013-01-15T20:00Z	&	-27	&	15	&	23	&	428	&	2013-01-19T16:47Z	\\
18	&	2013-03-12T10:36Z	&	2013-03-12T15:24Z	&	2013-03-12T18:00Z	&	42	&	8	&	52	&	734	&	2013-03-15T05:05Z	\\
19	&	2013-03-13T00:36Z	&	2013-03-13T06:13Z	&	2013-03-15T10:00Z	&	-23	&	-50	&	42	&	763	&	2013-03-16T01:50Z	\\
19	&	2013-03-15T07:12Z	&	2013-03-15T09:34Z	&	2013-03-15T10:00Z	&	-9	&	-5	&	55	&	1399	&	2013-03-17T05:28Z	\\
20	&	2013-04-11T07:24Z	&	2013-04-11T11:54Z	&	2013-04-11T12:00Z	&	-6	&	-17	&	48	&	743	&	2013-04-13T22:15Z	\\
21	&	2013-04-26T18:24Z	&	2013-04-27T04:37Z	&	2013-04-29T16:00Z	&	-32	&	8	&	30	&	493	&	2013-04-30T08:57Z	\\
22	&	2013-05-17T09:12Z	&	2013-05-17T11:22Z	&	2013-05-17T18:00Z	&	6	&	-24	&	40	&	1498	&	2013-05-19T22:21Z	\\
23	&	2013-05-22T08:48Z	&	2013-05-22T15:37Z	&	2013-05-22T20:00Z	&	9	&	49	&	55	&	1488	&	2013-05-24T17:35Z	\\
24	&	2013-06-28T02:00Z	&	2013-06-28T05:09Z	&	2013-06-28T16:00Z	&	-29	&	42	&	45	&	1063	&	2013-06-30T10:40Z	\\
25$^*$	&	2013-07-06T00:00Z	&	2013-07-06T06:24Z	&	2013-07-07T14:00Z	&	-15	&	-2	&	45	&	307	&	---	\\
25	&	2013-07-06T19:36Z	&	2013-07-06T22:05Z	&	2013-07-07T14:00Z	&	2	&	4	&	42	&	560	&	2013-07-09T19:58Z	\\
26	&	2014-01-04T21:22Z	&	2014-01-04T23:12Z	&	2014-01-06T18:00Z	&	-38	&	6	&	42	&	806	&	2014-01-07T14:25Z	\\
27	&	2014-01-07T18:24Z	&	2014-01-07T20:01Z	&	2014-01-08T00:00Z	&	-21	&	21	&	50	&	2048	&	2014-01-09T19:31Z	\\
28	&	2014-03-23T03:36Z	&	2014-03-23T08:34Z	&	2014-03-23T18:00Z	&	3	&	-53	&	47	&	768	&	2014-03-25T19:25Z	\\
29	&	2014-04-17T01:25Z	&	2014-04-17T07:07Z	&	2014-04-18T18:00Z	&	-26	&	15	&	38	&	529	&	2014-04-20T10:24Z	\\
29	&	2014-04-17T10:12Z	&	2014-04-18T16:14Z	&	2014-04-18T18:00Z	&	-18	&	17	&	43	&	1043	&	2014-04-20T10:24Z	\\
30	&	2014-08-15T17:48Z	&	2014-08-16T02:14Z	&	2014-08-16T16:00Z	&	12	&	2	&	38	&	438	&	2014-08-19T05:58Z	\\
31	&	2014-09-02T16:36Z	&	2014-09-02T22:17Z	&	2014-09-03T14:00Z	&	37	&	-9	&	37	&	708	&	2014-09-06T14:19Z	\\
32	&	2014-09-09T00:06Z	&	2014-09-09T04:05Z	&	2014-09-09T16:00Z	&	24	&	-23	&	43	&	767	&	2014-09-11T22:58Z	\\
33$^*$	&	2014-09-09T00:00Z	&	2014-09-09T04:05Z	&	2014-09-11T02:00Z	&	24	&	-23	&	43	&	767	&	2014-09-11T22:58Z	\\
33	&	2014-09-10T18:00Z	&	2014-09-10T20:16Z	&	2014-09-11T02:00Z	&	15	&	2	&	45	&	1343	&	2014-09-12T15:30Z	\\
34	&	2017-06-28T16:24Z	&	2017-06-28T22:34Z	&	2017-06-29T04:00Z	&	3	&	9	&	25	&	596	&	2017-07-01T16:26Z	\\
35	&	2017-07-14T01:25Z	&	2017-07-14T06:09Z	&	2017-07-14T08:00Z	&	-8	&	45	&	59	&	825	&	2017-07-16T05:14Z	\\
36	&	2017-09-04T20:28Z	&	2017-09-04T23:10Z	&	2017-09-05T02:00Z	&	-24	&	3	&	54	&	1323	&	2017-09-06T23:08Z	\\
37	&	2018-02-12T01:25Z	&	2018-02-12T05:13Z	&	2018-02-12T18:00Z	&	-11	&	20	&	39	&	933	&	2018-02-15T07:50Z	\\
38	&	2019-03-20T11:48Z	&	2019-03-20T16:40Z	&	2019-03-21T00:00Z	&	7	&	28	&	36	&	686	&	2019-03-24T20:43Z	\\
38$^*$	&	2019-03-20T09:15Z	&	2019-03-20T18:31Z	&	2019-03-21T00:00Z	&	-4	&	30	&	11	&	338	&	---	\\
38$^*$	&	2019-03-20T10:45Z	&	2019-03-20T20:20Z	&	2019-03-21T00:00Z	&	0	&	39	&	19	&	325	&	---	\\																		
\tableline
\multicolumn{9}{l}{$^*$  CME included in simulation but excluded from arrival time statistics.}\\
\multicolumn{9}{l}{$^{**}$ CME-associated shock observed by the ACE spacecraft.}\\
\multicolumn{9}{l}{$\dagger$  Observation time of latest magnetogram.  Applicable to single map-driven runs only.}\\
\multicolumn{9}{l}{$\ddagger$ CME event chosen for example ADAPT-WSA-ENLIL simulation with 12 realizations of ADAPT (item c).}
\end{tabular}
\end{center}
\end{table*}

The CME start time is determined from the first time the CME is visible in the {\it SOlar and Heliospheric Observatory} (SOHO: \citet{Domingo1995}) {\it Large Angle and Spectrometric Coronagraph Experiment} (LASCO: \citet{Brueckner1995}) or the {\it Solar TErrestrial RElations Observatory} (STEREO: \citet{Kaiser2008}) {\it Sun Earth Connection Coronal and Heliospheric Investigation} \citep[SECCHI:][]{howard2008} Ahead/Behind coronagraphs. The events and parameters for this study are also available from CCMC's public Space Weather Database Of Notifications, Knowledge, Information (DONKI; \href{https://ccmc.gsfc.nasa.gov/donki}{ccmc.gsfc.nasa.gov/donki}) database via an API and can be downloaded in text format:

\noindent{\href{https://kauai.ccmc.gsfc.nasa.gov/DONKI/WS/get/CMEAnalysis.txt?startDate=2012-01-01&endDate=2014-12-31&mostAccurateOnly=false&keyword=swpc_annex}{https://kauai.ccmc.gsfc.nasa.gov/DONKI/WS/get/CME Analysis.txt?startDate=2012-01-01\&endDate=2014-12-31 \&mostAccurateOnly=false\&keyword=swpc\_annex}} 

\noindent or JSON format:

\noindent{\href{https://kauai.ccmc.gsfc.nasa.gov/DONKI/WS/get/CMEAnalysis?startDate=2012-01-01&endDate=2014-12-31&mostAccurateOnly=false&keyword=swpc_annex}{https://kauai.ccmc.gsfc.nasa.gov/DONKI/WS/get/CME Analysis?startDate=2012-01-01\&endDate=2014-12-31 \&mostAccurateOnly=false\&keyword=swpc\_annex}} 

Methods and lessons learned from the CME Arrival Time and Impact Team \citep{Verbeke2018} were applied to this SWPC/CCMC project, and vice versa. We are also collaborating with the UK MetOffice who plan to perform ADAPT simulations for the same events and input parameters. Additionally, the SWPC/CCMC project's set of operational parameters for 38 events will be used as a validation test set by team members from the research community.  This opens up the possibility for other flux transport, coronal, and heliospheric models to test their performance using the same input parameters as the SWPC/CCMC project.

\section{Model Settings and Inputs}\label{settings}
\subsection{Replication}\label{rep}
SWPC selected seven different events for which full operational WSA (2.2) ENLIL (2.6.2) model inputs and outputs were available to provide to CCMC to test run replication.This 7 event subset is listed in Table \ref{tbl:subset} and was used to confirm that ENLIL versions 2.9e could replicate version 2.6.2 (operational prior to May 2019).  This was done for two reasons: (1) full model outputs were not available for all of the operational runs and (2) this allows us to create variations on the benchmark, such as using WSA 4.5 and {\sf{corobs=gongz}}.   Replication is necessary to form the first benchmark for item (a) as described in Section \ref{method}. 

\begin{figure*}[!p]
\includegraphics[width=0.5\textwidth]{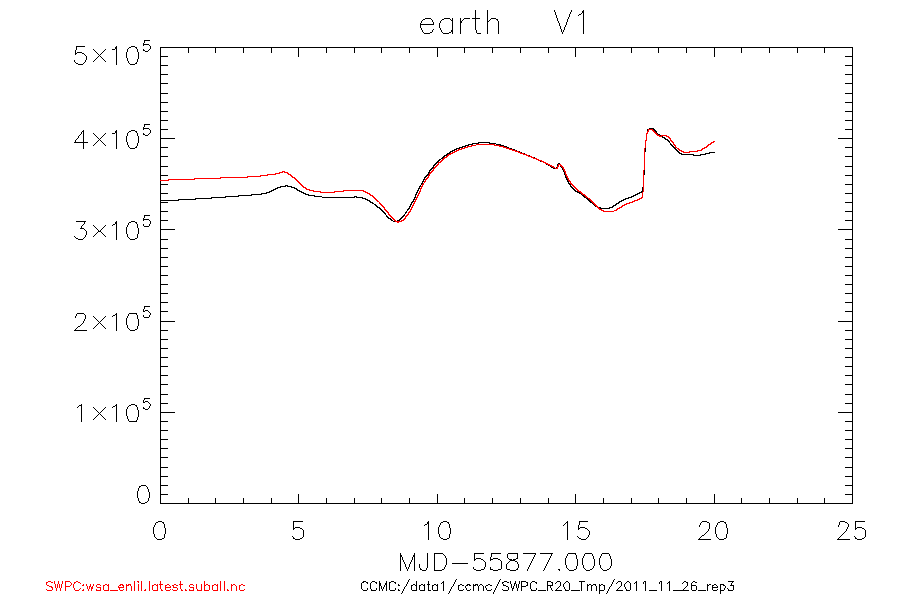}
\includegraphics[width=0.5\textwidth]{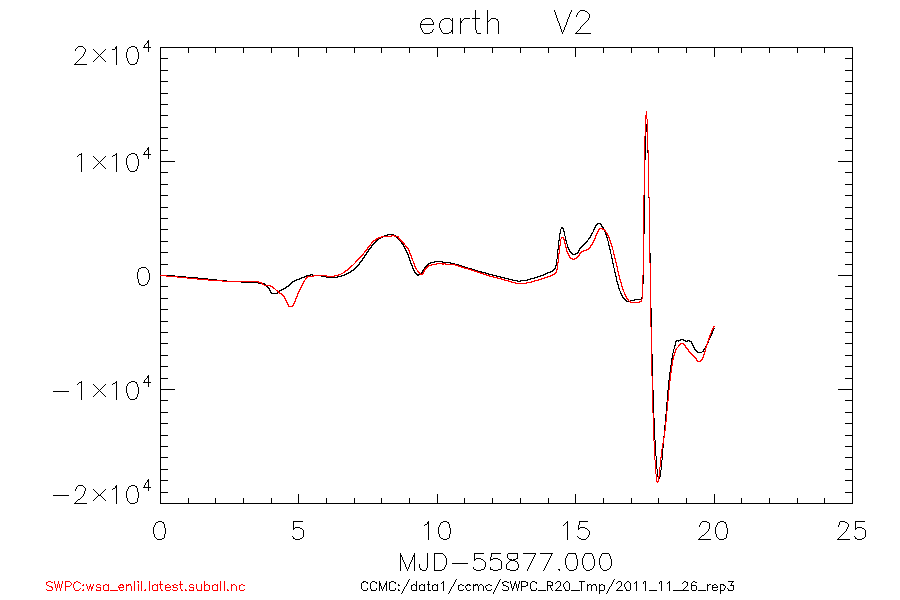}\\
\includegraphics[width=0.5\textwidth]{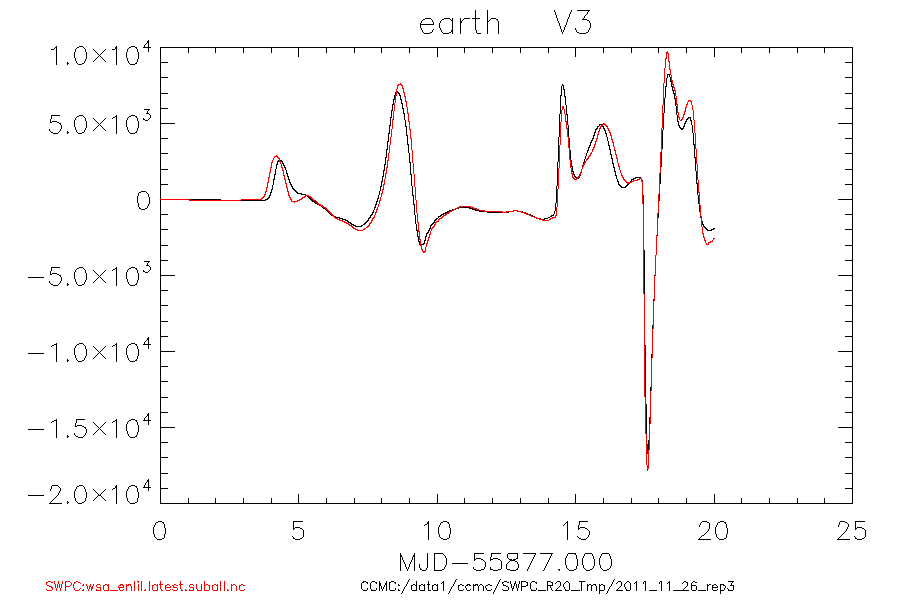}
\includegraphics[width=0.5\textwidth]{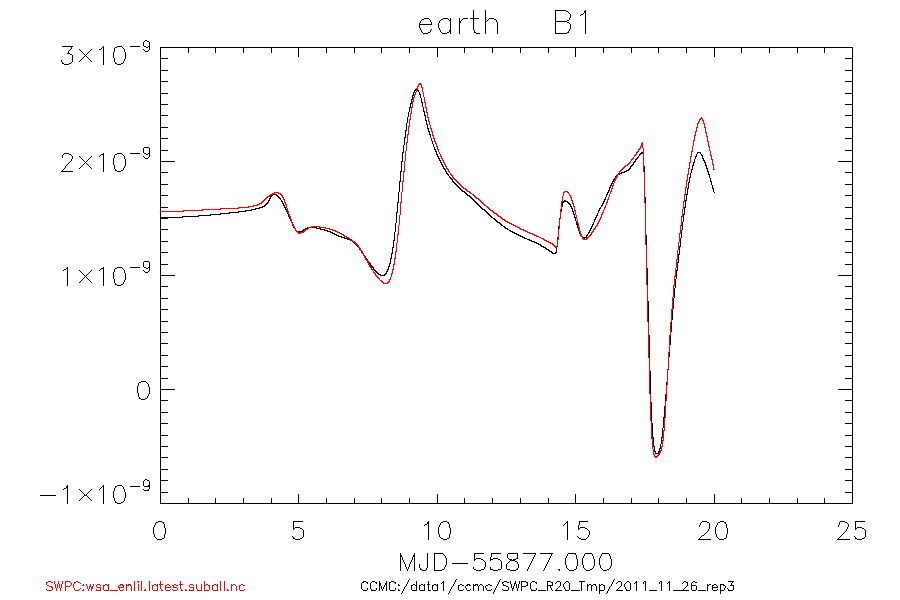}\\
\includegraphics[width=0.5\textwidth]{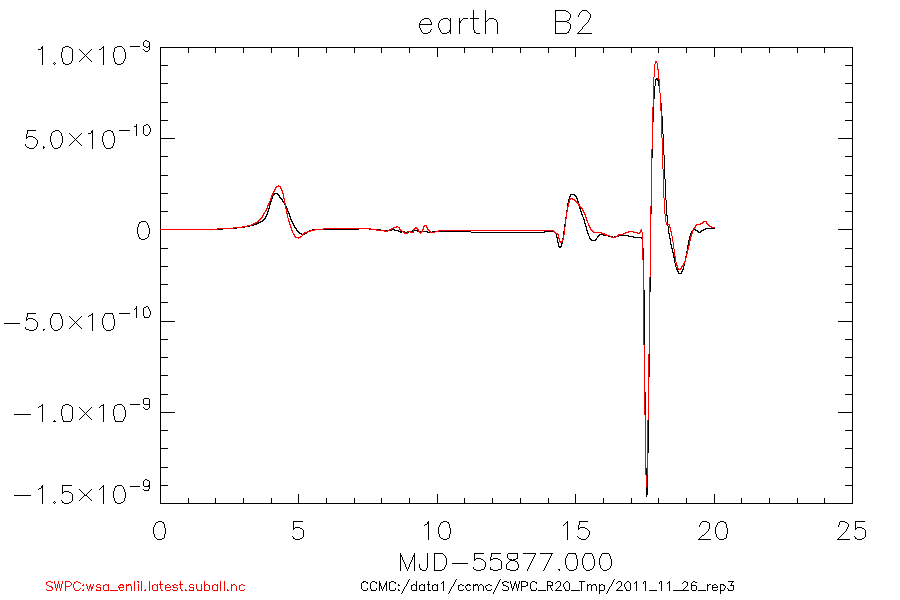}
\includegraphics[width=0.5\textwidth]{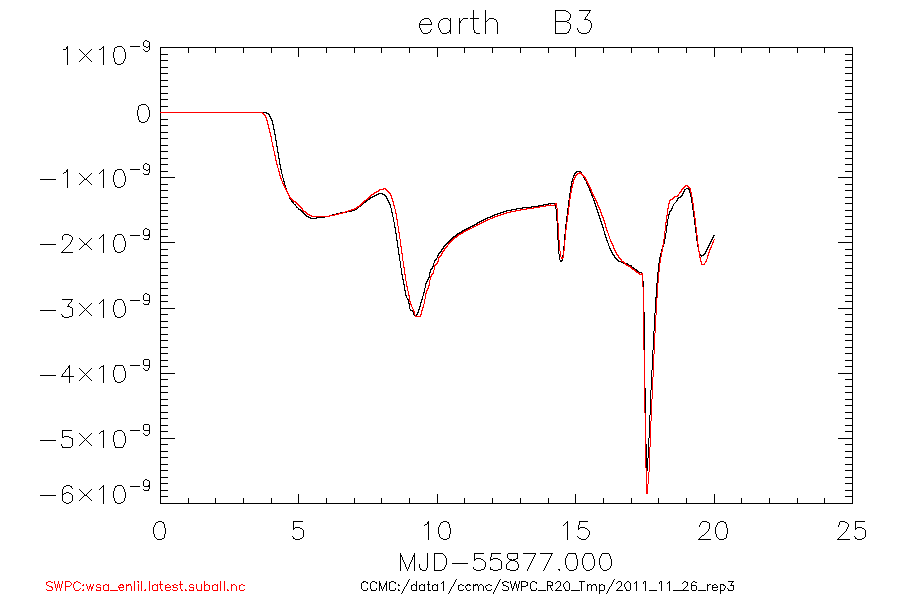}\\
\includegraphics[width=0.5\textwidth]{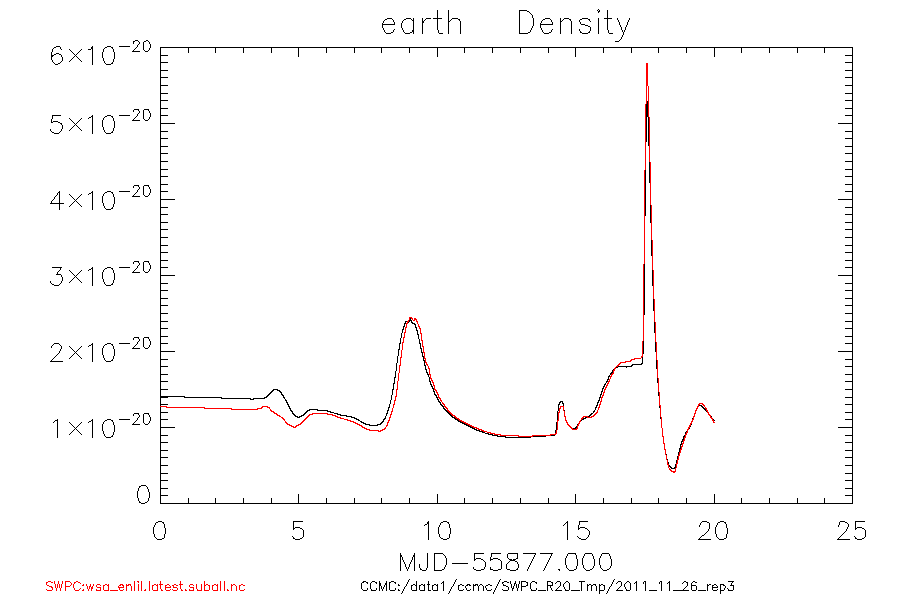}
\includegraphics[width=0.5\textwidth]{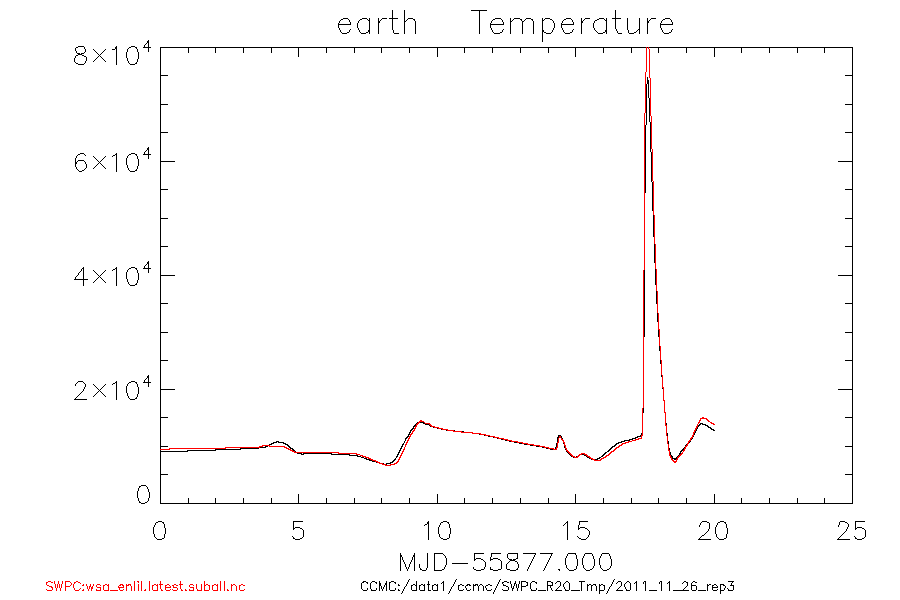}
\caption{Example of replication results at Earth for Event \#1 in Table \ref{tbl:subset}. \label{fig:rep}}
\end{figure*}

The ENLIL model settings for version 2.6.2 identified to be important to achieve replication with version 2.9e are listed in Table \ref{tbl:rep}. We tested our replication settings for all 7 events (\ref{tbl:subset}) and Figure \ref{fig:rep} shows comparisons between the SWPC provided operational model outputs (red) and the ENLIL version 2.9e replicated outputs (black) for MHD quantities at Earth for event \#1 (2011-11-22 12:21 UT and 2011-11-26 11:46 UT CMEs). These figures (and the ones for the other events) give us confidence that, for the purposes of CME arrival time, we are able to achieve adequate replication of SWPC versions of WSA 2.2 and ENLIL 2.6.2 (operational prior to May 2019).  This is further discussed in Section \ref{bench}, where we assess if the benchmark simulations correctly replicate the simulated CME arrival times provided by SWPC. 

However, we note that there are some replication nuances that we do not consider in this study, but would be important to achieve exact replication. Eric Adamson has been investigating differences  between the different ENLIL versions in how the variables {\sf vfast}, {\sf bfast}, {\sf dfast}, and {\sf tfast} are used for scaling and/or clipping WSA values. We also note that the general default value of {\sf bscl} for GONG-driven WSA-ENLIL simulations installed at CCMC, as provided by the model developer, is {\sf bscl}=4 or 5. Dusan Odstrcil suspects a miscommunication led to using {\sf bscl=1} in SWPC operations.  Both of the issues, and the time spent on achieving replication for this study, illustrate the importance of code documentation and version control moving forward.

\begin{table}[ht]
\tabletypesize{\scriptsize}
\caption{Seven events selected by SWPC to test ENLIL version 2.6.2 replication.\label{tbl:subset}}
\begin{center}
\begin{tabular}{lccccc}
\tableline
\tableline
Event & CME time at 21.5 R$_{\odot}$ &  Lat & Lon & $w/2$ & v \\
& {\scriptsize[UT]} & {\scriptsize[$^{\circ}$]} &  {\scriptsize[$^{\circ}$]} &  {\scriptsize[$^{\circ}$]} & \scriptsize{[km\,s$^{-1}$]} \\
\tableline
1$\dagger$ & 2011-11-22 12:21 &  12 & -46 & 34 & 559 \\
           & 2011-11-26 11:46 &   8 &  52 & 55 &  790\\
2$\dagger$ & 2012-01-23 05:50 &  29 &  17 & 52 & 1796\\
3          & 2012-03-13 19:48 &  18 &  55 & 51 & 1512\\
4$\dagger$ & 2012-06-13 20:00 & -36 & -33 & 49 &  597\\
           & 2012-06-14 16:57 & -27 &  -6 & 47 & 1177\\
5          & 2012-08-31 22:57 &   5 & -27 & 35 & 1002\\
6$\dagger$ & 2012-09-28 03:49 &  10 &  20 & 55 &  872\\
7          & 2012-11-20 17:40 &  19 &  24 & 47 &  664\\
\tableline
\multicolumn{6}{l}{$\dagger$ Resolution tests were also performed for these events.}
\end{tabular}
\end{center}
\end{table}

\begin{table*}[!p]
\tabletypesize{\scriptsize}
\caption{ENLIL version 2.9e model settings. Column 2: a3b2 ambient settings and numerical settings used to replicate version 2.6.2 (operational prior to May 2019). Column 3: a8b1 ambient settings and default numerical settings recommended for ENLIL version 2.9e by the model developer. Settings in column 2 were used for simulations (a)-(b) and 2(a)-2(b), and settings in column 3 were used for simulations 1(a)-1(b), 3(a)-3(b),  (d)-(e) (with {\sf{cormode}}=multi for all (b) simulations, and corobs set to the input observatory of each variation).\label{tbl:rep}}
\begin{center}
\def\arraystretch{1.1}
\begin{tabular}{lrrl}
\tableline
\tableline
Variable & \multicolumn{2}{r}{Value} & Description\\
\tableline
{{\sf\textbf{amb}}}	&	{\bf a3b2}	&	{\bf a8b1}	&	{\bf Ambient wind conditions setting:}\\
{\sf{nbrad}}	&	3	&	1	&	Magnetic field correction\\
{\sf{bfast}}	&	300	&	500	&	Radial magnetic field of fast stream (nT) \\
{\sf{bslow}}	&	0	&	400	&	Radial magnetic field of slow stream (nT)\\
{\sf{bmean}}	&	0	&	350	&	Radial magnetic field of mean stream (nT)\\
{\sf{bscl}}	&	1	&	5	&	Magnetic field scaling factor \\
{\sf{dfast}}	&	200	&	200	&	Number density of fast stream (cm$^{-3}$) \\
{\sf{dslow}}	&	2000	&	4000	&	Number density of slow stream (cm$^{-3}$) \\
{\sf{dmean}}	&	0	&	300	&	Number density of mean stream (cm$^{-3}$)  \\
{\sf{dscl}}	&	1	&	1	&	Number density scaling factor  \\
{\sf{tfast}}	&	0.8	&	1.5	&	Mean temperature of fast stream (MK)  \\
{\sf{tslow}}	&	0.1	&	0.1	&	Mean temperature of slow stream (MK)  \\
{\sf{tmean}}	&	0	&	0.5	&	Mean temperature of mean stream (MK)  \\
{\sf{tscl}}	&	1	&	1	&	Mean temperature scaling factor  \\
{\sf{vfast}}	&	675	&	700	&	Radial flow velocity of fast stream (km\,s$^{-1}$)  \\
{\sf{vslow}}	&	225	&	200	&	Radial flow velocity of slow stream (km\,s$^{-1}$)  \\
{\sf{vmean}}	&	0	&	450	&	Radial flow velocity of mean stream (km\,s$^{-1}$)  \\
{\sf{vrfast}}	&	25	&	25	&	Reduction of the maximum flow velocity (km\,s$^{-1}$)  \\
{\sf{vrslow}}	&	25	&	75	&	Reduction of the minimum flow velocity (km\,s$^{-1}$)  \\
{\sf{shift}}	&	8	&	8	&	Azimuthal shift at the inner boundary (deg)  \\
{\sf{nshift}}	&	1	&	1	&	Azimuthal shift at the inner boundary  \\
{\sf{xalpha}}	&	0	&	0.05	&	Fraction of alpha particles (rel to protons)  \\
{\sf{dvexp}}	&	2	&	2	&	Exponent in $NV^{dvexp}$ constant condition  \\
{\sf{nptot}}	&	0	&	0	&	0 if P$_{the}$ balance at boundary  \\
\tableline						
{\bf{\sf{numo}}}	&	{\bf mcp1umn1de}	&	{\bf mcp1va2d}	&	{\bf Numerical model setting}\\
{\sf{difbb}}	&	nodifbb	&	nodifbb	&	Magnetic field diffusion  \\
{\sf{cdifbb}}	&	1	&	1	&	Div(B) diffusion coefficient  \\
{\sf{cdifb}}	&	0.4	&	0.2	&	Div(B) diffusion coefficient\\
{\sf{ceclip}}	&	0.01	&	0.01	&	Clip the thermal energy below fraction of the total energy  \\
{\sf{energy}}	&	ethe	&	ethe	&	Energy equation (ethe | etot | etot2)\\
{\sf{vc}}	&	vc\_avr	&	vc\_max	&	Characteristic speed at cell interfaces \\
{\sf{vs}}	&	vs\_max	&	vs\_max	&	Signal speed at cell interfaces  \\
{\sf{bchalf}}	&	nobchalf	&	nobchalf	&	Boundary conditions at the half step  \\
{\sf{divb}}	&	nodivb	&	nodivb	&	Div(B) treatment  \\
{\sf{mftrace}}	&	mfeuler	&	mfrk4	&	IMF line tracing \\
{\sf{limiter}}	&	minmod	&	albada	&	Slope limiter\\
{\sf{flux}}	&	llf	&	hll	&	Numerical flux\\
{\sf{upwind}}	&	upwind	&	upwindp	&	Numerical upwinding\\
{\sf{gamma}} and heating:	&		&		&	\\
{\sf{heat}}	&	noheat	&	heat	&	Global heating  \\
{\sf{gamma}}	&	1.6666667	&	1.6666667	&	Ratio of specific heats  \\
{\sf{qheat}}	&	0	&	100000	&	Volumetric heating factor  \\
{\sf{nheat}}	&	0	&	2	&	Volumetric heating type  \\
\tableline						
{\bf Spatial grid and timing}	&		&		&	\\
{\sf{x1l}}	&	0.1	&	0.1	&	Inner boundary (AU) \\
{\sf{x1r}}	&	1.7	&	1.1	&	Outer boundary (AU) \\
{\sf{tfrom}}	&	-48	&	-48	&	Output from this time (h)\\
{\sf{tstart}}	&	-360	&	-180	&	Start computation at this time (h)\\
{\sf{tstop}}	&	120	&	120	&	Stop computation at this time (h)\\
{\sf{tstep}}	&	1	&	120	&	Output with this step in time (h)\\
{\sf{res}}	&	med	&	med	&	Numerical grid resolution\\
	&		&		&	\\
{\sf{cormode}}	&	single	&	single/multi	&	Coronal data mode\\
{\sf{corobs}}	&	gongb	&	gongb/gongz/adapt	&	Observatory name\\
\tableline
\end{tabular}
\end{center}
\end{table*}

\subsection{Spatial Resolution}\label{spatial}
\begin{figure*}
\includegraphics[width=0.5\textwidth]{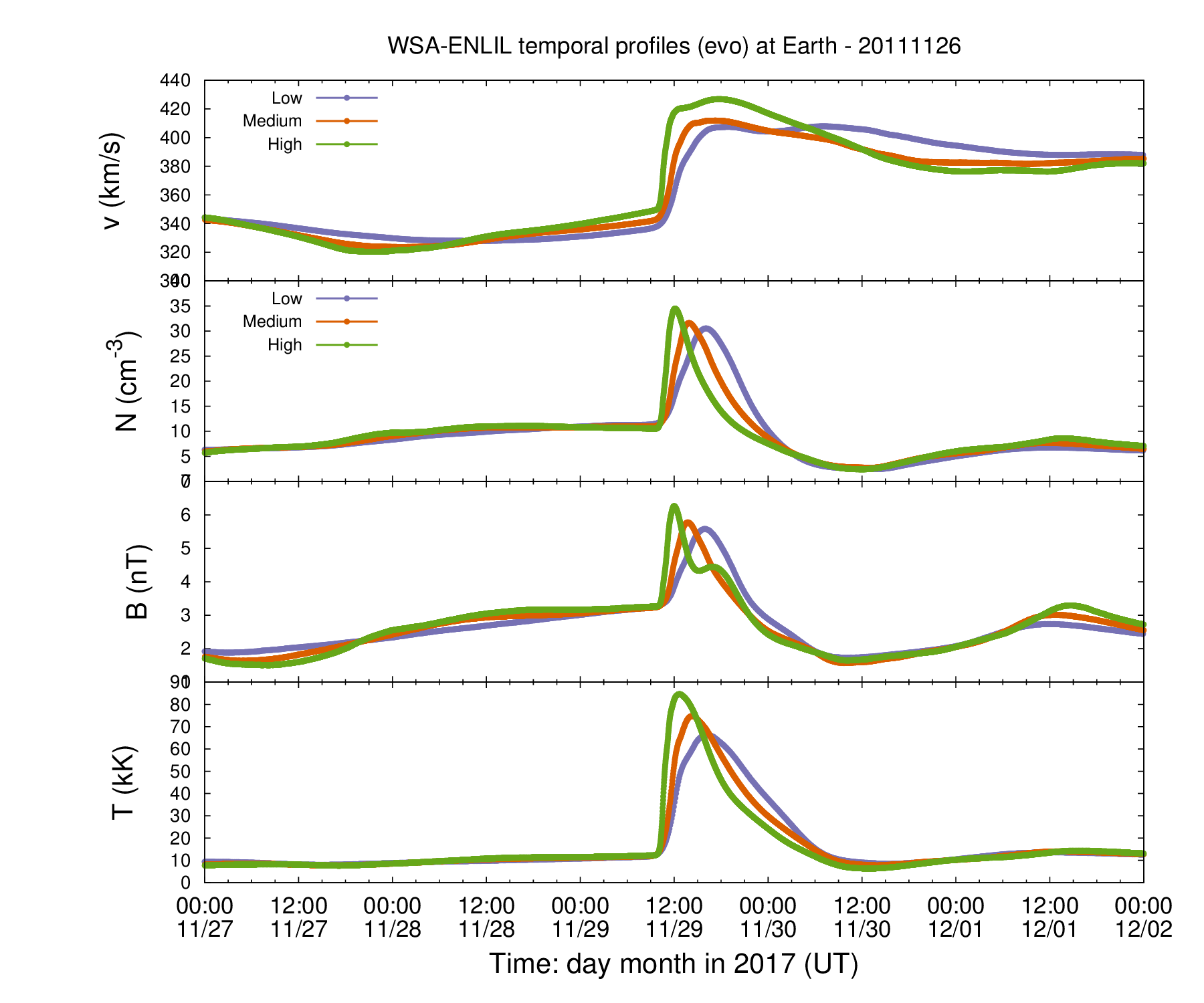}
\includegraphics[width=0.5\textwidth]{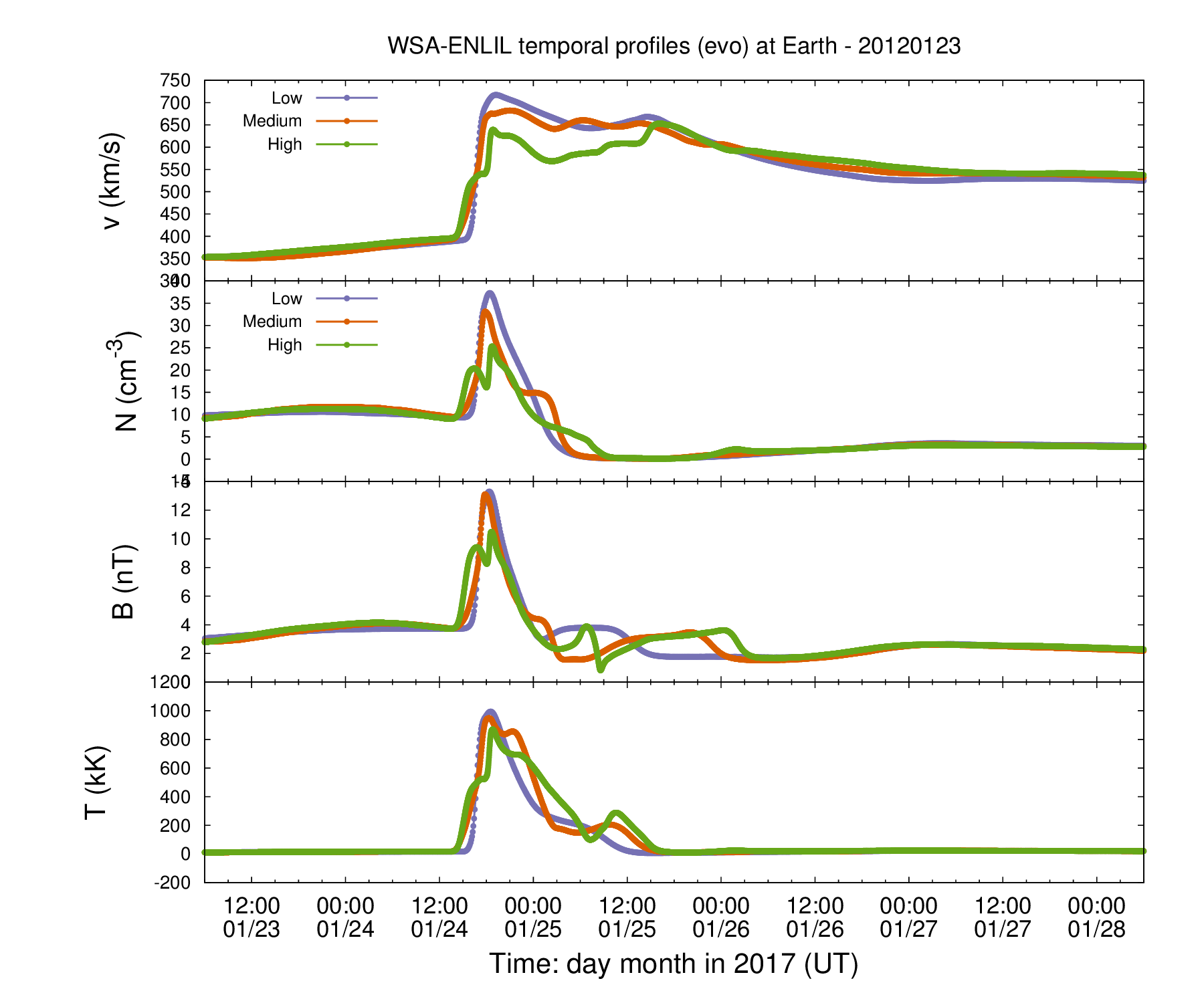}\\
\includegraphics[width=0.5\textwidth]{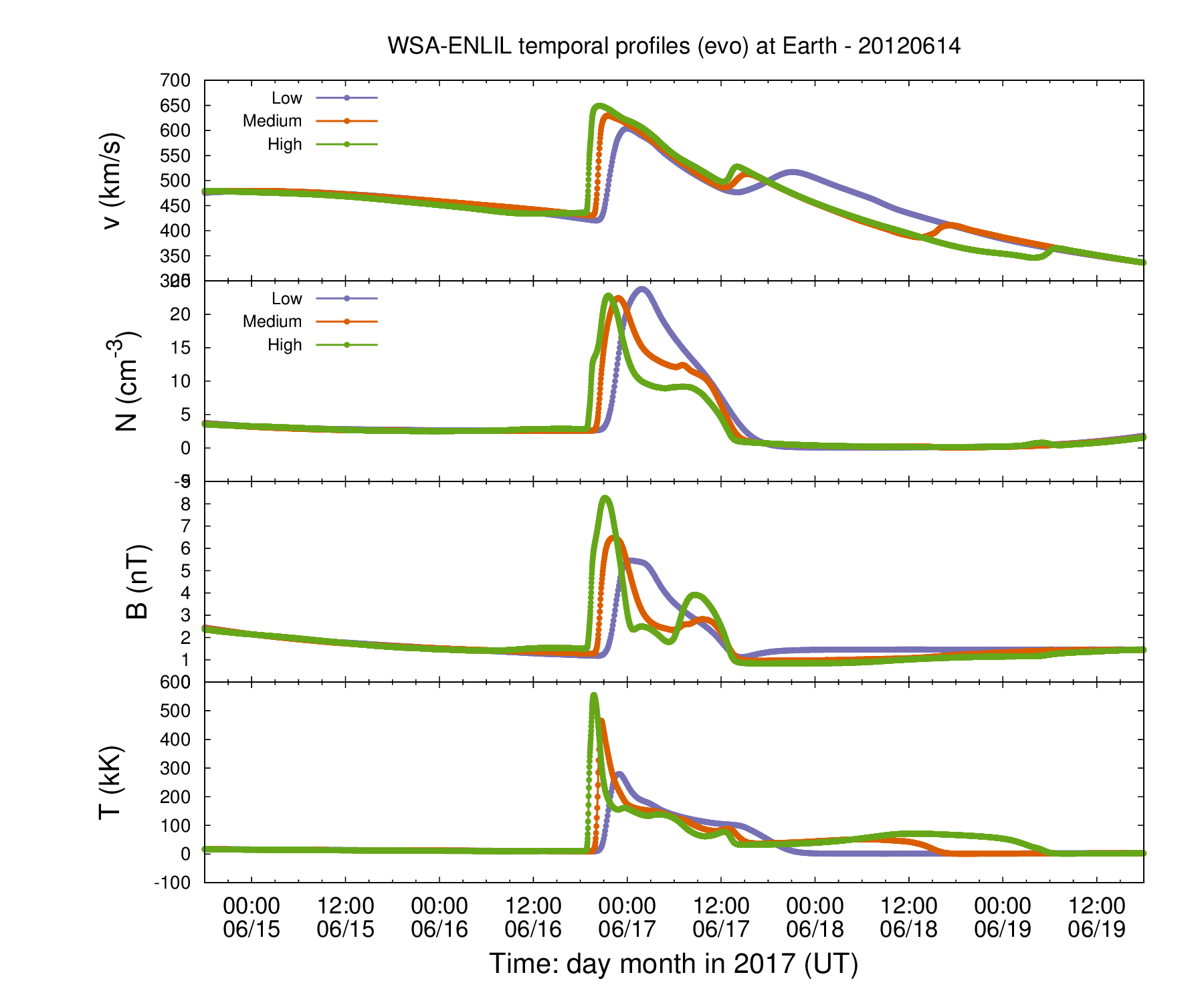}
\includegraphics[width=0.5\textwidth]{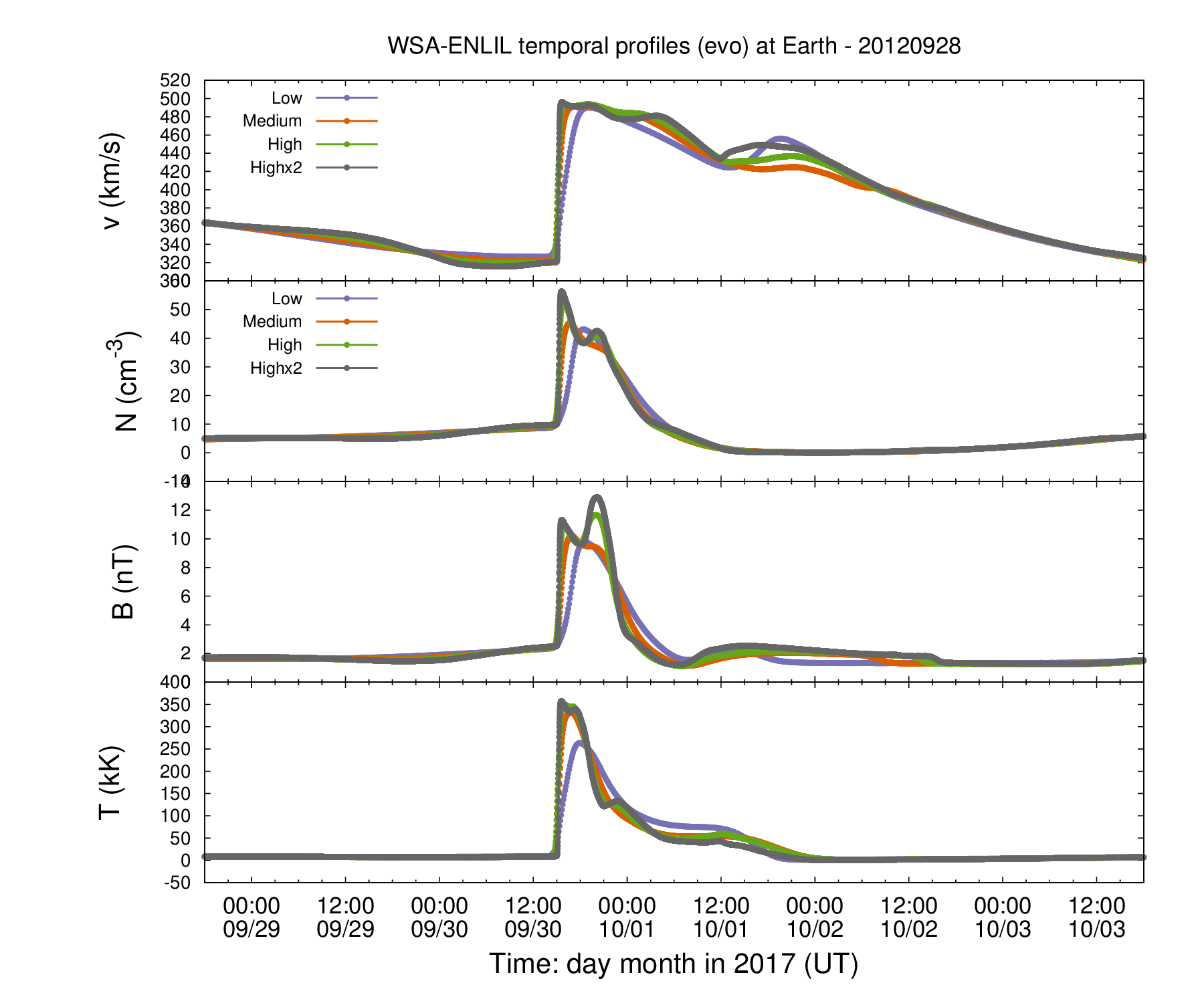}
\caption{Model output time-series of speed, density, magnetic field, and temperature for different resolution settings for the 4 events (of the 7 events listed in Table \ref{tbl:subset}). Low resolution--purple, medium resolution--orange, high resolution--green, high$\times$2 resolution--grey.\label{fig:res}}
\end{figure*}

CCMC performed ENLIL spatial resolution tests for four events (2011-11-22, 2012-01-23, 2012-06-13, 2012-09-28) of the replication event subset listed in Table \ref{tbl:subset} (marked with the $\dagger$ symbol). For these four events we performed ENLIL simulations at low 256$\times$30$\times$90 ($r,\theta,\phi$), medium 512$\times$60$\times$180, and high 1024$\times$120$\times$360 resolutions. We additionally simulated the 2011-11-22 event at high$\times$2 resolution 2048$\times$240$\times$720. Figure \ref{fig:res} shows the model output time-series of speed, density, magnetic field, and temperature at Earth for different spatial resolutions for each event.  SWPC inspected these resolution tests and chose to continue at medium resolution for the validation study of the 38 events (listed in Table \ref{tbl:events}), based on the sharpness in the CME arrival compared to low resolution, and the small difference in CME arrival compared to high resolution.

\subsection{Time resolution}\label{time}
SWPC decided to use a 1 hour time resolution for the 3D output time step, and 1-3 minute output at locations of interest.  This is the output setting for the current operational ENLIL version at SWPC.  We used this setting for stages (a)-(b), however in the interests of computation time and storage full 3D output was supressed (to the beginning and end of the run) for the run variations in stages 1(a-b), 2(a-b), 3(a-b) and ADAPT simulations in stages (d)-(e).  It is always possible to rerun these cases with full 3D output if desired.

\subsection{GONG Magnetogram Inputs to WSA}\label{magnetograms}
GONG acquires a 677nm magnetogram every minute at each currently day-lit observing site.  For each batch of ten at a given site, outliers are rejected and an average is made if there are three or more remaining.  These 10-minute averages are transferred from the remote GONG sites in near-real-time and are denoted by the data product code ``bqa'' (average of quickreduce magnetic field measurements).  Within the context of this CCMC/SWPC project, data products derived from this dataset have been referred to as ``GONGb''.

Because GONG was not originally designed for precise calibration and removal of non-solar magnetic field bias, a separate zero-point corrected (ZPC) ``zqa'' file is [normally] created from each ``bqa''.  The two are identical except for the subtraction of a planar zero-point correction.  The overall amplitude of the correction is determined from modeling of daily variations in the magnetic field measurements, for a given site, over the previous solar rotation period.  It is assumed that the solar mean field is common between sites and averages to zero over this period.  It is further assumed that the non-solar bias varies smoothly throughout each day as well as from one day to the next.  Optionally, the correction also incorporates non-zero orthogonal gradient components acquired from a direct fit to the corresponding ``bqa''.  Within the context of this SWPC/CCMC project, these ZPC data products have been referred to as ``GONGz''.

The validity of the above assumptions and the absence of higher-order structure in the magnetic field bias determines the accuracy of the zero-point corrections.  For example, the corrections are temporarily (i.e., for ~30 days) adversely impacted to a diminishing degree by certain hardware changes or unusually extended gaps in data acquisition.  Thus, while very few changes have been made to the ZPC software itself (one recent exception is noted below), changes in the underlying observations have impacted the nature of the ZPC data products.  

During the latter half of 2015, it began to be appreciated that GONG zero-point corrections made at that time of relatively high solar cycle activity might not have been as reliable as during the quiet Sun period when the underlying techniques were originally developed.  Direct assessment of the accuracy of these corrections is non-trivial; however, an extended analysis subsequently determined that noise associated with modulator transition was leaking into GONG observations and introducing additional magnetic field bias with complicated spatial distribution and daily variation.  Consequently, best-fit planar zero-point corrections potentially over-subtracted the same polar regions that disproportionately impact WSA-ENLIL modeling.  The so-called ``frame-exclusion'' modification to the GONG data acquisition system allowed for the acquisition of (``bqa'') magnetograms free of the identified modulator transition contamination.  This change went into effect at five of the six GONG sites from 2016-08-18 to 2016-09-01 and finally at the remaining Mauna Loa site on 2017-04-20.  Unfortunately, because the modification was to the acquisition of the data itself, there is no way to implement the change retroactively for previously acquired observations.

Eventually, the inclusion of the fitted gradients in the zero-point corrections was deemed no longer necessary following the implementation of the ``frame-exclusion'' modification to the GONG data acquisition system.  In order to avoid any unnecessary introduction of additional noise into the ``zqa'' data products, the ZPC software was modified to suppress subtraction of these gradients for all sites except Cerro Tololo (for which systematic daily variation of these gradients still warranted their removal).

The ``bqs'' and ``zqs'' GONG synoptic maps are derived, respectively, from heliographic remaps of the ``bqa'' and ``zqa'' magnetograms.  A pole-filler is utilized in the creation of both sets of synoptic maps, which has reportedly undergone no substantive changes.  However, the (non-ZPC  ``GONGb'') pole-filler additionally includes a correction for B-angle variations based on sinusoidal fits to polar data from approximately ten years ago.  The current validity of this correction has not been evaluated in recent years, since the ``bqs'' synoptic maps were not intended for continued operational usage.  ADAPT uses heliographic remaps of the GONG ``zqa'' magnetograms as input, from which it creates its own synoptic maps.

This list summarizes the GONG inputs used for all of the different model settings/versions in this study, where (a) is single map driven and (b) is time-dependent sequence of maps driven:
\begin{enumerate}
\item[(a-b).] GONGb WSA 2.2 ENLIL 2.6.2 a3b2 (operational prior to May 2019)
\item[1(a-b).] GONGb WSA 2.2 ENLIL 2.9e a8b1 (operational after May 2019)
\item[2(a-b).] GONGz WSA 4.5 ENLIL 2.6.2 a3b2
\item[3(a-b).] GONGz WSA 4.5 ENLIL 2.9e a8b1
\item[(c).] Single event time-dependent test of GONGz ADAPT WSA 4.5 ENLIL 2.9e a8b1
\item[(d-e)] GONGz ADAPT WSA 4.5 ENLIL 2.9e a8b1
\end{enumerate}

\begin{figure*}
\includegraphics[width=0.49\textwidth]{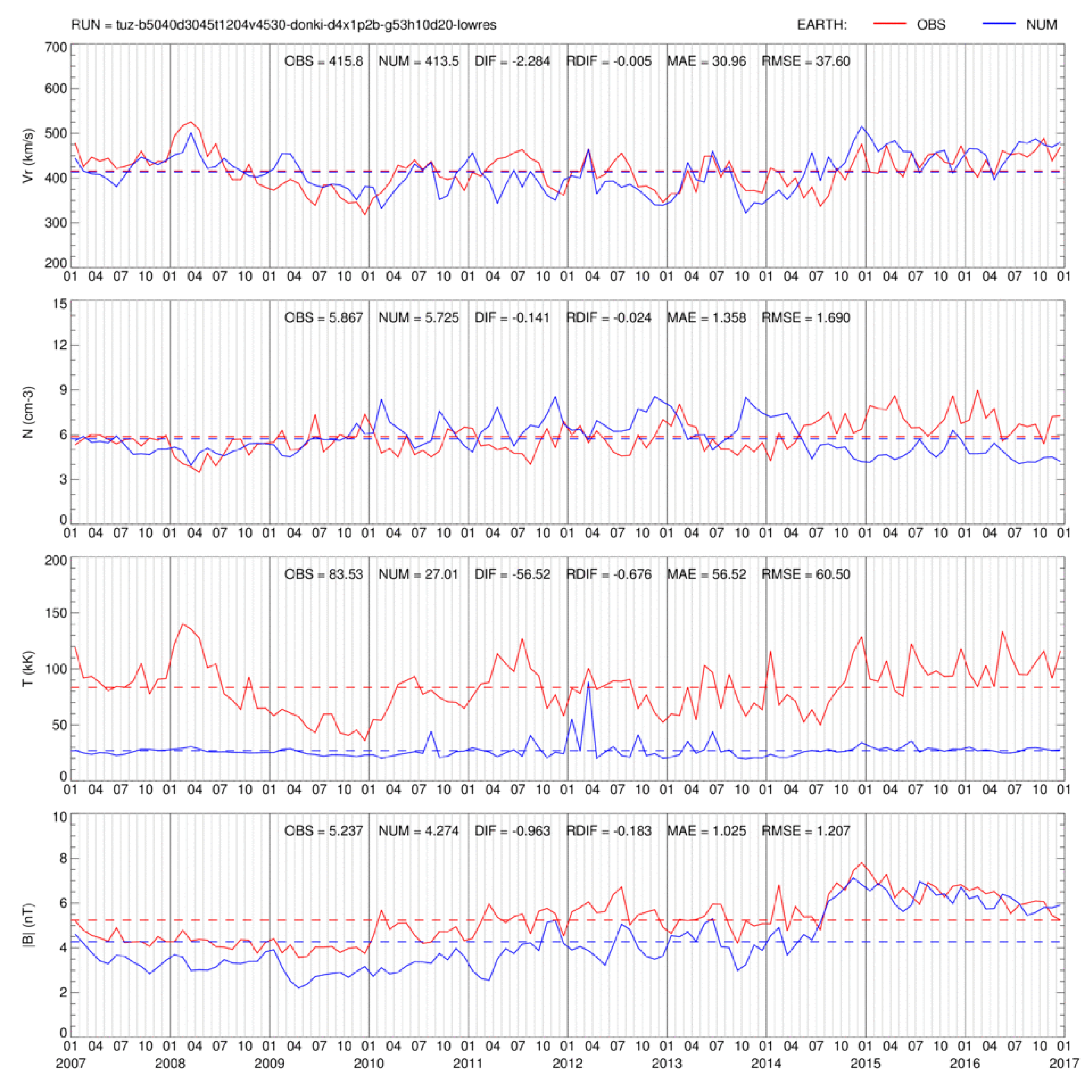}
\includegraphics[width=0.49\textwidth]{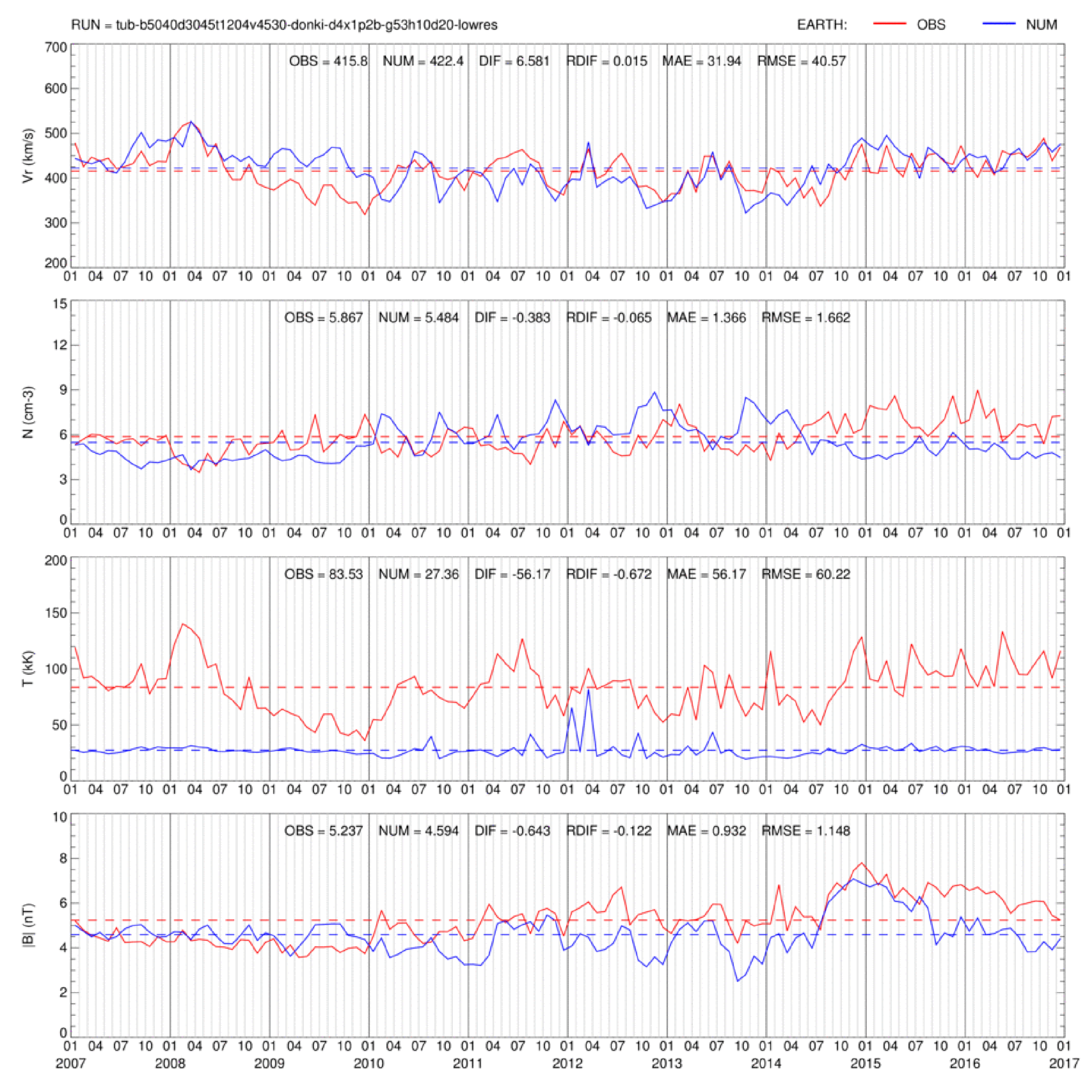}
\caption{Comparison of WSA version 2.2 and ENLIL version 2.9f speed, density, magnetic field, and temperature results (blue) at Earth when using zero-point corrected maps {\sf{corobs=gongz}} (left) and uncorrected maps (right) compared to OMNI observations (red).\label{fig:gongzgongb}}
\end{figure*}

Model developer Dusan Odstrcil has provided a comparison of the WSA (2.2) ENLIL background solar wind prediction time series at Earth for the 2007--2017 time period when using zero-point corrected maps ({\sf{gongz}}) or uncorrected maps ({\sf{gongb}}), both with a 24 hour input cadence.  Figures \ref{fig:gongzgongb} shows the differences in the model output for these two input options. Overall, the magnetic field magnitude is better captured when using the zero-point corrected maps. Note that, compared to version 2.2, WSA version 4.5 uses different coefficients in the velocity equation which are more appropriate for the more recent zero-point corrected GONGz observations. We discuss the impact of corrected maps on CME arrival time for the events in this study in Section \ref{results}.

For the benchmark runs (a), the simulations were driven by a single zero-point uncorrected GONGb map provided by SWPC which was used in their operations. The WSA velocity output file from each map that was used operationally to model the CME event in real-time was provided to CCMC, and was used as input for the benchmark simulations, and for all of the simulations in this study that were driven by GONGb WSA 2.2.  For time-dependent runs (b) a sequence of zero-point uncorrected GONGb maps were used.  Note the original maps provided by SWPC may have since been reprocessed since they were originally downloaded, and therefore may not be the same as one of the maps appearing in the time-dependent sequence. In November 2012, all of the ``bqs'' maps prior to 2012-11-04 17:54 UT were reprocessed, which could effect up to 14 of our events. However since this is just one map in the time-dependent sequence, it is likely a negligible effect. For time-dependent inputs, it is necessary to choose the time-interval for updating the magnetogram at the model inner boundary. Hourly daily-updated magnetograms are available after November 2012, and prior to this date they are available every 6 hours. For settings to be uniform for all runs, SWPC has chosen a magnetogram update frequency of 6 hours for all events. All of the runs for (a) and (b) and their variations in this report use a uniform 6 hour input time-cadence, and (c) uses a 2 hour time-cadence for ADAPT.

\section{Metrics}\label{metrics}
Starting in September 2017, SWPC and CCMC began discussing metrics. SWPC determined that the main metrics for this validation study will be the Mean Error (ME), Mean Absolute Error (MAE), and the Root Mean Square Error (RMSE). We define the CME arrival time prediction error $\Delta t$ for each particular forecast as

\begin{equation}\label{eqn:error}
\Delta t= t^{fcst} - t^{obs},
\end{equation}
where $t^{fcst}$ is the predicted CME arrival time and $t^{obs}$ is the observed CME arrival time and is computed for hit events. This follows the standard practice from atmospheric sciences \citep{Jolliffe2011} and the CME Arrival Time Scoreboard. A negative $\Delta t$ corresponds to a CME arrival observed later than predicted and a positive $\Delta t$, to an arrival observed earlier than predicted.

The observed arrival times $t^{obs}$ at ACE used for this validation study were provided by SWPC and are listed in the last column of Table \ref{tbl:events} and are also available in the DONKI database.  Note that we are comparing model outputs at the Earth trajectory location to ACE observations at L1, about 0.1 AU from the Earth along the Sun-Earth line.  With the medium resolution grid, it takes about 0.08--0.2 hours for CMEs 400--1000 km\,s$^{-1}$ to traverse one grid cell in the radial direction, and about 0.4--1 hours to traverse the grid cells between L1 and the Earth.  Therefore we do not expect the CME arrival time error $\Delta t$ from these runs to be less than 1 hour on average.

SWPC and CCMC agreed to use an existing CCMC algorithm (briefly described in \citet{Wold2018}) to derive the arrival time $t^{fcst}$ from the simulation time series that checks for when the time derivative of the simulated dynamic pressure exceeds 3 nPa. This detects the beginning of the rise in simulated speed and density related to the CME arrival.  SWPC generally determines the arrival time from the simulation from the mid-point of the rise visible in the speed. For the purposes of this project these two methods are not expected to produce very different results compared to the overall arrival time errors. Note that operationally, SWPC forecasters adjust the predicted arrival time $t^{fcst}$ based on uncertainties before issuing a forecast, and we do not consider these human-in-the-loop forecasts here.

In general, the dynamic pressure threshold algorithm detects the simulated CME arrival as expected.  However, in the process of performing this study, we found that the cases where the only the ``flank" of the CME is simulated to arrive at Earth, the algorithm does not detect the CME arrival.  Additionally, for one event (2012-04-02) the CME is simulated to have a head-on arrival but has a weak arrival signature that was not detected.  And finally, of the events that contain more than 2 CMEs, the 2nd CME arrival was not detected some of them. In the interests of keeping events that are not only head-on arrivals, we kept these ``flank" events in the validation study and instead experimented with decreasing the threshold in the algorithm to 0.5 nPa/hr and adding extra filter conditions.  However, even with the adjustments, the algorithm fails to detect arrivals for two events. This adjusted pressure algorithm with filters, developed by Jan Merka, consists of:
\begin{enumerate}
\item Search for CMEs by looking for increasing values of {\sf dp} (cloud tracer variable). Keep track of the maximum {\sf dp} for the CME and search for the next CME.
\item For each detected CME, search for the CME arrival for 6 hours before the {\sf dp} rise up until the maximum {\sf dp} for that CME.
\item The CME arrival time is detected when the time derivative of the dynamic pressure $P_{\rm dyn}= m_p n v^2$ exceeds the threshold of 0.5 nPa/hr, and the temperature and magnetic field derivatives are zero or positive.
\item Search for the next CME arrival time only after the derivative of the dynamic pressure is negative (below -0.0075 nPa/hr).
\item If no CME arrival is detected for CME(s) detected by {\sf dp}, repeat the arrival time search once more, using a lower threshold of 0.1 nPa.
\end{enumerate}
  
However, while these algorithm adjustments worked somewhat well for studies (a) and (b), we wanted a robust, automated method that will work for the entire project.  Especially when there are 12 arrivals to detect for each CME when using ADAPT inputs.  For this reason, we followed ENLIL model developer Dusan Odstrcil's recommendation of running the simulations on 2 ``blocks" ({\sf nblk}=2), wherein one simulation block contains the ambient simulation, and the other contains the ambient + CME(s) simulation. The main reason for multiple simulation block approach in ENLIL is to ensure that both the ambient and CME calculations run at identical time steps so no time interpolations involved, so that they can be directly compared.  The initial algorithm from Odstrcil checks for a 25\% increase between the two simulation blocks in the density time-series, and detects the arrival time as the inflection point after this criteria is reached.  This algorithm proved to be more robust than the pressure-threshold algorithm (using one simulation block), however, there were still issues detecting the 2nd CME in a few of the events.   We worked with Odstrcil to slightly adjust the {\sf nblk}=2 algorithm to the following:  
\begin{enumerate}
\item Search for the timestamp when the ratio of CME simulation density to the ambient simulation density is greater than 1.10: $n_{\rm CME}/n_{\rm ambient} > 1.10$.
\item Next, the inflection point is found immediately following the timestamp from step 1. This is recorded as the CME arrival time.
\item Find the maximum difference between CME and ambient simulation densities after arrival time detection.  Do not search for another CME arrival until difference between the CME and ambient simulation densities start to increase again after this: $n_{\rm CME} (t)/n_{\rm ambient} (t) > n_{\rm CME} (t-1)/n_{\rm ambient} (t-1)$
\item Before resuming the search after the timestamp found in step 3, first check that CME simulation density goes back down to the ambient simulation density ($n_{\rm CME}/n_{\rm ambient} \leq 1$).
\item Continue searching for the next CME arrival, starting at step 1.
\end{enumerate}

Finally, we settled on a simpler multi-block algorithm, by running all of the simulations on {\sf nblk}$=n_{\rm CMEs}+1$ blocks.  For our study, this means all of the single CME simulations were performed with {\sf nblk}=2, all of the double CME simulations were performed with {\sf nblk}=3, and the triple CME simulations were performed with {\sf nblk}=4.  For {\sf nblk}=3: block 2 is for the ambient simulation (no CMEs), block 1, the simulation for the first CME, and block 0, the simulation for the first and second CMEs together (all CMEs). Still, one 2 CME simulation had such a small density signature for the 2nd CME arrival, that we changed the algorithm to use the simulation temperature instead of density for all events. The drawback of using the {\sf nblk}=2, {\sf nblk}=3, and {\sf nblk}=4 methods is the extra computational resources and storage required. Note that in SWPC operations {\sf nblk}=1 is used. The final multi-block algorithm used for this study is as follows: 

For the 1st CME:
\begin{enumerate}
\item Search for the timestamp when the ratio of CME 1+2 simulation temperature (block 0) to the ambient simulation temperature is greater than 1.10: $T_{\rm CME1+2}/T_{\rm ambient} > 1.10$.
\item Next, the inflection point is found immediately following the timestamp from step 1. This is recorded as the 1st CME arrival time.
\end{enumerate}

For the 2nd CME (if {\sf nblk}=3):
\begin{enumerate}
\setcounter{enumi}{3}
\item Search for the timestamp when the ratio of CME 2 simulation temperature to the ambient simulation temperature is greater than 1.10: ($T_{\rm CME1+2}-T_{\rm CME1}+T_{\rm ambient})/T_{\rm ambient} > 1.10$.
\item Next, the inflection point is found immediately following the timestamp from step 3. This is recorded as the 2nd CME arrival time.
\end{enumerate}

The work described above has shown that it is non-trivial to reliably, automatically detect CME arrivals in the model output time-series using any of these algorithms (pressure threshold, {\sf nblk}=2 density, multi-block temperature).  For the time being, we have chosen to use the multi-block temperature algorithm for the validation performed in this report, which we have demonstrated will automatically work for all of the events.  This algorithm uses an increase of 10\% of the temperature ratio which was successful for all of the events except for when it was reduced in events: event 5 2012-04-02 02:24 UT  by 5\% for (d) realization 11, event 7 2012-05-17 01:30 UT  by 5\% for (e) realization 3, by 5\% for 2b, and by 2\% for 2a and 3a.

Now that $t^{obs}$ and $t^{fcst}$ have been defined for computing the CME arrival time error $\Delta t$, next we outline the metrics. The first metric is the Mean Error (ME), given by
\begin{equation}
ME = Bias = \frac{1}{N} \sum_{i=1}^N \Delta t_i
\end{equation}
where $N$ is the total number of hit events in the validation set consisting of each event $i$. The Mean Error is a way of quantifying the bias, as it describes  the models tendency to consistently predict early or late arrival with respect to observations. A negative bias corresponds to on average early predicted arrivals, while a positive bias corresponds to on average late predicted arrivals.  

The Mean Absolute Error (MAE) is defined as
\begin{equation}
MAE = \frac{1}{N} \sum_{i=1}^N |\Delta t_i|.
\end{equation}
While the ME is a measure of the model's bias, it is not adequate to measure the forecasting skill of a model, since negative errors can compensate positive errors. By taking the absolute value of the errors and measuring the distance between the observed and forecast values, we can overcome this. The MAE is very similar to the Mean Square Error (MSE) and the Root Mean Square Error (RMSE), but does have some differences. In practice, it is also more resistant to outlier errors. 

The Root Mean Square Error (RMSE) is a second order moment, and is given by
\begin{equation}
RMSE  = \sqrt{ \frac{1}{N} \sum_{i=1}^N (\Delta t_i)^2. }.
\end{equation}
Compared to the MAE, the RMSE gives more weight to larger errors, due to the errors being squared. 

SWPC and CCMC also discussed measuring forecast performance for other quantities such as comparing the ICME sheath observed mean or max plasma quantities to the simulated quantities, or the cloud duration (using the {\sf dp} cloud tracer variable).  More discussion is needed in this regard on how to make such comparisons and what to compare. Currently, this is outside the scope of this project, however, all of the simulations from this study will be available and we expect these comparisons could be made in follow-up studies by the community.

\section{Results}\label{results}
\subsection{Single Map Simulations (a)}\label{bench}
To complete item (a), GONGb WSA 2.2 ENLIL 2.6.2 a3b2 (operations prior to May 2019), a set of benchmark simulations were performed for the 38 events in this study.  To create the benchmark, we used the operational WSA velocity files and the CME parameters listed in Table \ref{tbl:events} as input and replicated ENLIL version 2.6.2 using version 2.9e, as discussed in Section \ref{rep}. Next, we compared the SWPC-provided manually identified simulated arrival time from the operational runs with the multi-block temperature algorithm identified simulated arrival time from the benchmark runs (see Section \ref{metrics}). Comparing these two simulated arrival times (algorithm identified arrival time$-$manually identified arrival time) yields a mean absolute error of MAE$_{\rm algorithm-manual}$=1.2 hours.  These errors were determined to be reasonably small to assume that replication was successful, and that the difference in general is due to the algorithm detecting first inflection point of the rise due to the CME arrival, compared to the SWPC manual method of using the mid-point of the rise.  We note that there are two outliers in which the benchmark time is very different (events 5 and 14), however we do not anticipate this impacting our results since all conclusions are drawn from examining differences in error compared to the replicated benchmark and not the original operational benchmark.  A full list of the benchmark simulation arrival times and prediction errors for each event are listed in Table \ref{tbl:errors}.  

Figures \ref{fig:time-series}-\ref{fig:time-series3} show the simulated and observed speed at Earth for each of the 38 events.  The SWPC provided simulated arrival times (magenta vertical lines) and those detected by the algorithm (dark purple vertical lines) are shown for comparison on the same plot. The benchmark simulation (a) and arrival times are plotted in dark purple, and the thin lines (of the same color) show the ambient simulation (with no CME). The one hour average OMNI data is plotted in black. Simulation variations 1(a), 2(a), and 3(a) are also over-plotted here in dark red, dark blue and dark green colors respectively. Similar plots with all of the plasma parameters (speed, density, magnetic field, and temperature) time series and plots zoomed in near the CME arrival time (for better clarity) are available to download from the project webpage.

We computed basic validation metrics (as discussed in Section \ref{metrics} for the benchmark simulations compared to the ACE observed arrival times provided by SWPC.  When considering all  of events we found a mean error of ME=-0.9 hours, a mean absolute error of MAE=6.6 hours, and root mean-squared error of RMSE=9.1 hours.  These values are shown in Table \ref{tbl:results} as (a) Benchmark, along with errors when using the SWPC provided manually identified arrival time from the original set of operational simulations.  This table also shows the results for all of the run variations and results divided by year, 2012--2014 when the proper implementation of ZPC for GONGz was not possible, 2017--2019 when GONGz was more accurate, and all of the events together.  Note that the sample size for 2017--2019 only contains 5 events, therefore it difficult to draw statistically significant conclusions, but we report the trends observed for this sample nonetheless.  Another interesting feature of this period is that the CME arrival time error from SWPC operations was much higher (MAE=18.8 hours) than their average, likely due to the difficulty in deriving accurate CME input parameters for these events. This period suffers from the lack of STEREO-B data due a communication loss since September 2014, and longitude of STEREO-A was between -134 to -98 degrees (Heliocentric Earth Equatorial). The error bars are 95\% confidence intervals computed using a bootstrap technique with 10,000 samples using replacement.

To complete the simulations for variation 1(a), GONGb WSA 2.2 ENLIL 2.9e a8b1, we used the same operational WSA velocity files as (a) but used ENLIL version 2.9e with default settings recommended by the model developer and ambient settings ``a8b1''.  These settings are listed in column 3 of Table \ref{tbl:rep}. Errors for each event for simulations 1(a-b) are listed in Table \ref{tbl:errors1}.  For all events, we found a mean error of ME=-1.3 hours, a mean absolute error of MAE=7.3 hours, and root mean-squared error of RMSE=9.9 hours, an increase in CME arrival time error compared to (a) for all time periods.

For variation 2(a), GONGz WSA 4.5 ENLIL 2.6.2 a3b2, we created WSA velocity files using GONGz magnetograms as input and use the same ENLIL version as (a) which was replicating ENLIL version 2.6.2 using version 2.9e.  Note that the GONGb and and GONGz magnetogram products do not have identical timestamps, therefore we chose the timestamp of the GONGz input magnetogram which was closest to the GONGb magnetogram that was used operationally. Errors for each event for simulations 2(a-b) are listed in Table \ref{tbl:errors2}. For all events, we found a mean error of ME=-0.6 hours, a mean absolute error of MAE=7.7 hours, and root mean-squared error of RMSE=10.4 hours, an overall increase in CME arrival time error compared to (a) and 1(a). However, for the period 2017--2019 there is a decrease in error by 2.7 hours compared to (a), reflecting that the reliably corrected GONGz magnetograms decreased the the CME arrival time error.

Finally for variation 3(a), GONGz WSA 4.5 ENLIL 2.9e a8b1, we used the same WSA velocity files from GONGz magnetograms as 2(a) but used ENLIL version 2.9e with default settings recommended by the model developer and ambient settings ``a8b1'' (see Table \ref{tbl:rep}).  This configuration represents the latest model versions and settings of all of the models and input data. Errors for each event for simulations 3(a-b) are listed in Table \ref{tbl:errors3}. For all events, we found a mean error of ME=-1.2 hours, a mean absolute error of MAE=9.0 hours, and root mean-squared error of RMSE=11.7 hours, an overall increase in CME arrival time error compared to (a), 1(a), and 2(a).  Note that for the period 2017--2019 we did not find a similar decrease in arrival time error as was found for 2(a), instead the error remained almost the same, increasing by 0.2 hours compared to (a).  The overall increase in error for variation 3(a) is likely due to the ambient settings (a8b1) not being as appropriate for the period 2012--2014.  The negligible difference in error for 2017--2019 is puzzling but may arise from the ambient parameters or 2.9e model settings.

\begin{table*}[!p]
\tabletypesize{\scriptsize}
\caption{CME arrival time error validation results for all of the simulations in this study. ME=Mean Error, MAE=Mean Absolution Error, RMSE=Root Mean Square Error, TD=Time Dependent.\label{tbl:results}}
  \def\arraystretch{1.8}
\begin{tabular}{lrrr|rrr|rrr}
\tableline
\tableline
& \multicolumn{3}{c|}{2012--2014} &    \multicolumn{3}{c|}{2017--2019} & \multicolumn{3}{c}{2012--2019} \\
& \multicolumn{3}{c|}{(33 events; 37 arrivals)} &    \multicolumn{3}{c|}{(5 events/arrivals)} & \multicolumn{3}{c}{(38; 42 arrivals)} \\
Simulation & ME &  MAE & RMSE & ME &  MAE & RMSE & ME &  MAE & RMSE \\
\tableline
\multicolumn{4}{l|}{\bf GONG WSA 2.2 ENLIL 2.6.2 a3b2}  & & & & & & \\
SWPC Operational$\dagger$ & $-0.9^{+1.8}_{-1.8}$ & $4.5^{+1.1}_{-1.0}$ & $5.6^{+1.0}_{-1.1}$ & $-3.5^{+18.0}_{-18.0}$ & $18.8^{+7.3}_{-8.7}$ & $20.8^{+5.4}_{-7.8}$ & $-1.2^{+2.6}_{-2.7}$ & $6.2^{+2.1}_{-1.8}$ & $8.9^{+2.8}_{-3.0}$\\
(a) Benchmark & $-0.8^{+2.0}_{-1.9}$ & $5.1^{+1.1}_{-1.1}$ & $6.1^{+1.0}_{-1.1}$ & $-1.7^{+17.5}_{-17.5}$ & $17.6^{+8.4}_{-9.6}$ & $20.4^{+6.1}_{-8.5}$ & $-0.9^{+2.7}_{-2.7}$ & $6.6^{+2.0}_{-1.7}$ & $9.1^{+2.8}_{-2.9}$\\
(b) Time dependent & $-0.3^{+2.2}_{-2.1}$ & $5.5^{+1.2}_{-1.2}$ & $6.6^{+1.2}_{-1.2}$ & $+0.5^{+15.9}_{-16.3}$ & $16.2^{+8.8}_{-9.3}$ & $19.3^{+6.5}_{-9.4}$ & $-0.2^{+2.7}_{-2.8}$ & $6.8^{+2.0}_{-1.7}$ & $9.1^{+2.7}_{-2.7}$\\
\multicolumn{1}{r}{MAE$_{\rm TD - bench}$} && $+0.4^{+0.8}_{-0.7}$ &&& $-1.4^{+4.4}_{-5.0}$ &&& $+0.2^{+0.9}_{-0.9}$ &\\
\tableline
\multicolumn{4}{l|}{\bf GONG WSA 2.2 ENLIL 2.9e a8b1}  & & & & & & \\
1(a) Single map & $-1.1^{+2.4}_{-2.3}$ & $5.8^{+1.5}_{-1.4}$ & $7.3^{+1.6}_{-1.6}$ & $-2.3^{+17.9}_{-17.9}$ & $18.6^{+8.1}_{-8.5}$ & $20.9^{+6.0}_{-8.7}$ & $-1.3^{+3.0}_{-2.9}$ & $7.3^{+2.2}_{-1.9}$ & $9.9^{+2.7}_{-2.8}$\\
1(b) Time-dependent & $-0.6^{+2.5}_{-2.4}$ & $6.4^{+1.4}_{-1.3}$ & $7.7^{+1.3}_{-1.4}$ & $-0.6^{+17.6}_{-17.6}$ & $18.3^{+8.1}_{-8.8}$ & $20.7^{+6.7}_{-8.9}$ & $-0.6^{+3.0}_{-3.1}$ & $7.8^{+2.0}_{-1.8}$ & $10.2^{+2.7}_{-2.6}$\\
\multicolumn{1}{r}{MAE$_{\rm TD - Single}$} && $+0.6^{+1.0}_{-1.0}$ &&& $-0.3^{+3.9}_{-4.2}$ &&& $+0.5^{+1.0}_{-1.0}$ &\\
\tableline
\multicolumn{4}{l|}{\bf GONGz WSA 4.5 ENLIL 2.6.2 a3b2}  & & & & & & \\
2(a) Single map & $-0.5^{+3.1}_{-2.8}$ & $6.7^{+2.1}_{-1.8}$ & $9.1^{+3.1}_{-2.8}$ & $-1.1^{+14.4}_{-14.4}$ & $14.9^{+6.9}_{-7.4}$ & $17.0^{+5.8}_{-7.2}$ & $-0.6^{+3.2}_{-3.1}$ & $7.7^{+2.2}_{-2.0}$ & $10.4^{+2.8}_{-2.9}$\\
2(b) Time-dependent & $-1.1^{+2.7}_{-2.6}$ & $6.5^{+1.7}_{-1.6}$ & $8.3^{+2.1}_{-2.0}$ & $-1.7^{+13.3}_{-13.3}$ & $14.6^{+4.9}_{-4.7}$ & $15.6^{+4.5}_{-5.1}$ & $-1.2^{+2.9}_{-2.8}$ & $7.5^{+1.8}_{-1.7}$ & $9.5^{+2.0}_{-2.1}$\\
\multicolumn{1}{r}{MAE$_{\rm TD - Single}$} && $-0.2^{+0.9}_{-0.9}$ &&& $-0.3^{+2.3}_{-2.3}$ &&& $-0.2^{+0.8}_{-0.8}$ &\\
\tableline
\multicolumn{4}{l|}{\bf GONGz WSA 4.5 ENLIL 2.9e a8b1}  & & & & & & \\
3(a) Single map & $-0.9^{+3.4}_{-3.1}$ & $7.8^{+2.2}_{-1.9}$ & $10.1^{+3.3}_{-2.8}$ & $-3.3^{+16.6}_{-16.6}$ & $17.8^{+8.1}_{-6.9}$ & $19.8^{+6.8}_{-7.9}$ & $-1.2^{+3.6}_{-3.5}$ & $9.0^{+2.4}_{-2.1}$ & $11.7^{+3.1}_{-3.1}$\\
3(b) Time-dependent & $-1.2^{+3.1}_{-2.9}$ & $7.6^{+1.9}_{-1.7}$ & $9.4^{+2.4}_{-2.3}$ & $-2.1^{+13.9}_{-13.9}$ & $15.5^{+3.8}_{-4.3}$ & $16.1^{+3.4}_{-4.2}$ & $-1.3^{+3.2}_{-3.1}$ & $8.6^{+1.9}_{-1.8}$ & $10.4^{+2.2}_{-2.1}$\\
\multicolumn{1}{r}{MAE$_{\rm TD - Single}$} && $-0.2^{+0.9}_{-0.9}$ &&& $-2.3^{+3.7}_{-4.4}$ &&& $-0.4^{+0.9}_{-1.0}$ &\\
\tableline
\multicolumn{4}{l|}{\bf GONGz ADAPT WSA 4.5 ENLIL 2.9e a8b1 (median)}  & & & & & & \\
(d) Single map & $-2.7^{+2.8}_{-2.8}$ & $7.5^{+1.7}_{-1.6}$ & $9.1^{+1.7}_{-1.9}$ & $-0.1^{+13.9}_{-13.9}$ & $14.4^{+4.9}_{-6.2}$ & $15.8^{+3.6}_{-5.0}$ & $-2.4^{+3.0}_{-3.0}$ & $8.3^{+1.8}_{-1.7}$ & $10.1^{+1.8}_{-1.9}$\\
(e) Time-dependent & $-0.5^{+2.8}_{-2.8}$ & $6.8^{+1.8}_{-1.7}$ & $8.7^{+1.9}_{-2.0}$ & $-4.2^{+11.2}_{-10.7}$ & $11.8^{+5.3}_{-5.2}$ & $13.4^{+4.9}_{-6.1}$ & $-0.9^{+2.8}_{-2.8}$ & $7.4^{+1.8}_{-1.7}$ & $9.4^{+1.9}_{-2.0}$\\
\multicolumn{1}{r}{MAE$_{\rm TD - Single}$} && $-0.7^{+1.4}_{-1.3}$ &&& $-2.6^{+4.2}_{-6.6}$ &&& $-0.9^{+1.4}_{-1.4}$ &\\
\tableline
\tableline
\end{tabular}

\end{table*}

\begin{figure*}[!ph]
\includegraphics[width=1.0\textwidth]{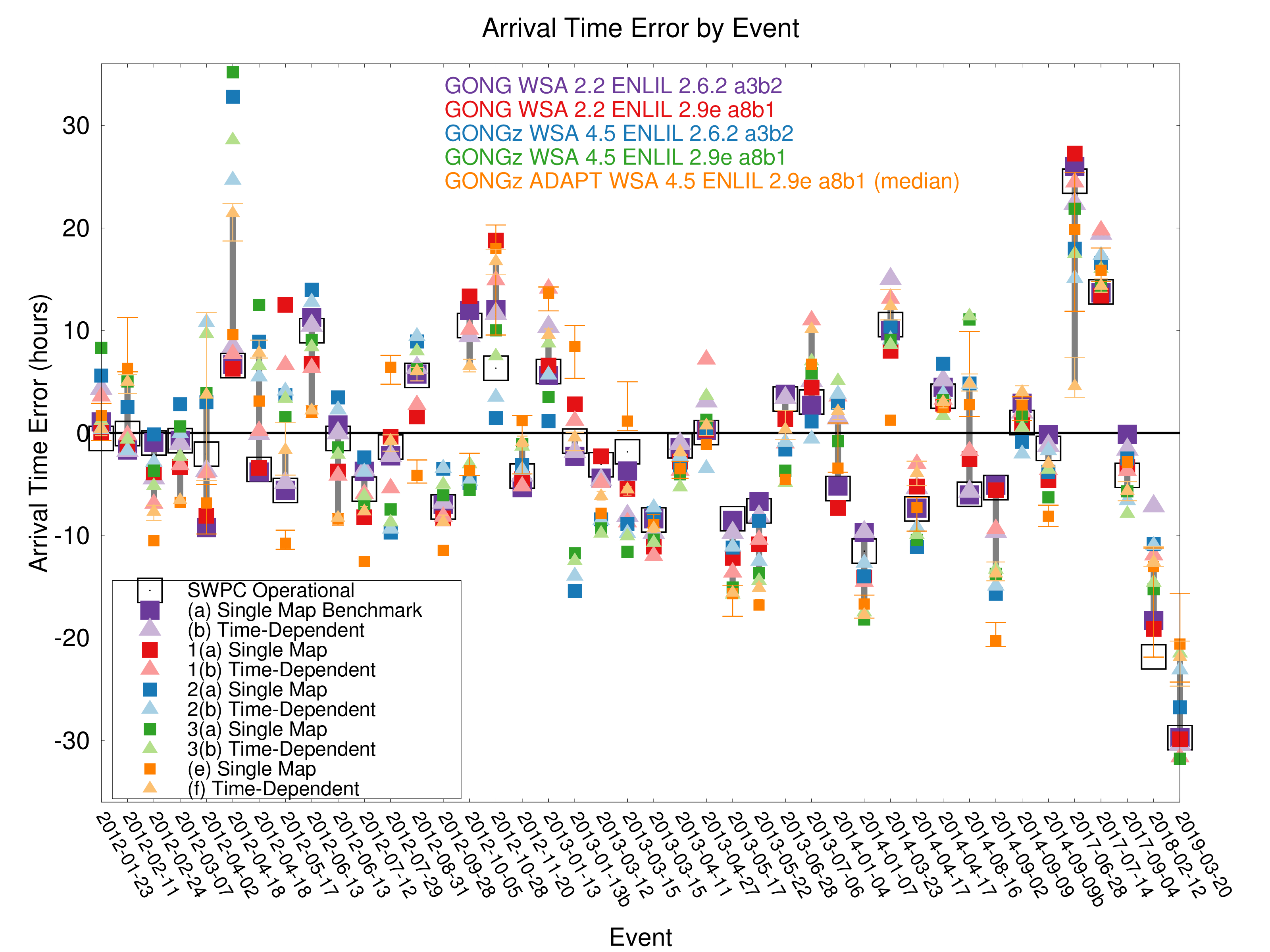}
\caption{Arrival time error for each event and simulation setting (the color for each setting is defined in the legend). Symbols:  hollow black squares=SWPC operational arrival times, squares=benchmark single-map driven arrival times, triangles=time-dependent arrival times.  The grey bars show the distance between the (a) single map benchmark and (e) ADAPT time-dependent map driven arrival times, showing the greatest improvement in arrival time error. The error bars show the range of arrival time errors from the ADAPT ensembles.\label{fig:byevent}}
\end{figure*}

\begin{figure*}[!ph]
\includegraphics[width=1.0\textwidth]{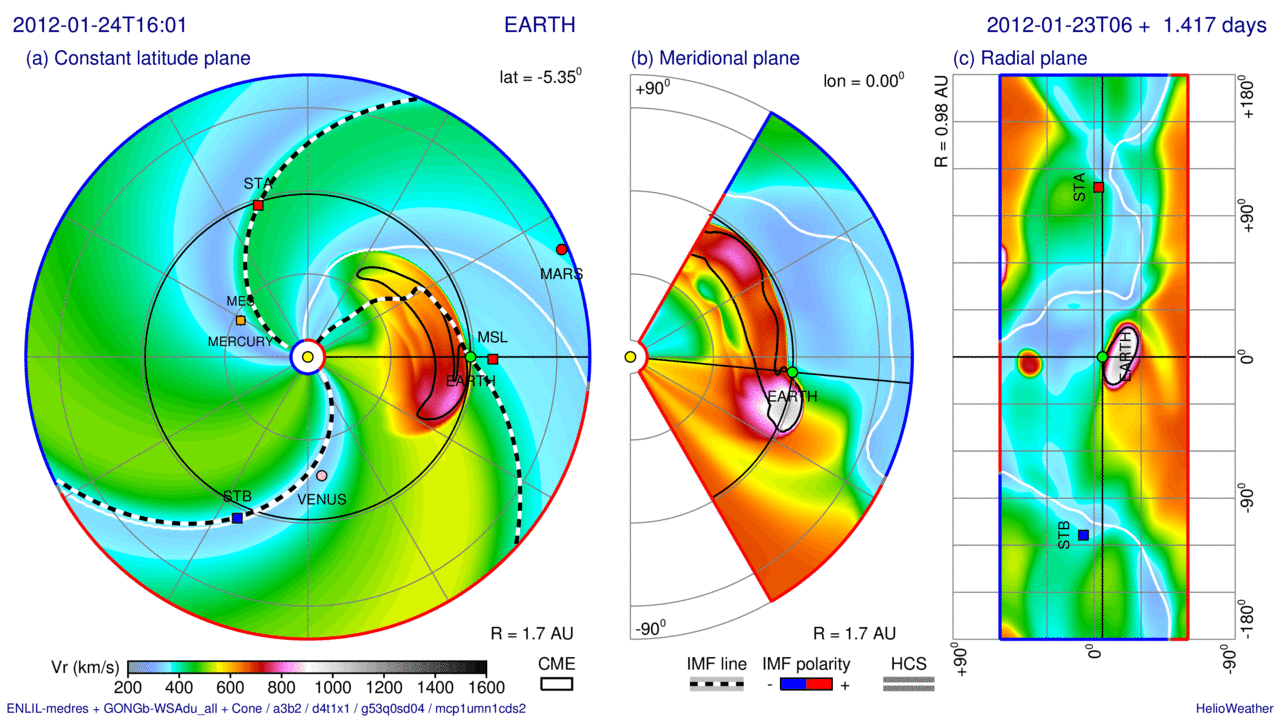}
\includegraphics[width=1.0\textwidth]{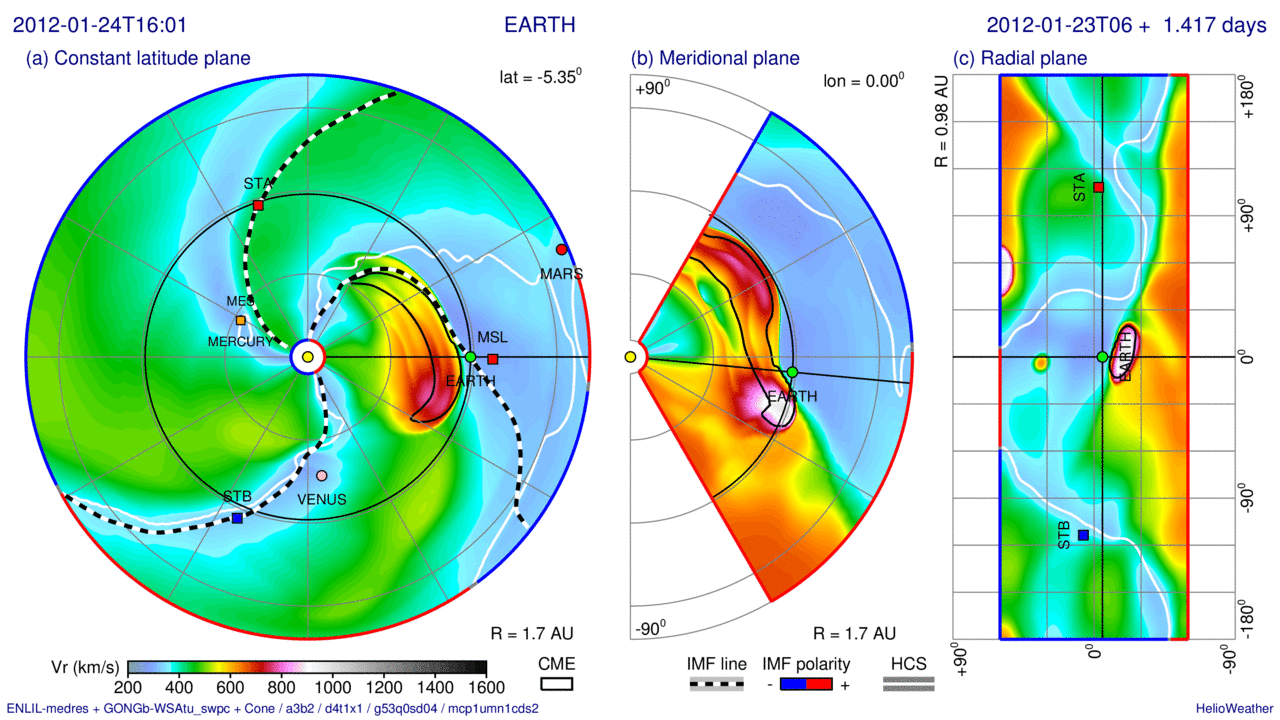}
\caption{GONG-WSA-ENLIL radial velocity contour plots for event 1 (2012-01-23 04:00 UT) showing the (panel a) constant Earth HEEQ latitude plane, (panel b) meridional plane of Earth, and (panel c) 1 AU sphere in cylindrical. The single map driven benchmark (a) simulation (top) shows the CME arrival around 2012-01-24 16:01 UT, while the CME has not yet arrived in the time-dependent driven (b) simulation (bottom) at the same timestamp. For this event, the CME propagates just inside the front of a high speed stream that has a slower speed in the time-dependent driven simulation, causing the CME to arrive later compared to the benchmark.\label{fig:diff1}}
\end{figure*}

\begin{figure*}[!ph]
\includegraphics[width=1.0\textwidth]{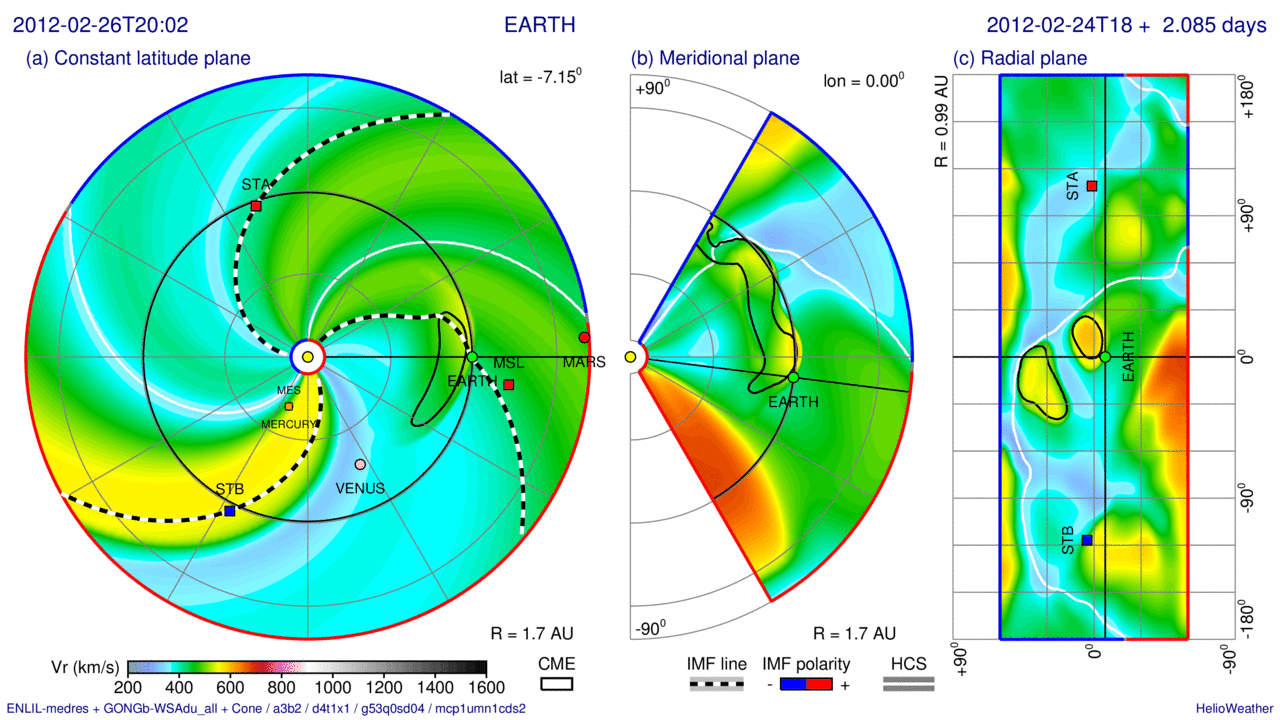}
\includegraphics[width=1.0\textwidth]{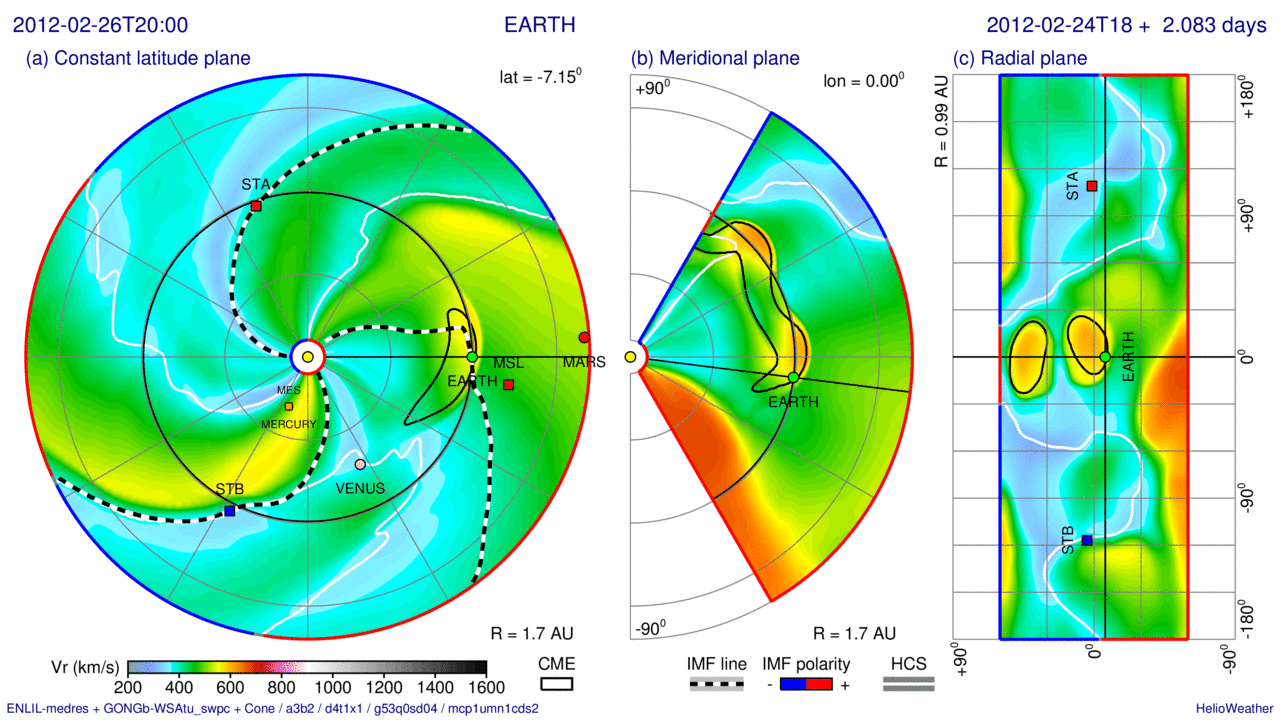}
\caption{GONG-WSA-ENLIL radial velocity contour plots for event 3 (2012-02-24 03:46 UT) in the same format as Figure \ref{fig:diff1}. The single map driven benchmark (a) simulation (top) shows the CME arrival around 2012-02-26 22:02 UT, while the CME has already arrived in the time-dependent driven (b) simulation (bottom) at the same timestamp. For this event, the CME propagates behind a high speed stream that has a higher speed in the time-dependent driven simulation, causing the CME to arrive earlier compared to the benchmark.\label{fig:diff3}}
\end{figure*}

\subsection{Time-dependent Simulations (b)}\label{td}
The same benchmark model settings (Table \ref{tbl:rep}) were applied to the time-dependent runs (b), except for the setting of {\sf{cormode=multi}} up until the simulation {\sf rundate} (CME time at 21.5 R$_{\odot}$, after relaxation).  With this setting, a time-sequence of WSA velocity output maps from a time-sequence of GONG magnetograms with a 6 hour time cadence are used to create the simulation inner boundary.  To emulate the real-time operational environment, the maps are updated until the simulation start date {\sf{rundate}}, after which the final map does not update but continues to co-rotate. Interpolation between velocity maps is performed within ENLIL at each numerical timestep using a co-rotating weighted linear interpolation method for the computational cells at the inner boundary ($\theta,\phi$).  For each numerical timestep, the co-rotating speed in $\phi$ of the immediately preceding and following maps is used to determine ($\theta,\phi$) grid cell location on each map, and an area-weighting procedure is used to calculate the interpolated values.

Figures \ref{fig:time-series}-\ref{fig:time-series3} show the simulation time-series for the benchmark single map-driven runs (a) and time-dependent map-driven runs (b) for all 38 events. The simulated benchmark time-series are plotted in dark purple, time-dependent runs in light purple, and the one hour average OMNI data is plotted in black. The magenta vertical lines show the SWPC provided simulated arrival times, the light purple vertical lines show those detected by the multi-block algorithm for the time-dependent runs, and the black vertical lines show the ACE observed arrival times (provided by SWPC).  Simulation variations 1(b), 2(b), and 3(b) are also over-plotted here in light red, light blue, and light green colors respectively.  The realizations corresponding to the median arrival from the ADAPT simulations (d) and (e) are also plotted here in dark and light orange. Overall, the effects of time-dependence show more variation in the background solar wind compared to the benchmark single-map driven background solar wind.  

Similar to the benchmark runs, we also compute validation metrics for the (b) time-dependent runs compared to the ACE observed arrival times provided by SWPC.  We found a mean error of ME=-0.2 hours, a mean absolute error of MAE=6.8 hours, and root mean-squared error of RMSE=9.1 hours. These metrics are shown in Table \ref{tbl:results} as (b) Time-dependent.  The error for each event is also plotted in Figure \ref{fig:byevent} as light purple triangles and can be compared to the error from the (a) single map run represented as dark purple squares. On average there is a slight increase in error: MAE$_{\rm TD - bench}$=0.2$^{+0.9}_{-0.9}$ hours.

To understand why the time-dependent CME arrival time errors sometimes increased and sometimes decreased we carefully studied the 3D output and time series at Earth for all of the simulations (a-b).  For all events, the difference in CME arrival time was due to the time-dependent simulations creating different background conditions for the CME to propagate through, with the largest arrival time differences being due to CME interactions with high speed streams. We show two examples of this in Figures \ref{fig:diff1} and \ref{fig:diff3} for event 1 (2012-01-23 04:00 UT) and event 3 (2012-02-24 03:46 UT). The figures show the WSA-ENLIL radial velocity contour plot of the CMEs for the (a) constant Earth HEEQ latitude plane, (b) meridional plane of Earth, and (c) 1 AU sphere in cylindrical projection. In general the time-dependent conditions are more variable for both the slow and fast wind, however an increase (or decrease) in the fast wind has the effect of the CME arriving sooner (or later).  For example in event 1, Figure \ref{fig:diff1}  shows the single map driven benchmark (a) simulation (top) with the CME arrival around 2012-01-24 16:01 UT, while the CME has not yet arrived in the time-dependent driven (b) simulation (bottom) at the same timestamp. In event 1, the CME propagates just inside the front of a high speed stream that has a slower speed in the time-dependent driven simulation, causing the CME to arrive later compared to the benchmark.  For another example in event 3, Figure \ref{fig:diff3} the CME arrival is around 2012-02-26 22:02 UT in the single map driven benchmark (a) simulation, while the CME has already arrived in the time-dependent driven (b) simulation at the same timestamp. In event 3, the CME propagates behind a high speed stream that has a higher speed in the time-dependent driven simulation, causing the CME to arrive earlier compared to the benchmark.  As illustrated in these examples, the entire simulated CME cloud does not need to interact with the high speed stream to be impacted, in many cases only part of the CME cloud speeds up or slows down due to the interaction, and this part may pass over the Earth. 

Variations 1(b), 2(b), 3(b) were set up in a similar manner as described in the previous Section \ref{bench} except a time dependent sequence of maps was used at the inner boundary.  For all events, variation 1(b) was found to have ME=-0.6 hours, MAE=7.8 hours, and RMSE=10.2 hours, an increase in CME arrival time error compared to (b) for all time periods.  Variation 2(b) resulted in ME=-1.2 hours, MAE=7.5 hours, and RMSE=9.5 hours, similar performance to 1(b), however there is a decrease of 1.6 hours in the CME arrival time error for 2017--2019.  Finally for variation 3(b) we have ME=-1.3 hours, MAE=8.6 hours, and RMSE=10.4 hours, an overall error increase compared to (b), 1(b), and 2(b).  However for 2017--2019 the error decreased by 2.7 hours for 3(b) compared to (b) (unlike 3(a) which  did not show such error decrease).


\section{ADAPT}\label{year2}
\subsection{ADAPT-driven Test Simulation (c)}\label{adapttest}
\begin{figure*}[ht]
\begin{center}
\includegraphics[width=0.49\textwidth]{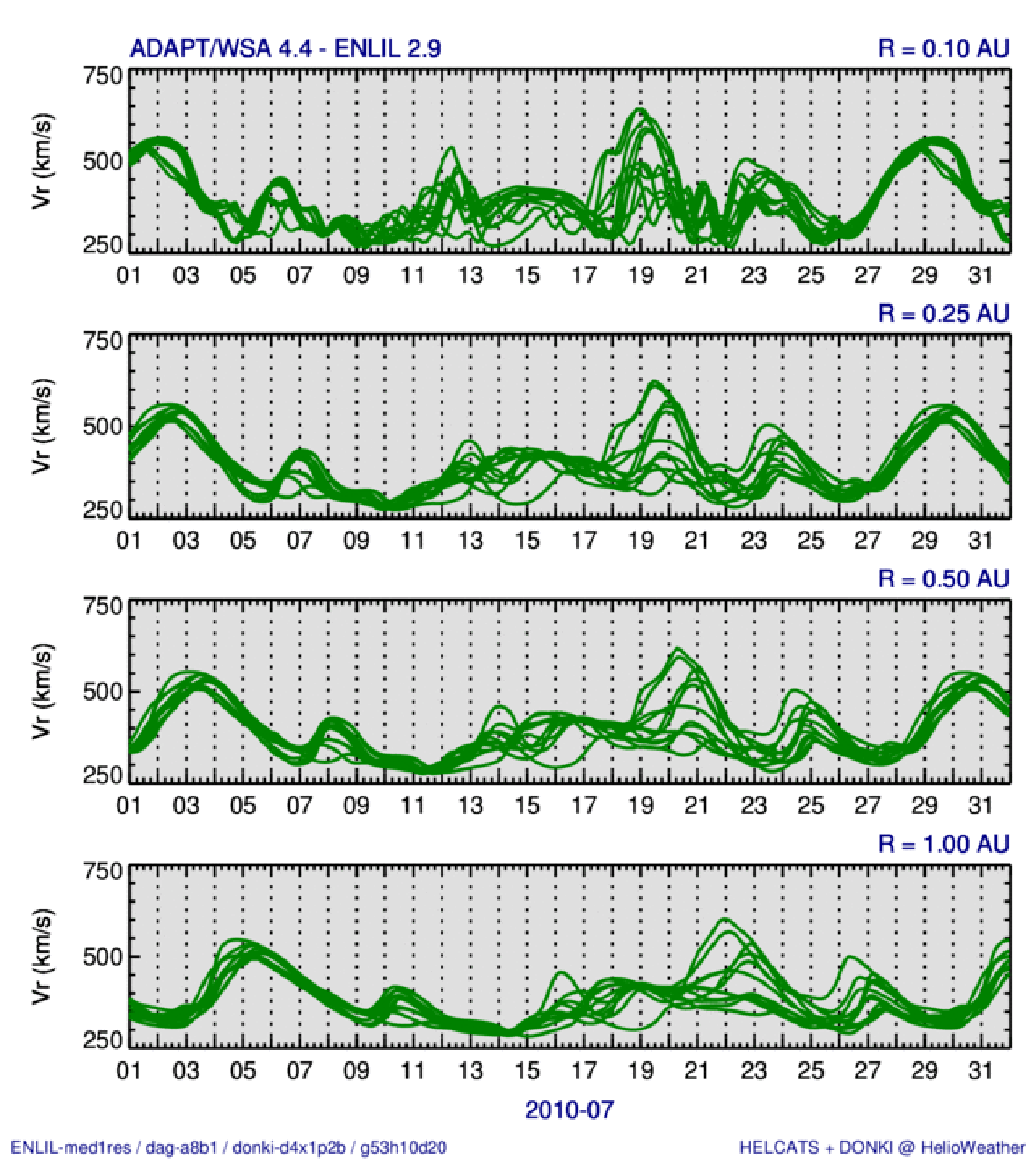}
\includegraphics[width=0.49\textwidth]{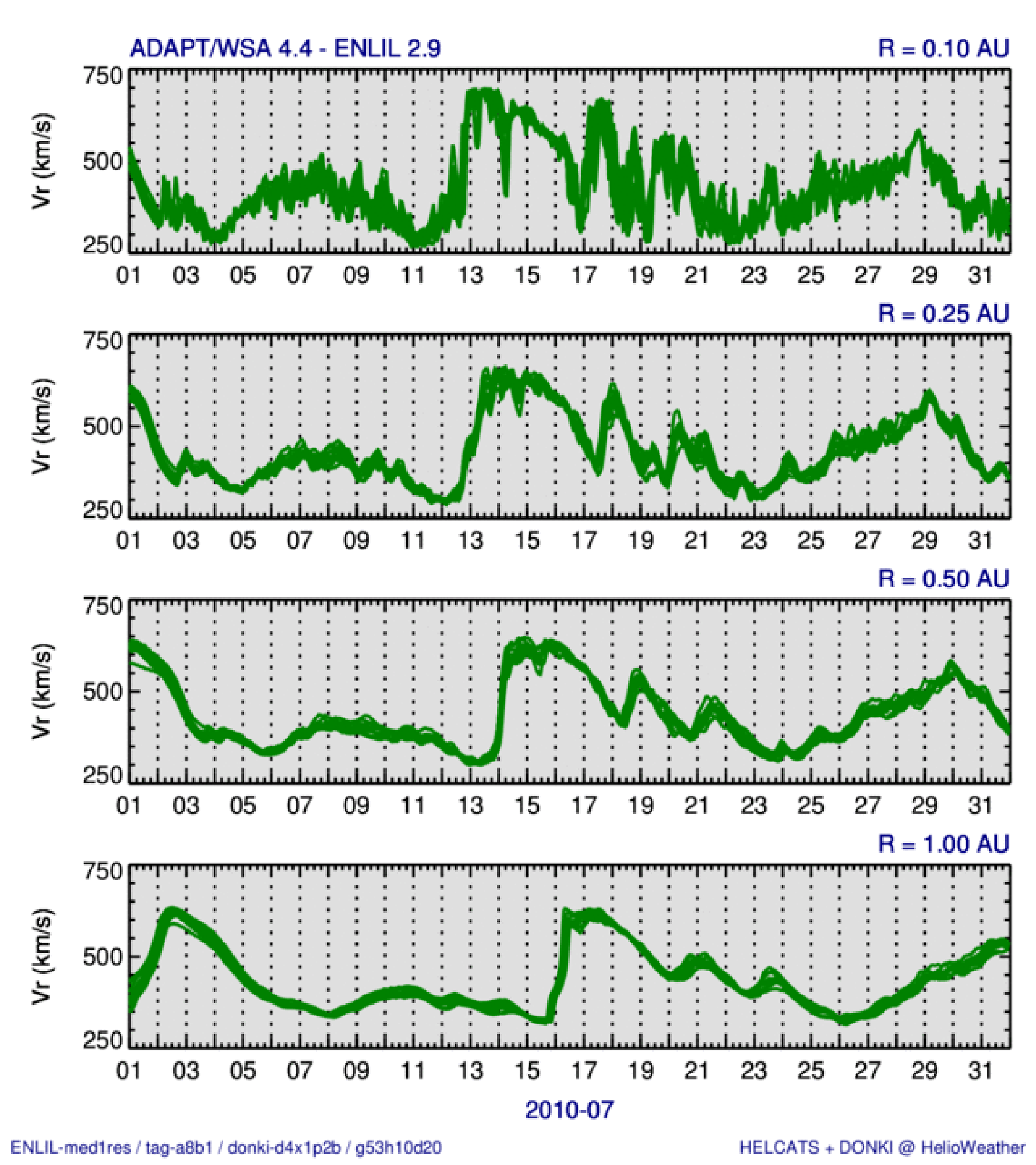}
\end{center}
\caption{Twelve ADAPT-WSA-ENLIL radial velocity model outputs at different radial distances (0.1, 0.25, 0.5, 1 AU) for July 2010 using a 24 hour ADAPT-WSA input cadence (left) or 2 hour input cadence (right).\label{fig:adaptamb}}
\end{figure*}

Before embarking on simulating a CME event we worked with model developers Nick Arge, Carl Henney, and Dusan Odstrcil to perform a one month long test run of the ambient solar wind for July--August 2010. This process allowed us to work out any code input/output issues between the models, and test model settings. Figure \ref{fig:adaptamb} provided by Dusan Odstrcil shows the 12 resulting ENLIL radial velocity model outputs at different radial distances (0.1, 0.25, 0.5, 1 AU) for July 2010 using a 24 hour ADAPT-WSA input cadence (left) or 2 hour input cadence (right). Our main lesson from this exercise was that running ENLIL at low resolution had the effects of averaging out the variation provided by the 12 ADAPT realizations due to interpolation at the inner boundary.   Moving forward, ENLIL runs using ADAPT should be performed at medium resolution or higher.  Secondly, we found that using a higher input time-cadence of the ADAPT-WSA at the ENLIL inner boundary can have the effect of smoothing out differences in the realizations.

\begin{figure*}[!ph]
\begin{center}
\includegraphics[width=0.75\textwidth]{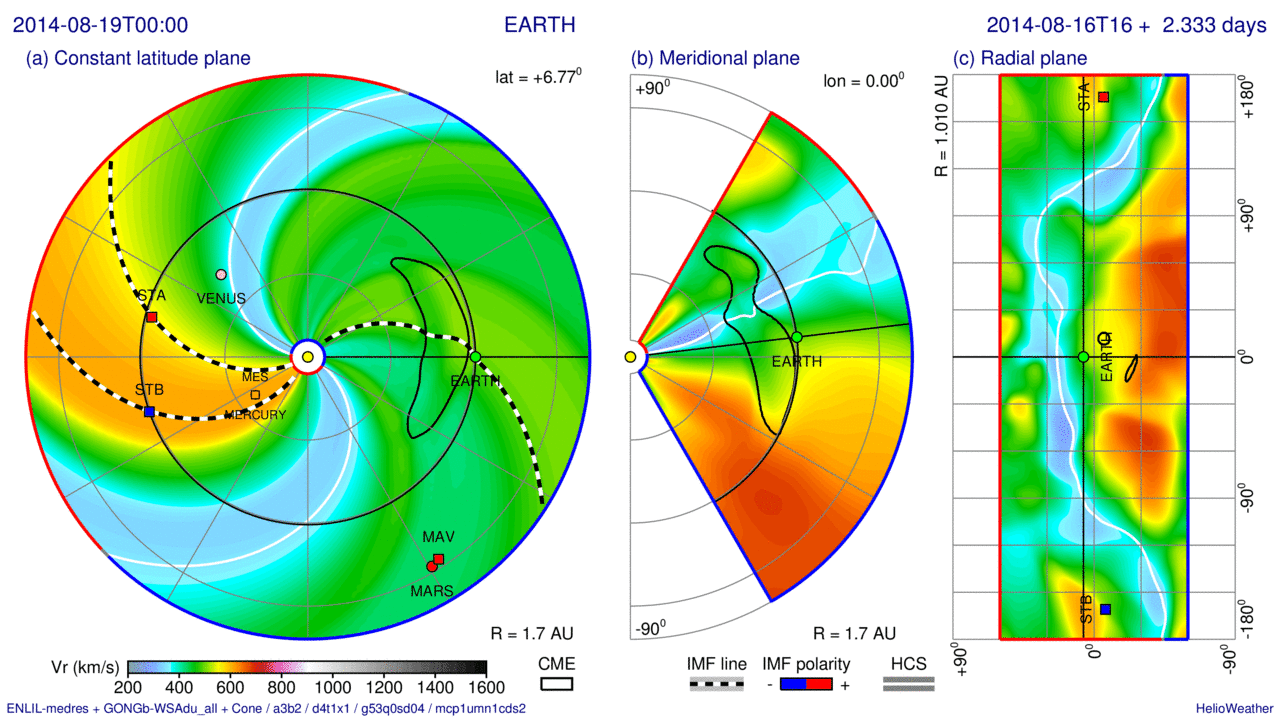}\\
\includegraphics[width=0.75\textwidth]{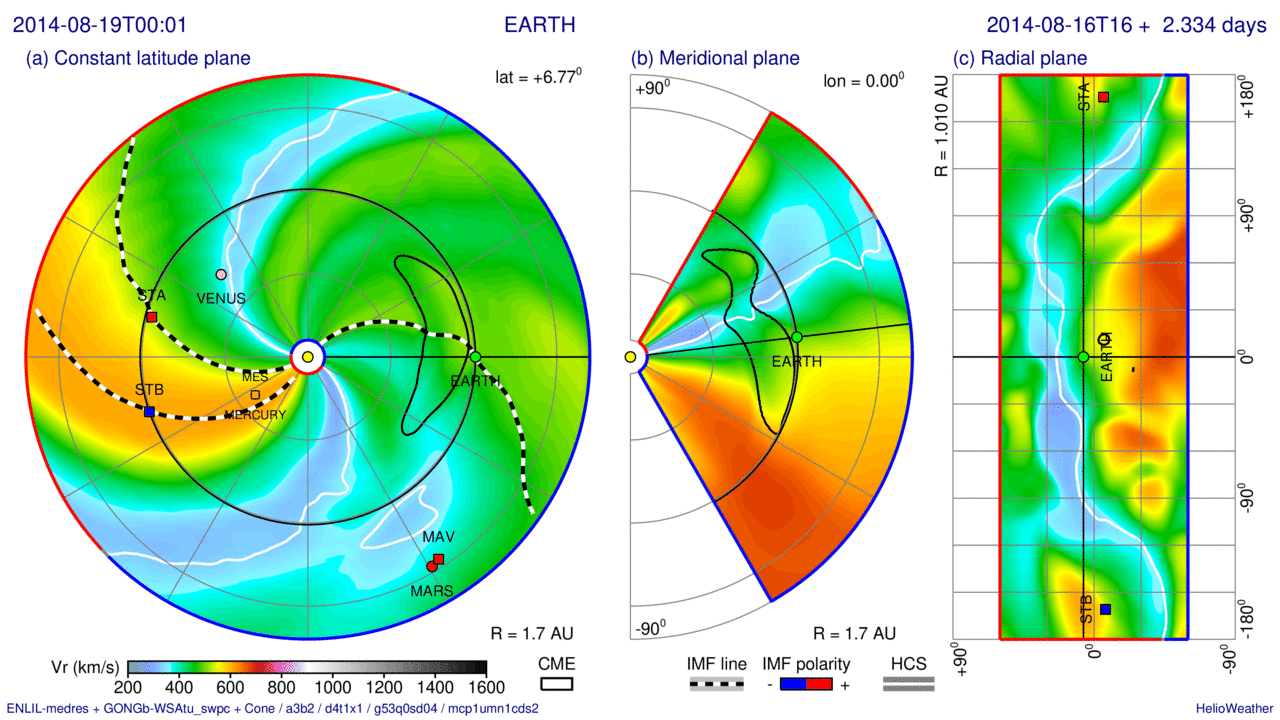}\\
\includegraphics[width=0.75\textwidth]{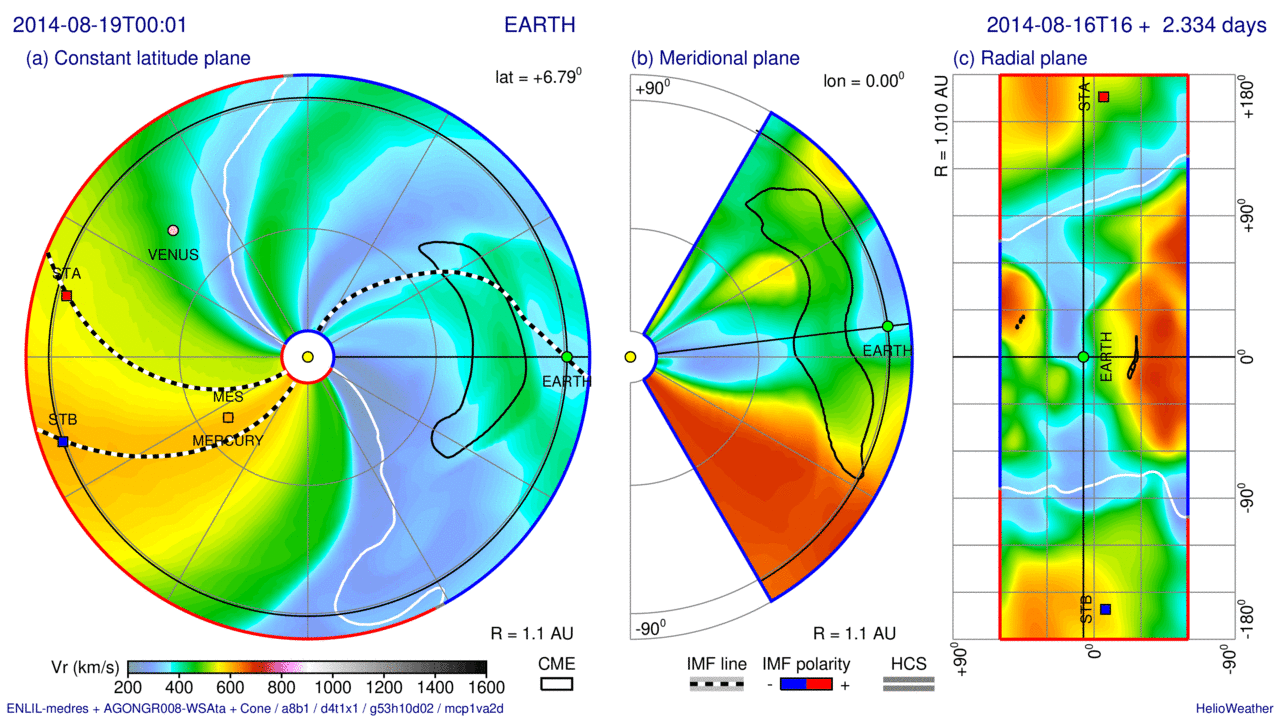}\\
\end{center}
\caption{Event 30: GONG-WSA-ENLIL radial velocity contour plots in the same format as Figure \ref{fig:diff1}. The top figure shows the single map-driven case, the middle figure shows the time-dependent map-driven case, and the bottom figure shows the time-dependent ADAPT map-driven case for realization R008.\label{fig:contour}}
\end{figure*}

Next we began work on item (c), GONGz ADAPT WSA 4.5 ENLIL 2.9e a8b1, a time-dependent test run for a single event using a time-dependent sequence of GONG maps driving ADAPT, WSA version 4.5 and ENLIL version 2.9e with {\sf{amb=a8b1}} settings.  For this test run SWPC selected the 2014-08-15 17:48 UT CME event.  Figure \ref{fig:contour}(bottom) shows the WSA-ENLIL radial velocity contour plot of the CME 2014-08-19 00:00 UT.  The figure shows that this slow (438 km s$^{-1}$) CME becomes embedded in slow background solar wind that is faster than the CME itself.  By comparison, the difference in the background solar wind can be seen in the top and middle panels show the contour plots for the benchmark (a) and time-dependent (b) simulations respectively.

\begin{figure*}[ht]
\includegraphics[width=0.5\textwidth]{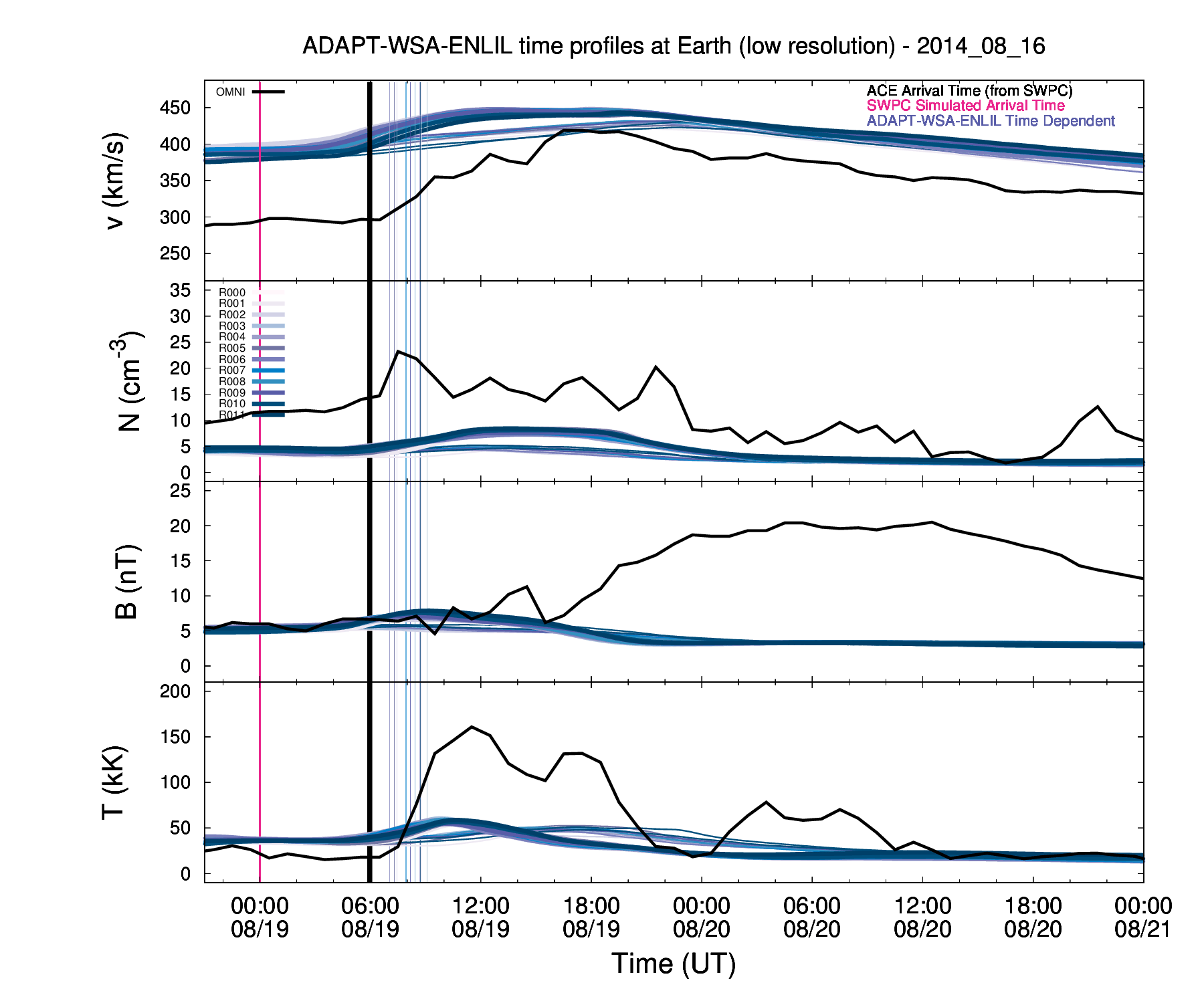}
\includegraphics[width=0.5\textwidth]{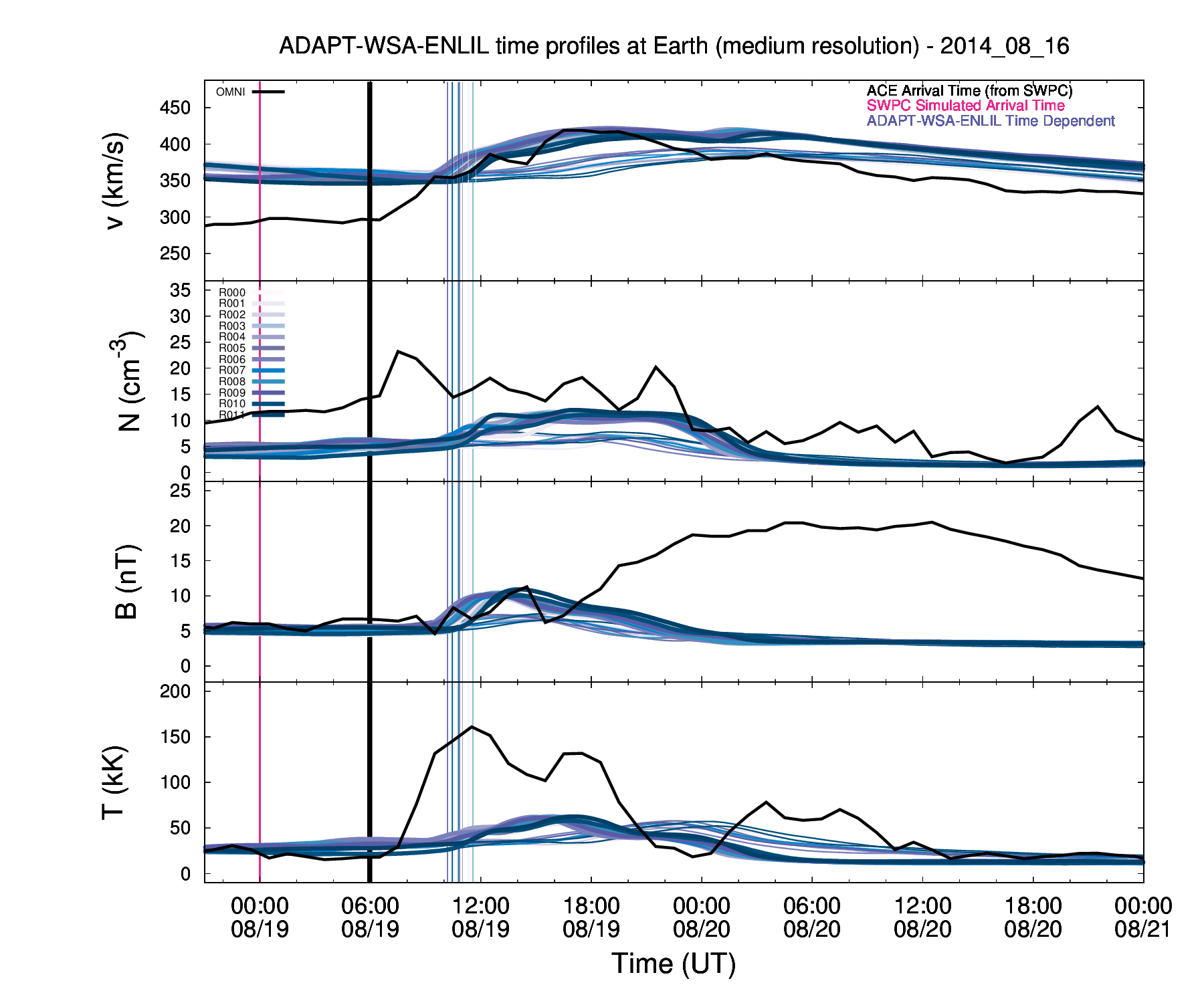}
\caption{GONGz ADAPT WSA 4.5 ENLIL 2.9e a8b1 (c) simulation results for the single event on 2014-08-15, at low resolution (left) and medium resolution (right). \label{fig:adaptmulti}}
\end{figure*}

We performed 12 ADAPT-WSA-ENLIL simulations (one for each realization) for the 2014-08-15 17:48 UT CME event, using a 2 hour ADAPT-WSA cadence at the ENLIL inner boundary. The time-dependent driven ENLIL results at Earth for each realization are plotted in shades of purple in Figure \ref{fig:adaptmulti} (right), and the thin lines show the ambient simulation (with no CME). Due to the slow speed of this CME, there is not a strong arrival signature in the time-series. The spread in CME arrival times ranges from 2014-08-19 10:10 to 11:44 UT (1.5 hours) with a median arrival time of 2014-08-19 10:46 UT. This gives an average arrival time error of $\Delta t$=+4.8 hours.  We also performed the same simulation at low resolution, shown in Figure \ref{fig:adaptmulti} (left) with a spread in CME arrival time range from 2014-08-19 07:03 to 2014-08-19T09:04 UT (2 hours) and a median of 2014-08-19 08:26 UT giving an error of $\Delta t$=+2.6 hours.  The low resolution simulation has the effect of smoothing out the arrival, arriving earlier, however the medium resolution simulation captures the speed increase at Earth due to the CME more accurately.  Both simulations predict the background solar wind to be 100 km/s higher than observed.  We attempted to perform a similar test at high resolution but the model developer Nick Arge needed to make corrections to the WSA coefficients to reflect the higher resolution and those results were not available in time for this report, but will completed in the future.   Another question the model developers are investigating is the  convergence of the spherical harmonic expansion which may be an issue at 1 degree resolution.

\begin{figure*}[ht]
\includegraphics[width=.8\textwidth]{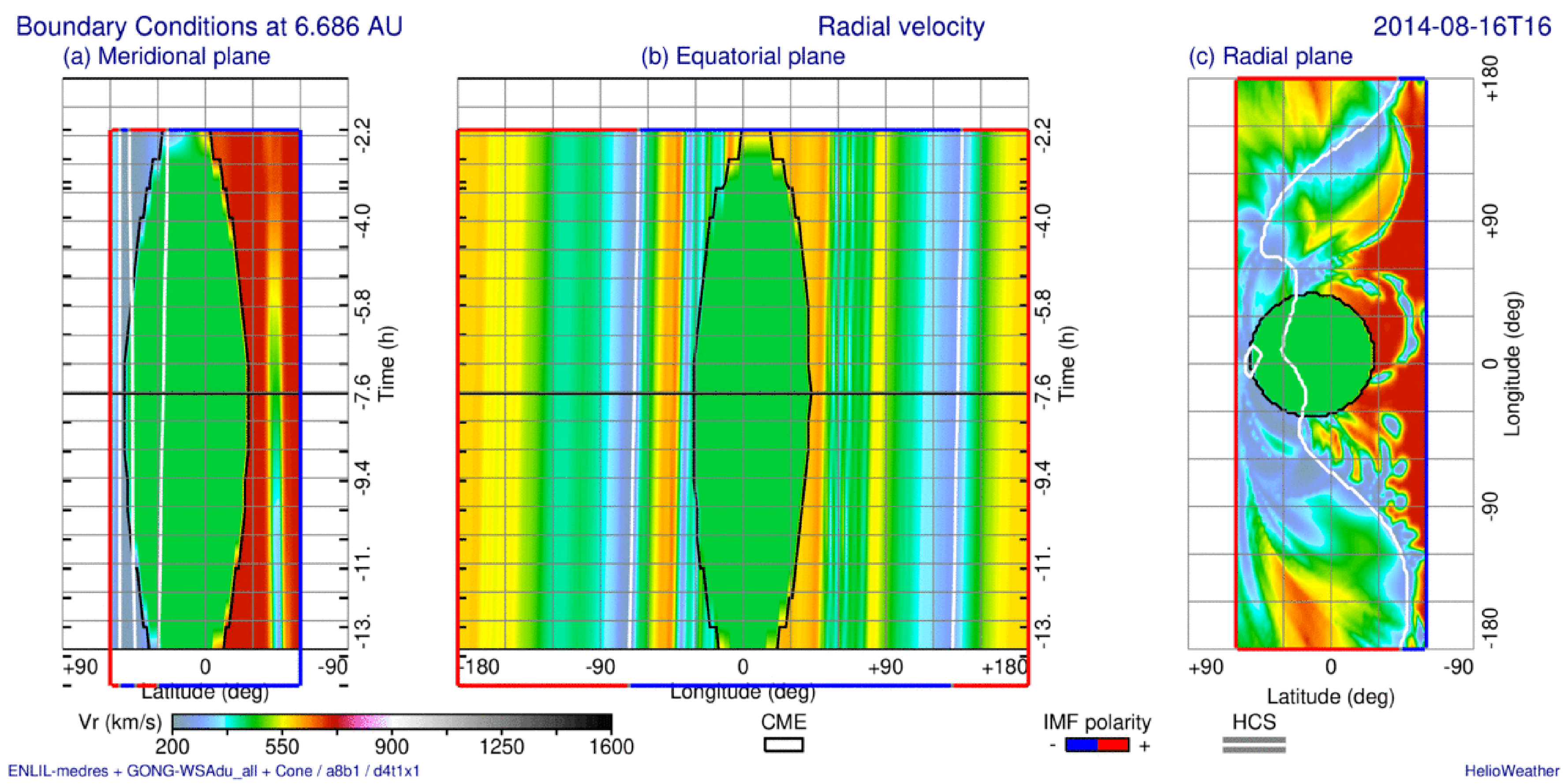}\\
\includegraphics[width=.8\textwidth]{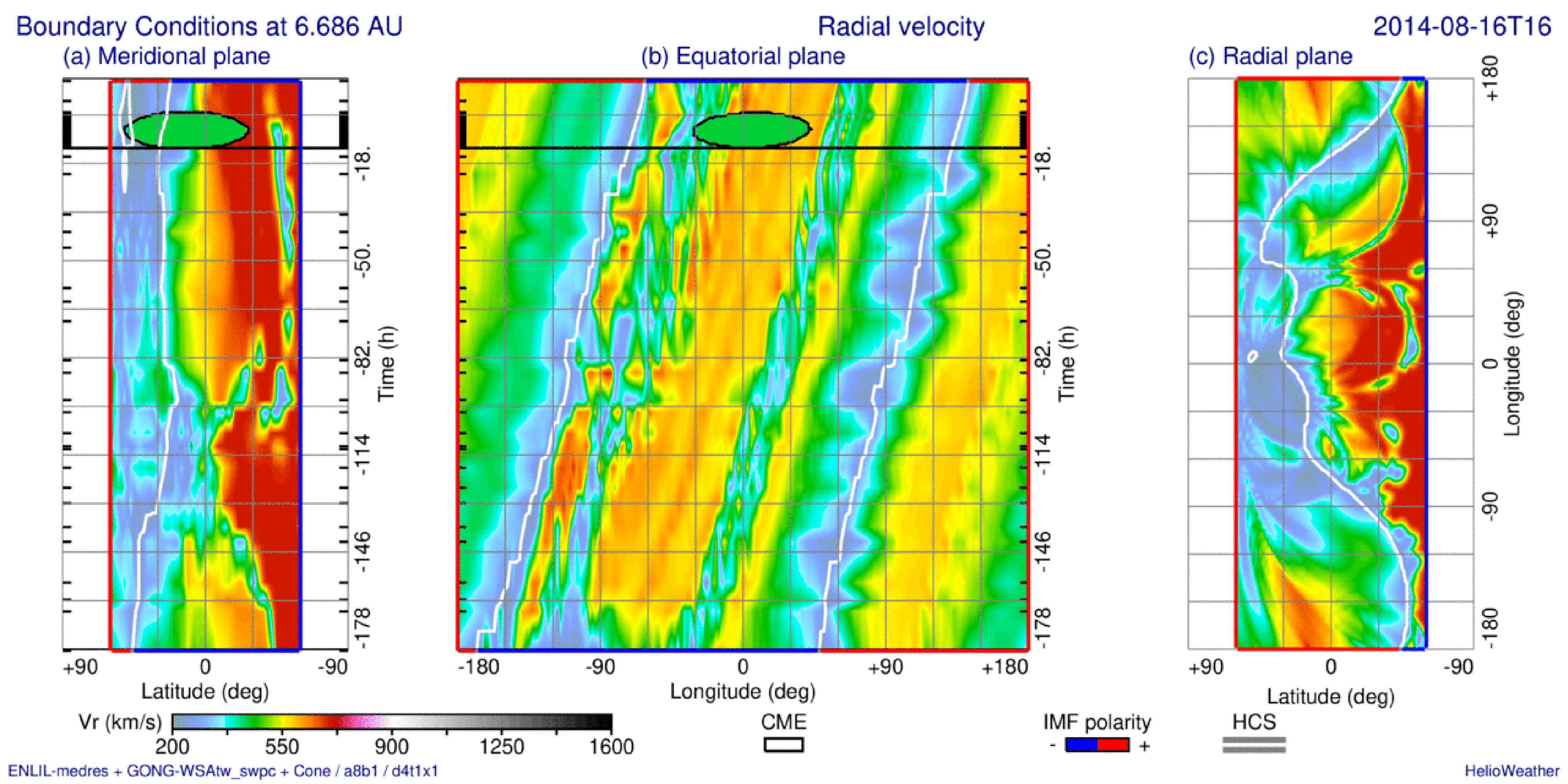}\\
\includegraphics[width=.8\textwidth]{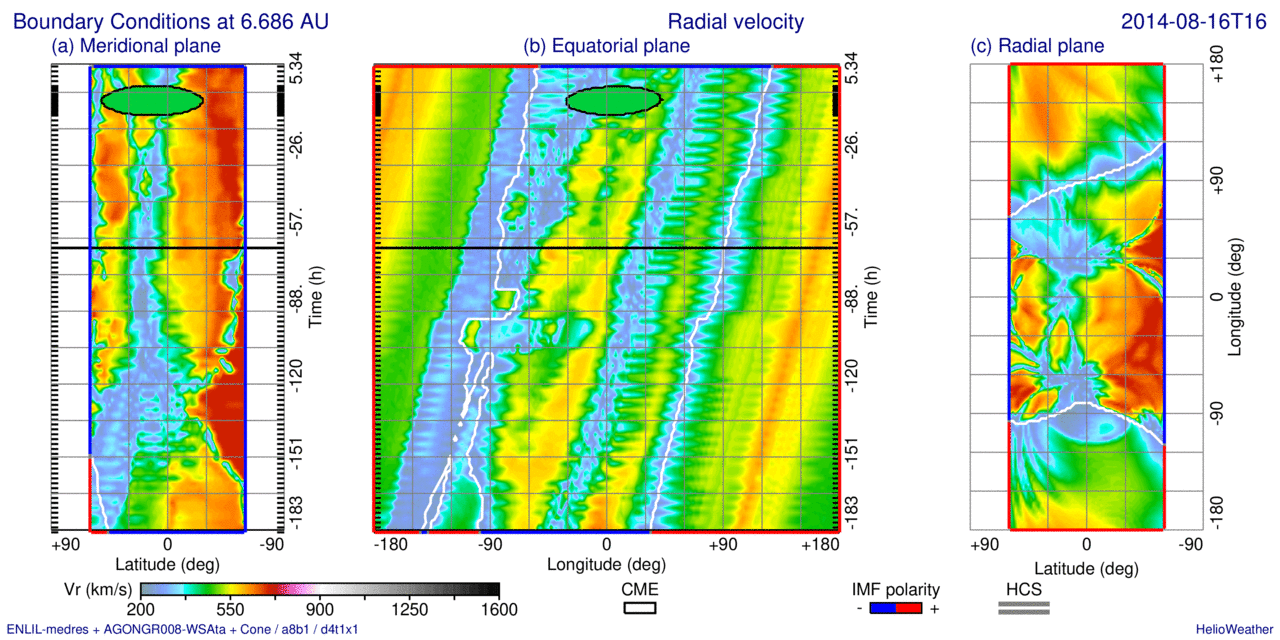}\\
\caption{Event 30 panels a and b: Meridonal and equatorial slices of the ENLIL inner boundary as a function of time.  The top figure shows the single map-driven case, the middle figure shows the time-dependent map-driven case, and the bottom figure shows the time-dependent ADAPT map-driven case for realization R008.  The black horizontal lines indicate the timestamp relative to the run date in the top right hand corner of the panel (c), the radial plane at 21.5 solar radii. This figure illustrates that for the single map-driven case (bottom), the background map is fixed and rotated through time during CME insertion, whereas for the time-dependent sequence of maps (middle, bottom), the CME is inserted into a clearly time-varying background.\label{fig:bnd}}
\end{figure*}

\begin{figure*}[ht]
\includegraphics[width=.5\textwidth]{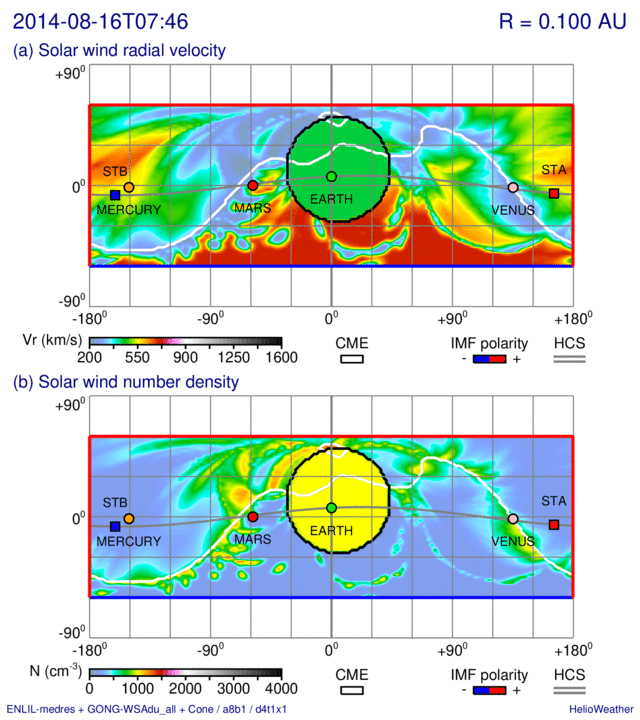}
\includegraphics[width=.5\textwidth]{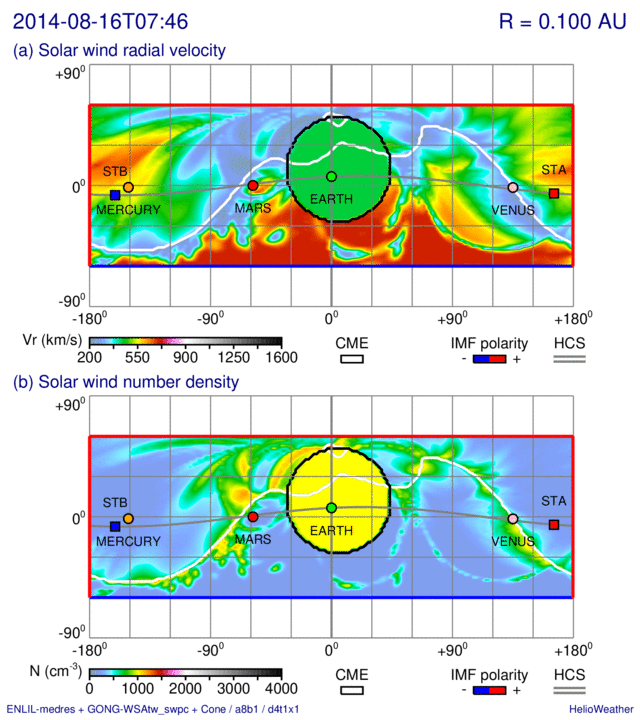}\\
\includegraphics[width=.5\textwidth]{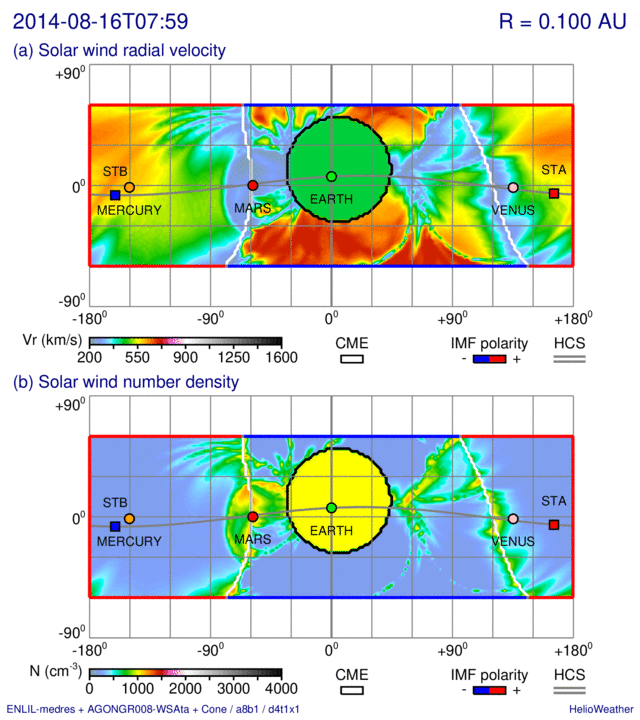}\\
\caption{Event 30: velocity (panel a) and density (panel b) at the radial distance of 21.5 solar radii at the ENLIL inner boundary around 2014-08-16 08:00 UT. The top figure shows the single map-driven case, the middle figure shows the time-dependent map-driven case, and the bottom figure shows the time-dependent ADAPT map-driven case for realization R008. \label{fig:bnd2}}
\end{figure*}

To illustrate the difference in the ENLIL inner boundary conditions when using time-dependent conditions ADAPT, for this event we show equatorial slices of the ENLIL inner boundary as a function of time in panel (b) of Figure \ref{fig:bnd}.  The top figure shows the single map-driven case, the middle figure shows the time-dependent map-driven case, and the bottom figure shows the time-dependent ADAPT map-driven case for realization R008.  This figure illustrates that for the single map-driven case (bottom), the background map is fixed and rotated through time during CME insertion, whereas for the time-dependent sequence of maps (middle, bottom), the CME is inserted into a clearly time-varying background. In Figure \ref{fig:bnd2} we show the velocity and density for the radial plane at the inner boundary of ENLIL for the same three simulations during CME insertion.

\subsection{Single map and Time-dependent ADAPT (d-e)}
Finally, we performed simulations for the GONGz ADAPT WSA 4.5 ENLIL 2.9e a8b1 case.  An archive of WSA 4.5 velocity maps was created using ADAPT inputs with a cadence of 2 hours. For the single map driven simulations (d), the timestamp closest to the operational magnetogram from (a) was selected.   For each event we performed 12 ADAPT-WSA-ENLIL simulations with single map (d) and time-dependent (e) inputs.  In order to compare the CME arrival time errors for earlier simulations to these ensemble simulations, we analyzed the errors for the realization with the median CME arrival.   In future work, we plan to perform validation for the full ensemble, and provide analysis for the highest ranking realization in terms of reproducing background solar wind at Earth up until the simulation start date.

\begin{figure*}
\includegraphics[width=0.33\textwidth]{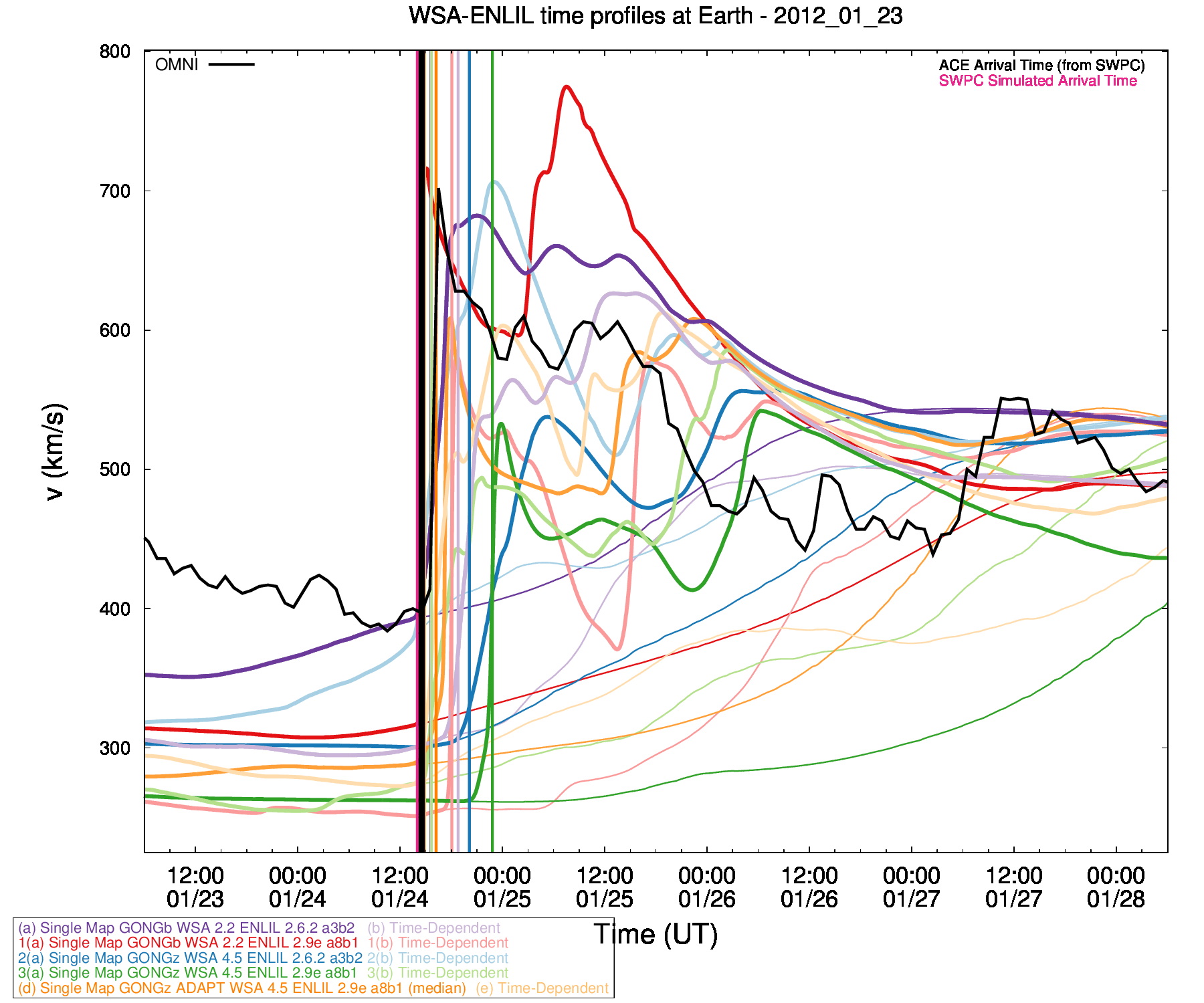}
\includegraphics[width=0.33\textwidth]{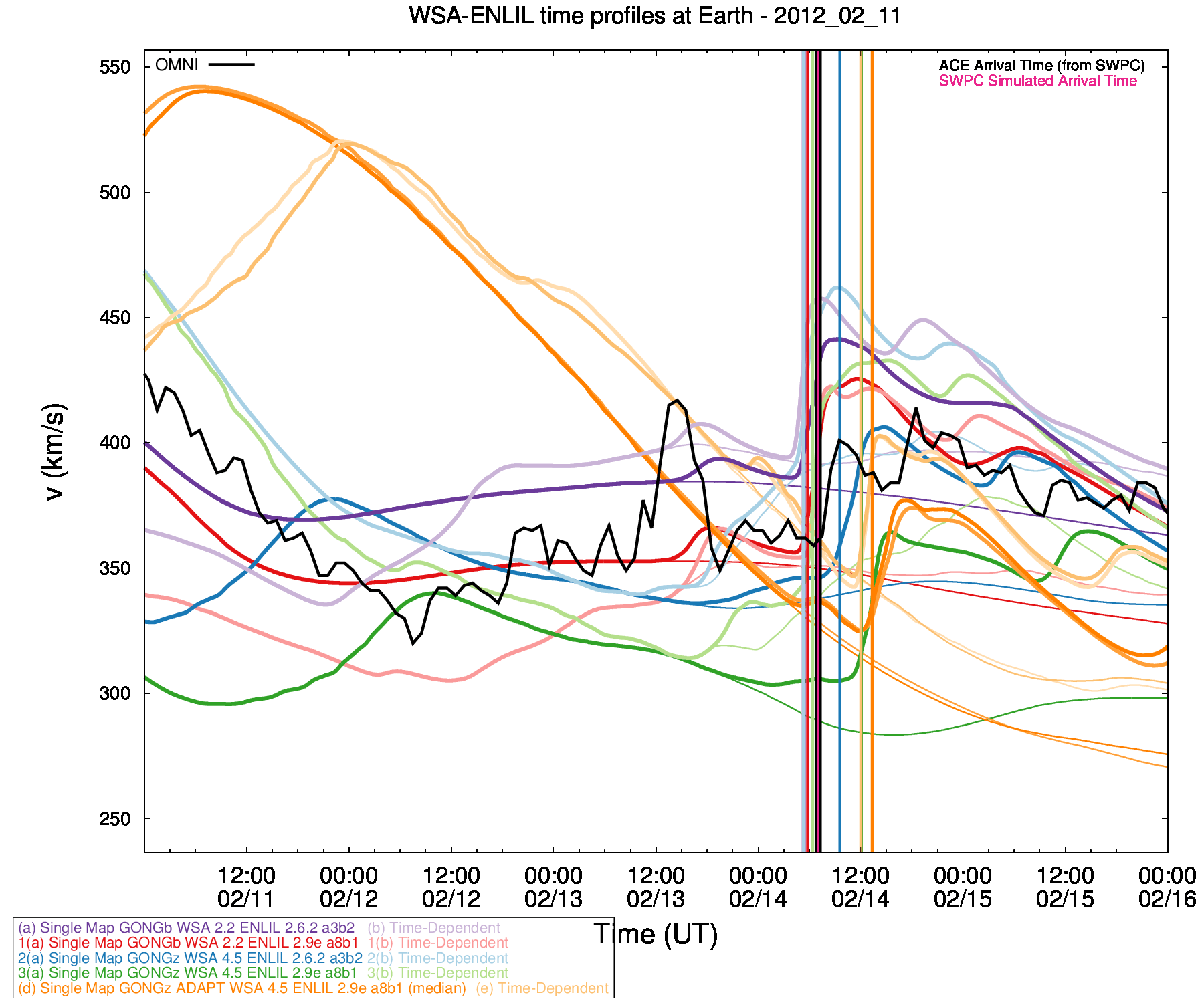}
\includegraphics[width=0.33\textwidth]{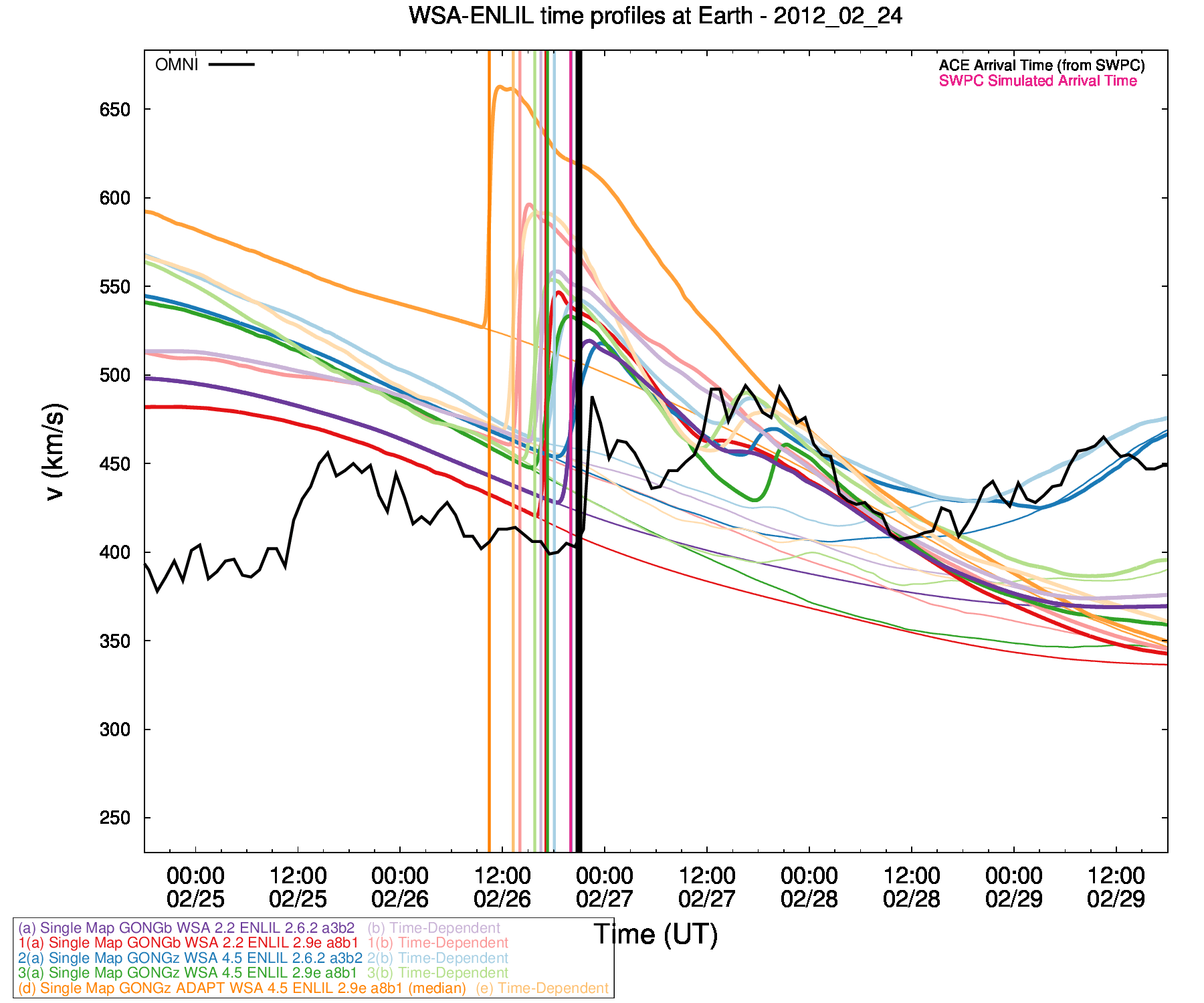}\\
\includegraphics[width=0.33\textwidth]{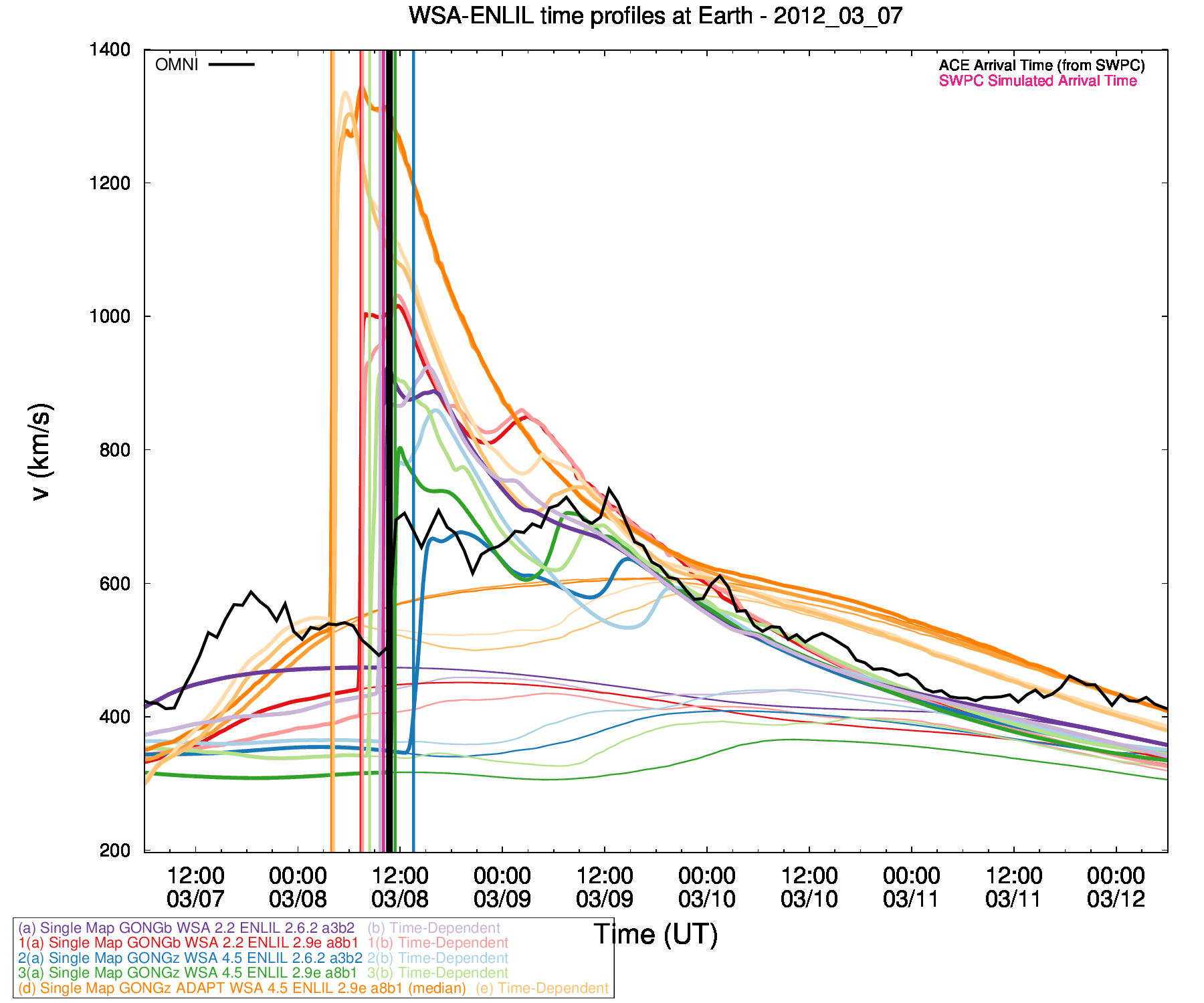}
\includegraphics[width=0.33\textwidth]{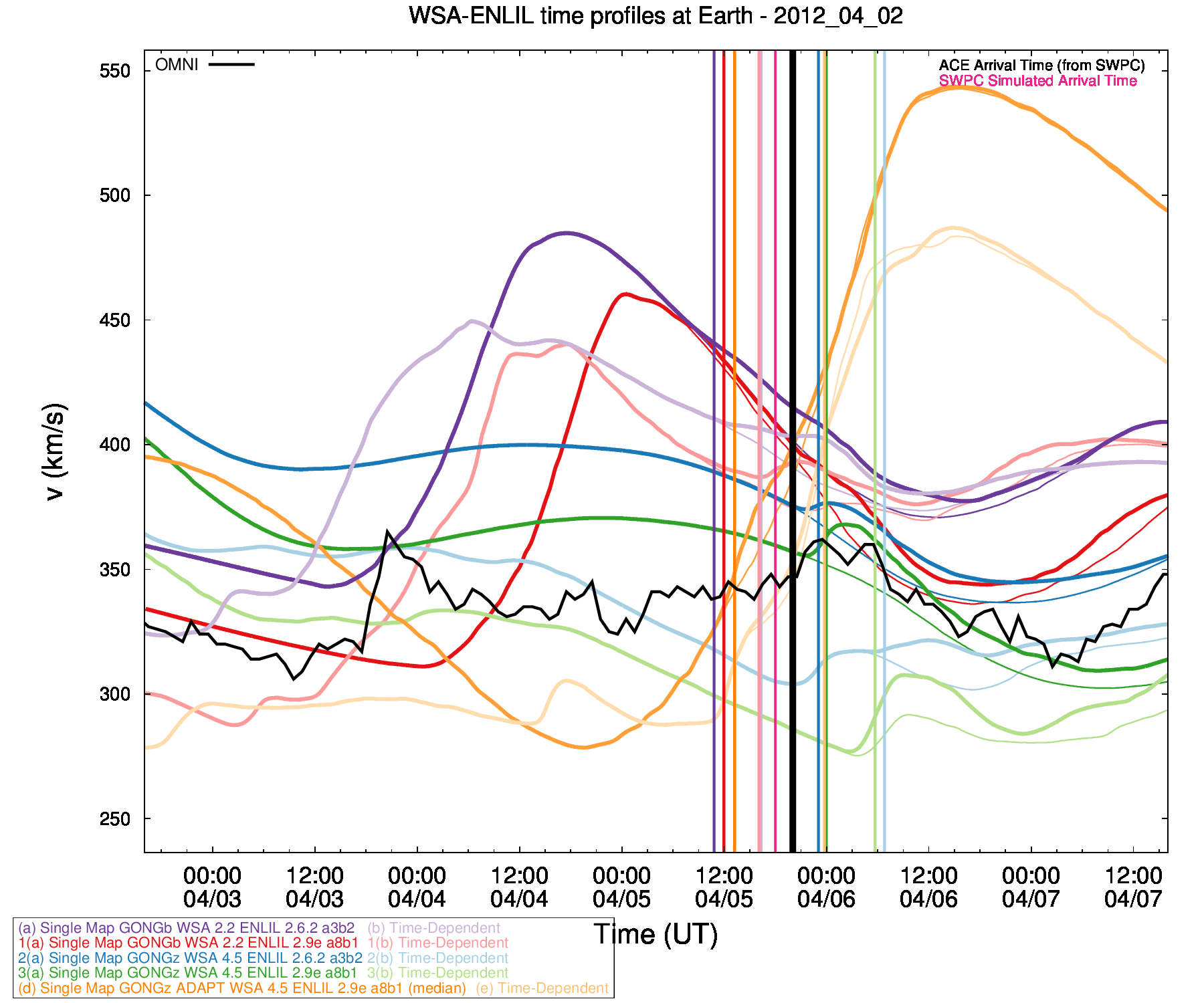}
\includegraphics[width=0.33\textwidth]{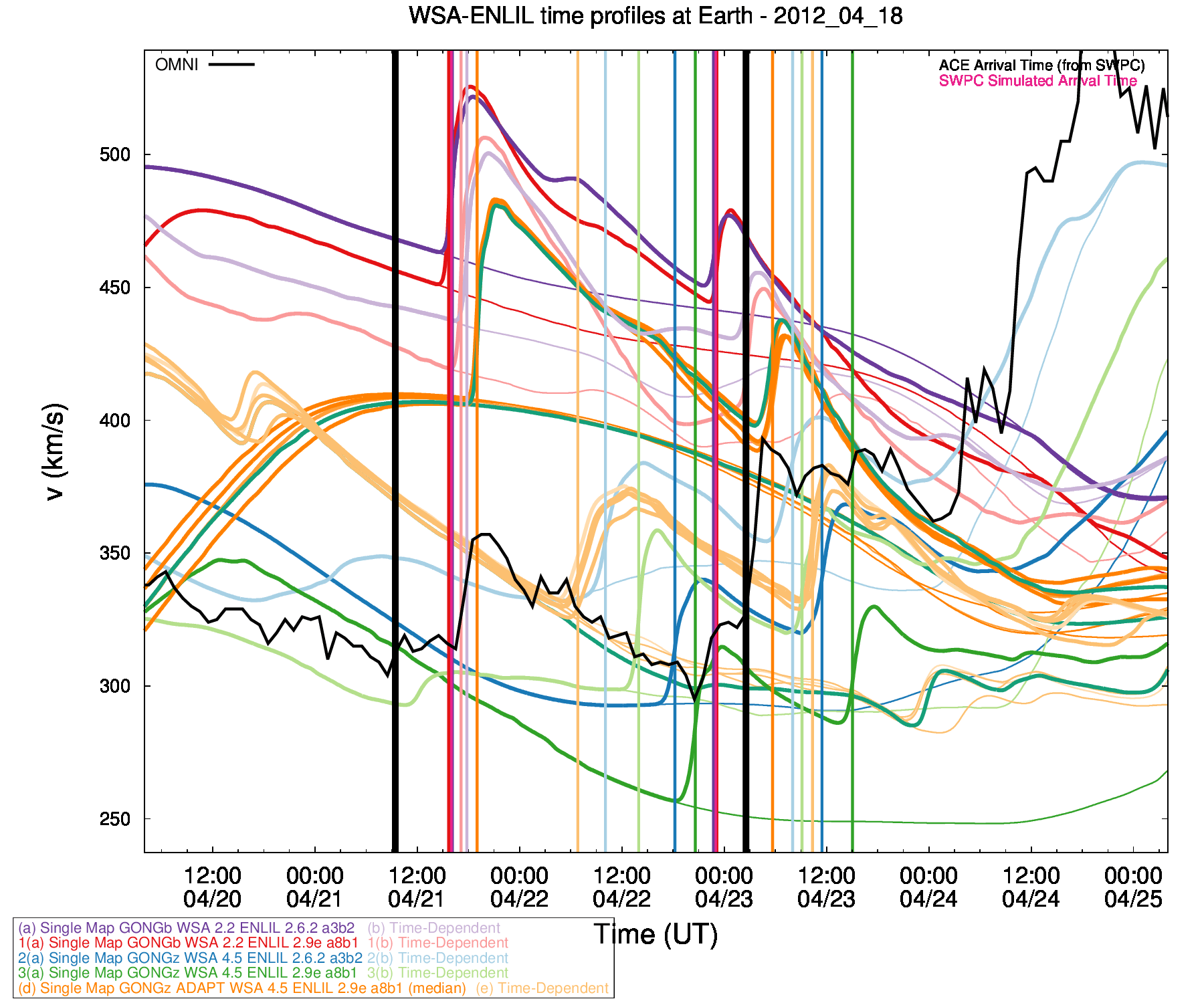}\\
\includegraphics[width=0.33\textwidth]{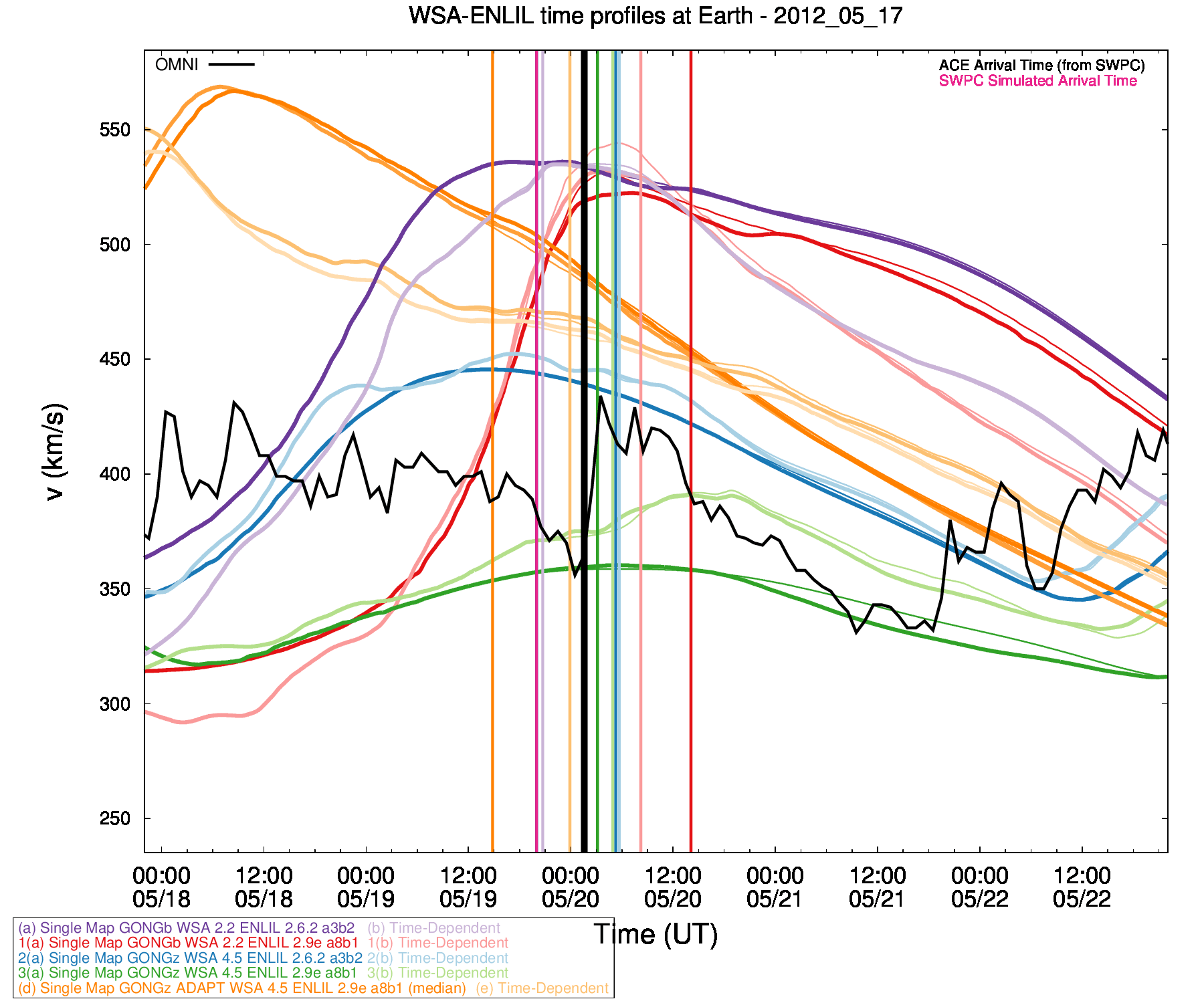}
\includegraphics[width=0.33\textwidth]{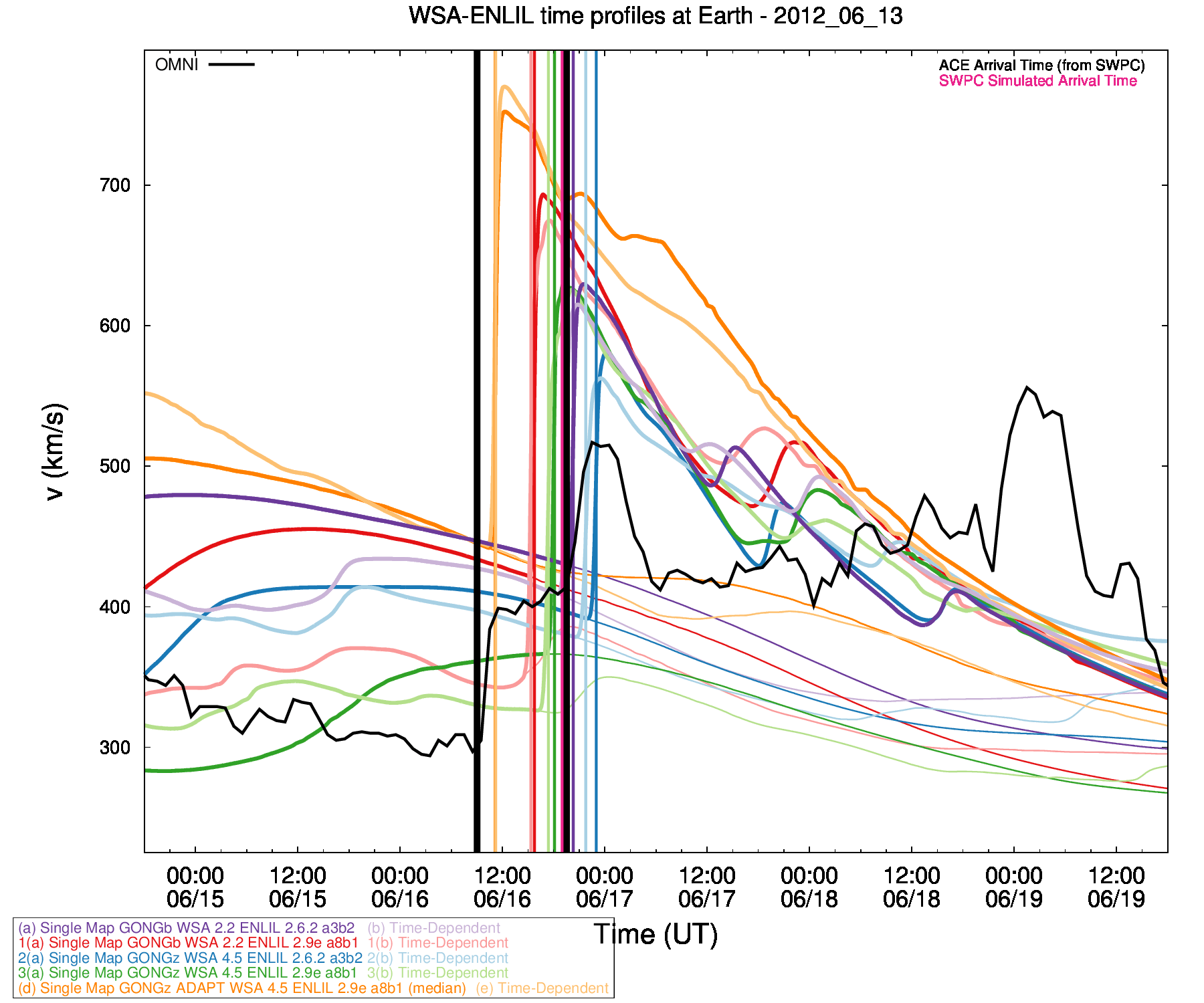}
\includegraphics[width=0.33\textwidth]{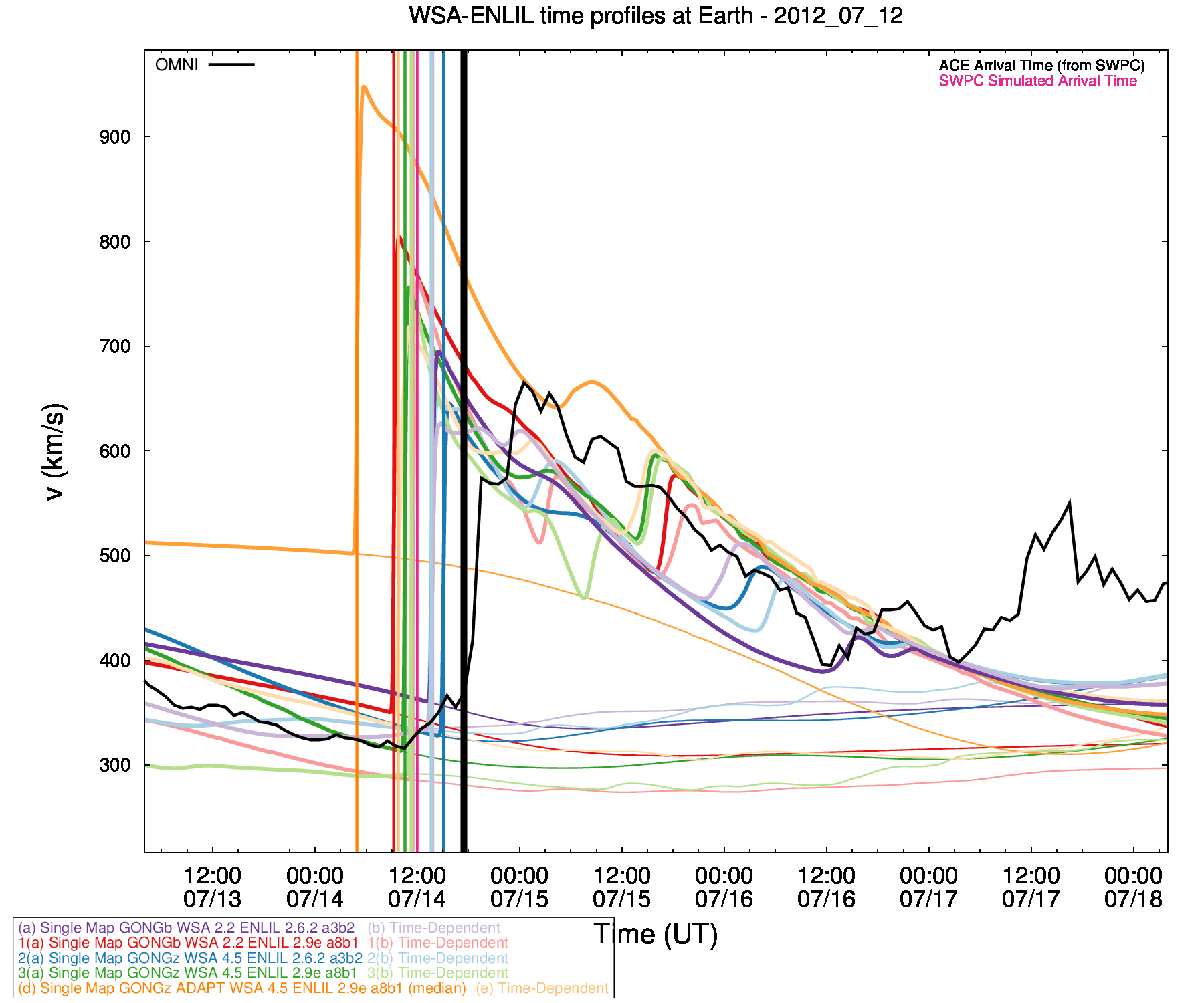}\\
\includegraphics[width=0.33\textwidth]{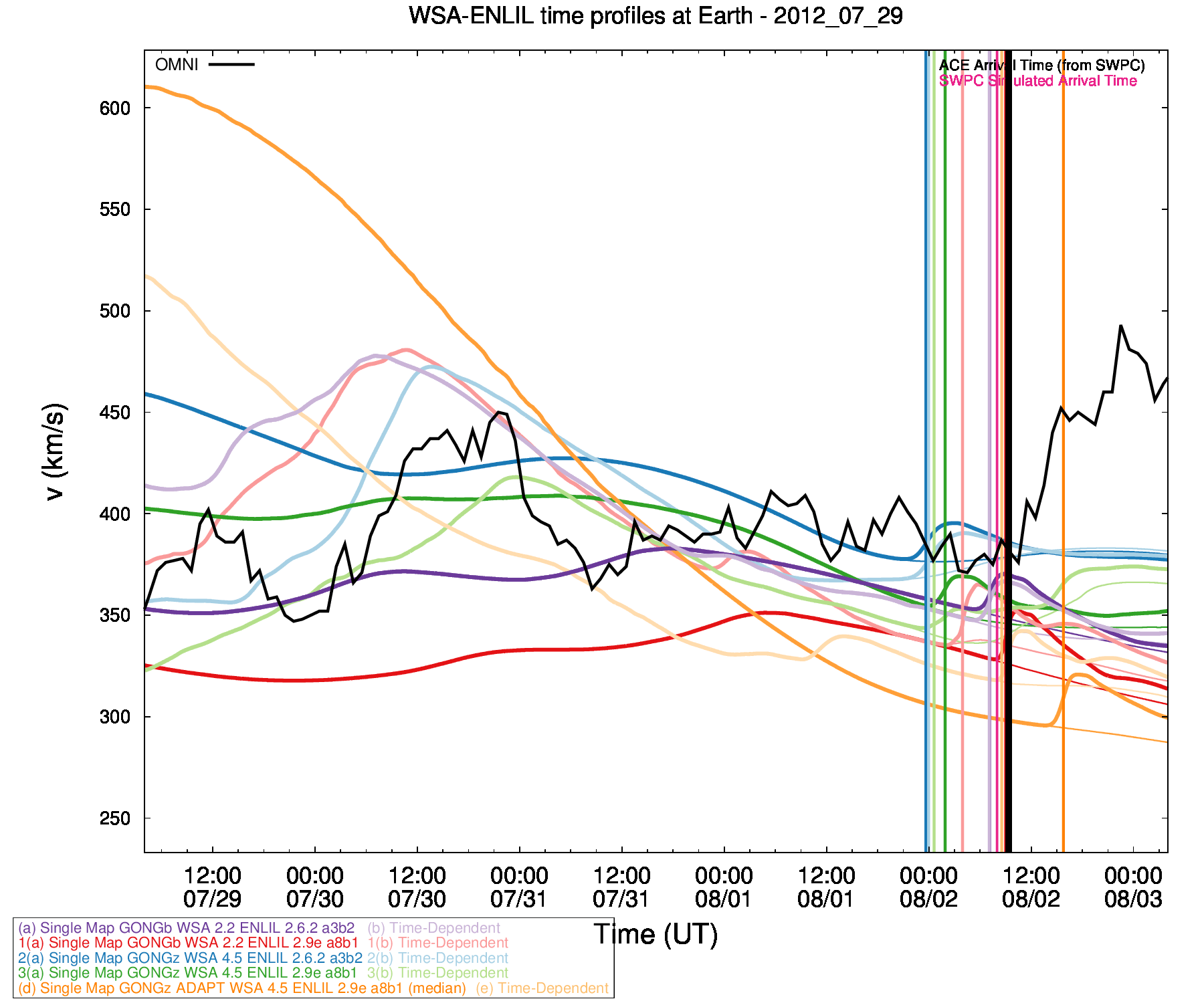}
\includegraphics[width=0.33\textwidth]{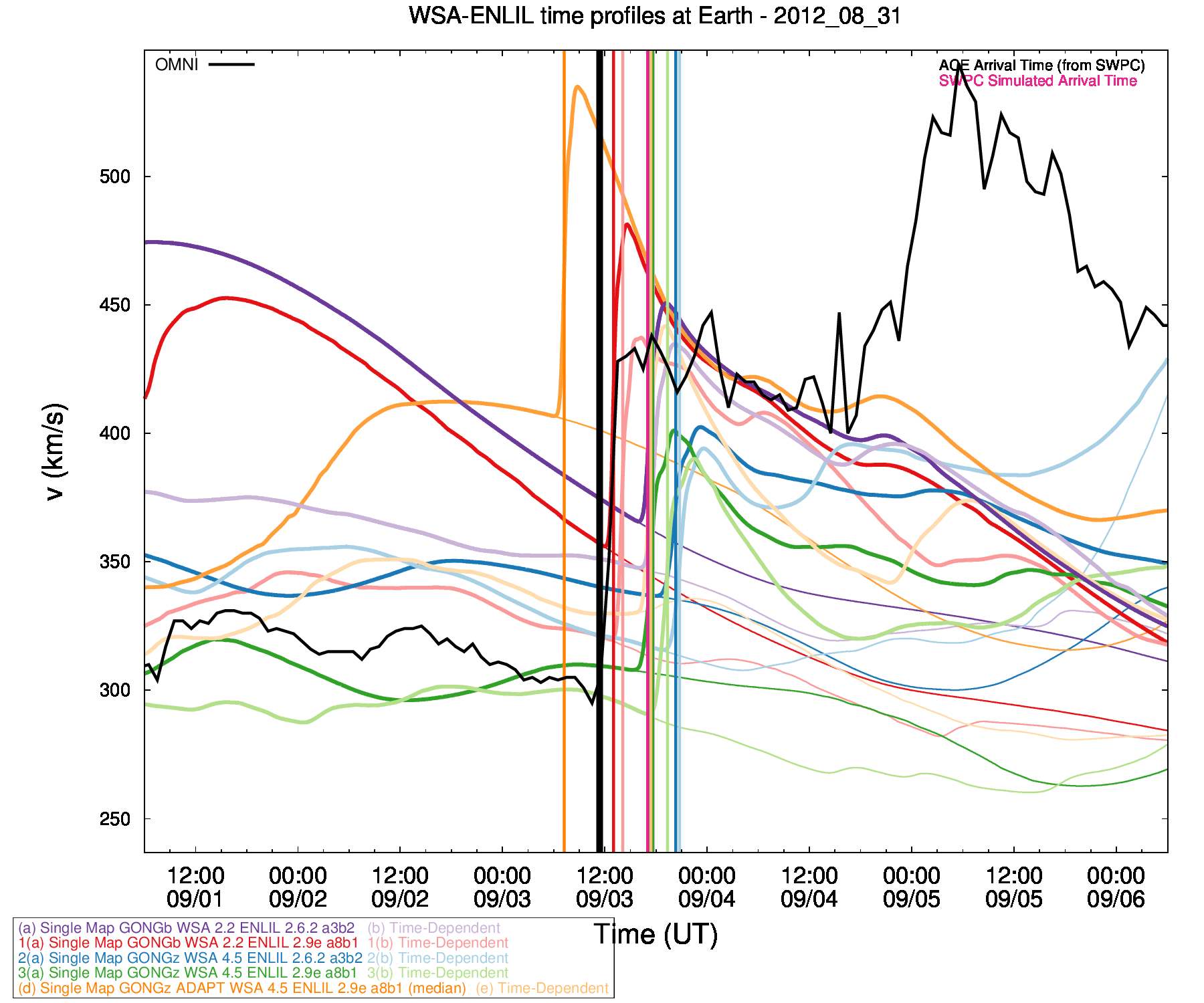}
\includegraphics[width=0.33\textwidth]{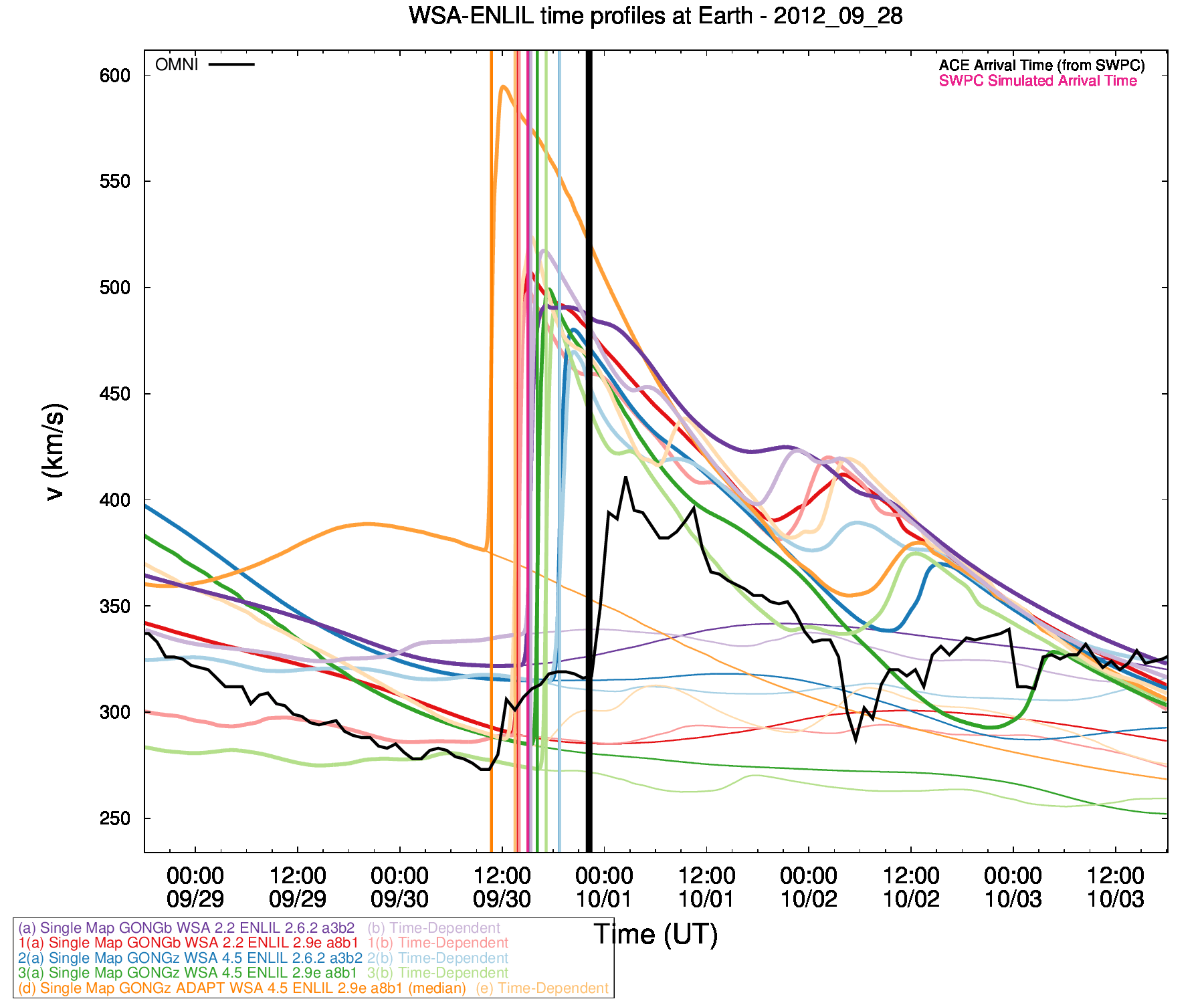}\\
\caption{Simulated and observed speed for the benchmark single map-driven runs (Section \ref{method}, task a) and time-dependent map-driven runs (Section \ref{method}, task b) for all 38 events. The simulated benchmark time-series are plotted in purple, time-dependent runs in light purple, and the hourly averaged OMNI data is plotted in black. The magenta vertical lines show the SWPC provided simulated arrival times, the purple and light purple vertical lines show those detected by the dynamic pressure algorithm, and the black vertical lines show the ACE observed arrival times.  Also over-plotted here are the single map and time-dependent results for simulation variations 1(a-b): dark and light red, 2(a-b): dark and light blue, 3(a-b) dark and light green. (d-e) in dark and light orange (median for each arrival).  In all cases the thin lines show the ambient simulation (with no CME). Similar plots with all of the plasma parameters (speed, density, magnetic field, and temperature) time series and plots zoomed in near the CME arrival time (for better clarity) are available to download from the project webpage.\label{fig:time-series}}
\end{figure*}

\begin{figure*}
\includegraphics[width=0.33\textwidth]{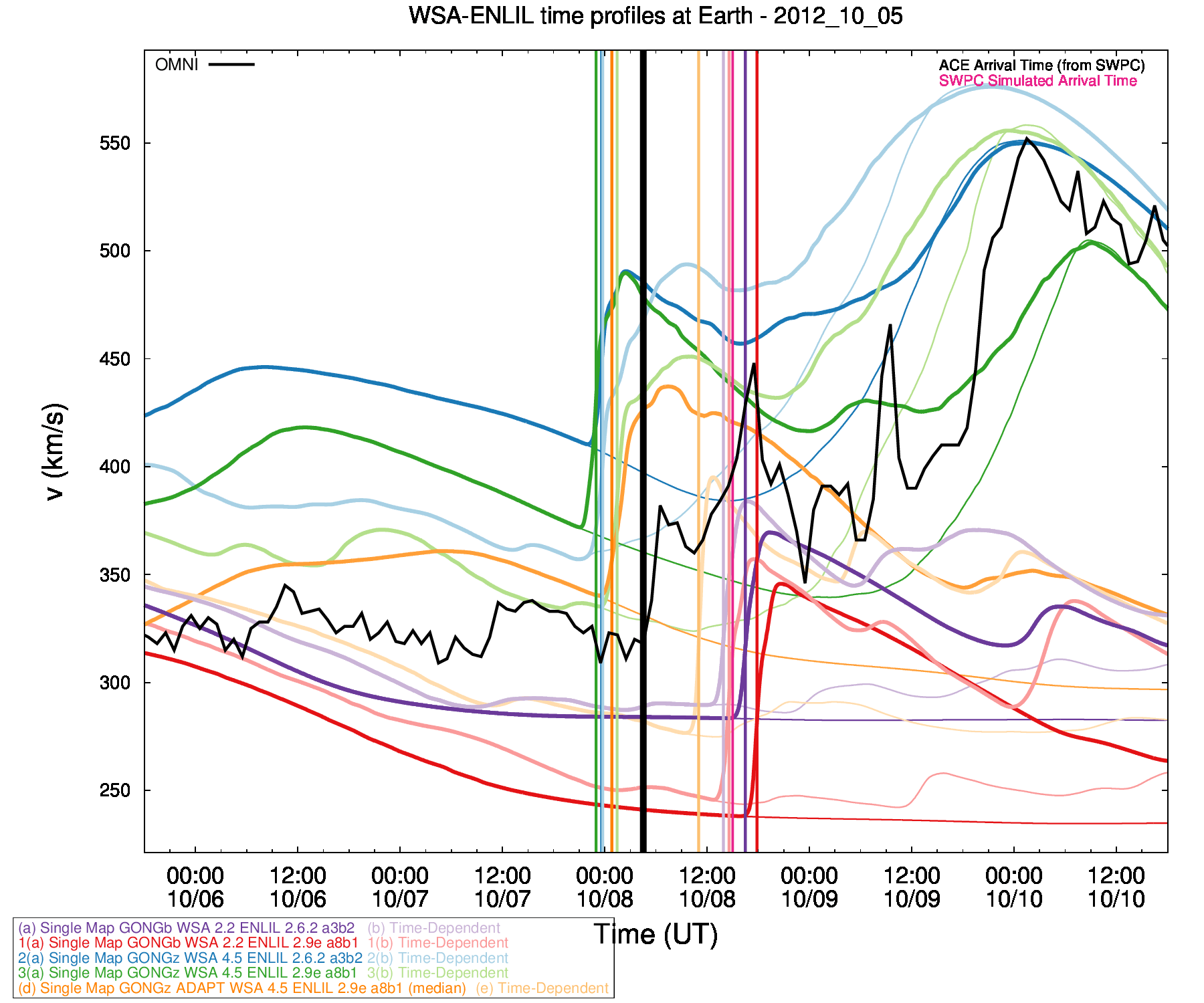}
\includegraphics[width=0.33\textwidth]{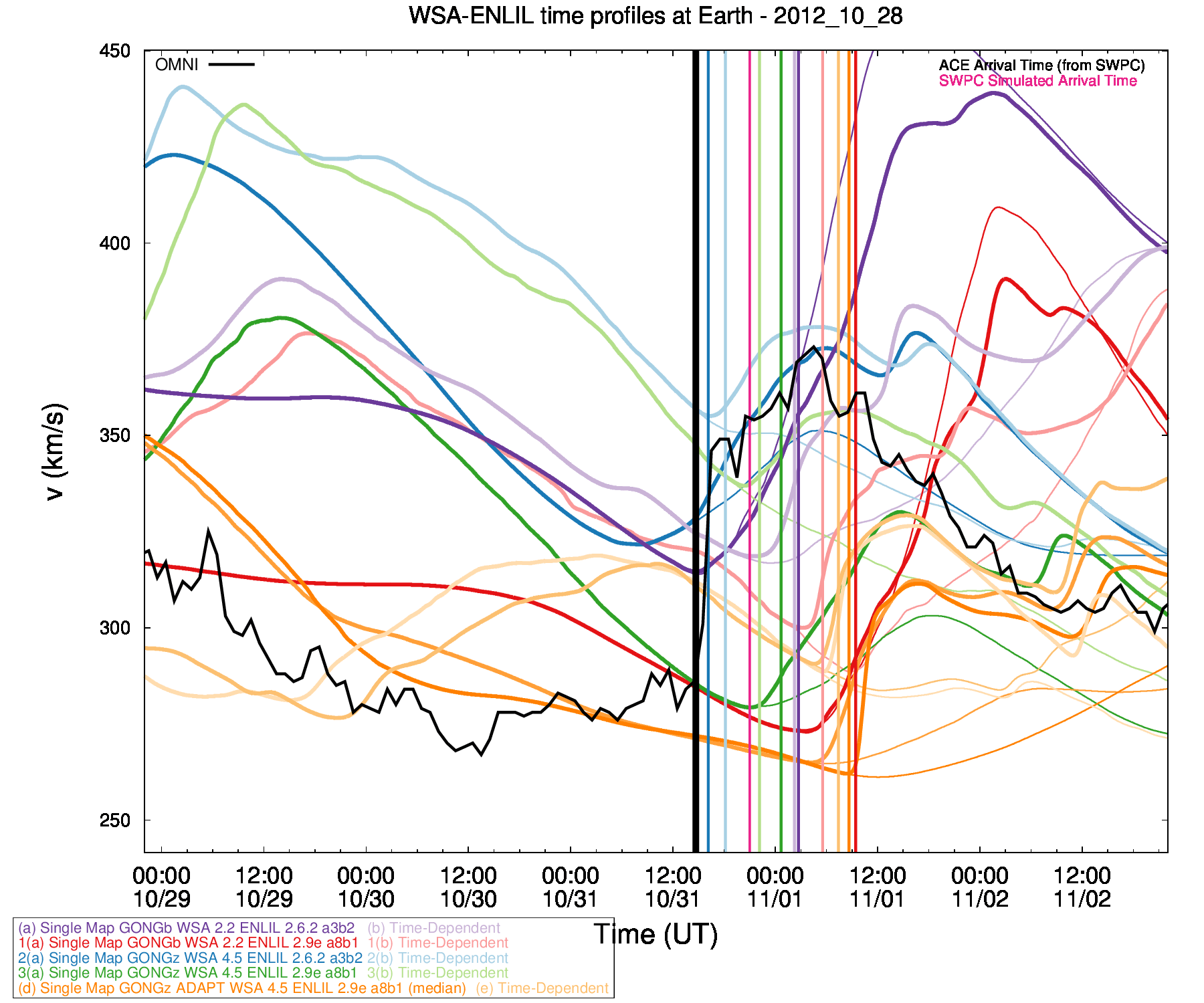}
\includegraphics[width=0.33\textwidth]{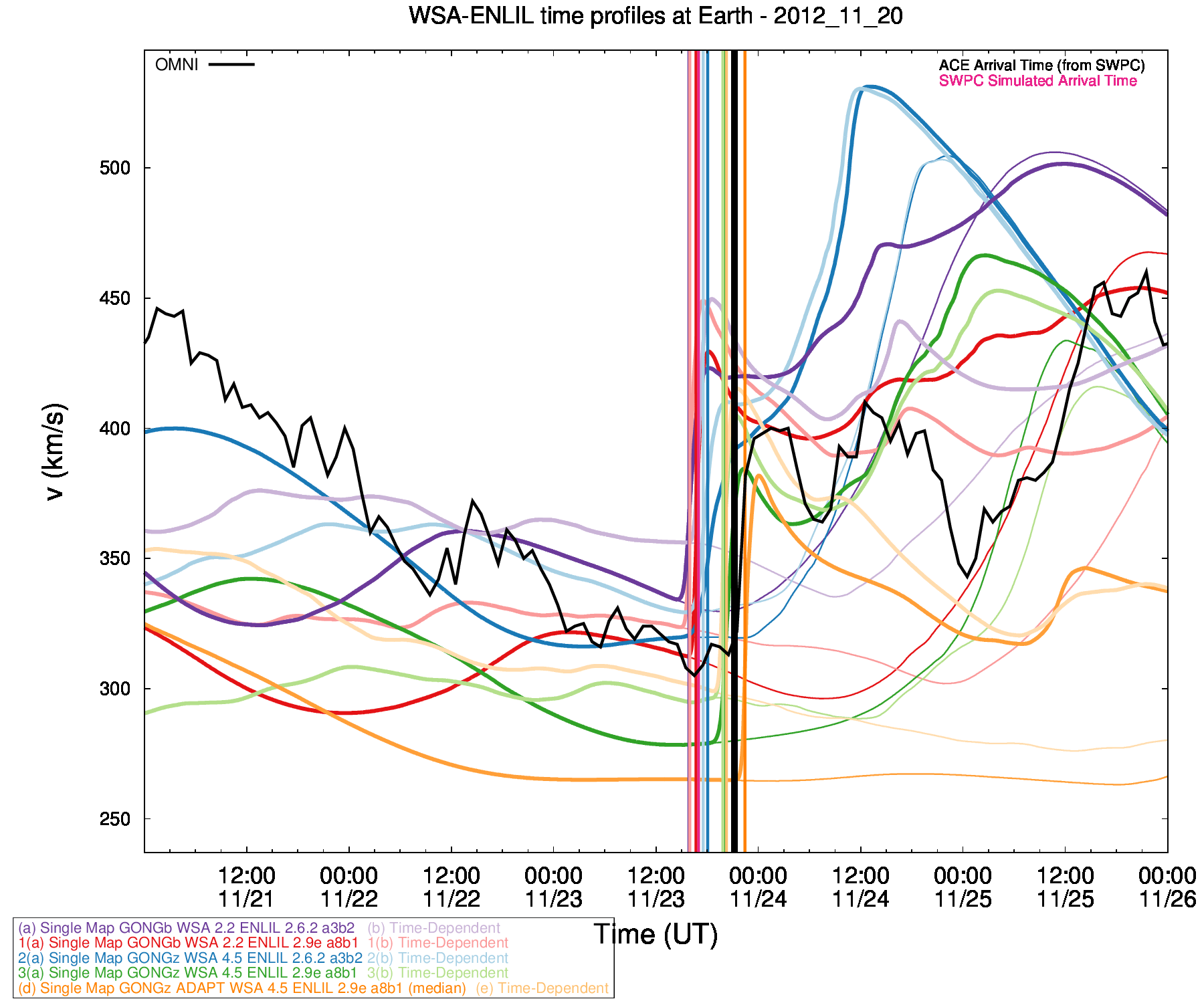}\\
\includegraphics[width=0.33\textwidth]{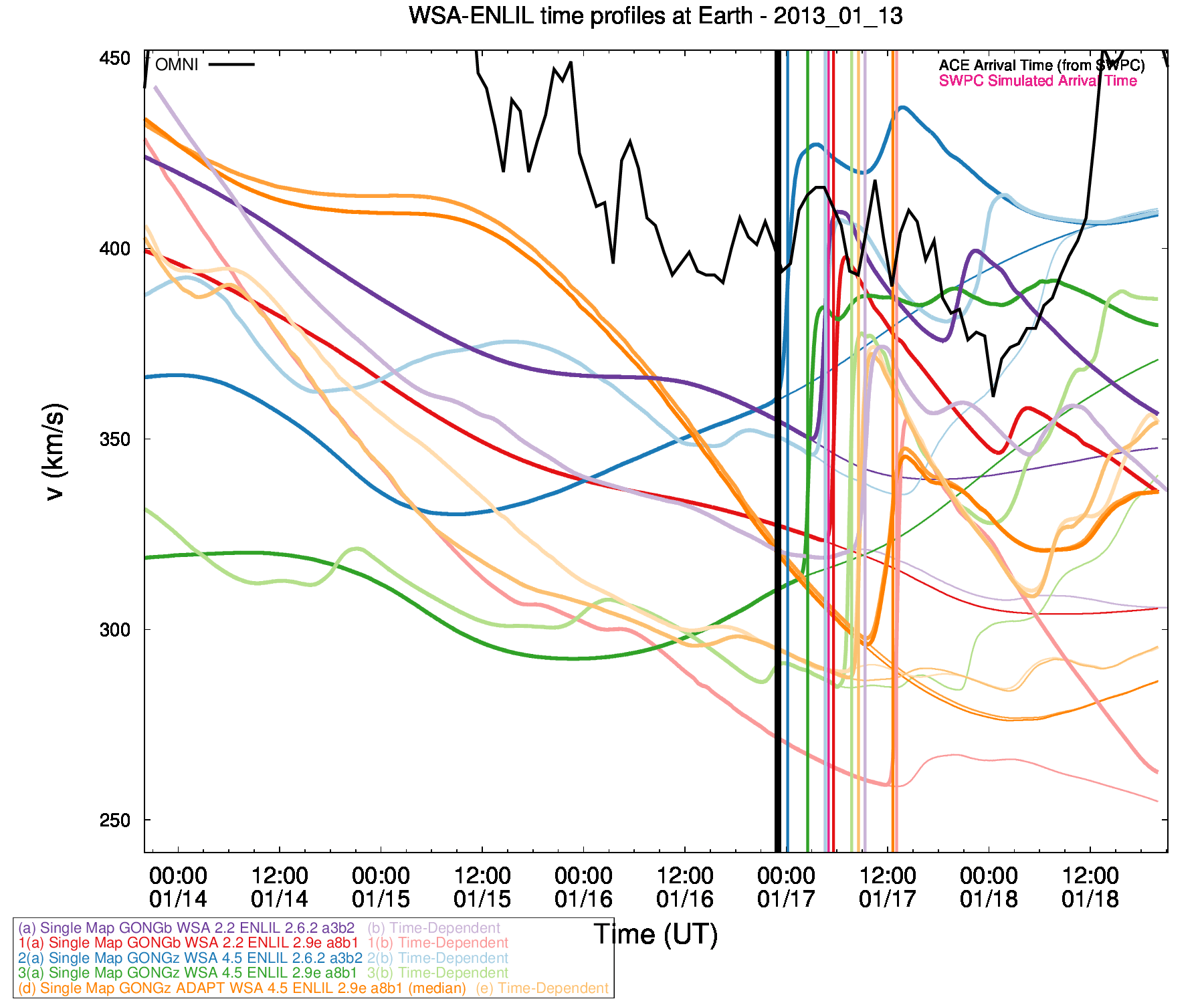}
\includegraphics[width=0.33\textwidth]{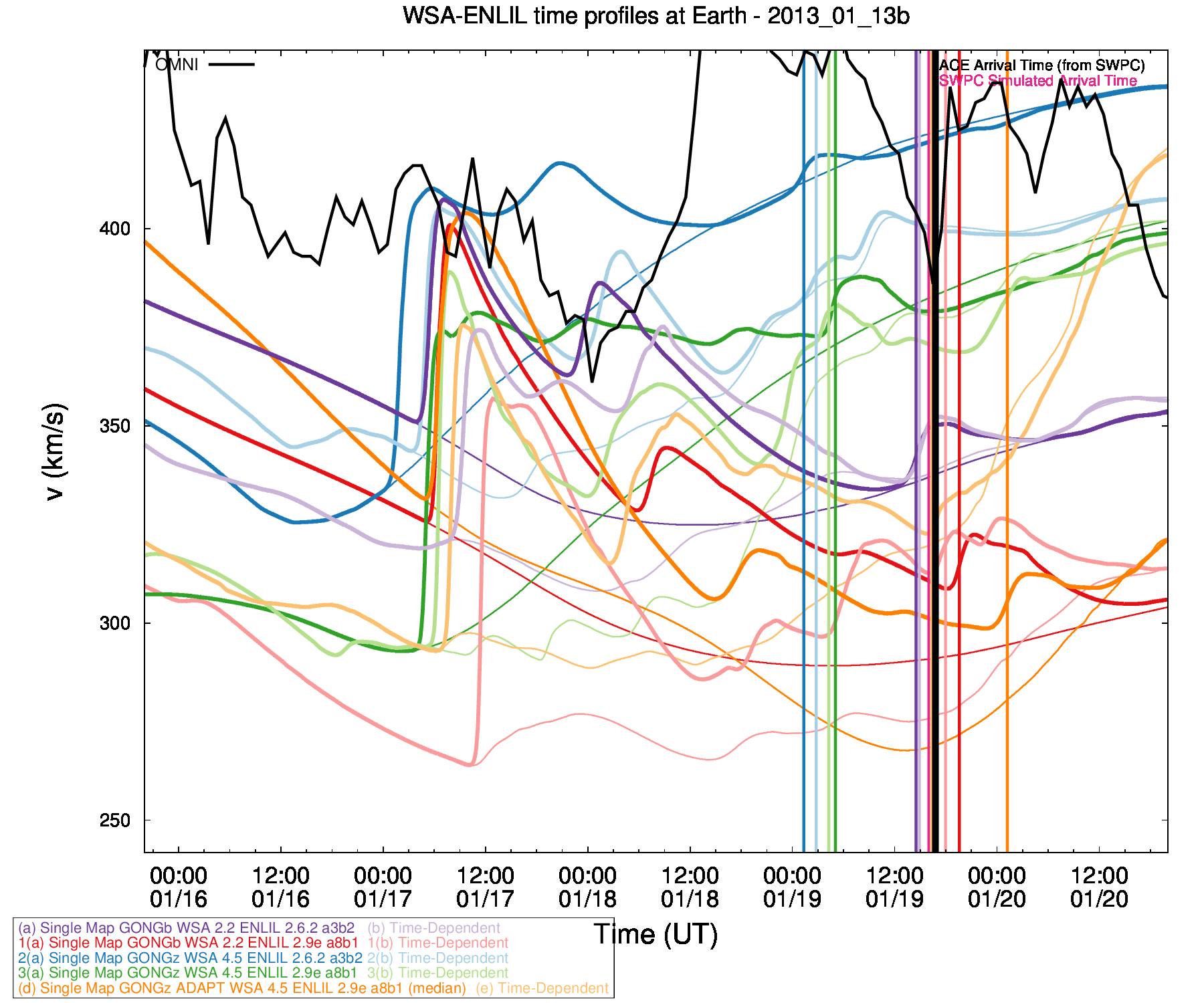}
\includegraphics[width=0.33\textwidth]{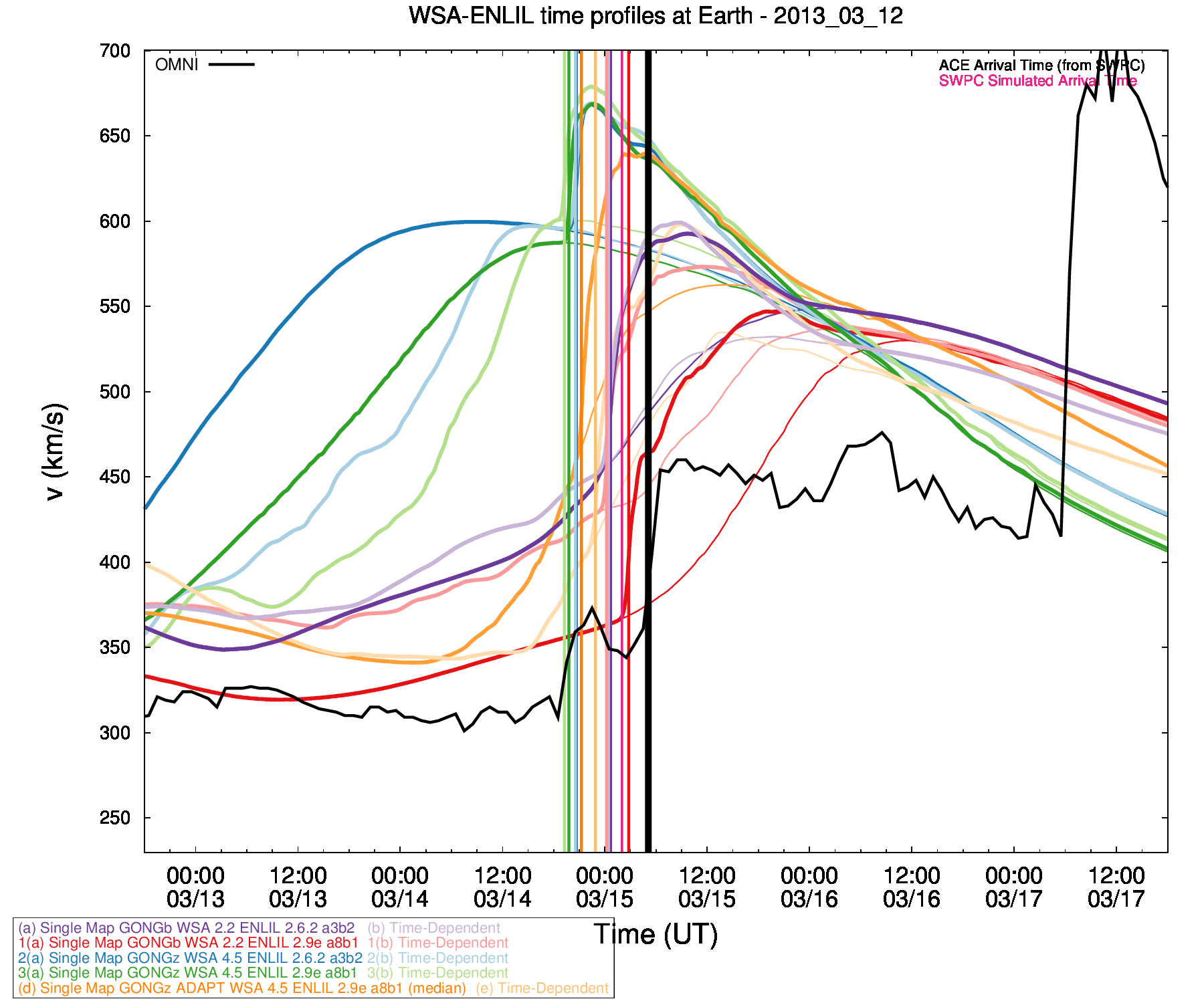}\\
\includegraphics[width=0.33\textwidth]{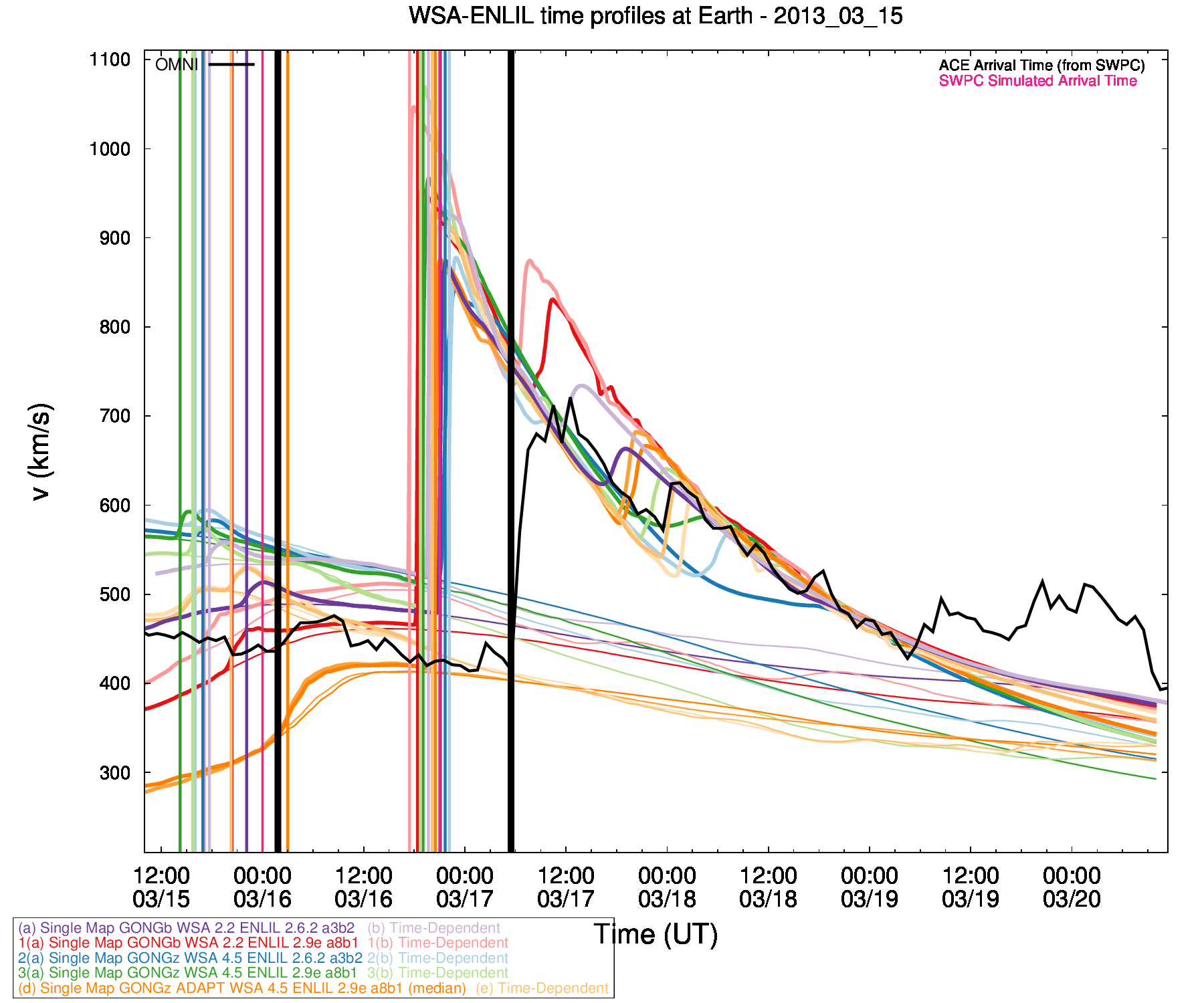}
\includegraphics[width=0.33\textwidth]{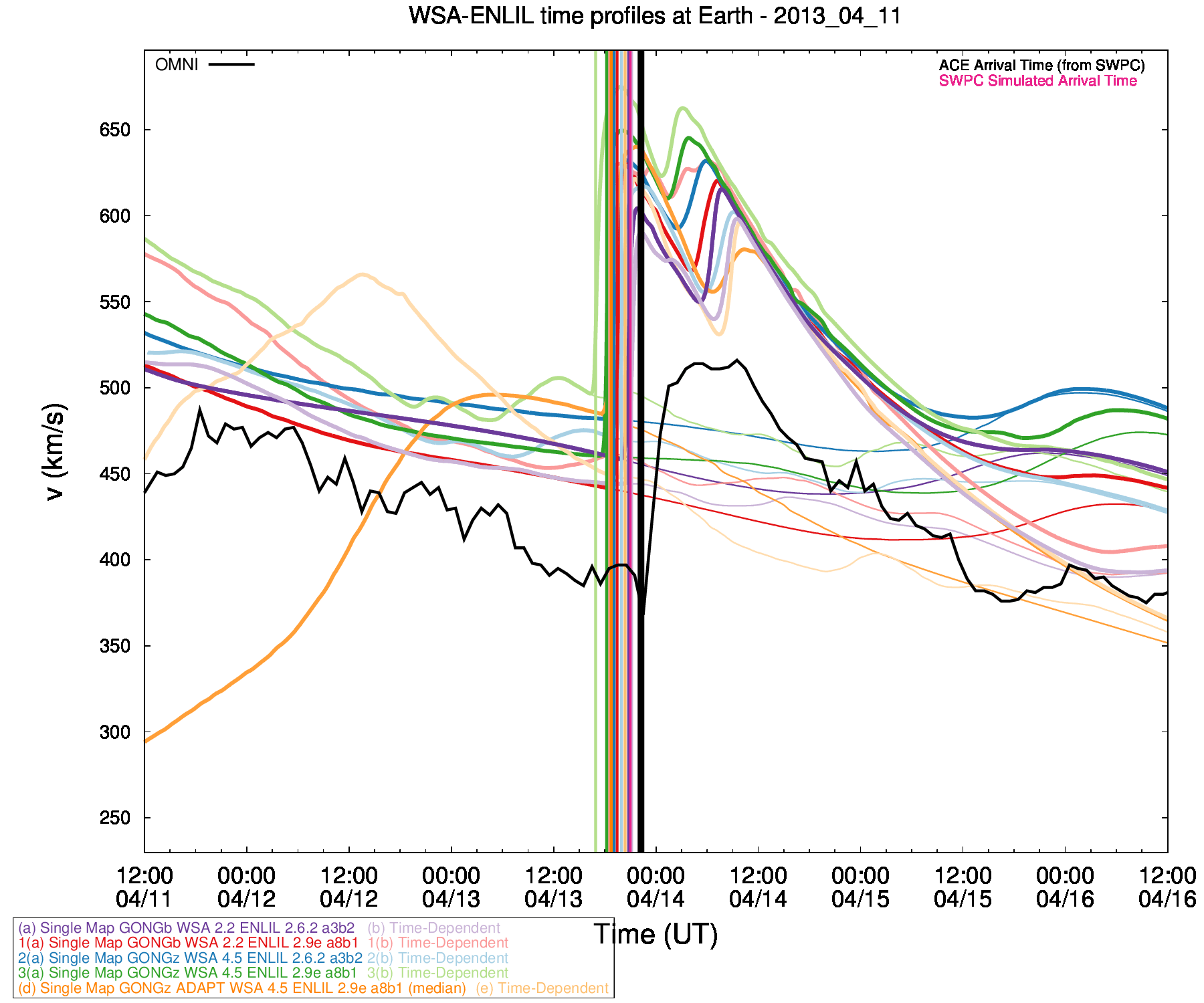}
\includegraphics[width=0.33\textwidth]{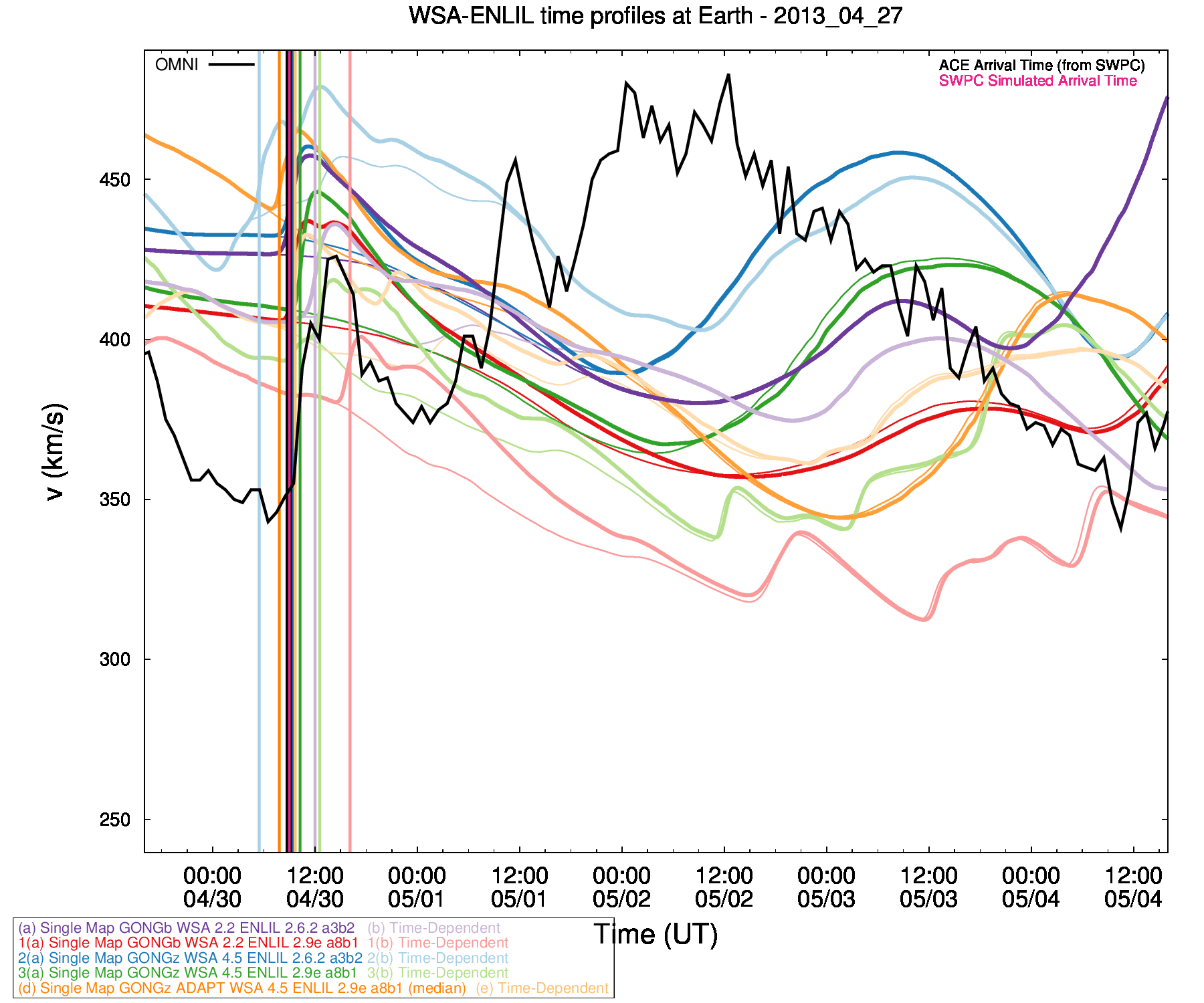}\\
\includegraphics[width=0.33\textwidth]{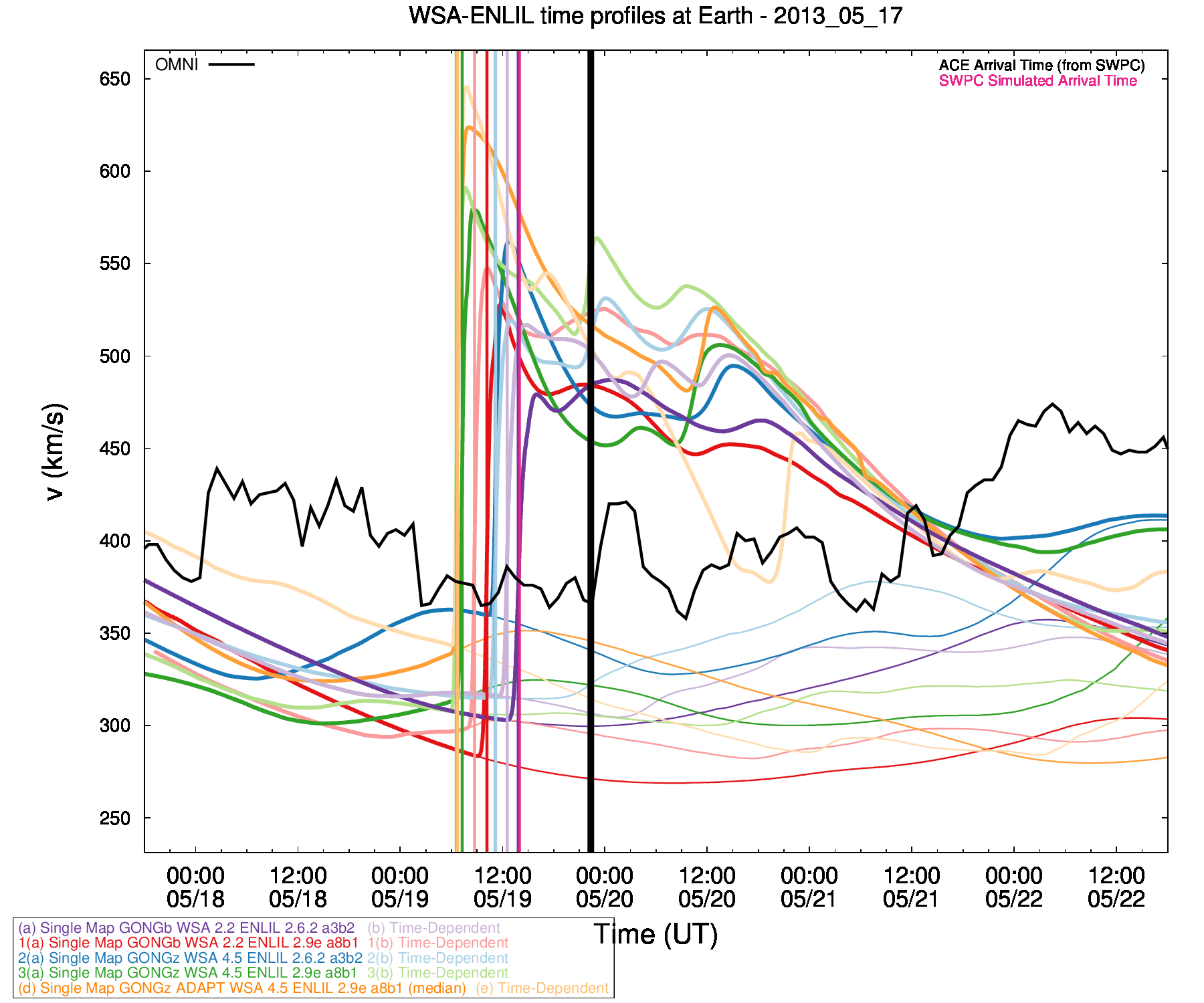}
\includegraphics[width=0.33\textwidth]{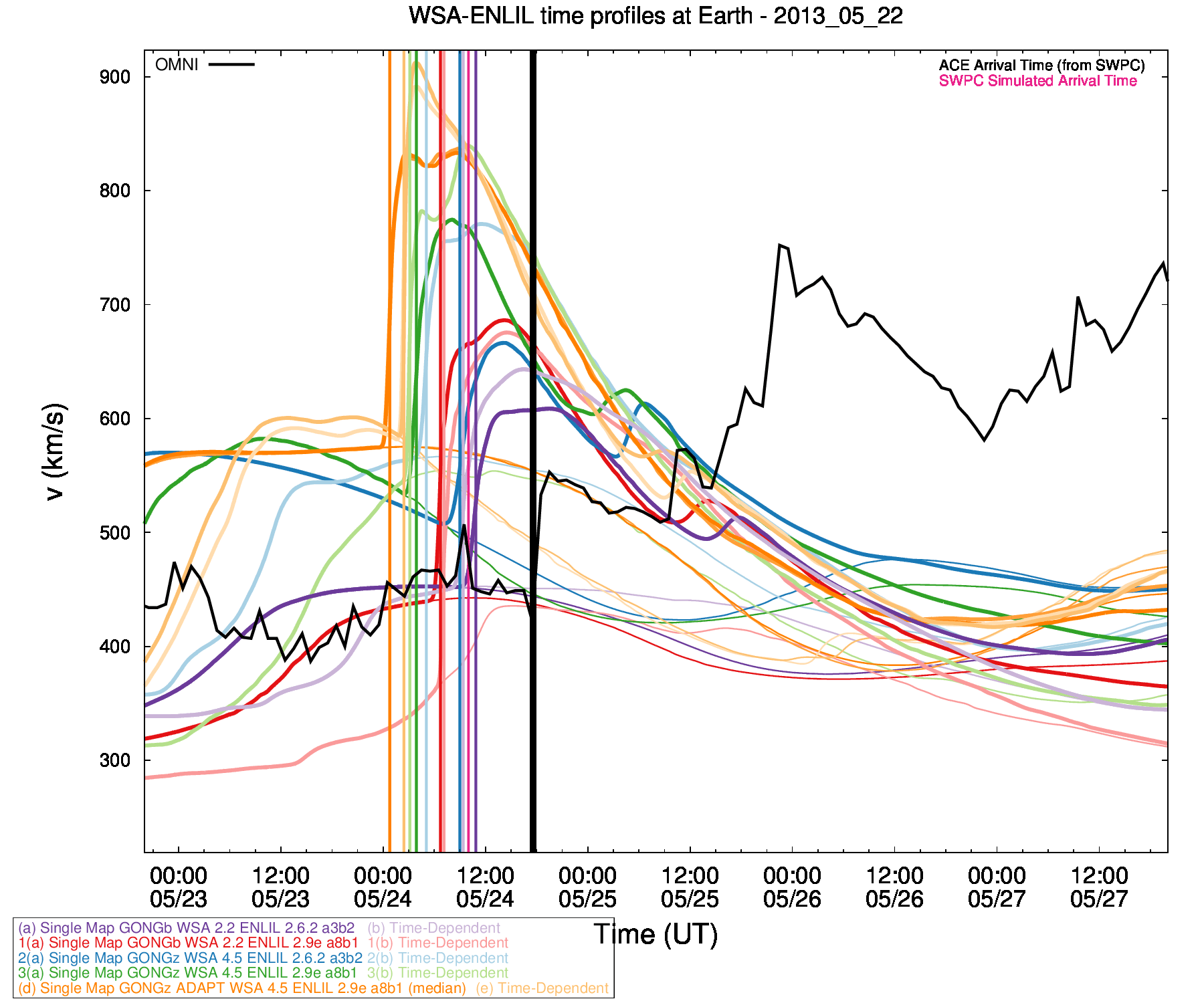}
\includegraphics[width=0.33\textwidth]{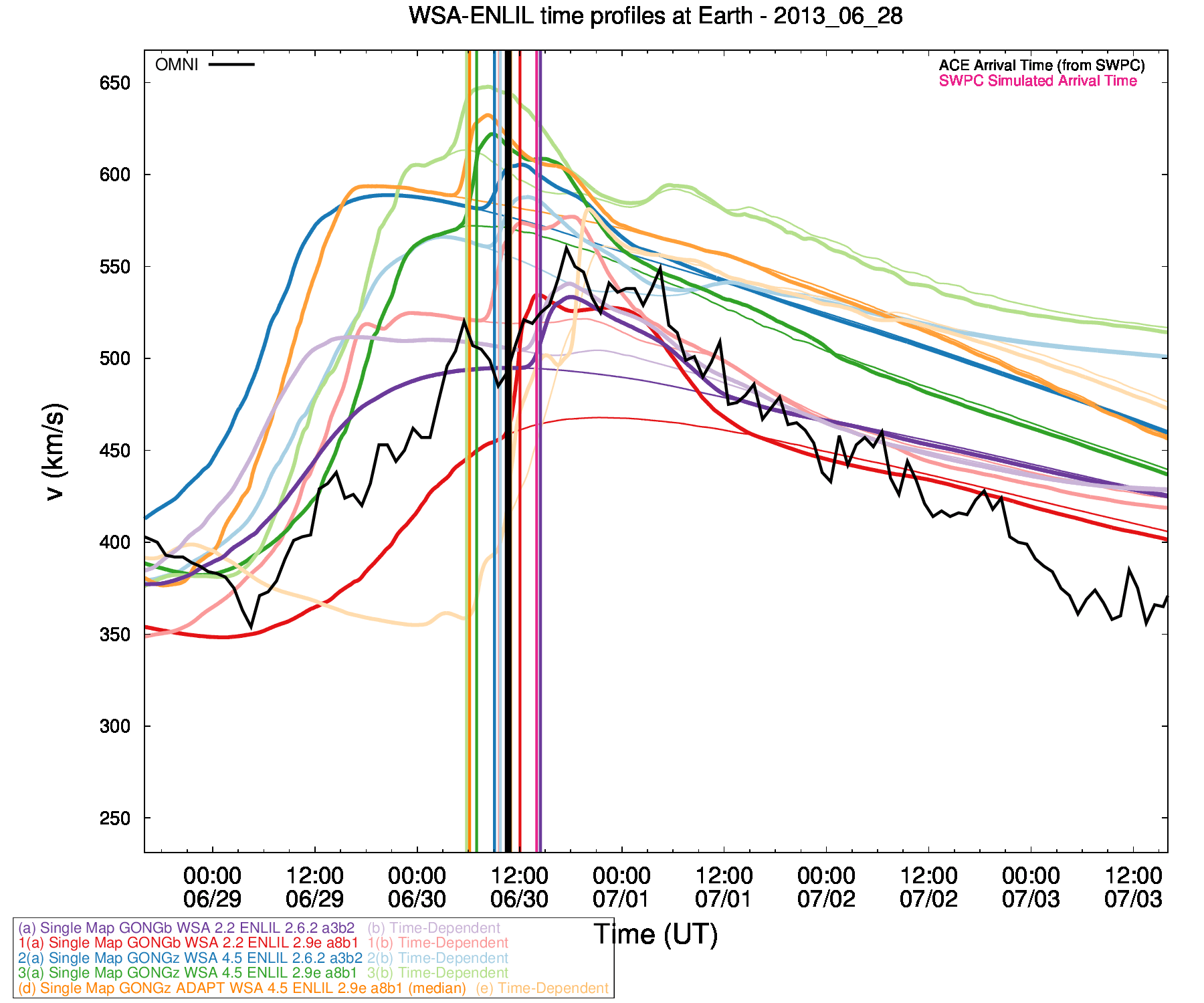}\\
\caption{Figure \ref{fig:time-series} continued.\label{fig:time-series2}}
\end{figure*}

\begin{figure*}
\includegraphics[width=0.33\textwidth]{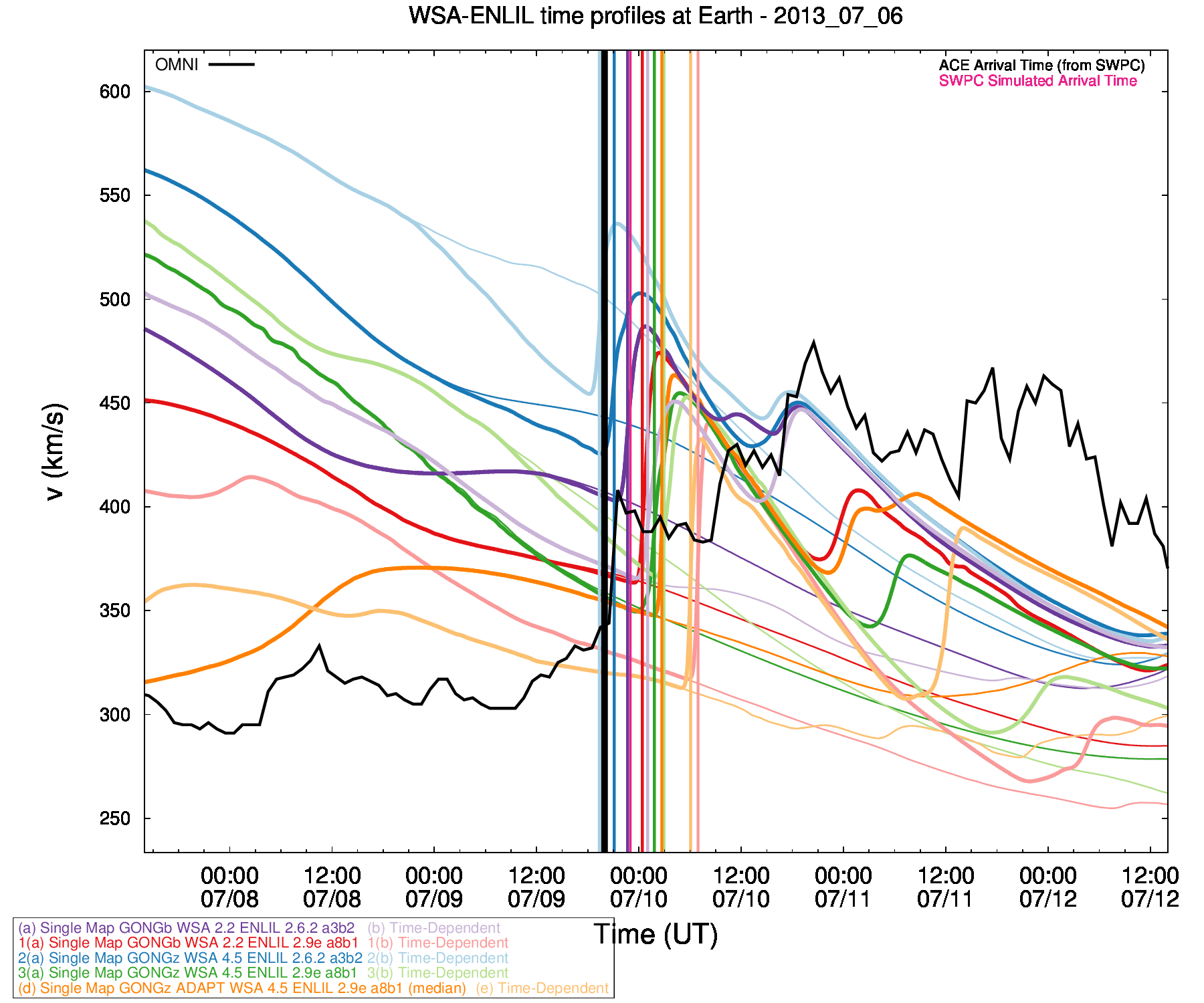}
\includegraphics[width=0.33\textwidth]{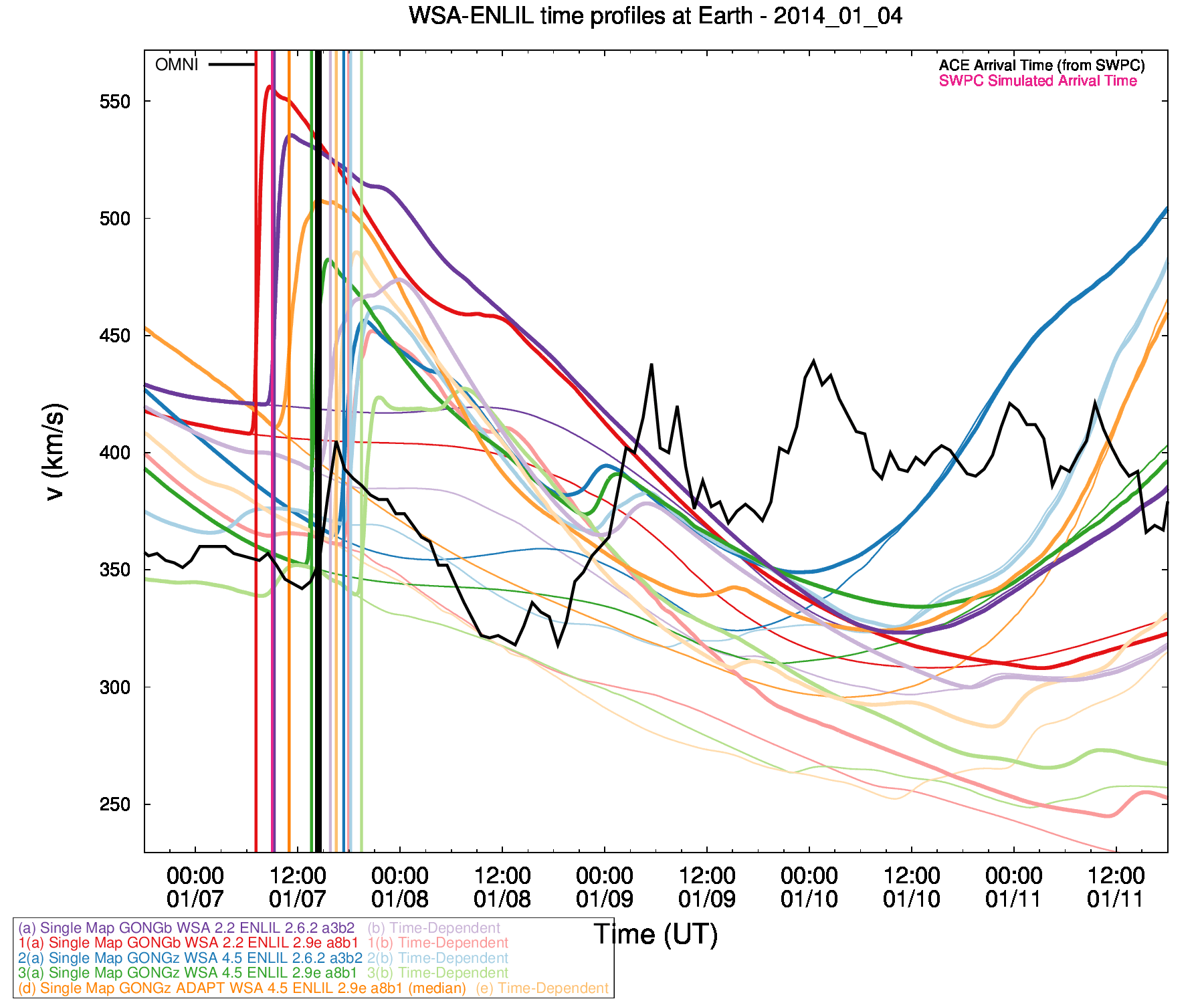}
\includegraphics[width=0.33\textwidth]{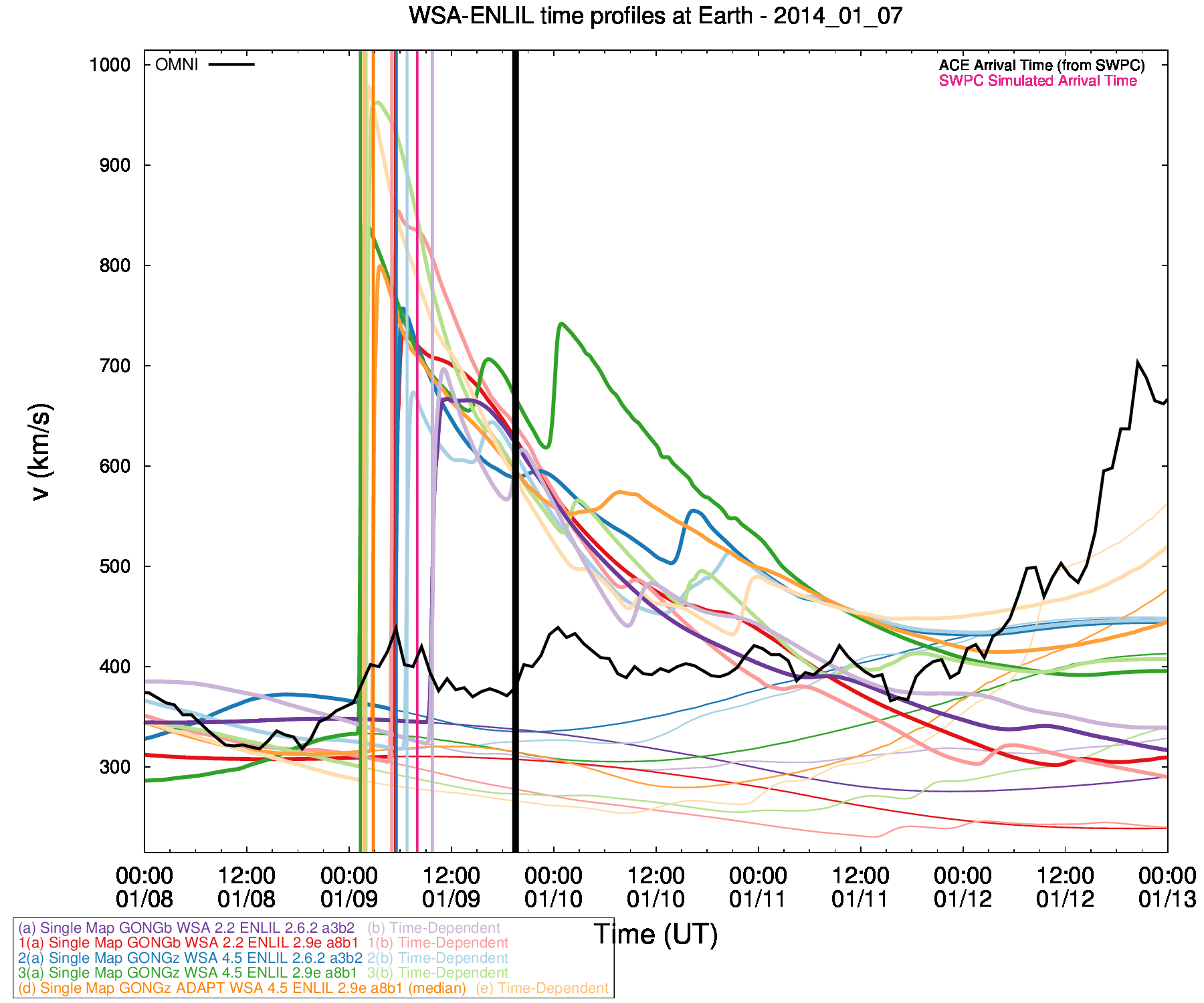}\\
\includegraphics[width=0.33\textwidth]{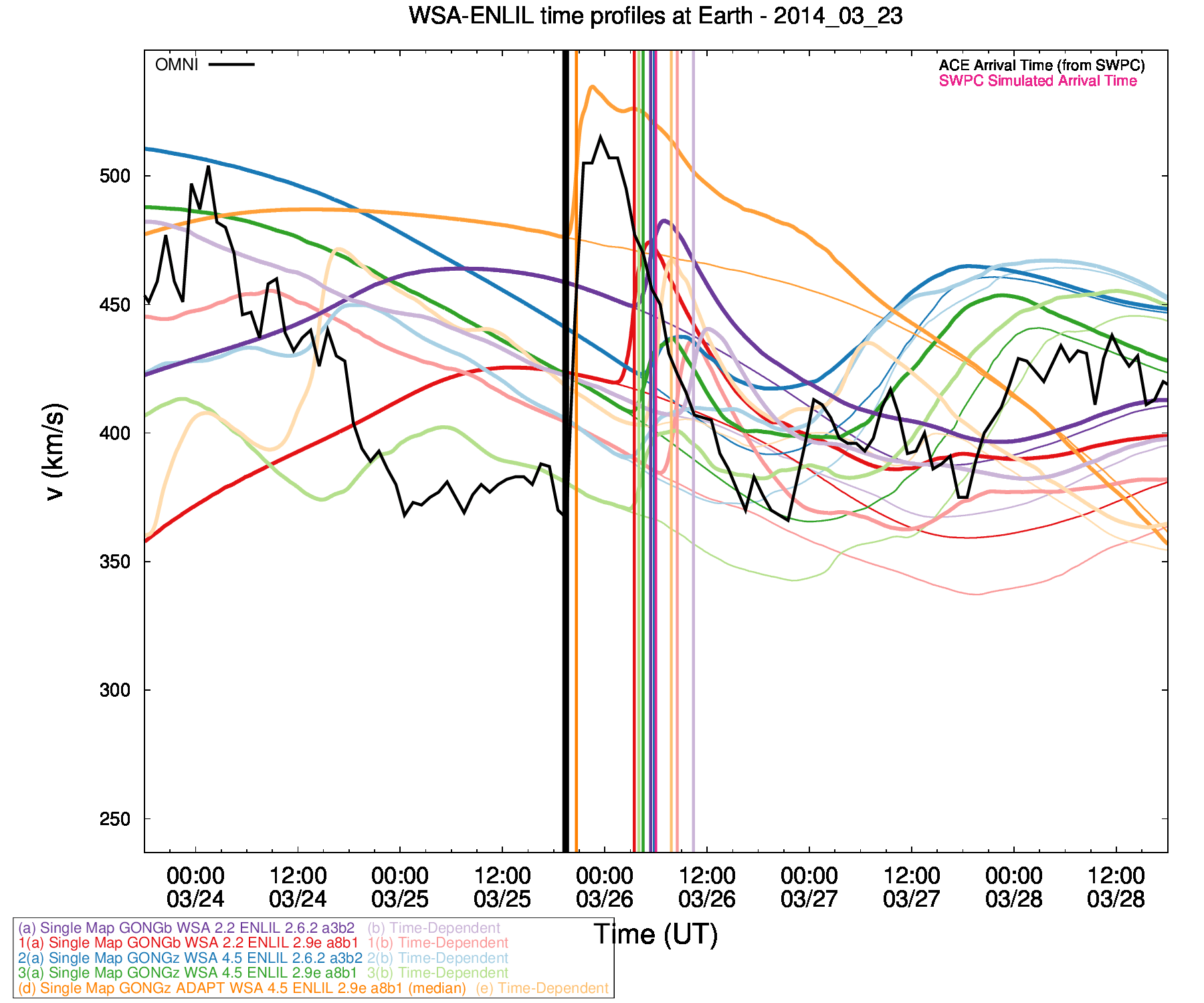}
\includegraphics[width=0.33\textwidth]{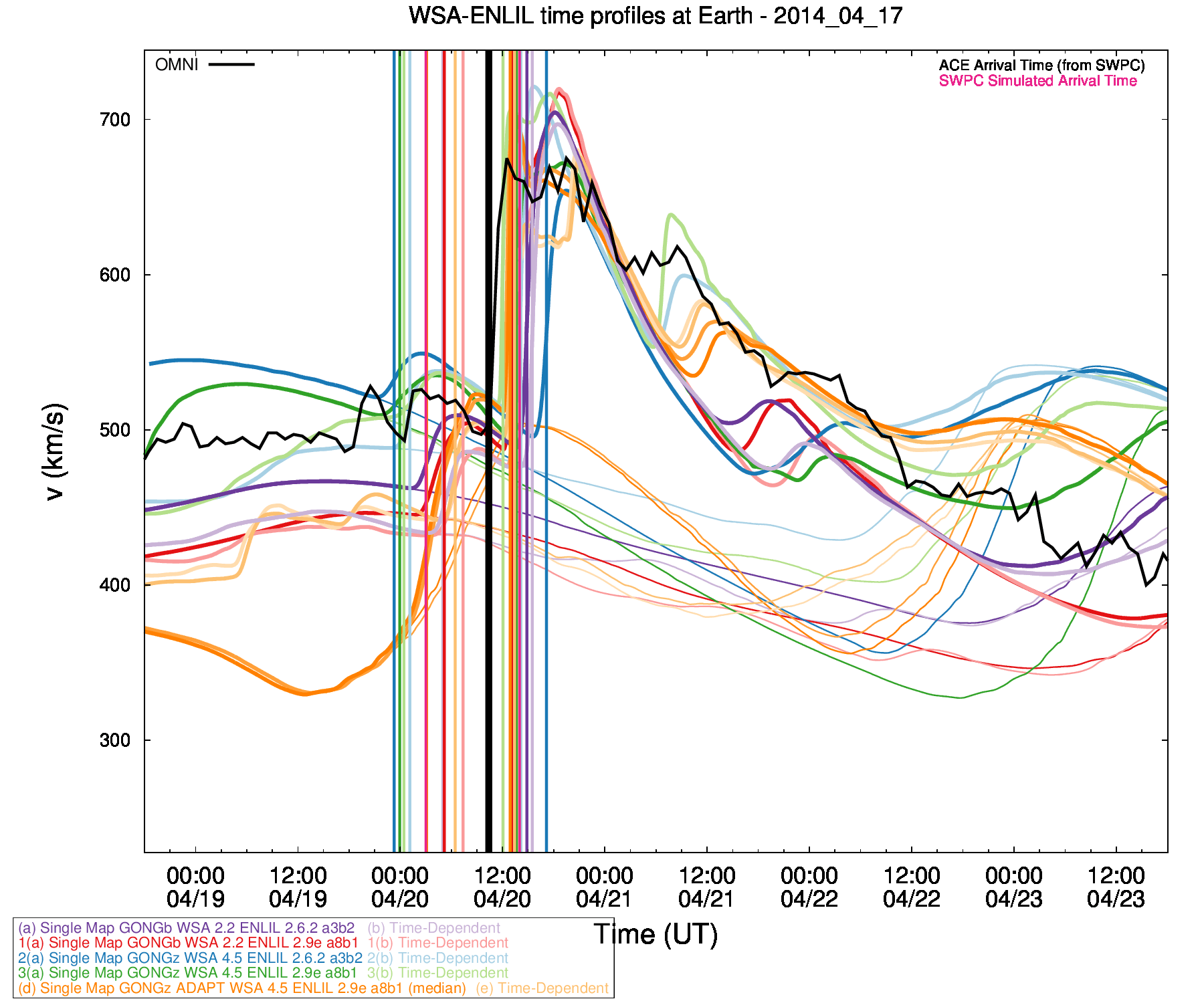}
\includegraphics[width=0.33\textwidth]{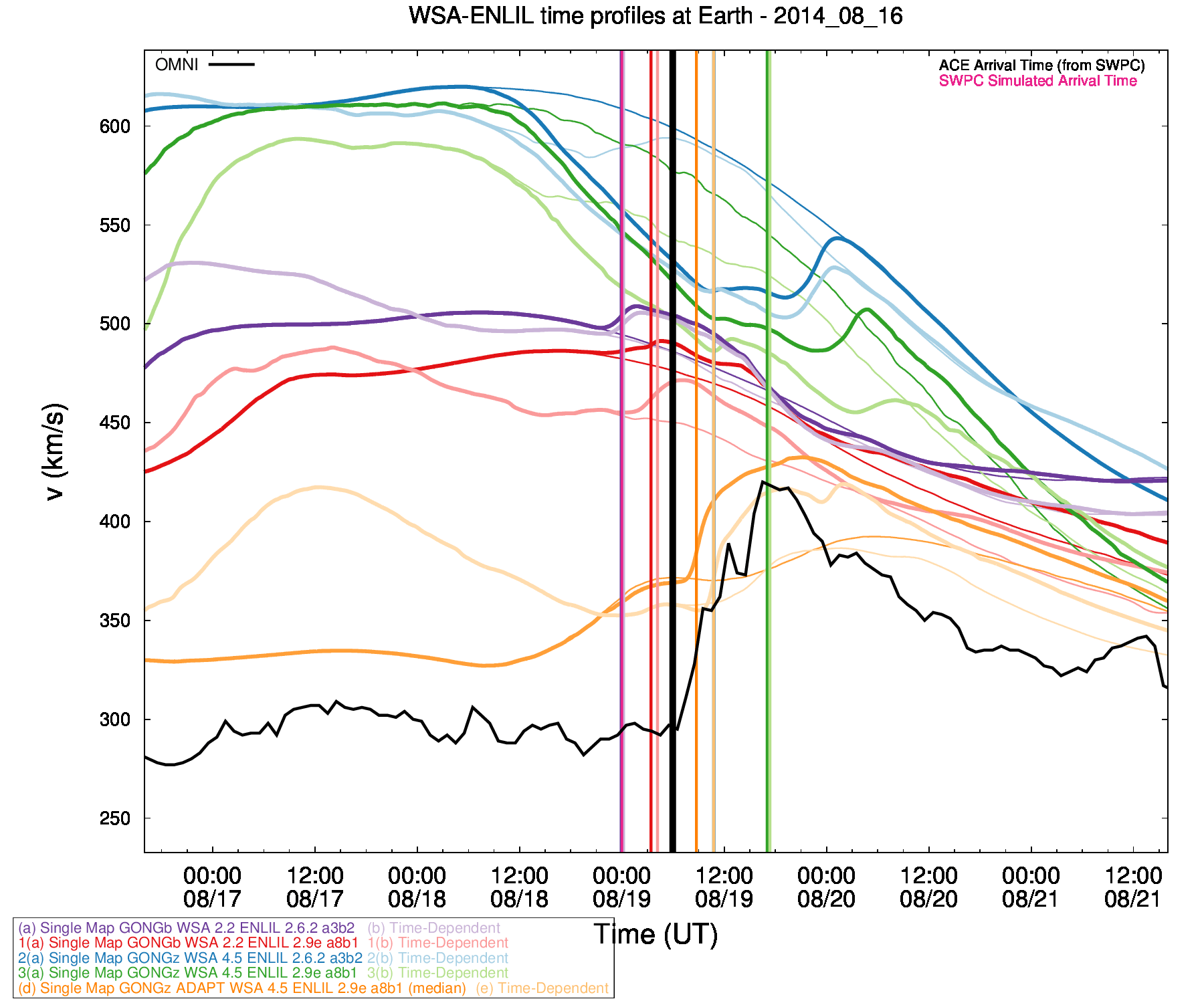}\\
\includegraphics[width=0.33\textwidth]{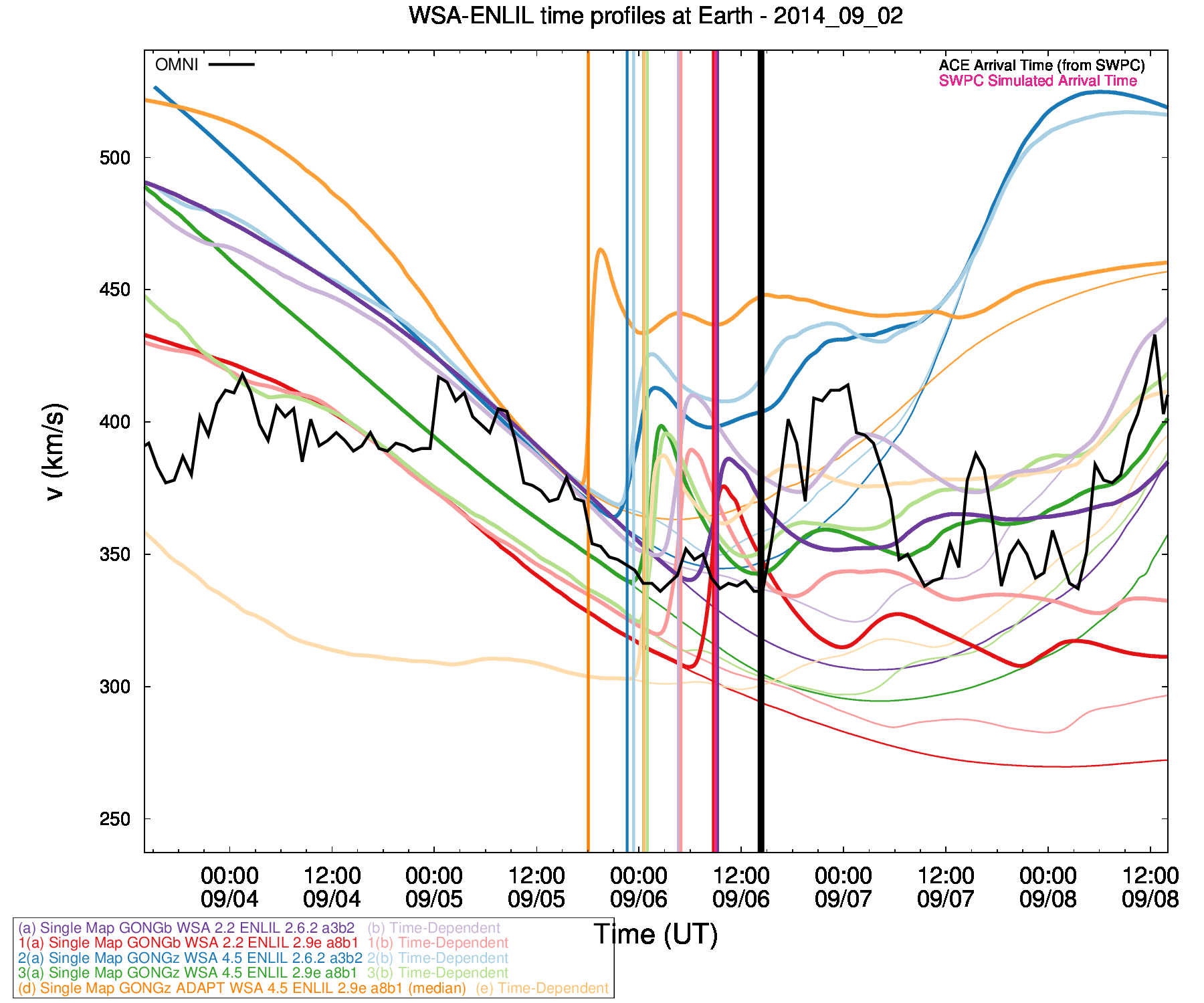}
\includegraphics[width=0.33\textwidth]{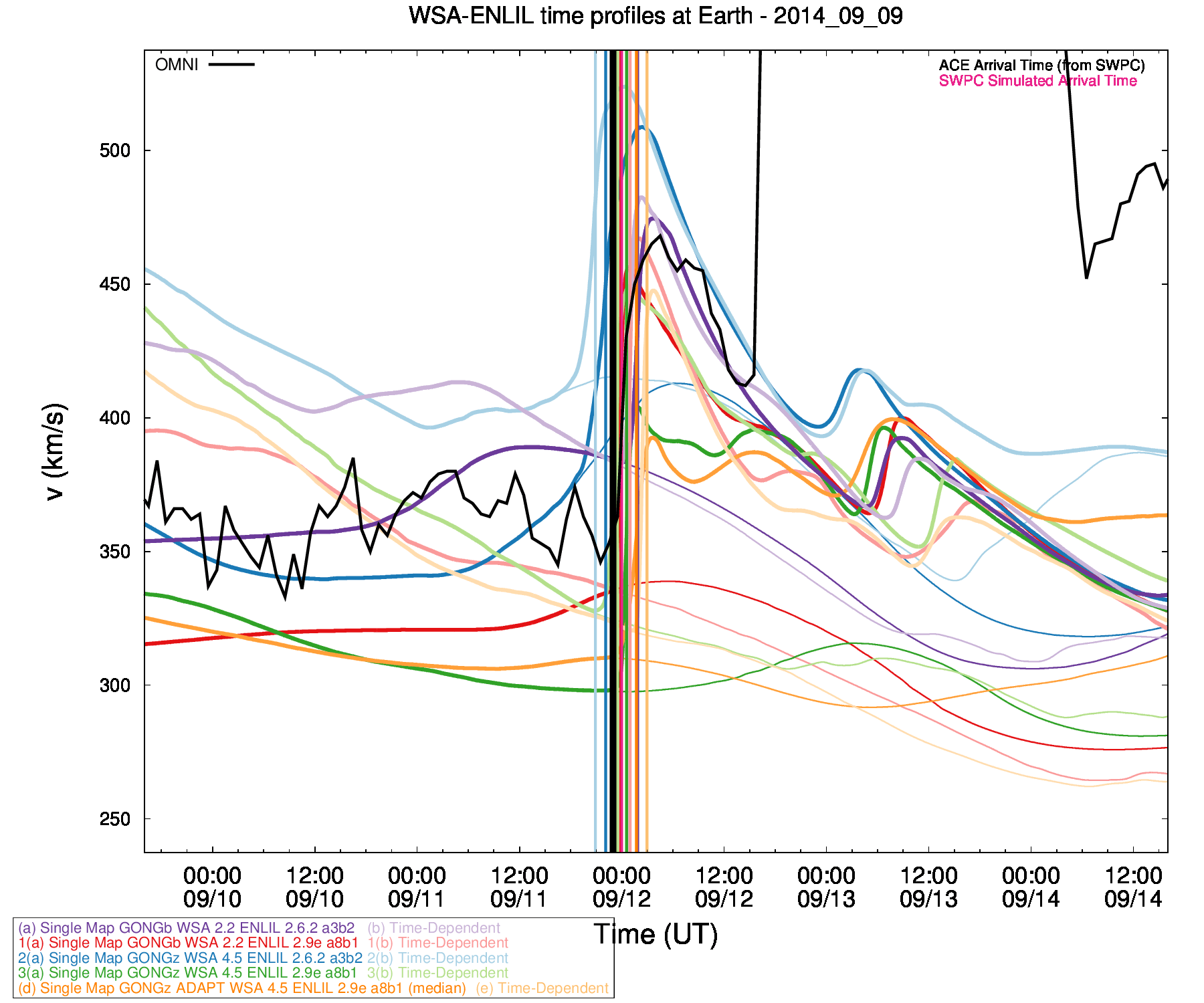}
\includegraphics[width=0.33\textwidth]{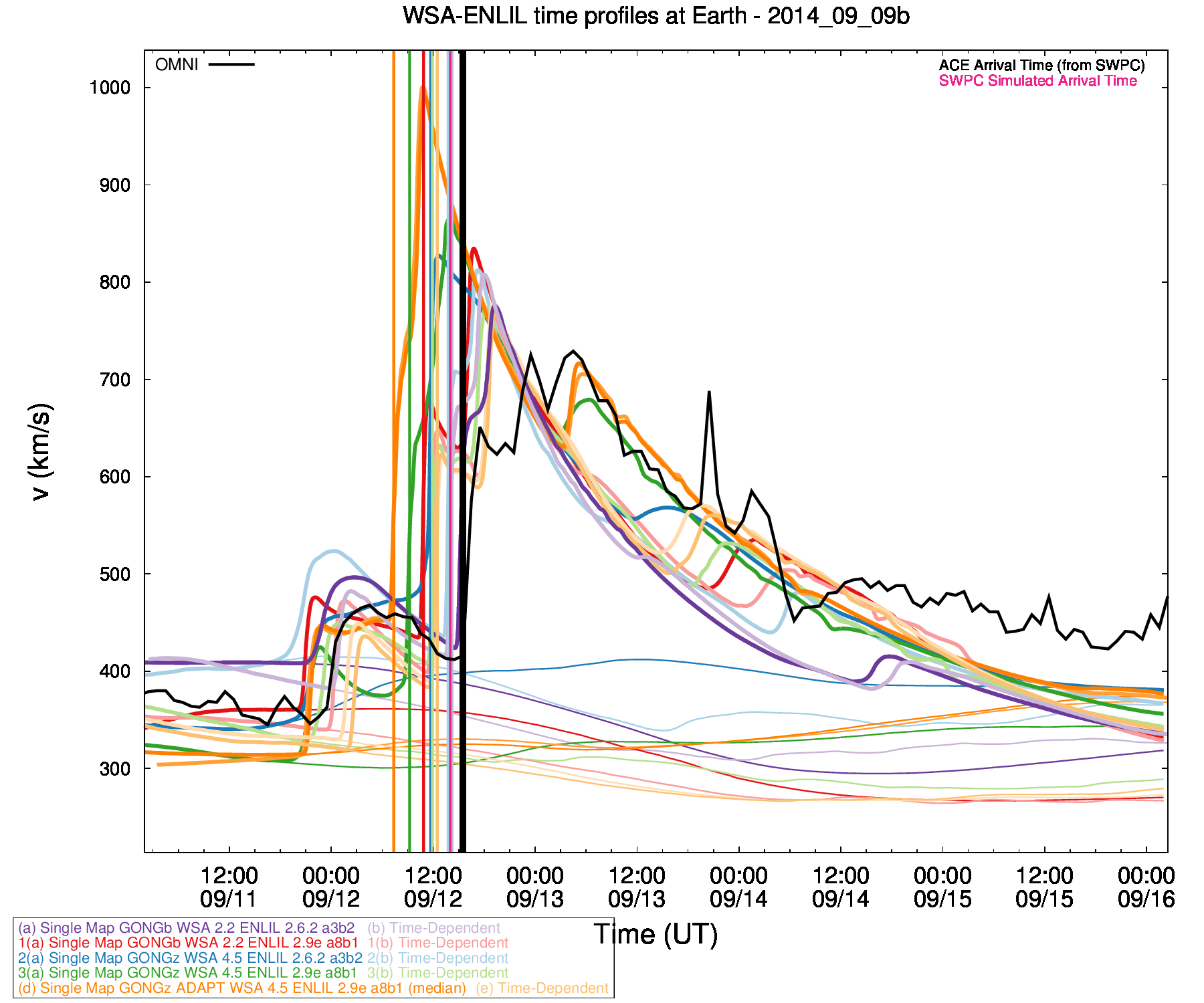}\\
\includegraphics[width=0.33\textwidth]{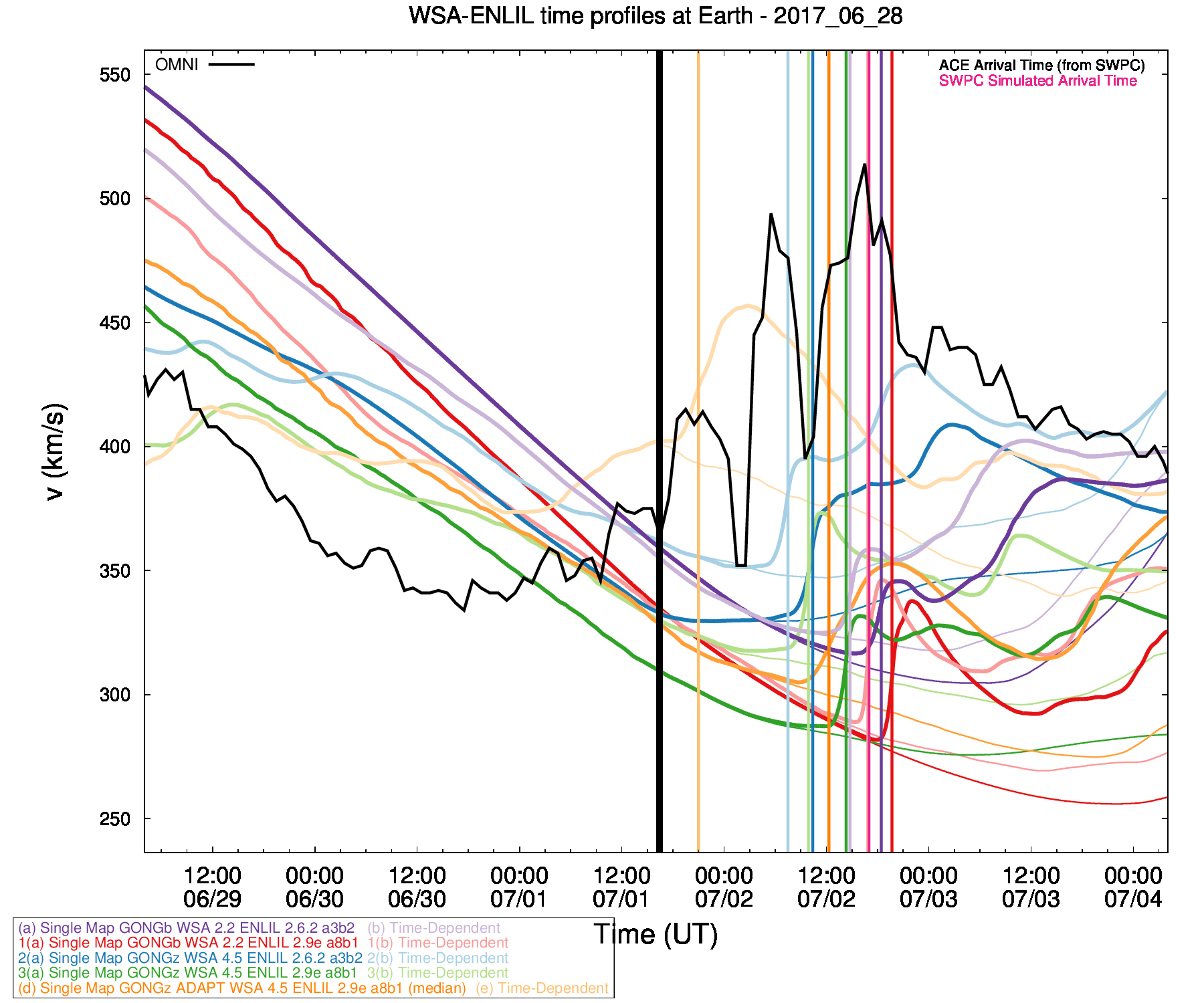}
\includegraphics[width=0.33\textwidth]{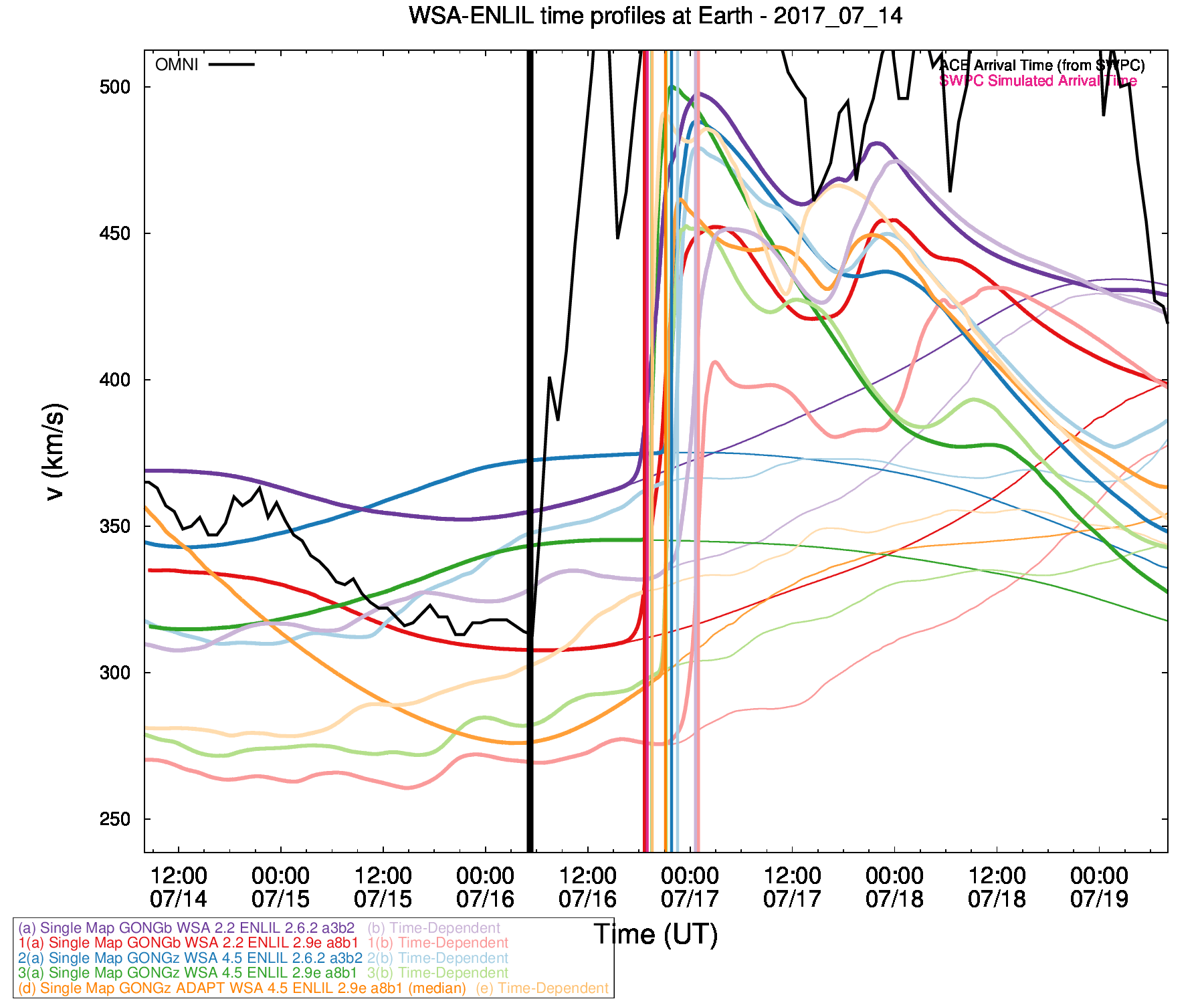}
\includegraphics[width=0.33\textwidth]{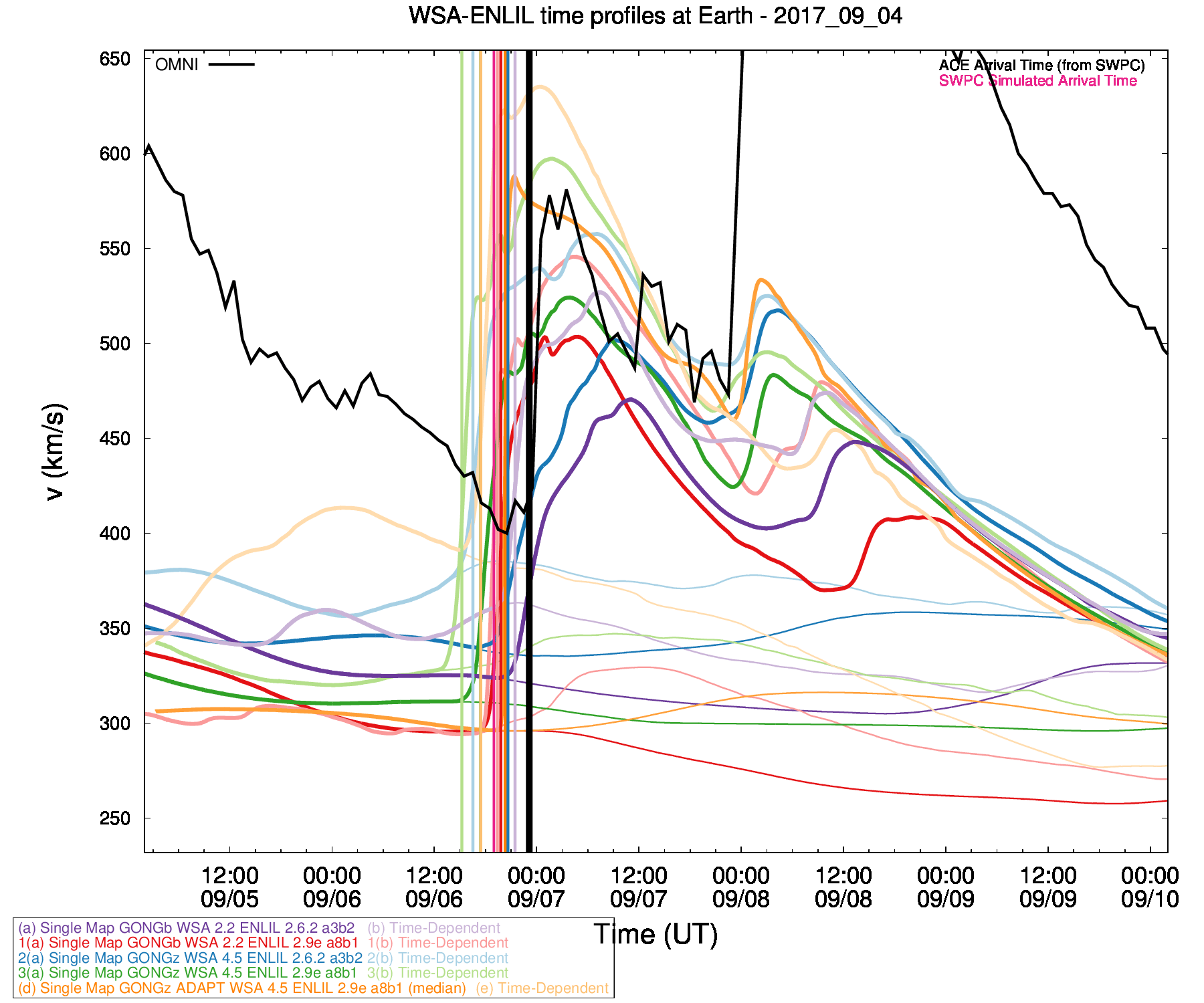}\\
\includegraphics[width=0.33\textwidth]{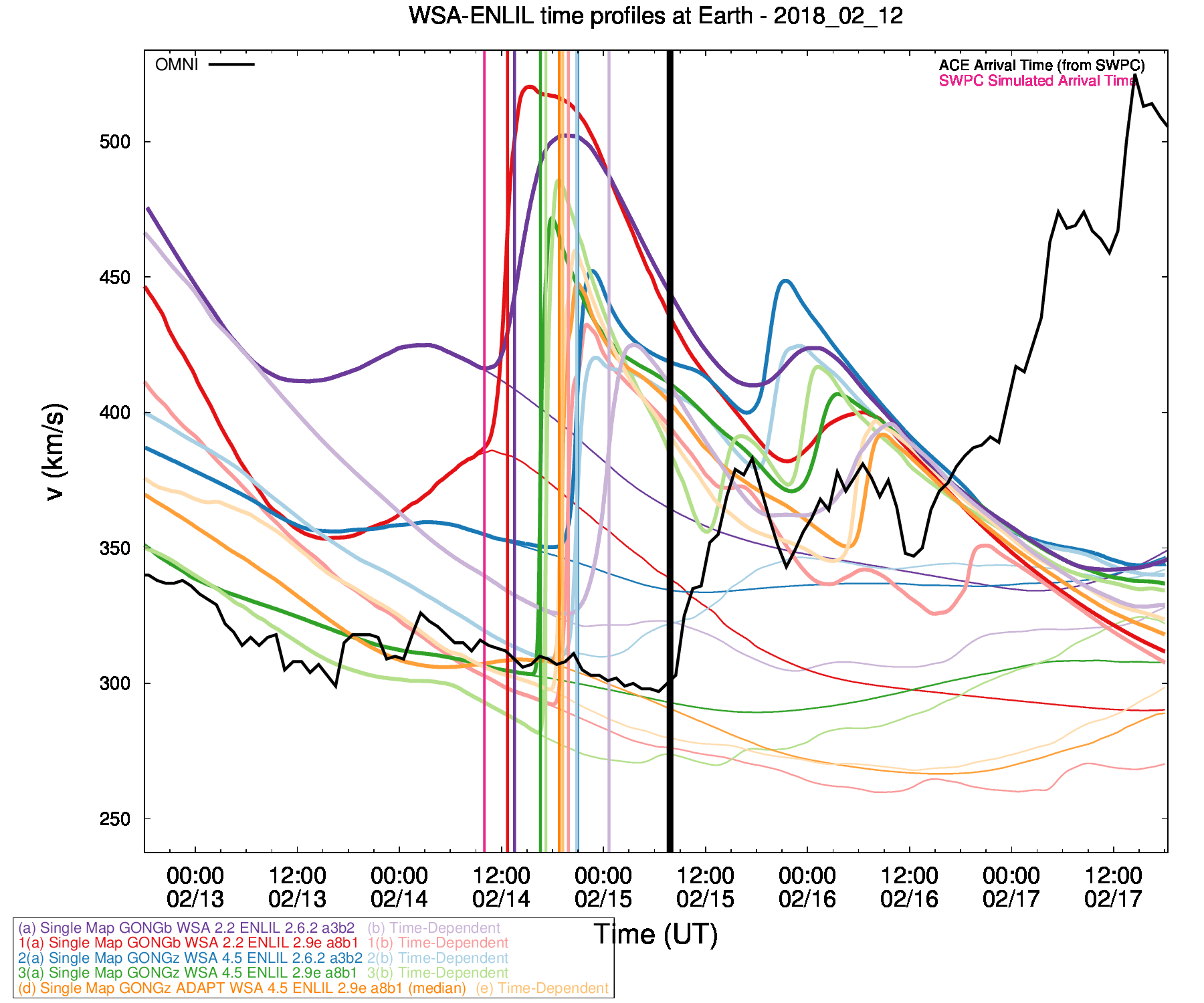}
\includegraphics[width=0.33\textwidth]{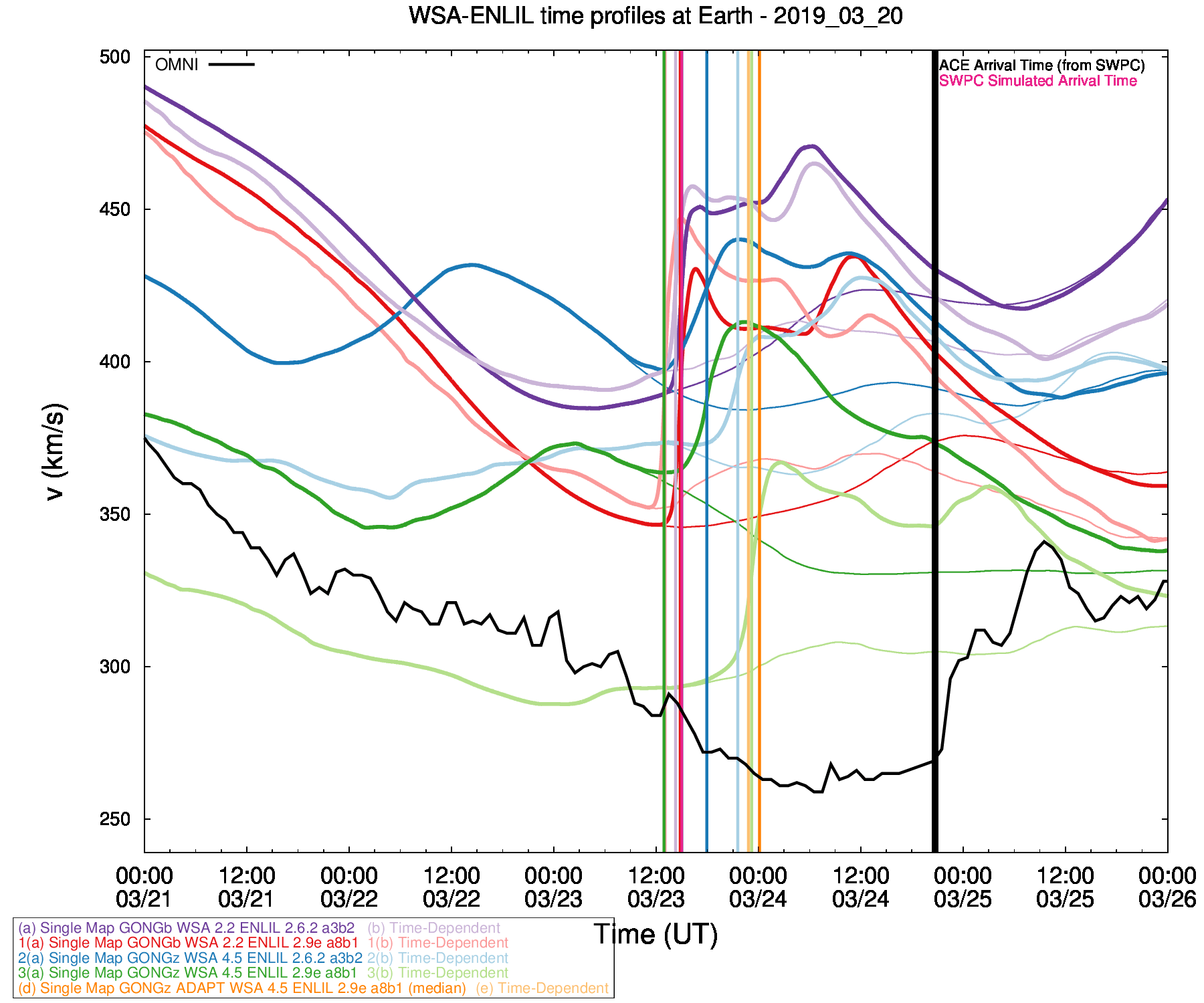}
\caption{Figure \ref{fig:time-series2} continued.\label{fig:time-series3}}
\end{figure*}

\begin{figure*}
\includegraphics[width=0.33\textwidth]{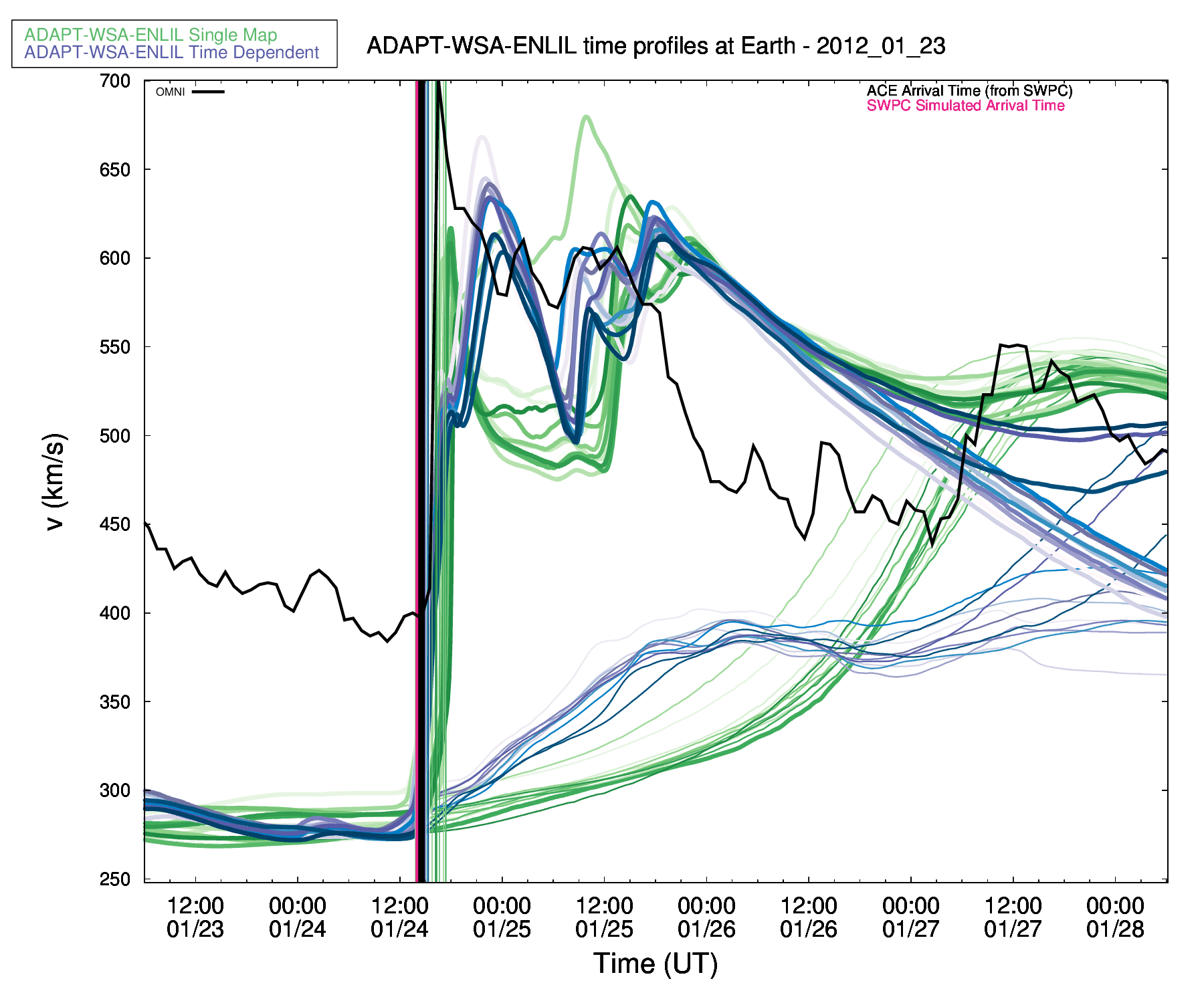}
\includegraphics[width=0.33\textwidth]{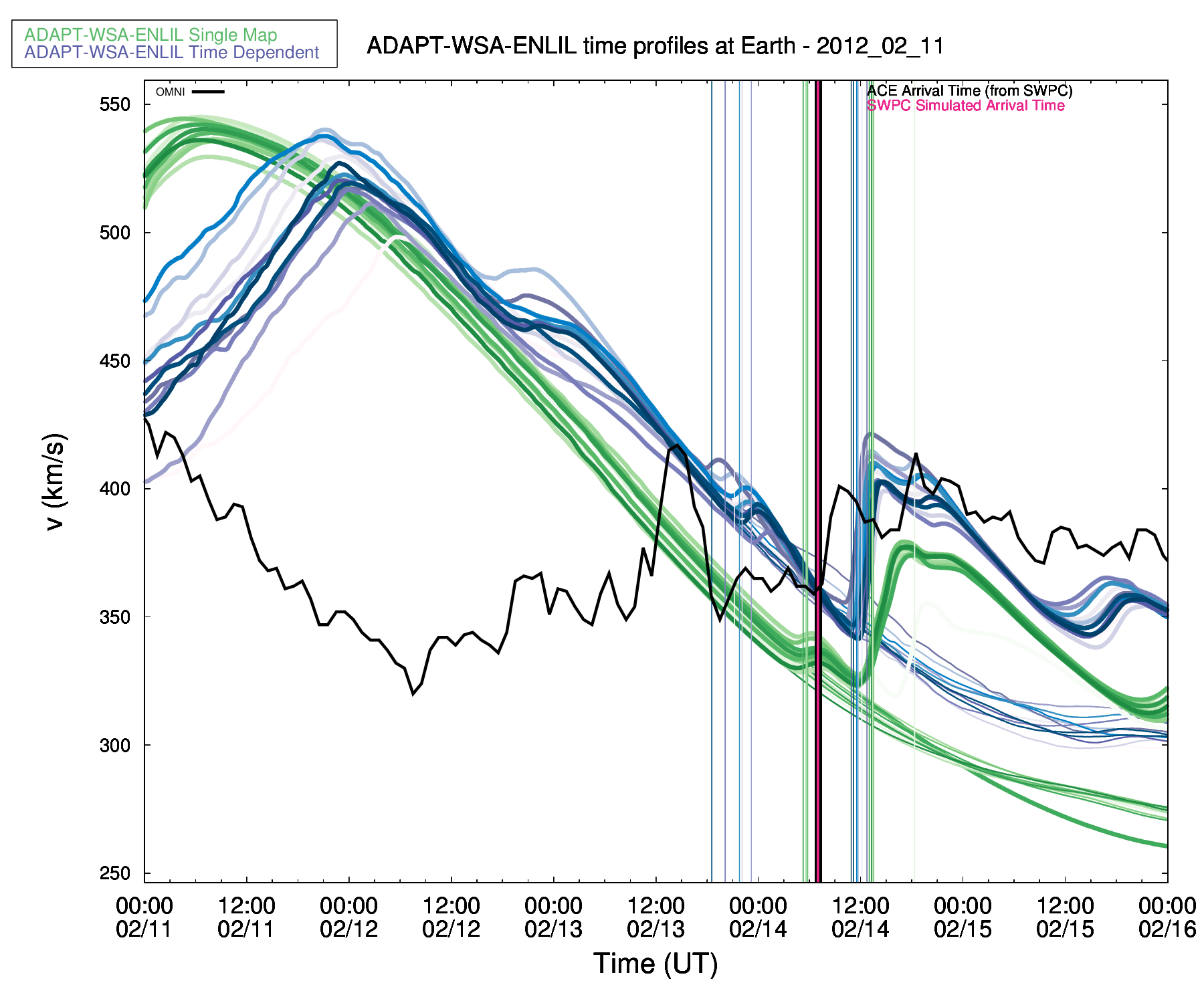}
\includegraphics[width=0.33\textwidth]{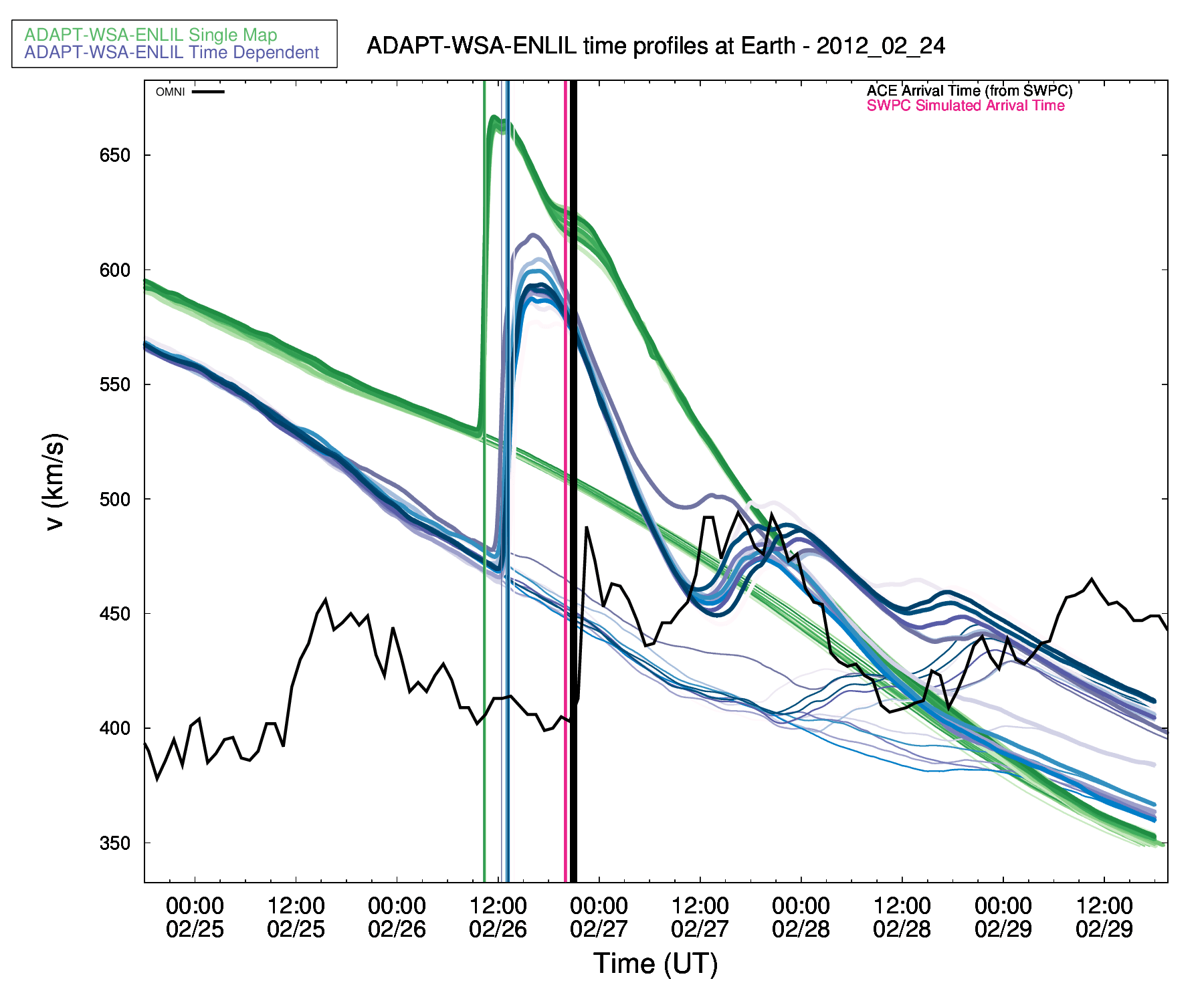}\\
\includegraphics[width=0.33\textwidth]{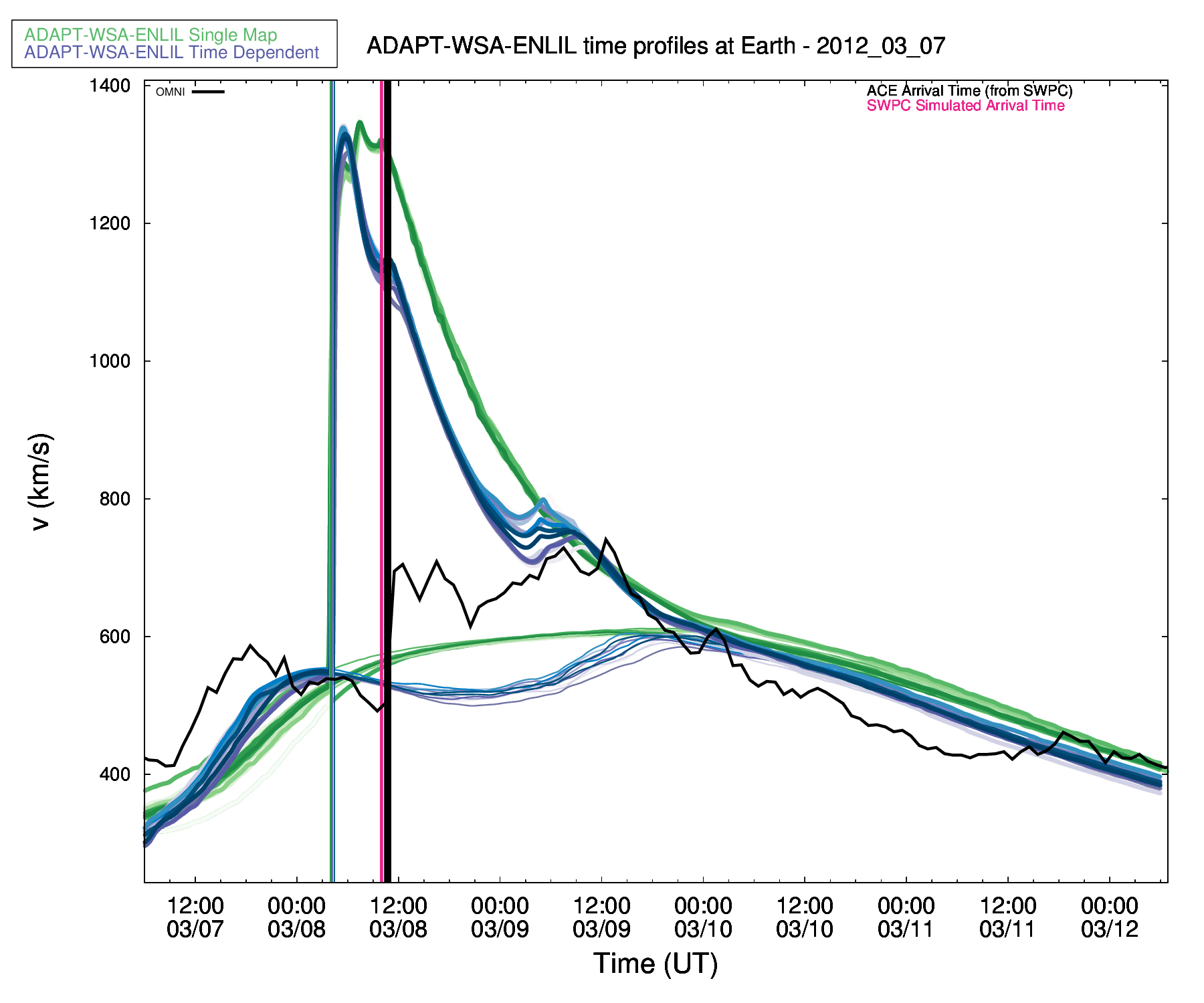}
\includegraphics[width=0.33\textwidth]{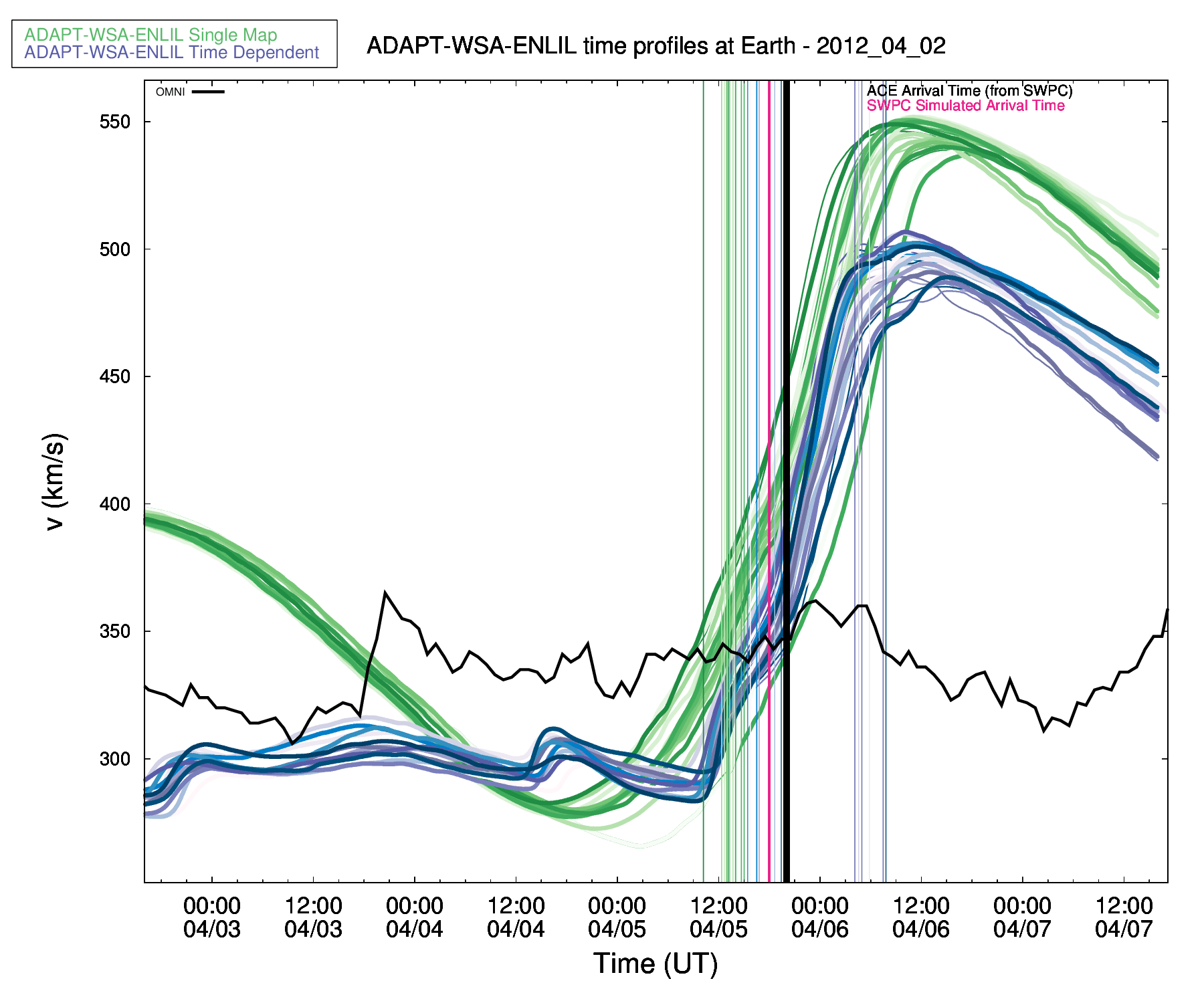}
\includegraphics[width=0.33\textwidth]{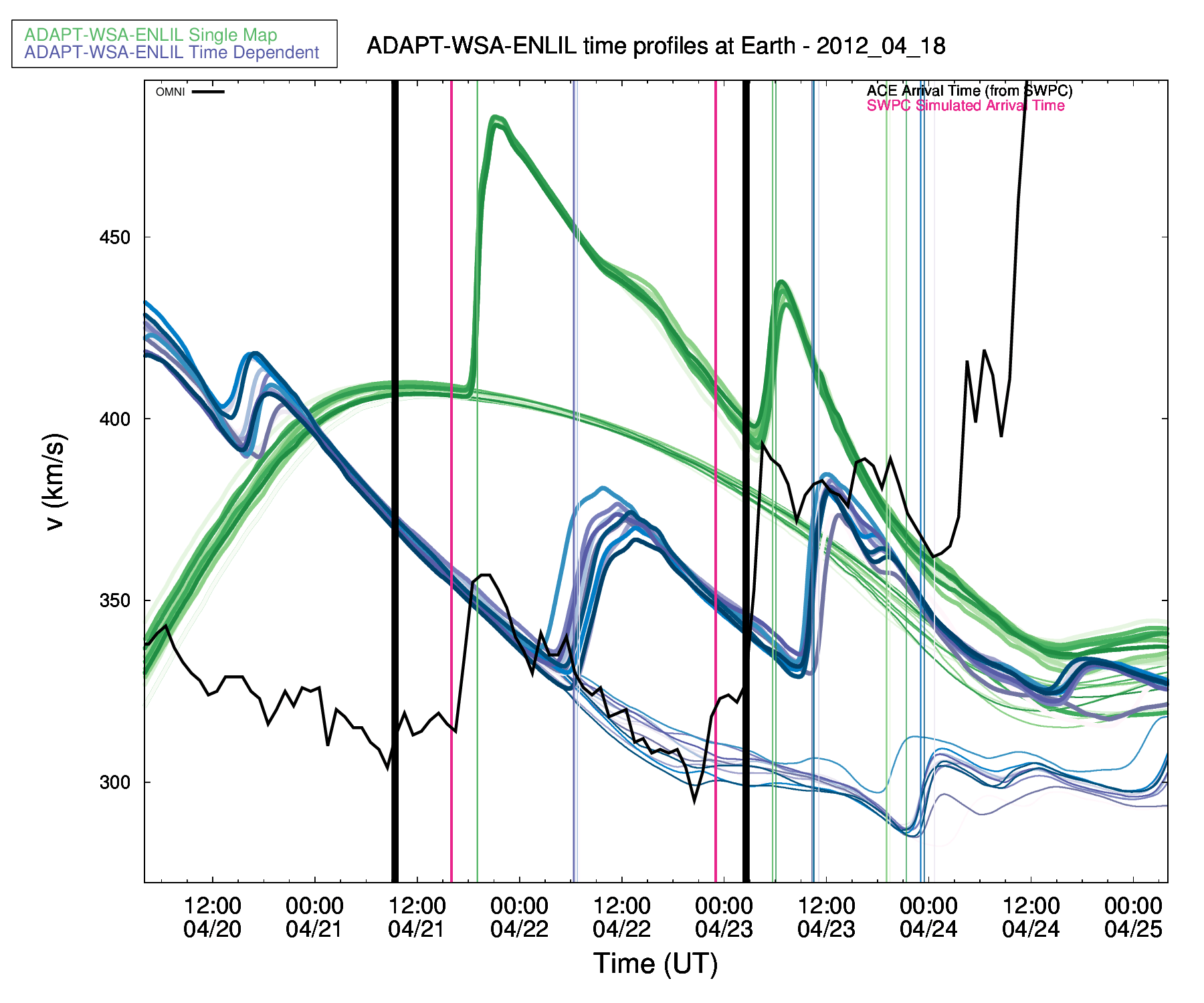}\\
\includegraphics[width=0.33\textwidth]{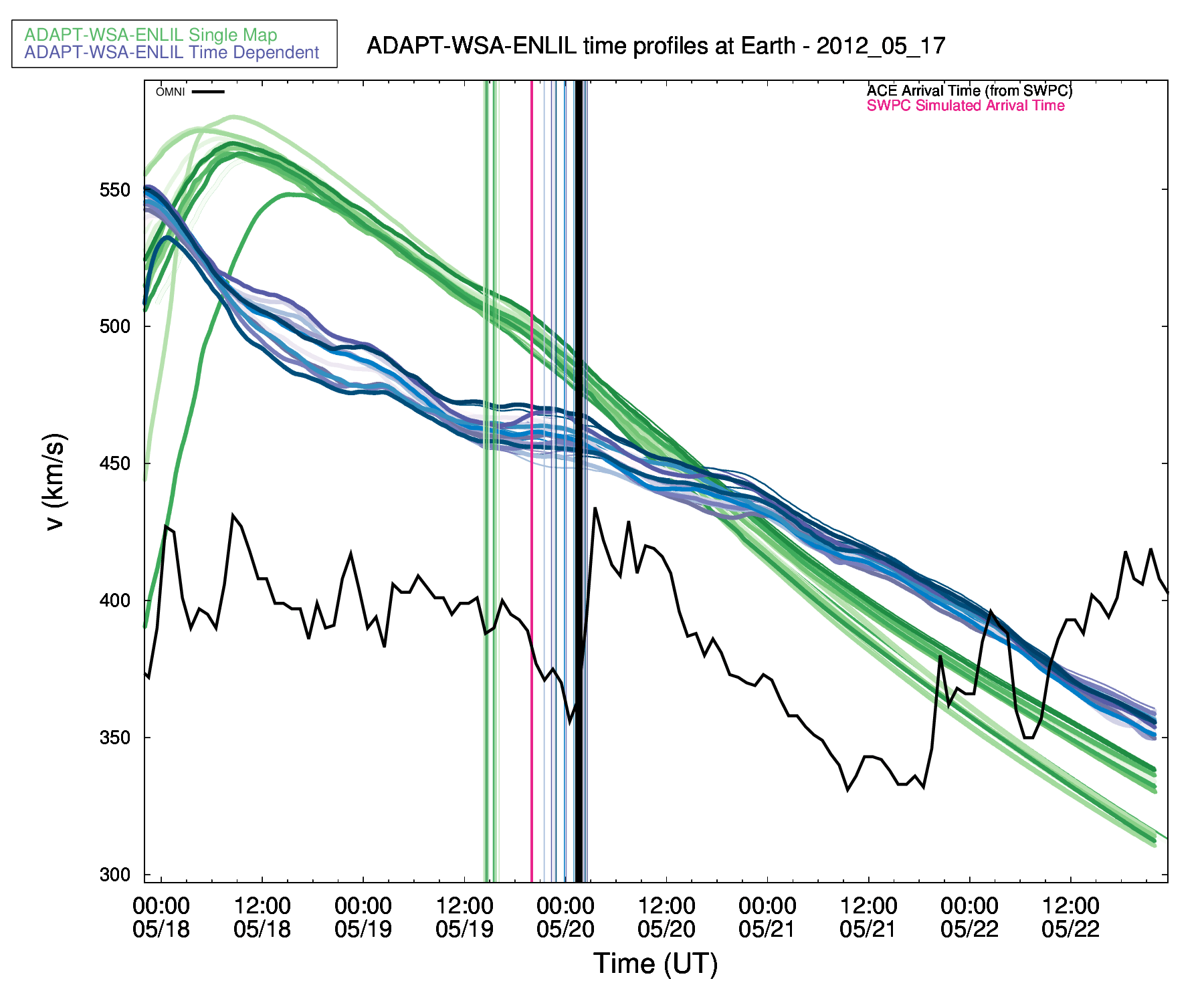}
\includegraphics[width=0.33\textwidth]{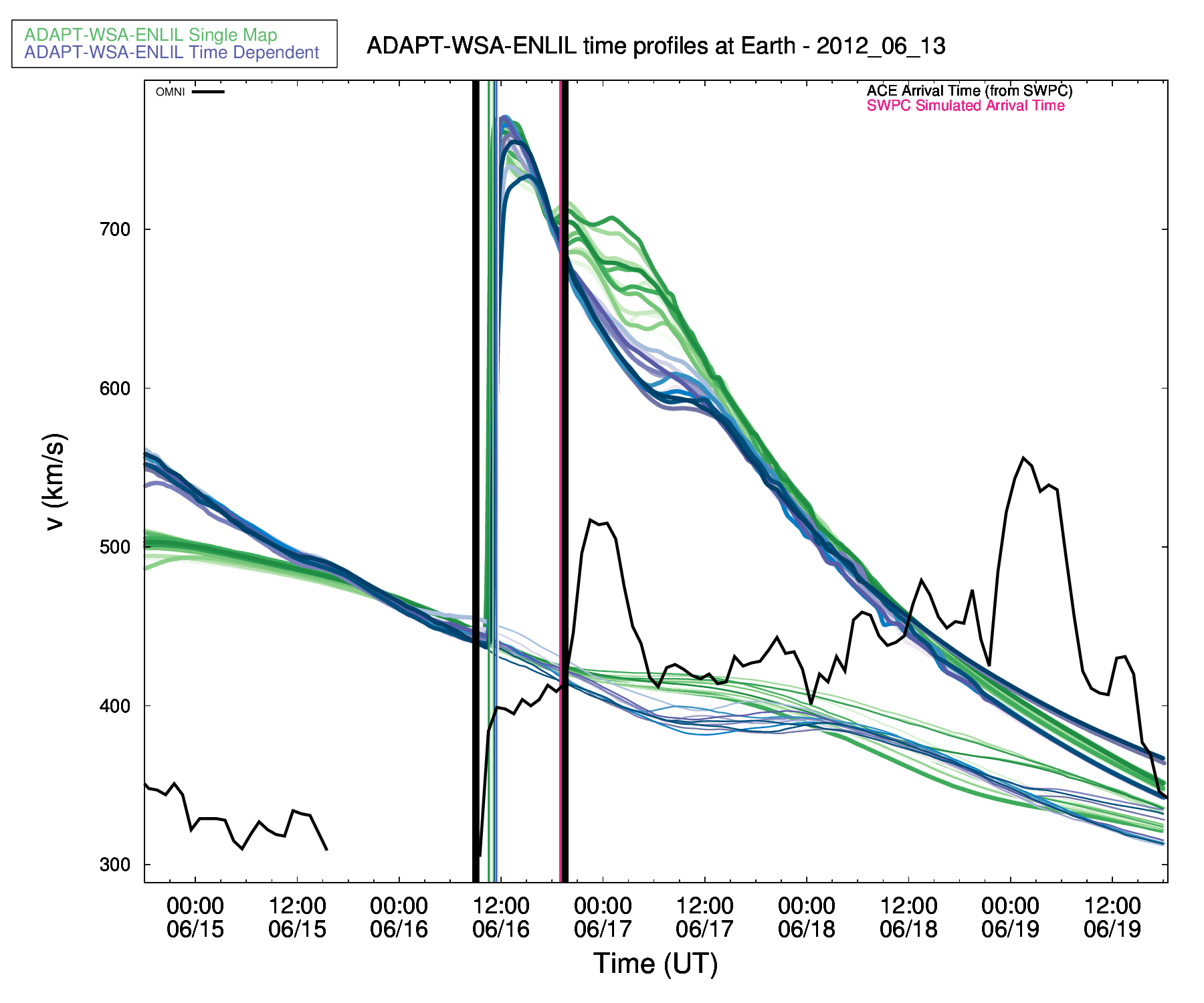}
\includegraphics[width=0.33\textwidth]{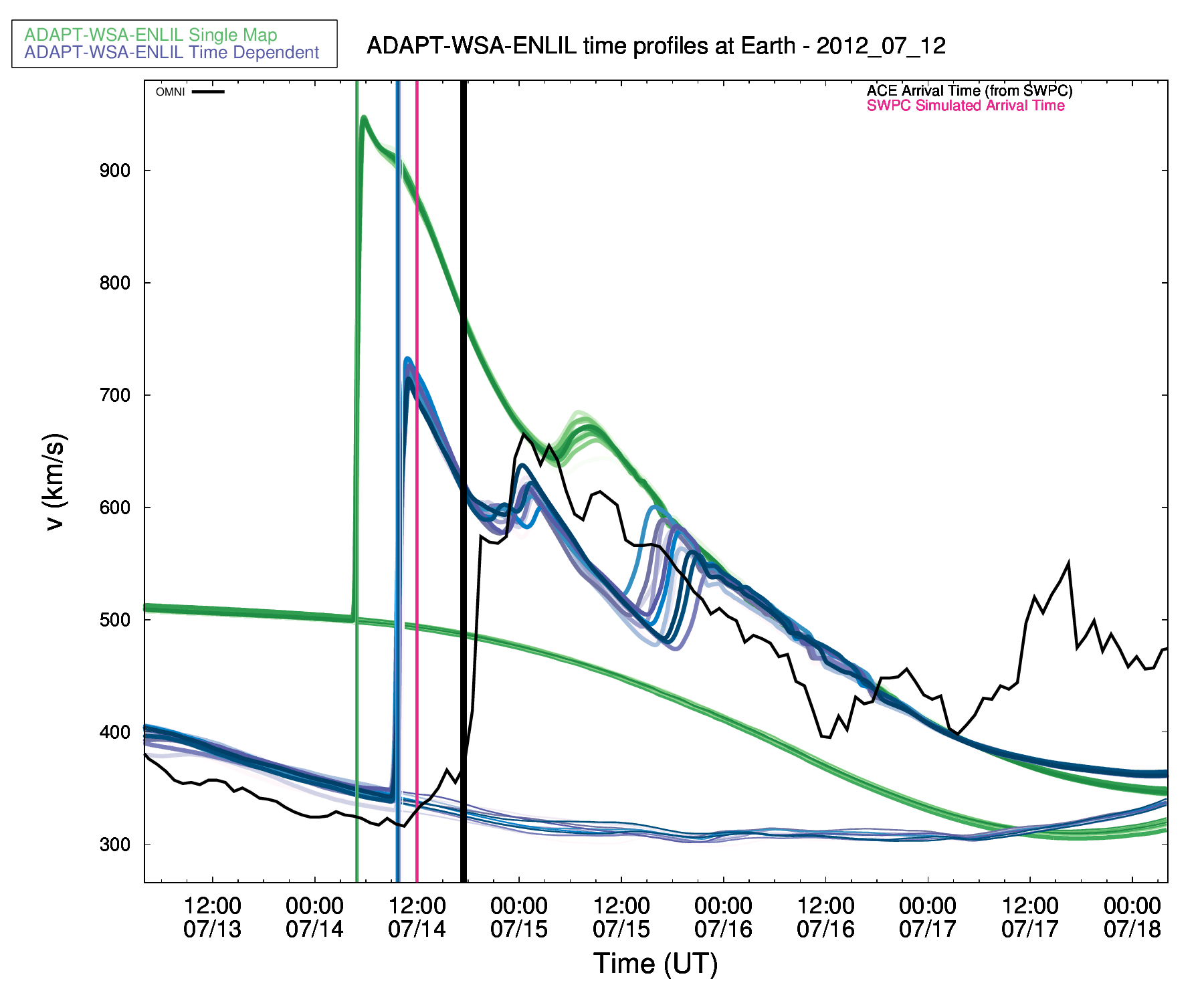}\\
\includegraphics[width=0.33\textwidth]{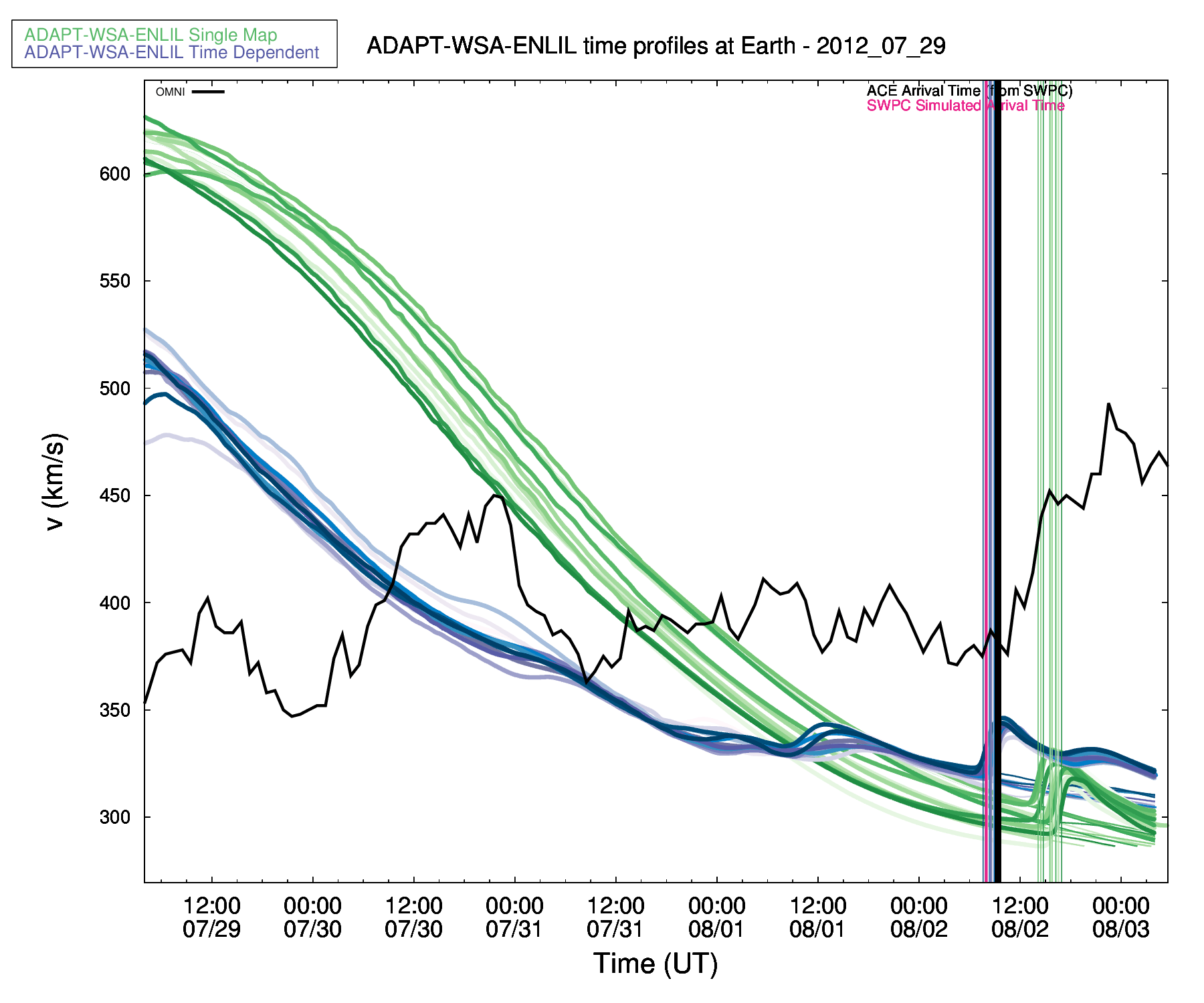}
\includegraphics[width=0.33\textwidth]{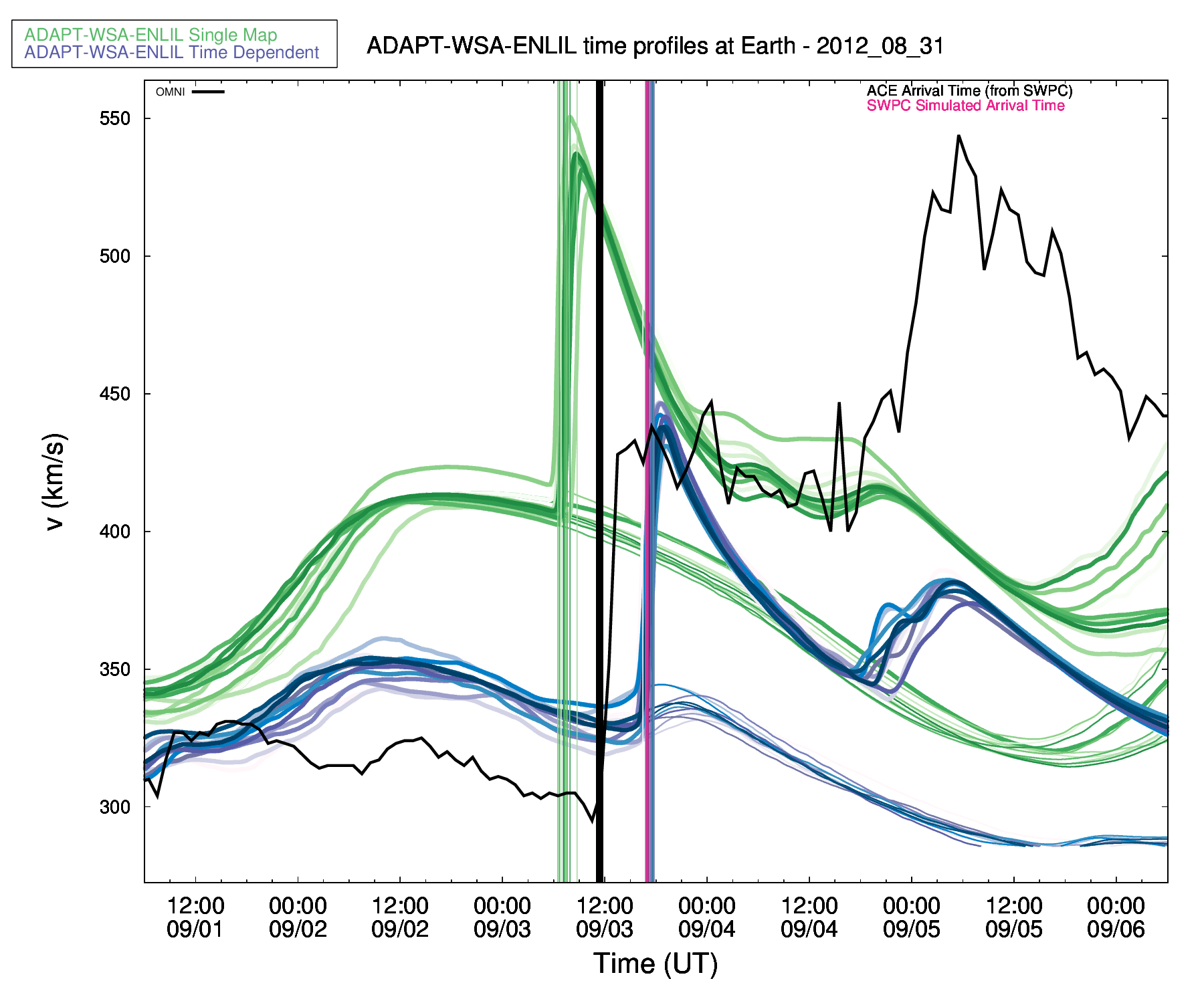}
\includegraphics[width=0.33\textwidth]{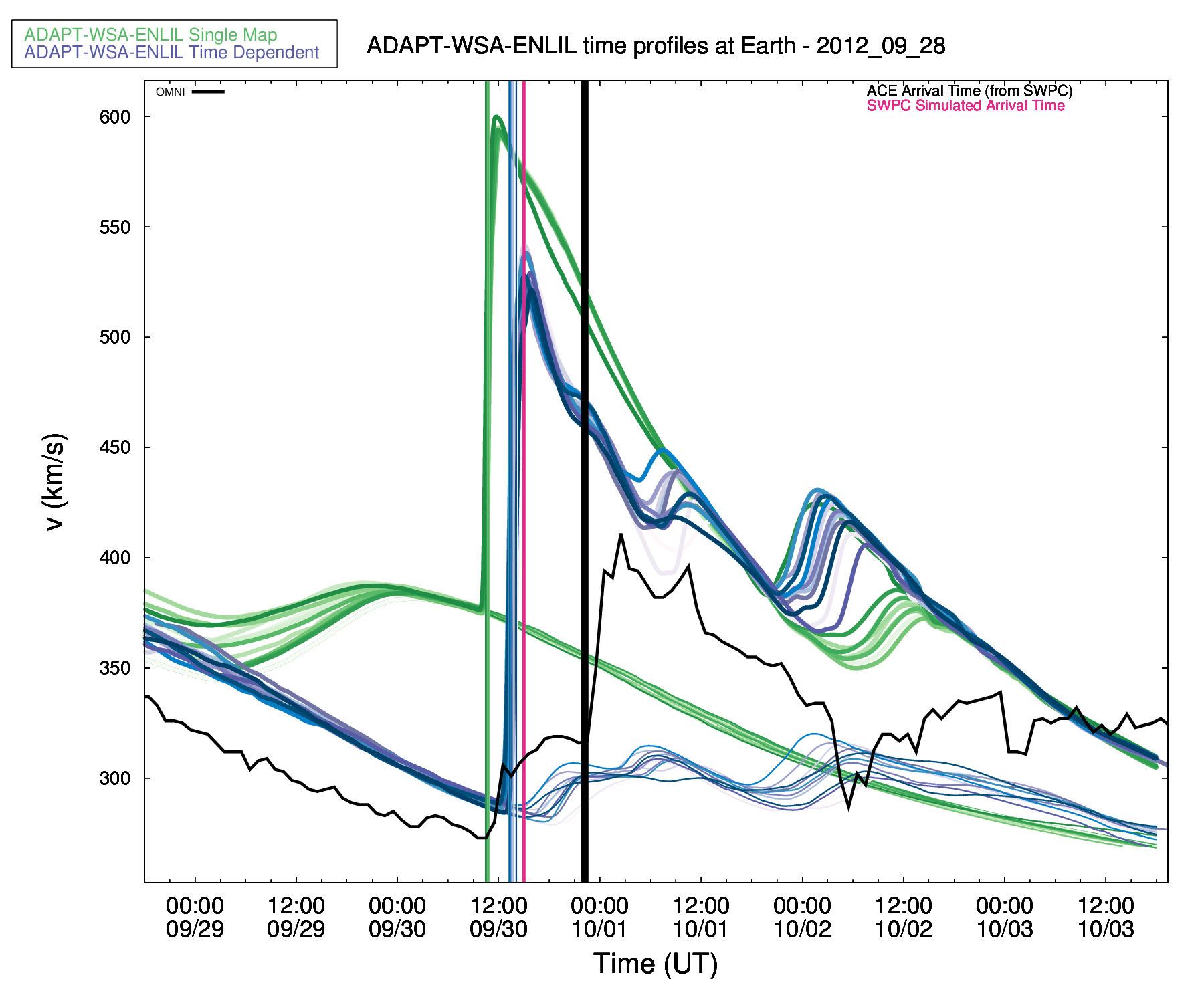}\\
\caption{Simulation results in a similar format to Figure \ref{fig:time-series} for ADAPT-WSA-ENLIL single map (greens) and time dependent map (purples) input with the respective colored vertical lines indicating simulated arrivals.\label{fig:adapt-time-series}}
\end{figure*}

\begin{figure*}
\includegraphics[width=0.33\textwidth]{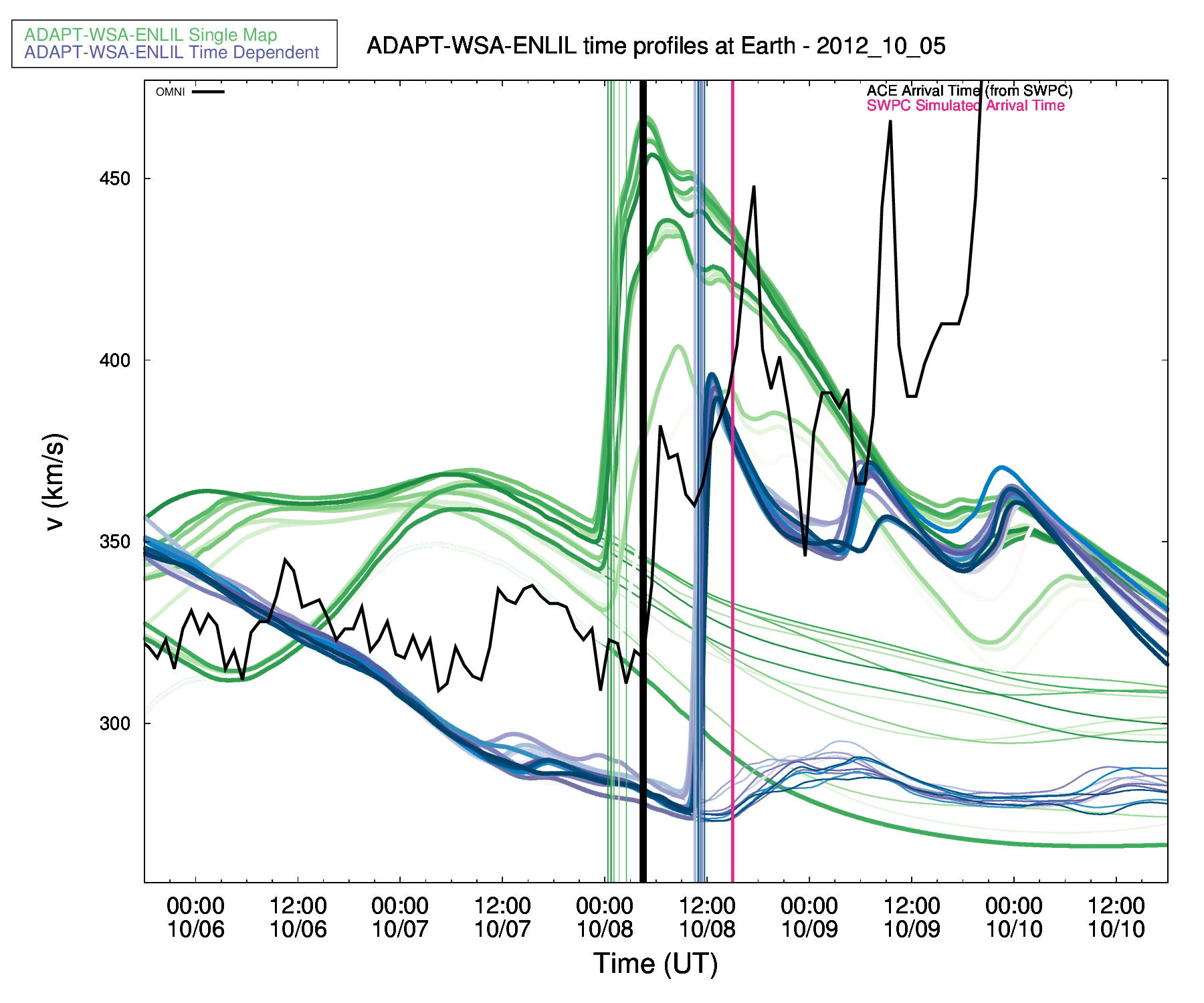}
\includegraphics[width=0.33\textwidth]{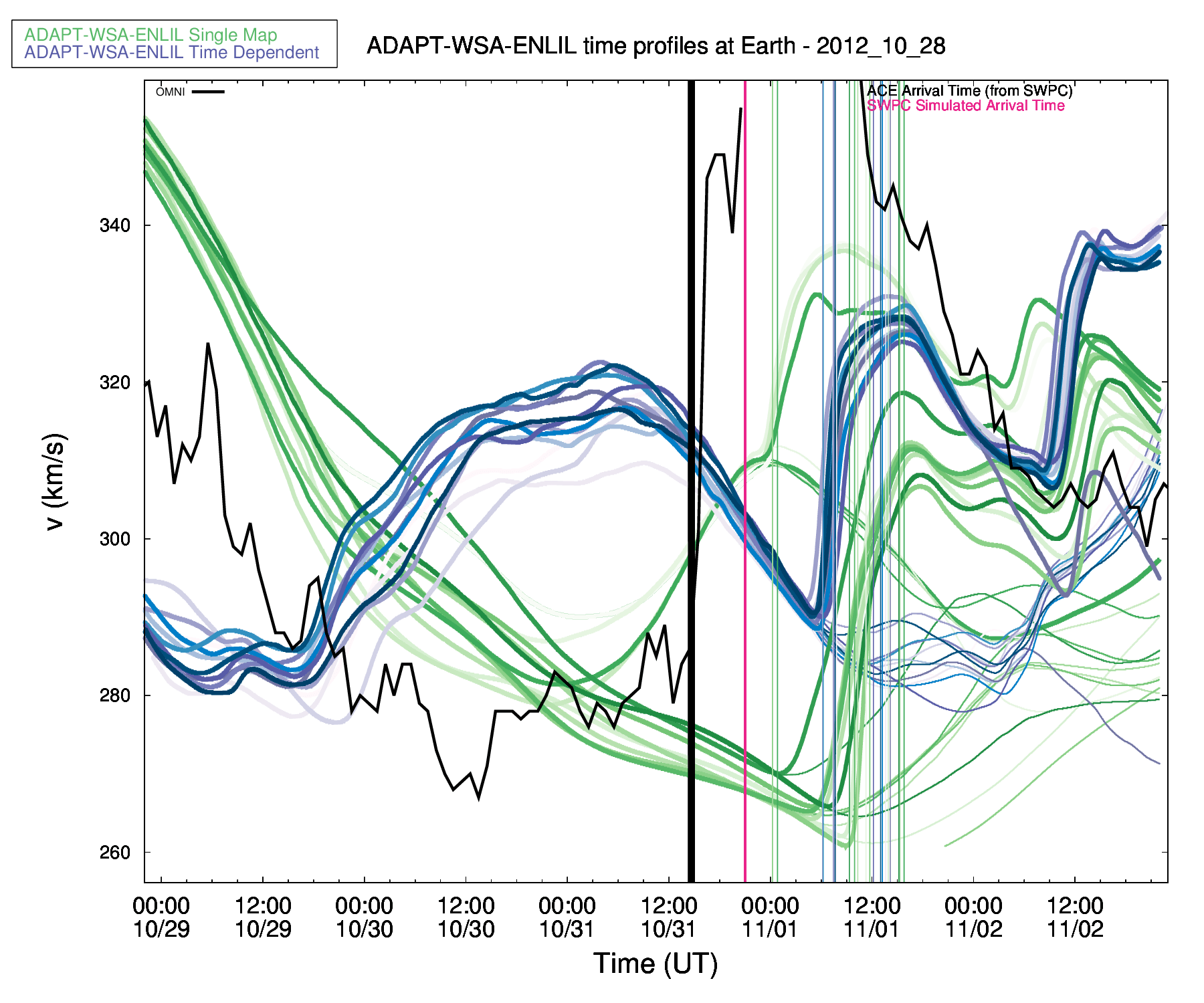}
\includegraphics[width=0.33\textwidth]{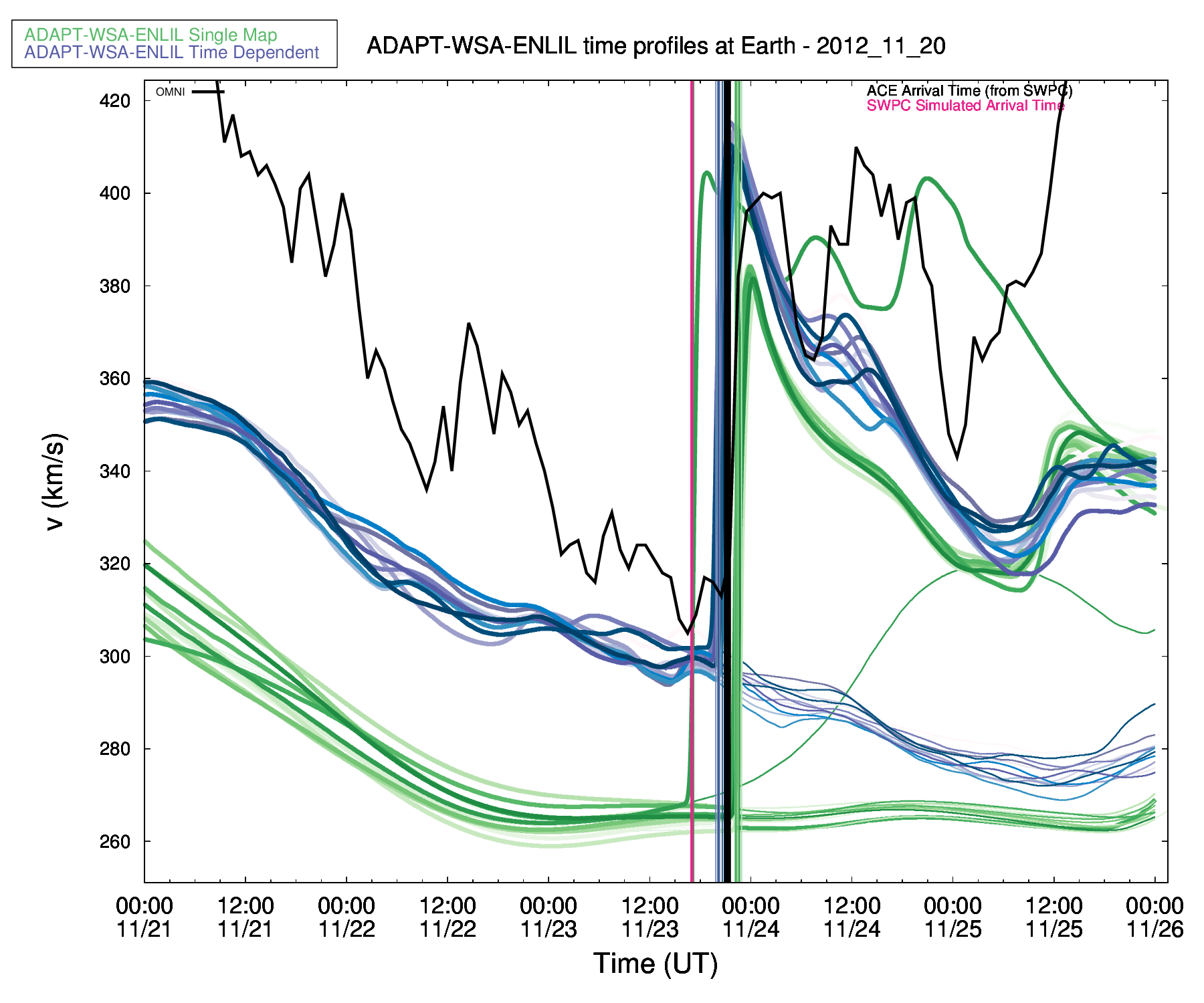}\\
\includegraphics[width=0.33\textwidth]{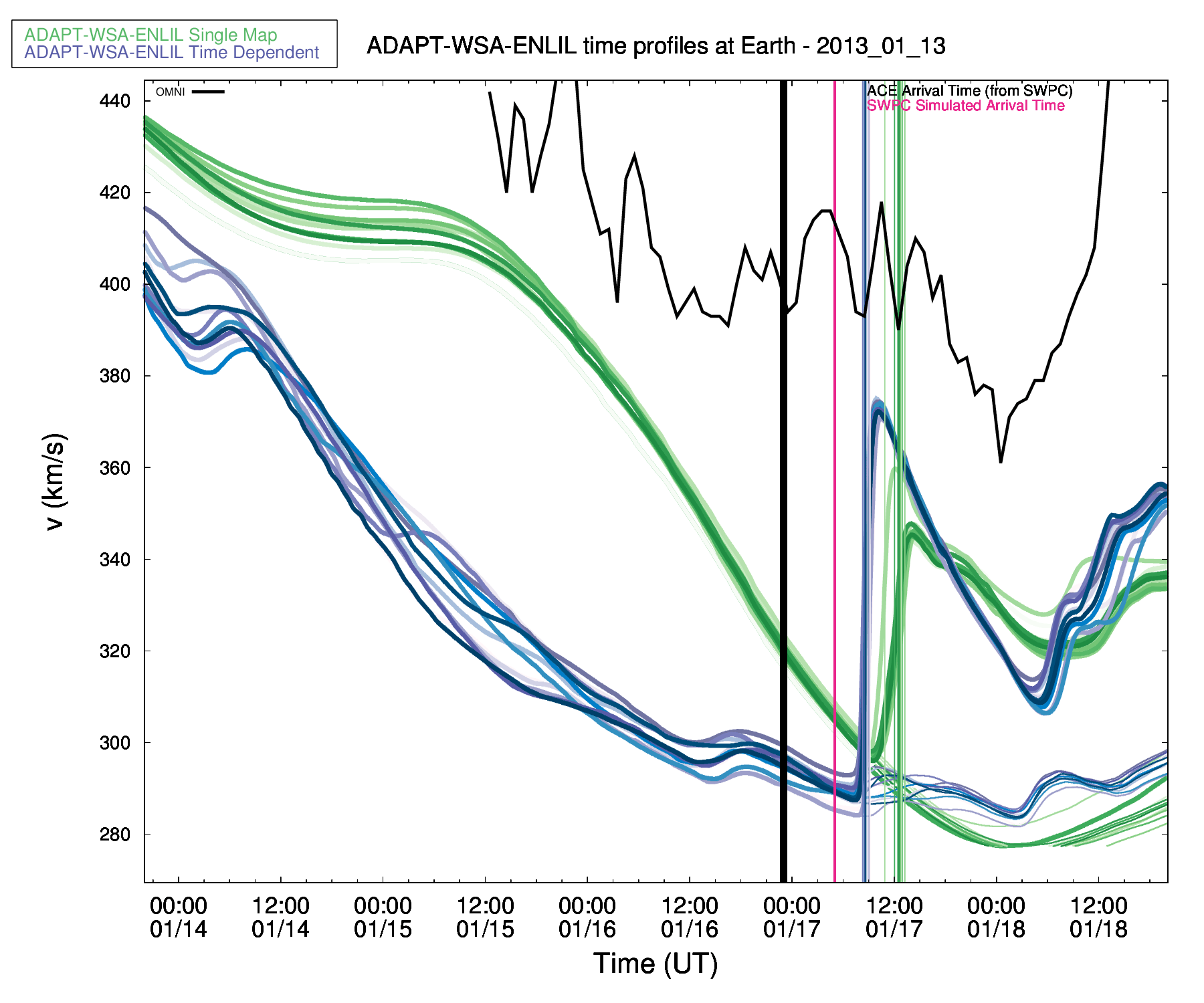}
\includegraphics[width=0.33\textwidth]{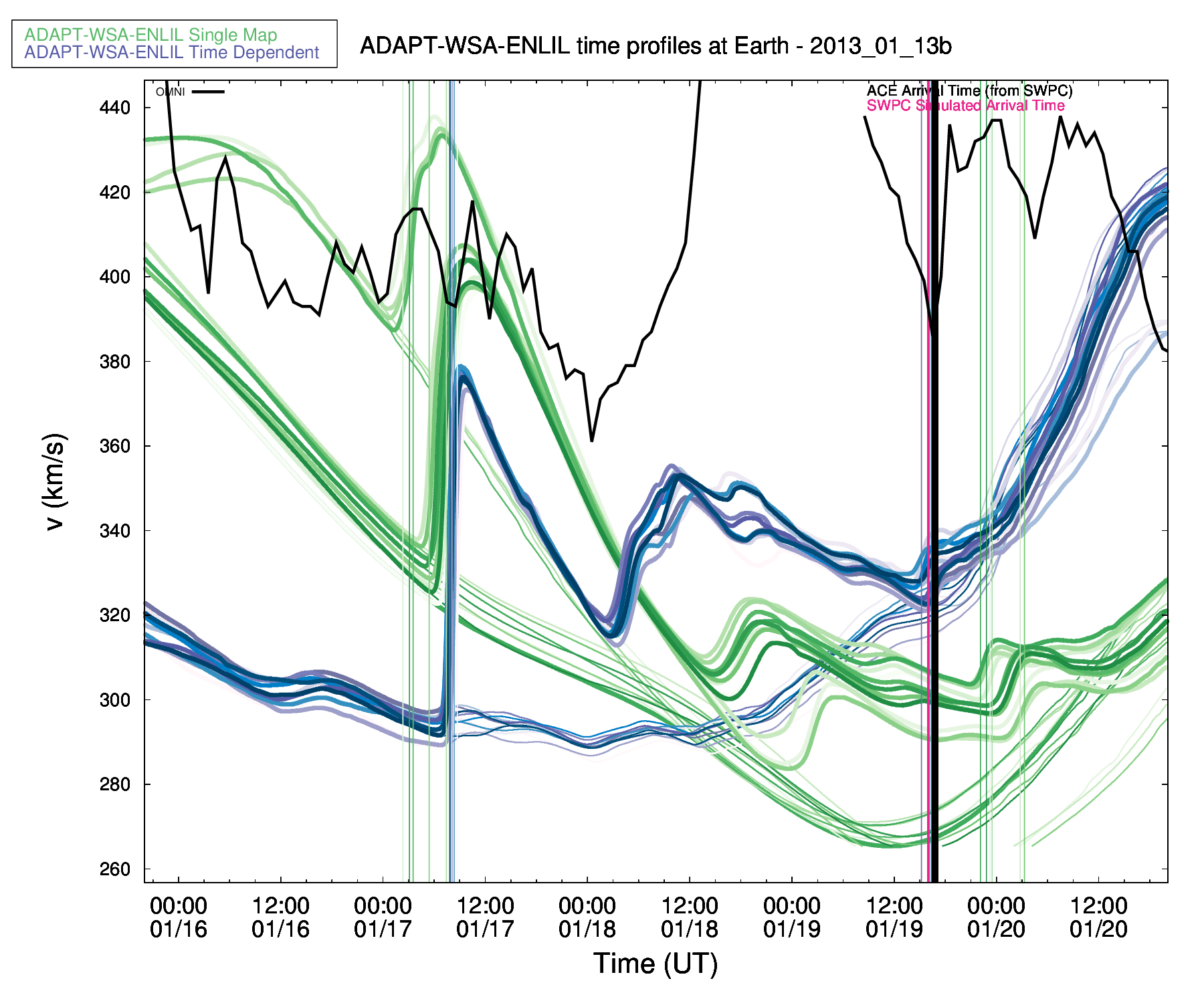}
\includegraphics[width=0.33\textwidth]{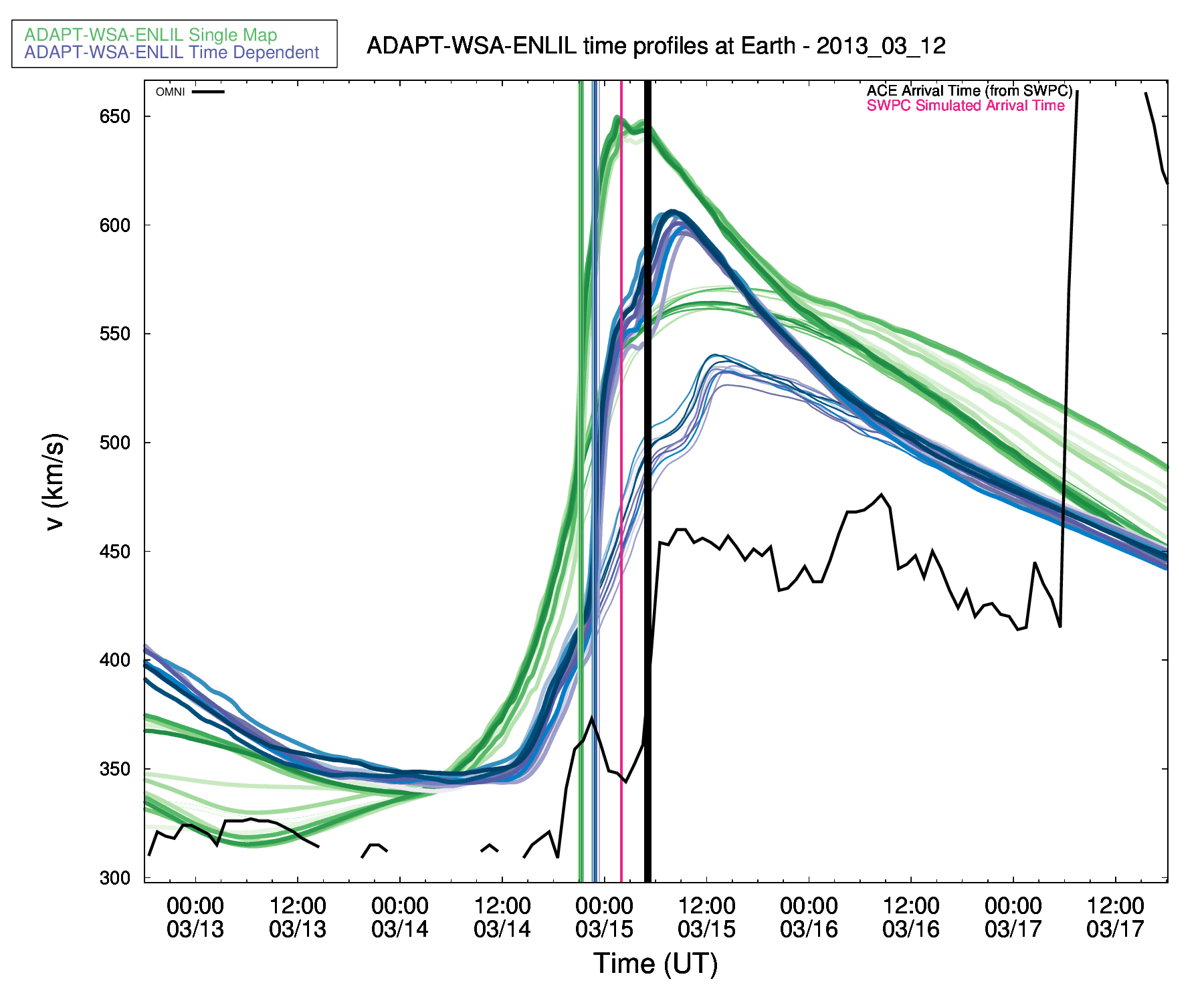}\\
\includegraphics[width=0.33\textwidth]{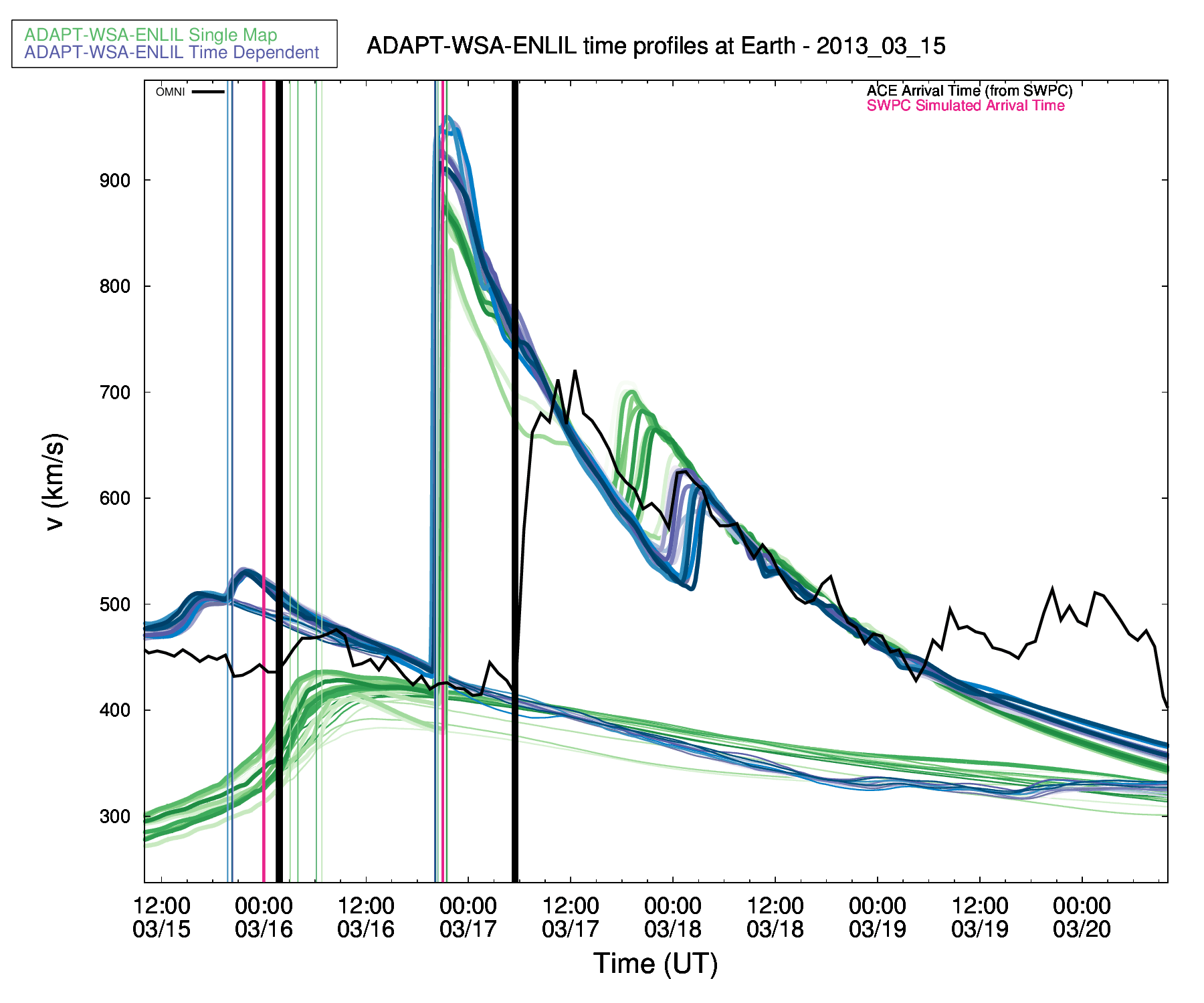}
\includegraphics[width=0.33\textwidth]{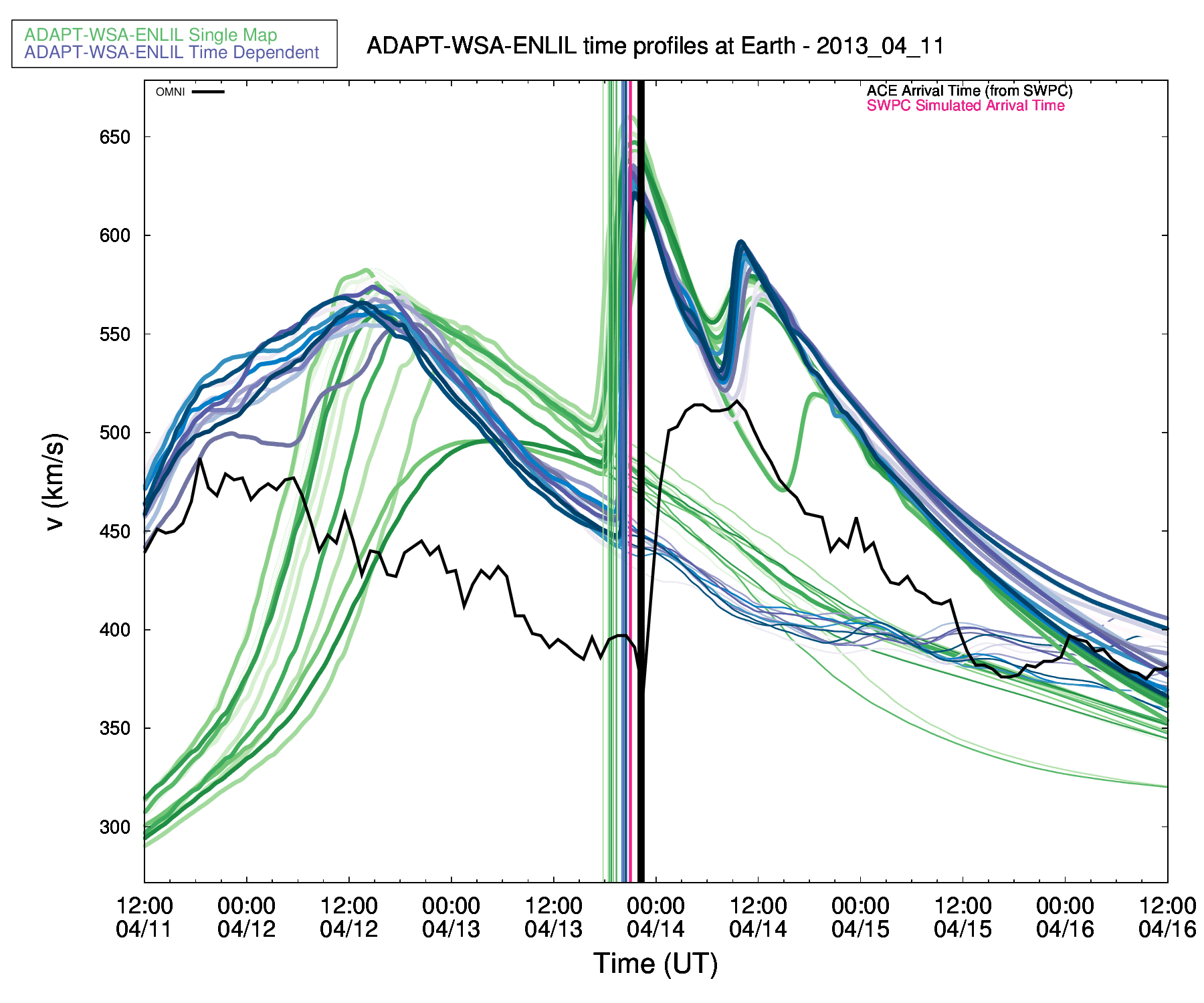}
\includegraphics[width=0.33\textwidth]{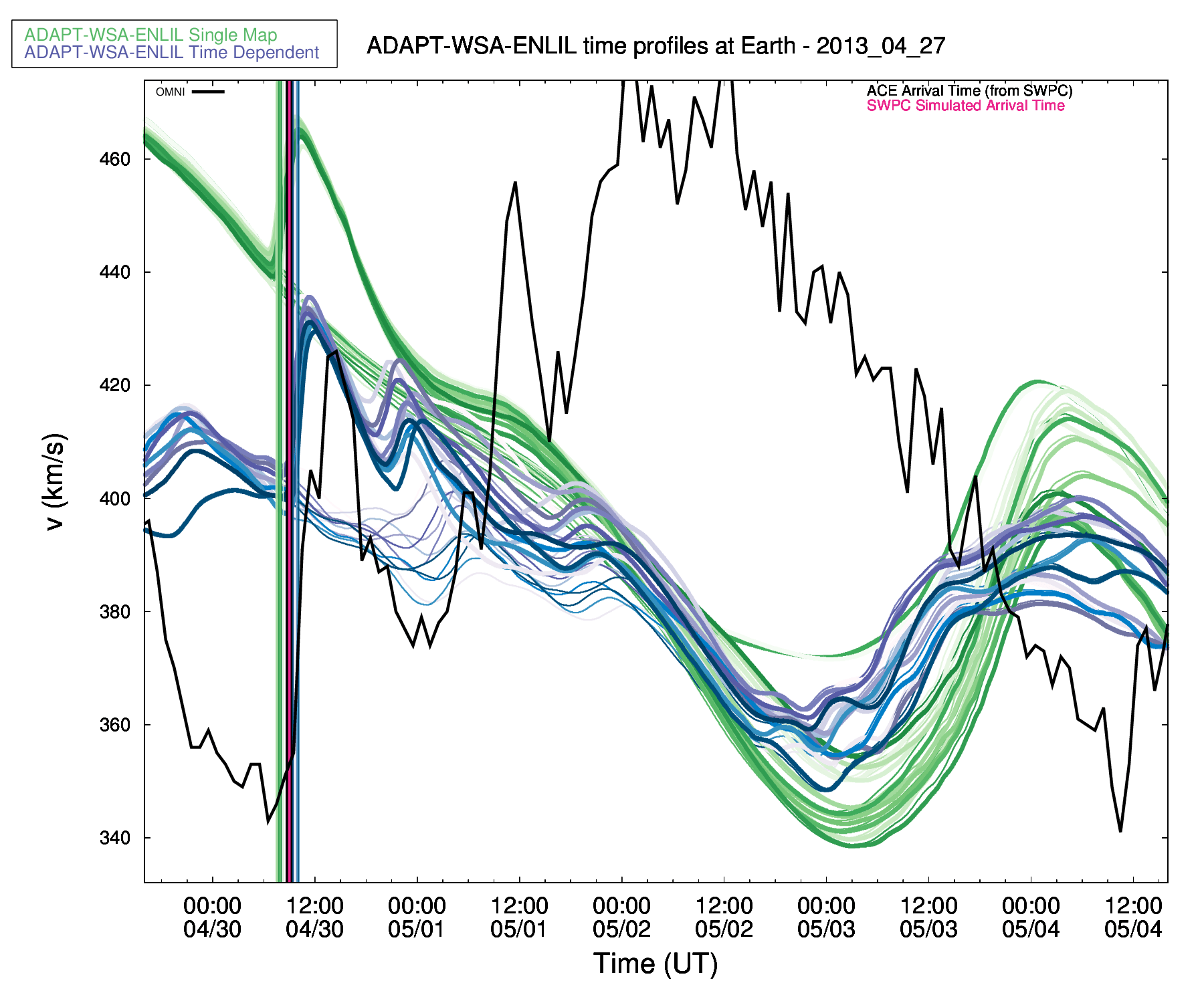}\\
\includegraphics[width=0.33\textwidth]{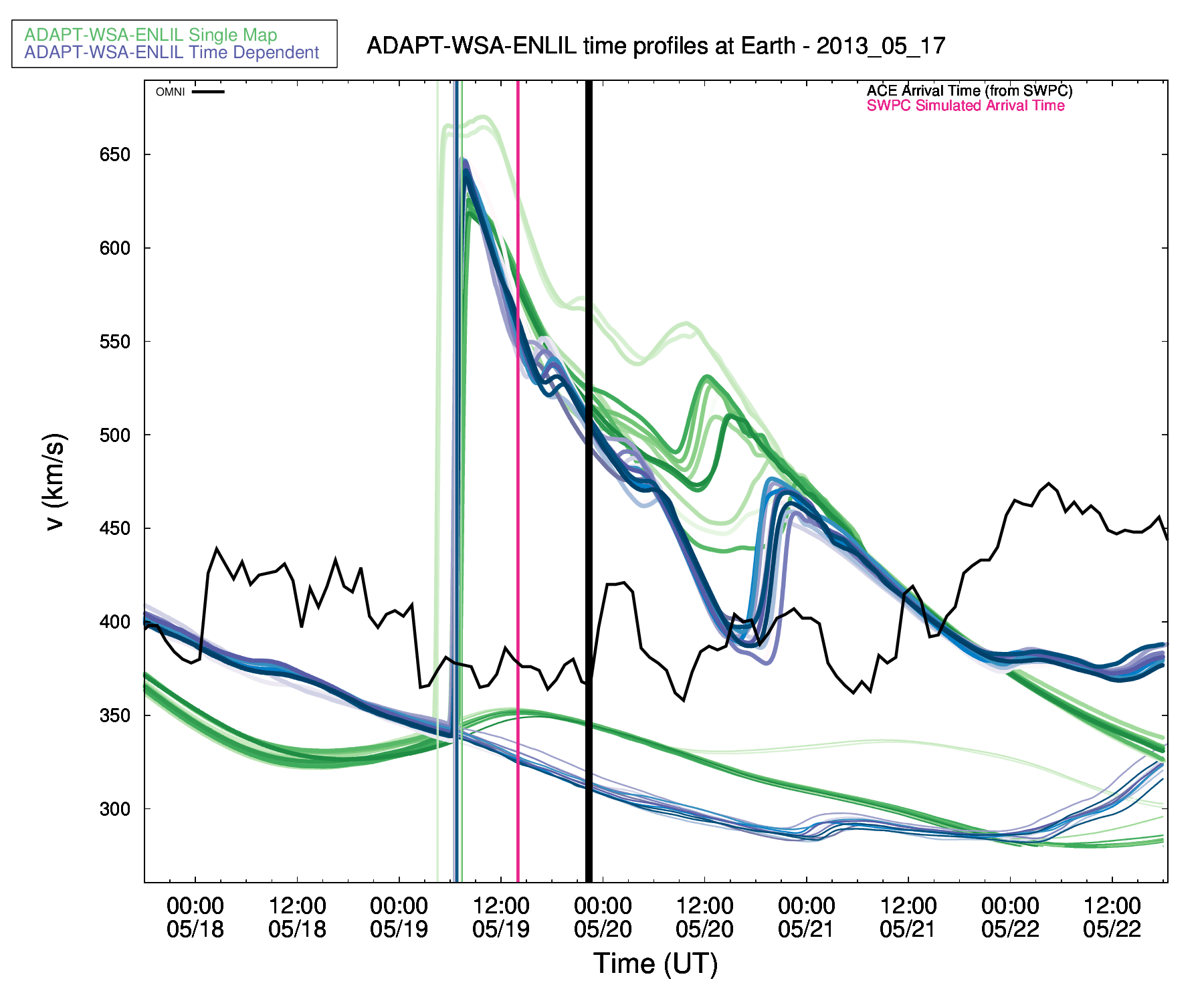}
\includegraphics[width=0.33\textwidth]{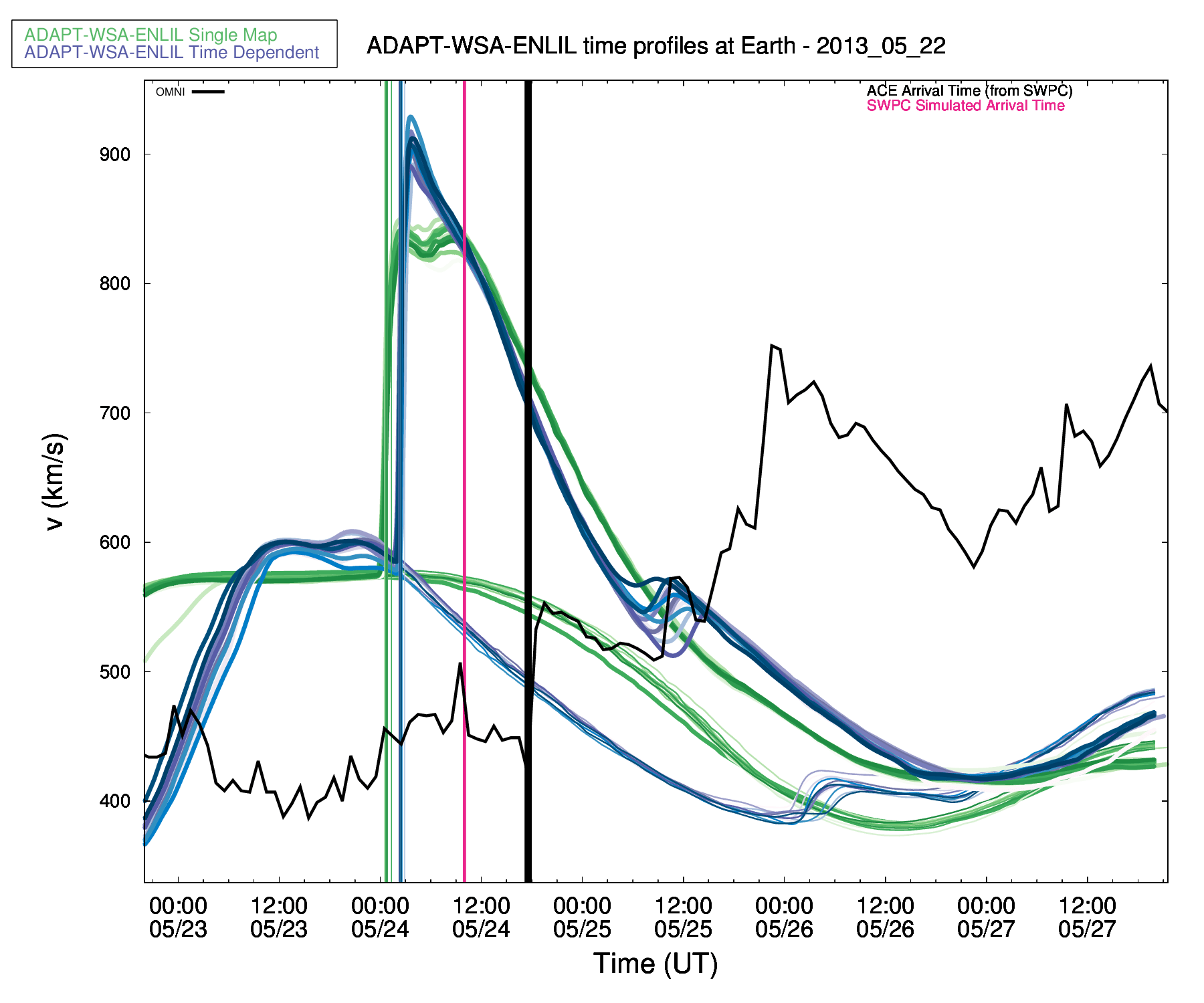}
\includegraphics[width=0.33\textwidth]{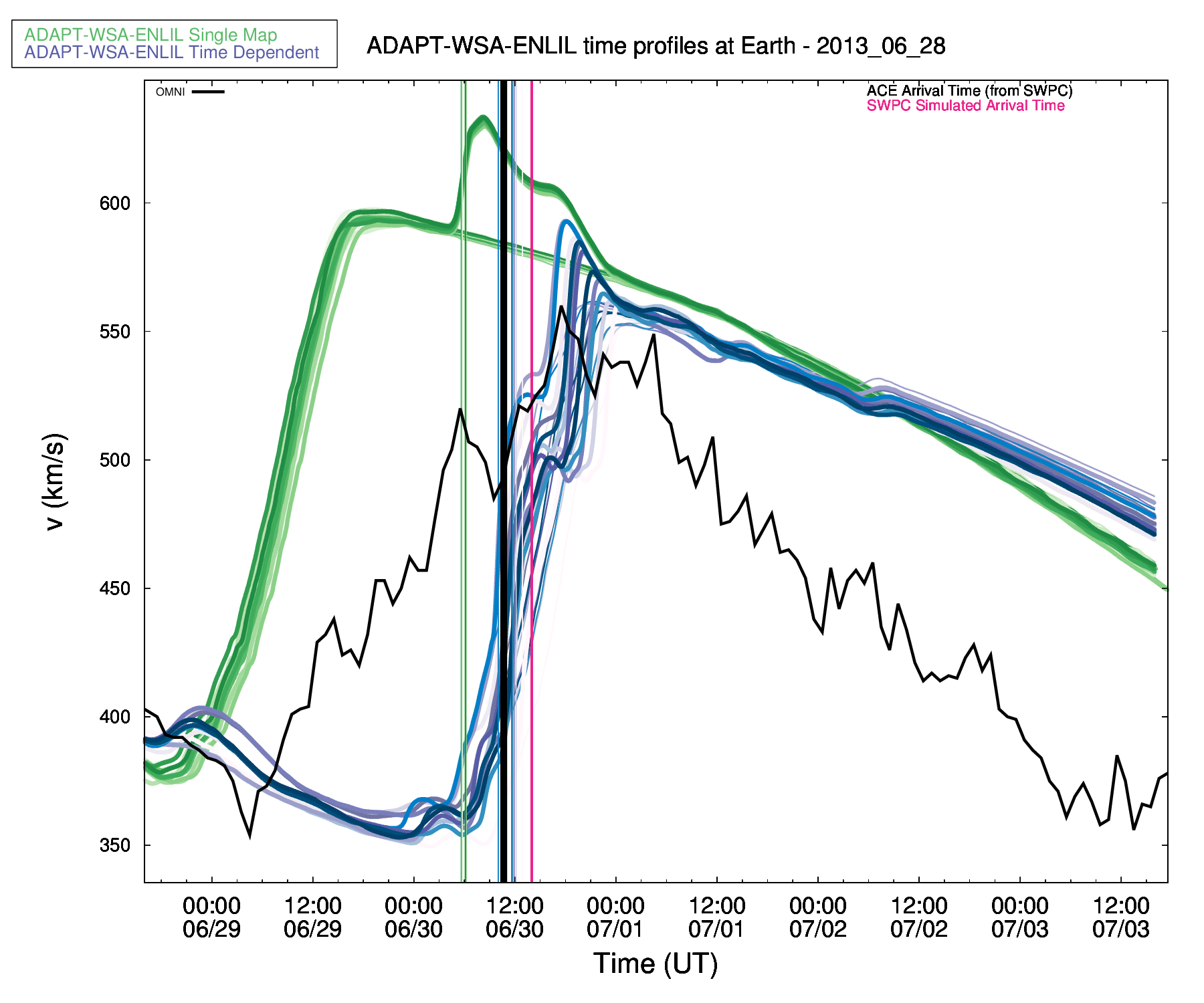}\\
\caption{Figure \ref{fig:adapt-time-series} continued.\label{fig:adapt-time-series2}}
\end{figure*}

\begin{figure*}
\includegraphics[width=0.33\textwidth]{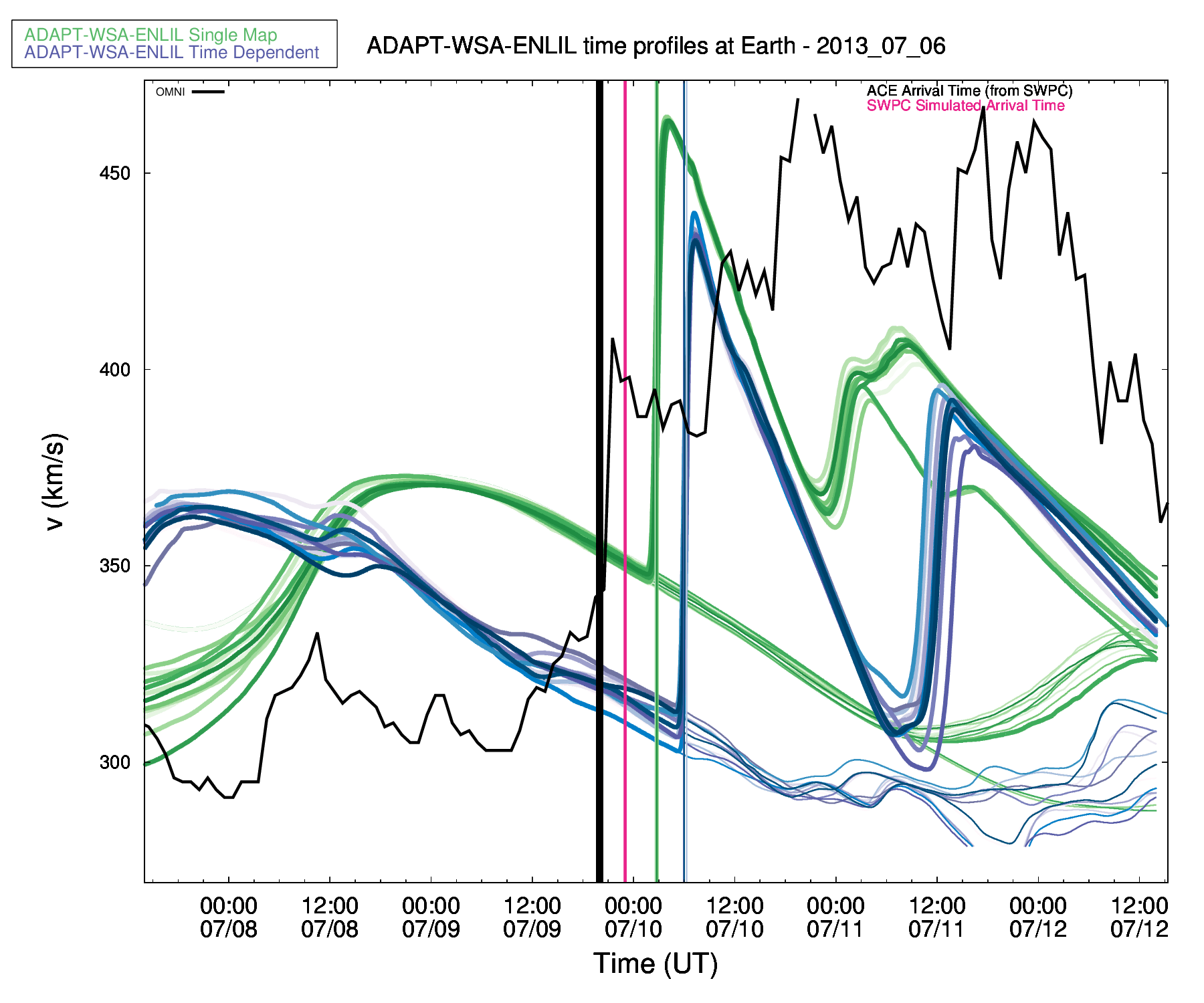}
\includegraphics[width=0.33\textwidth]{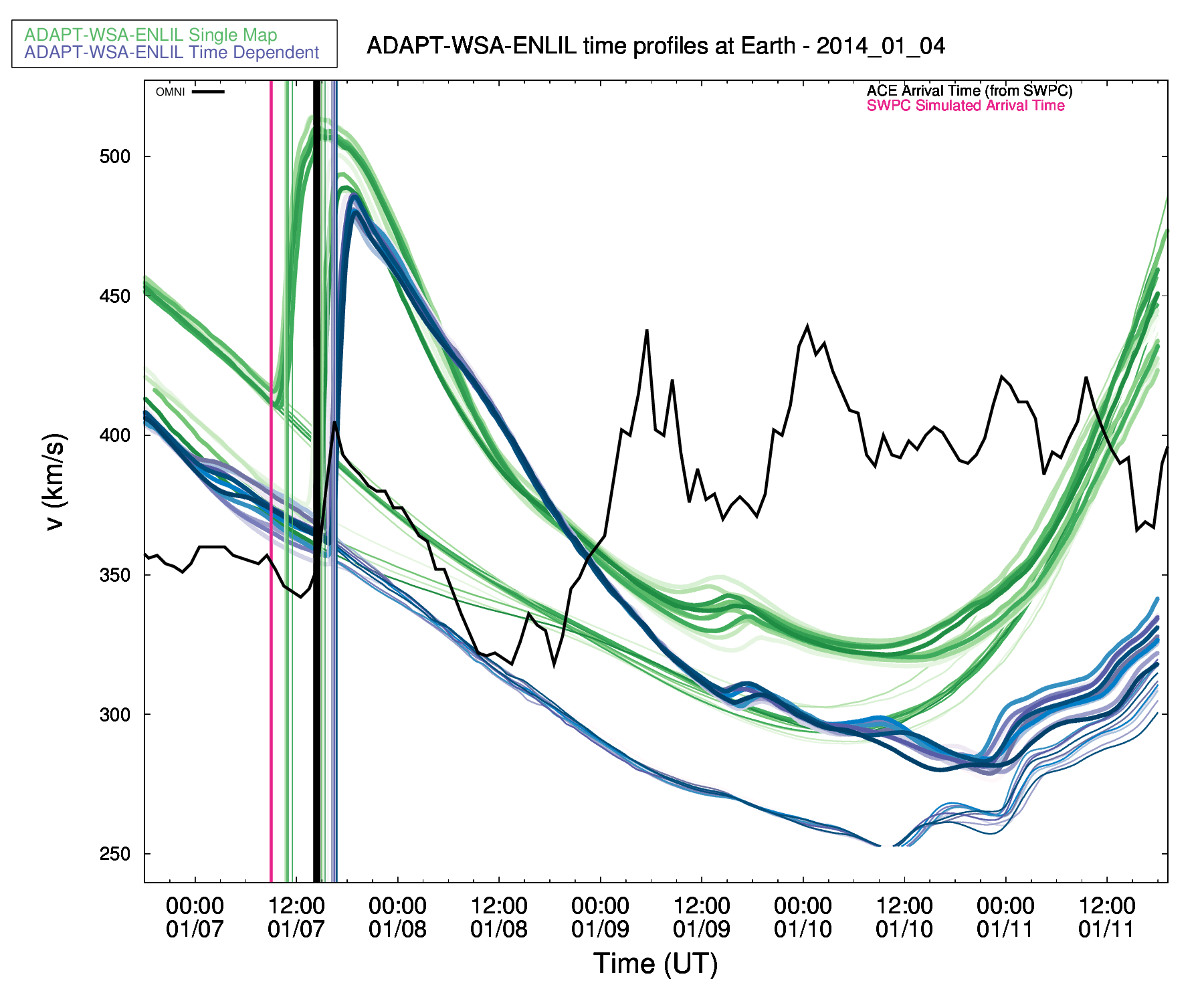}
\includegraphics[width=0.33\textwidth]{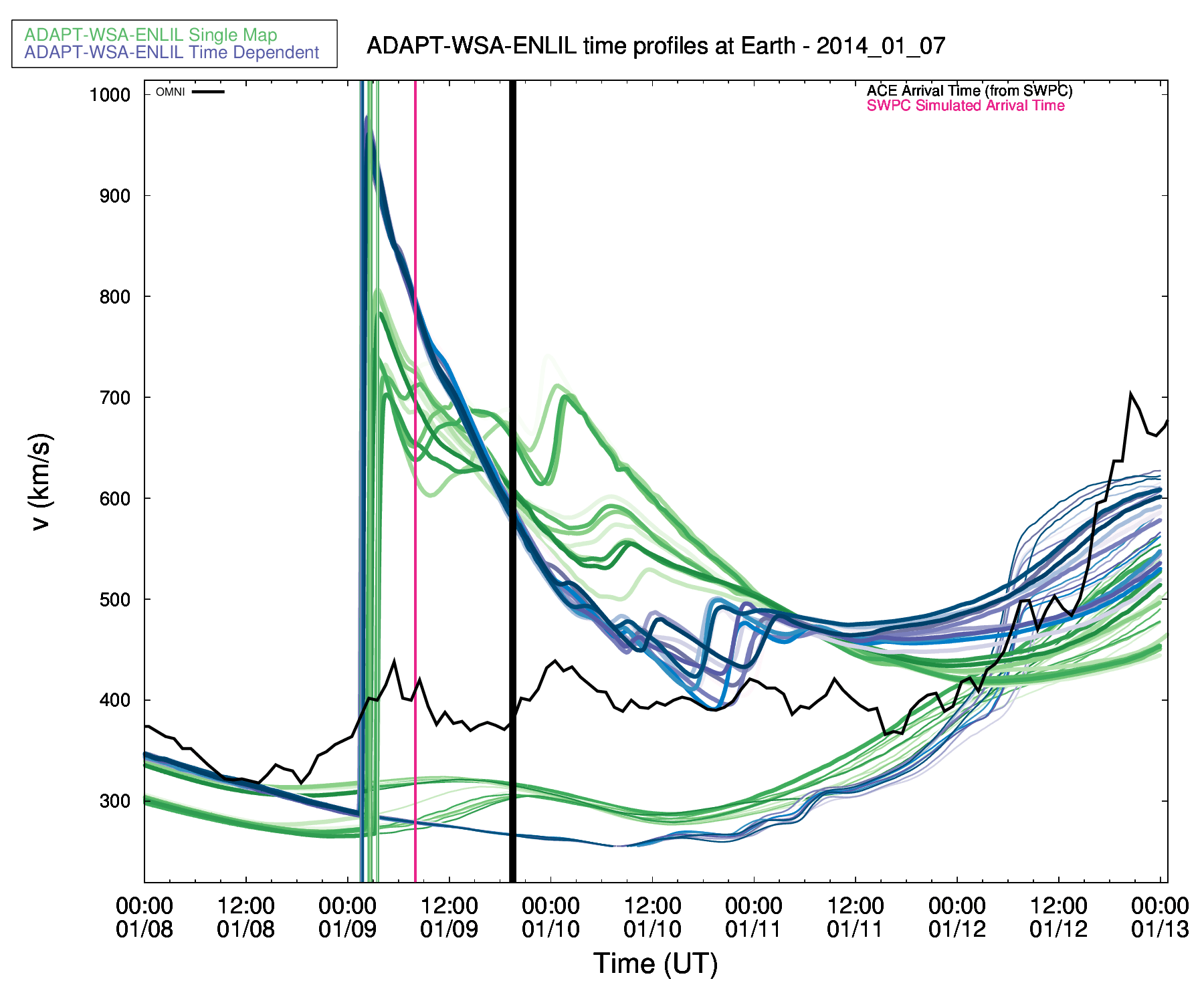}\\
\includegraphics[width=0.33\textwidth]{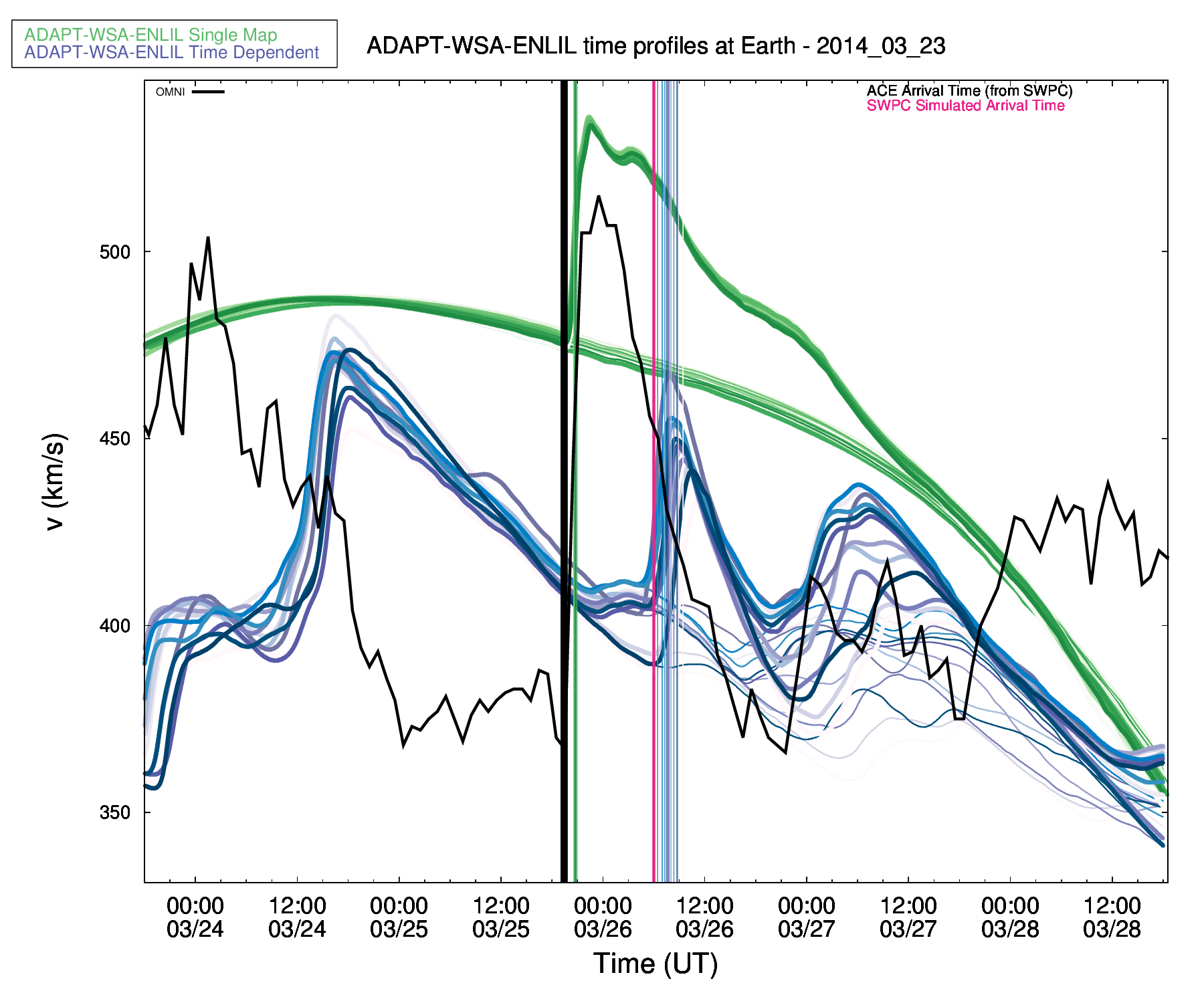}
\includegraphics[width=0.33\textwidth]{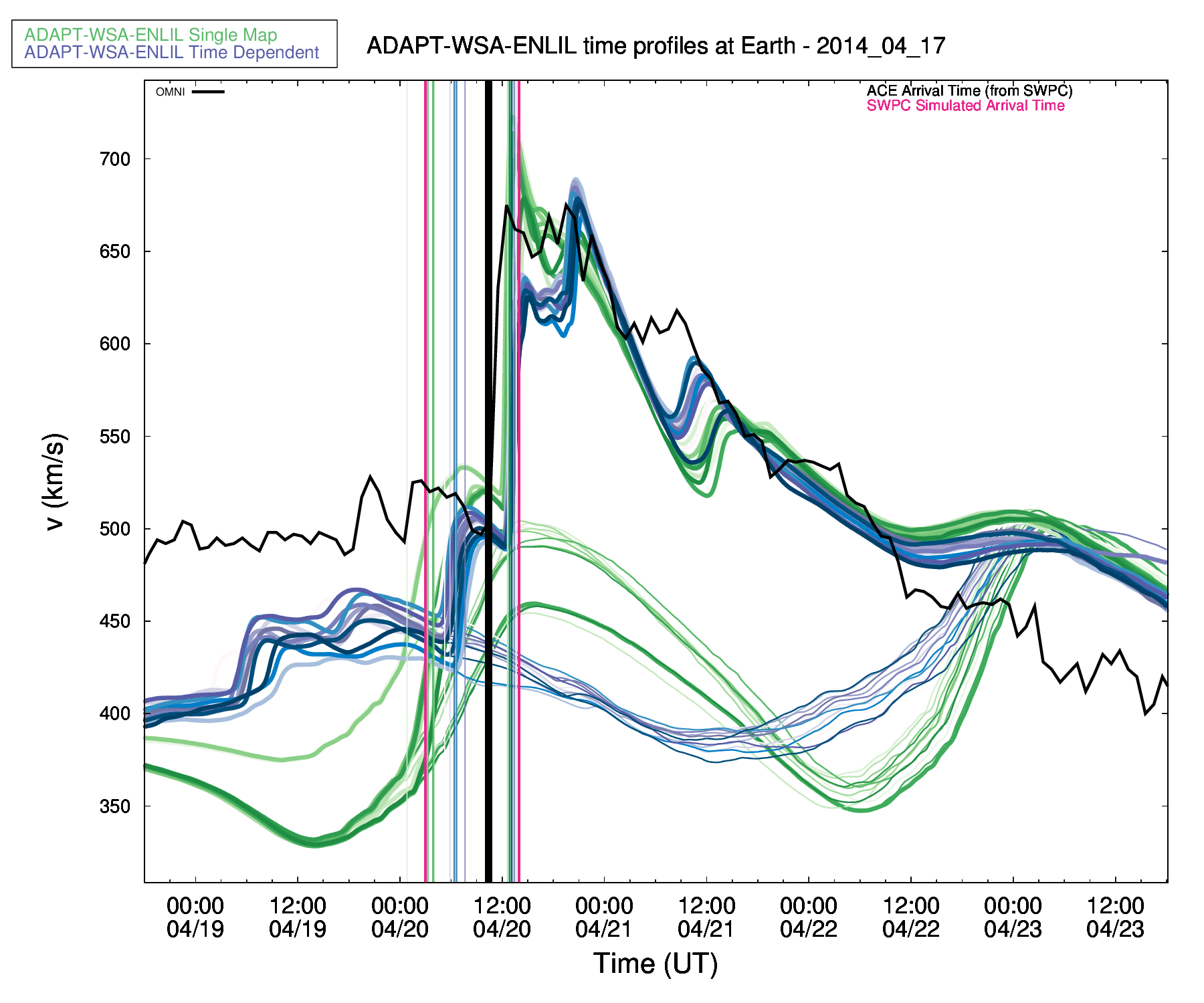}
\includegraphics[width=0.33\textwidth]{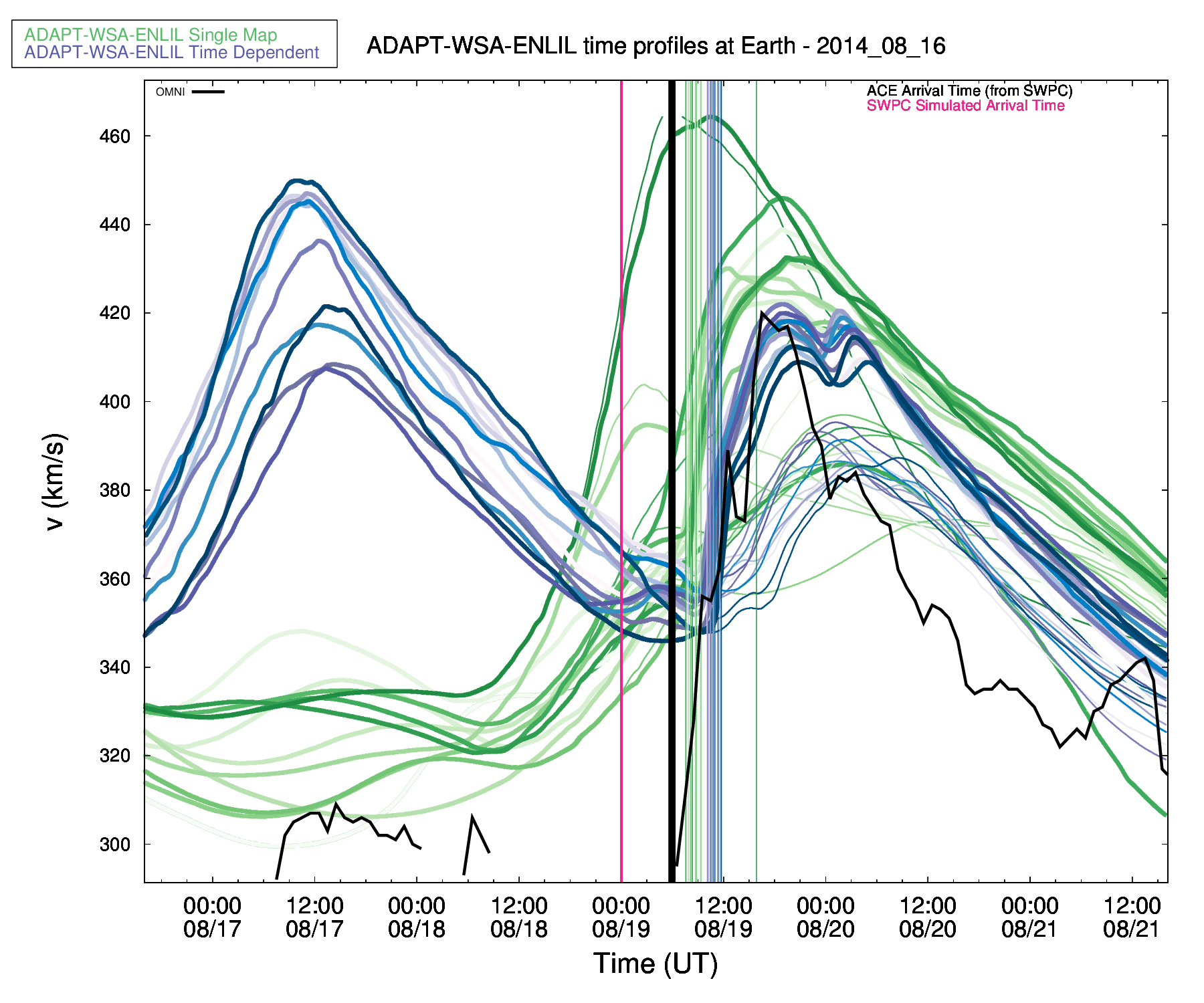}\\
\includegraphics[width=0.33\textwidth]{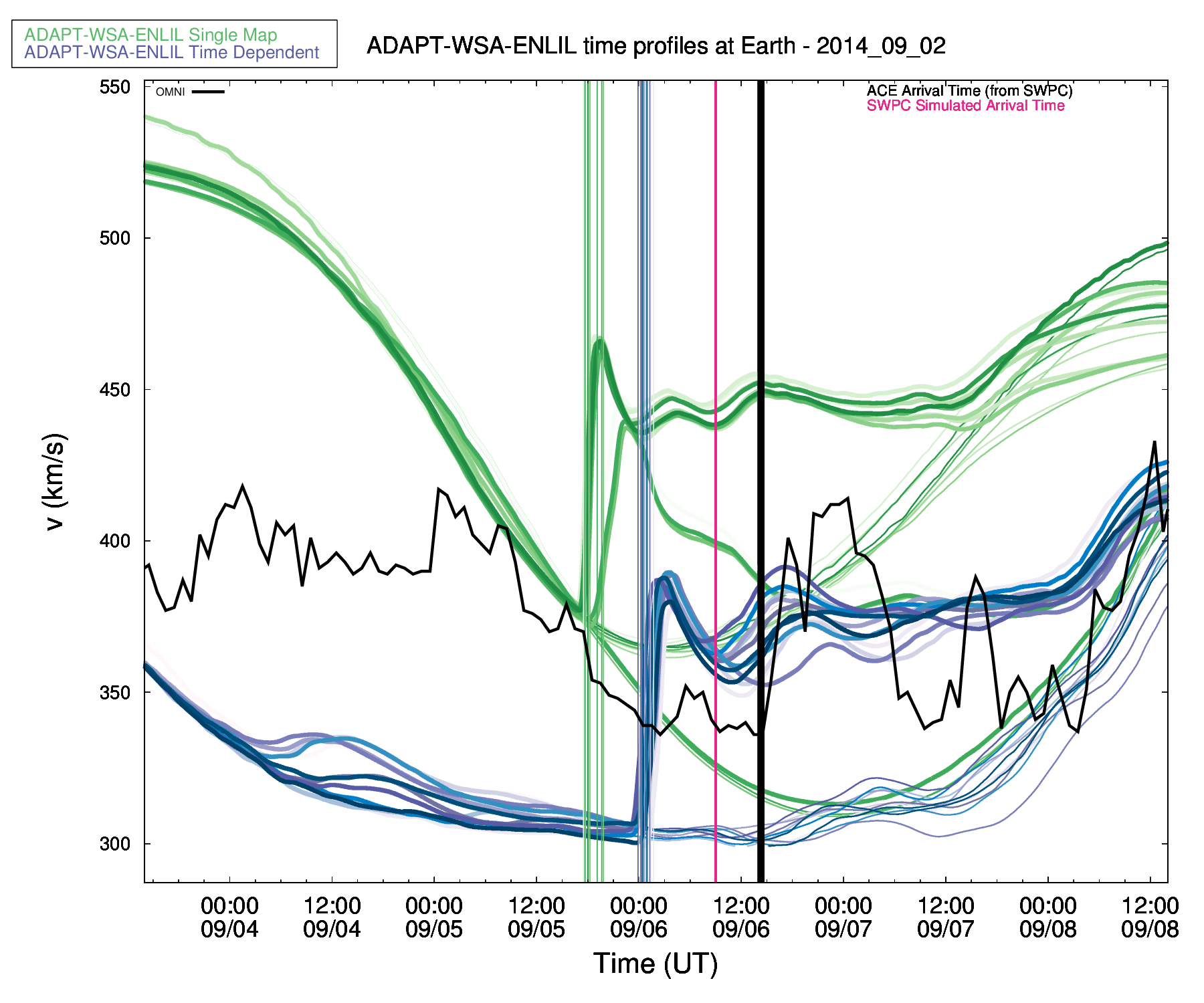}
\includegraphics[width=0.33\textwidth]{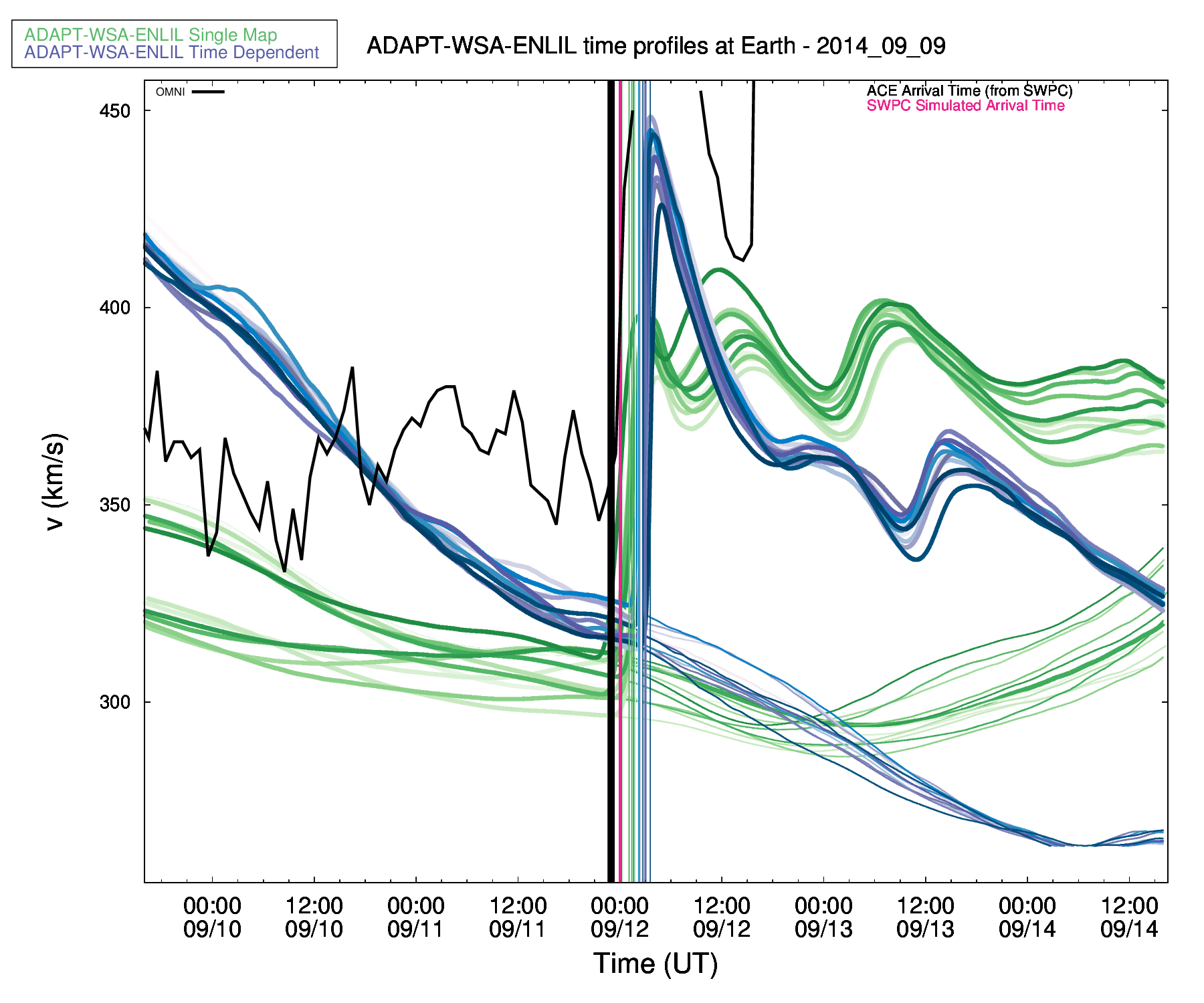}
\includegraphics[width=0.33\textwidth]{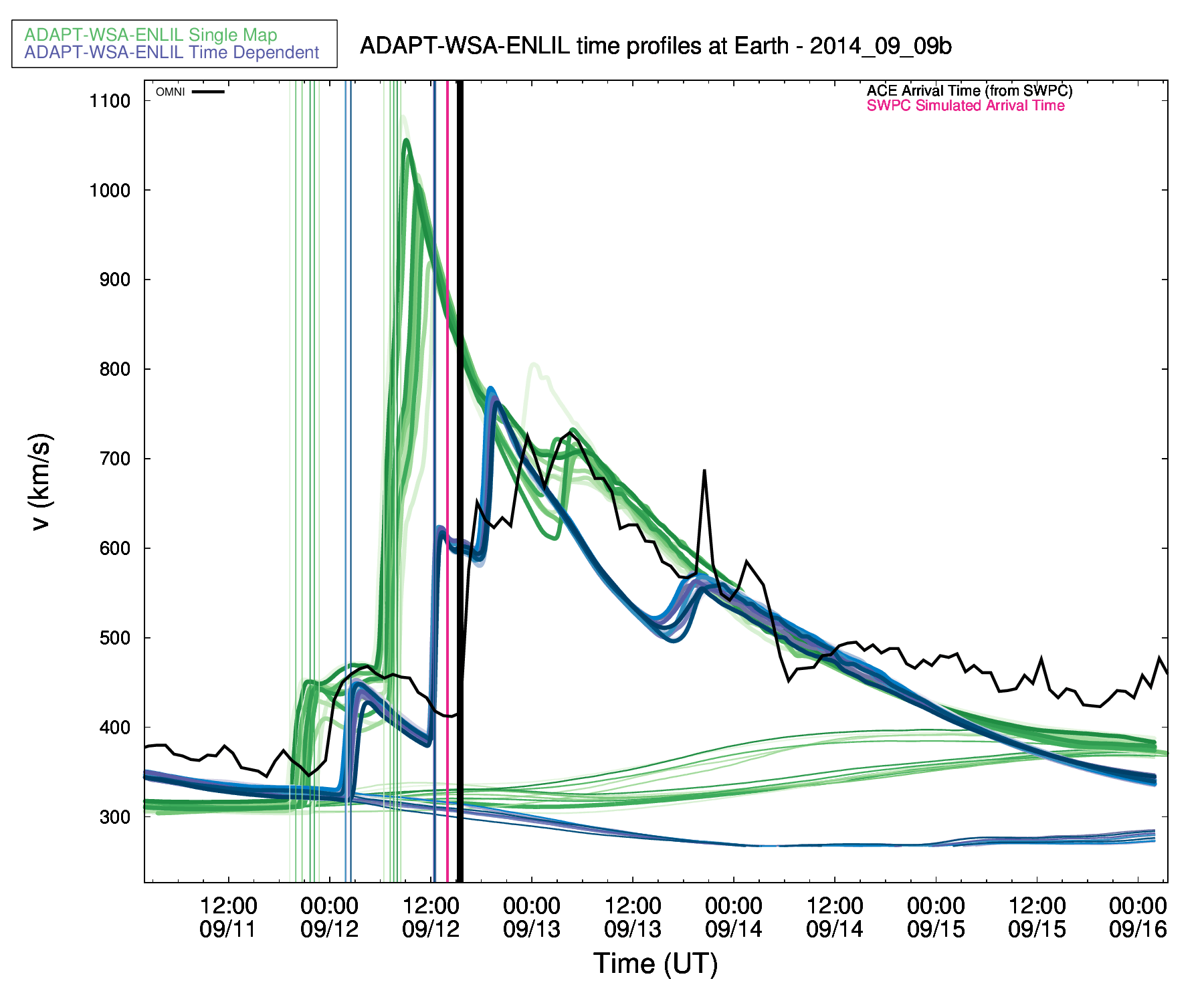}\\
\includegraphics[width=0.33\textwidth]{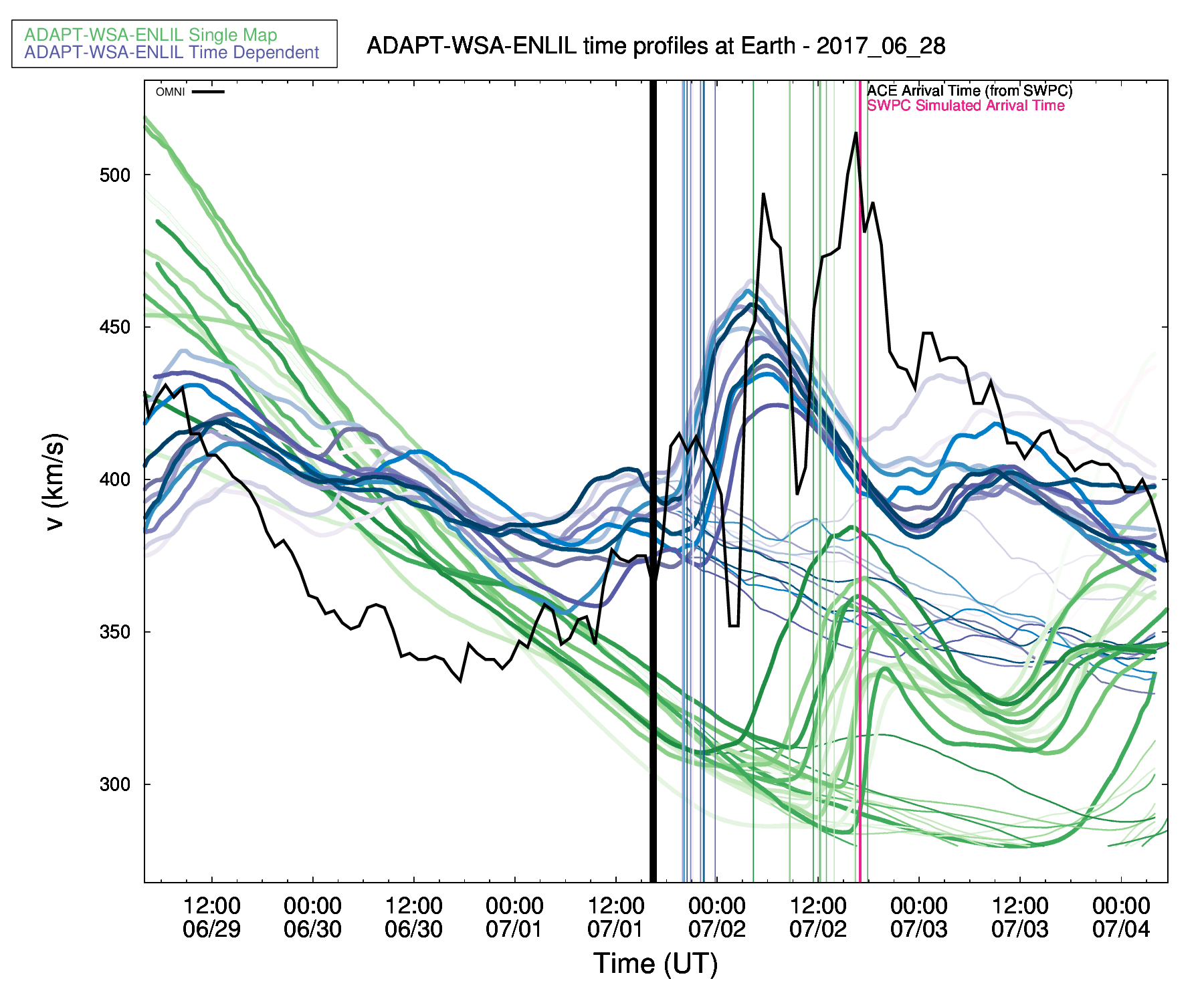}
\includegraphics[width=0.33\textwidth]{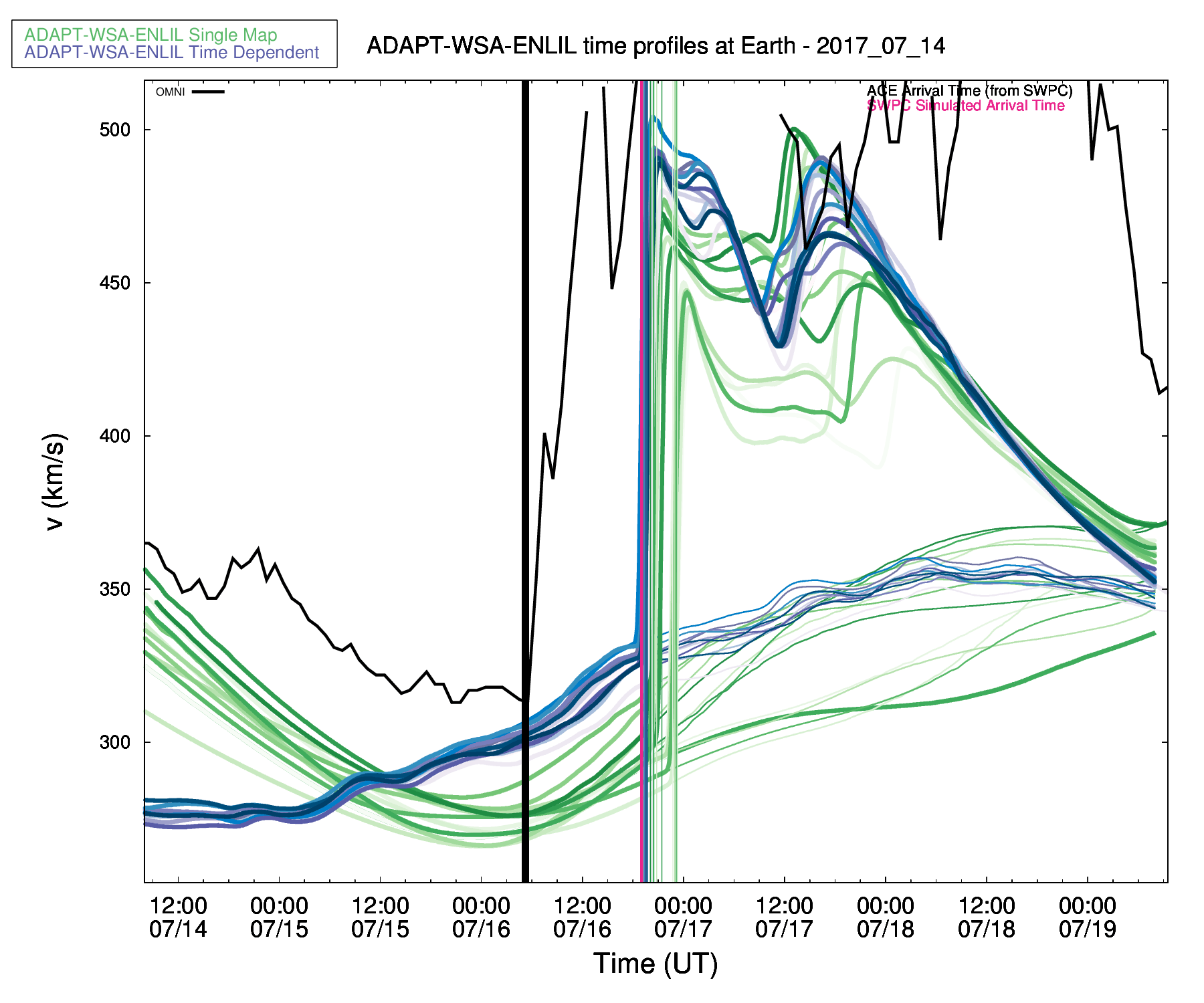}
\includegraphics[width=0.33\textwidth]{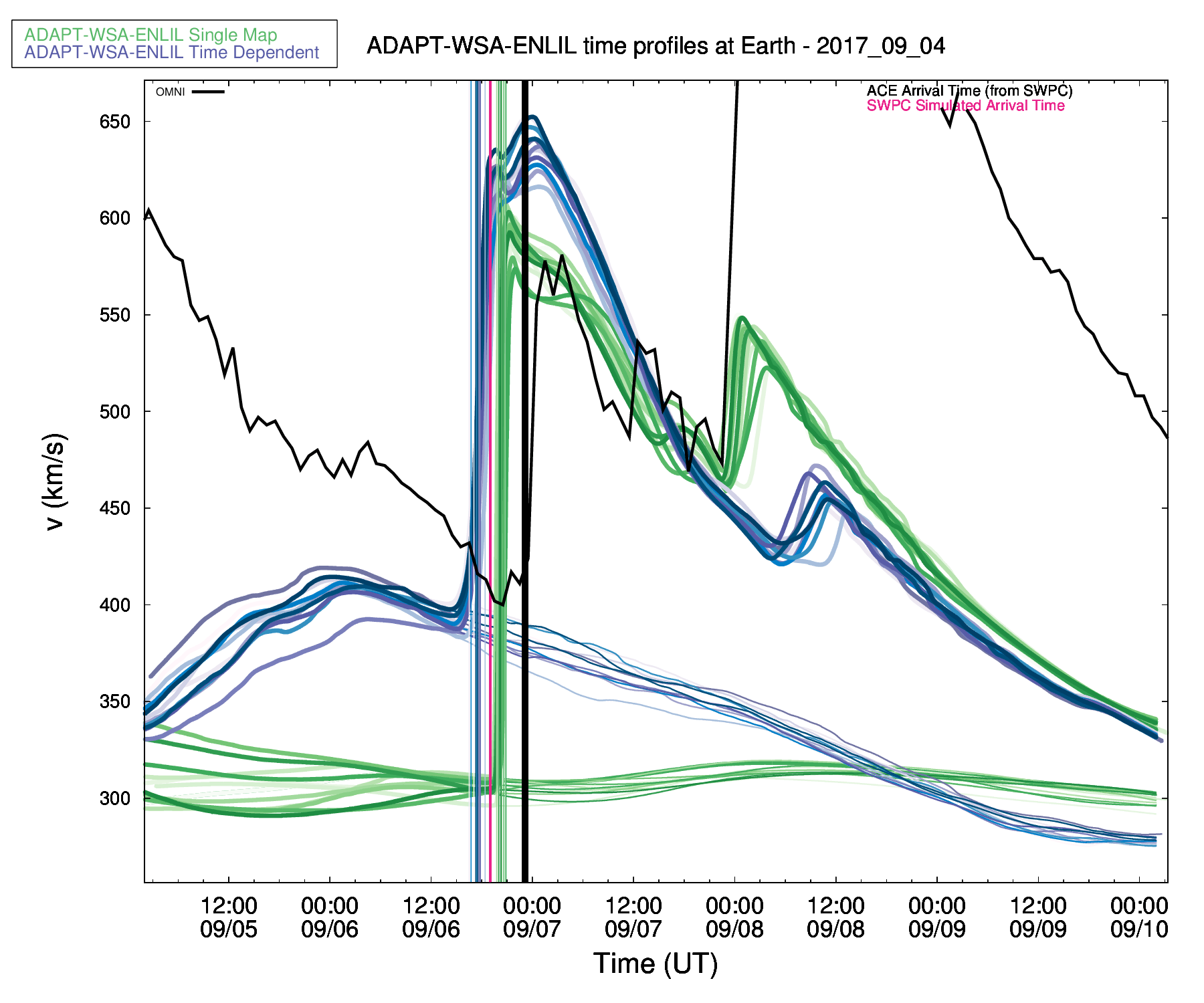}\\
\includegraphics[width=0.33\textwidth]{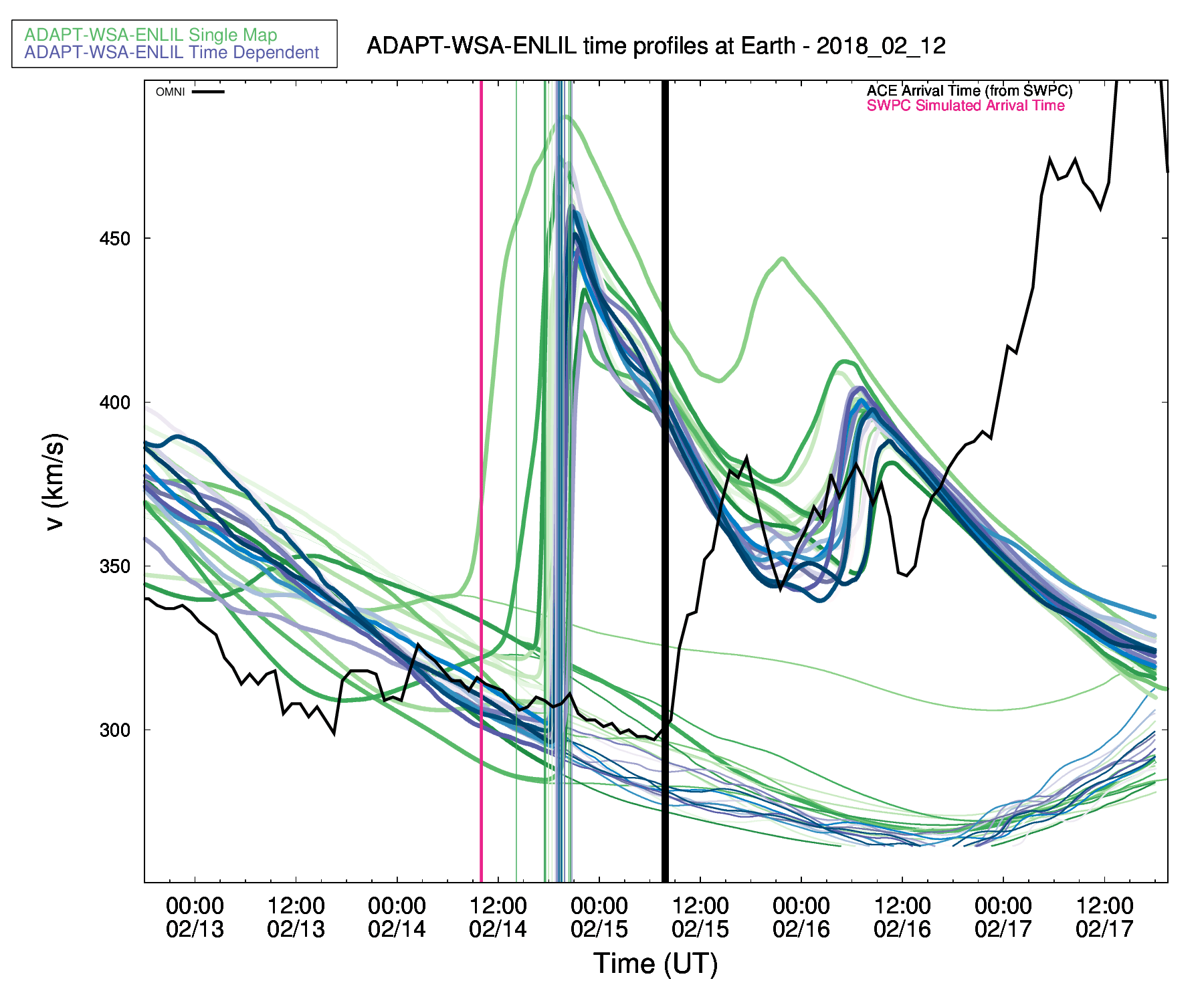}
\includegraphics[width=0.33\textwidth]{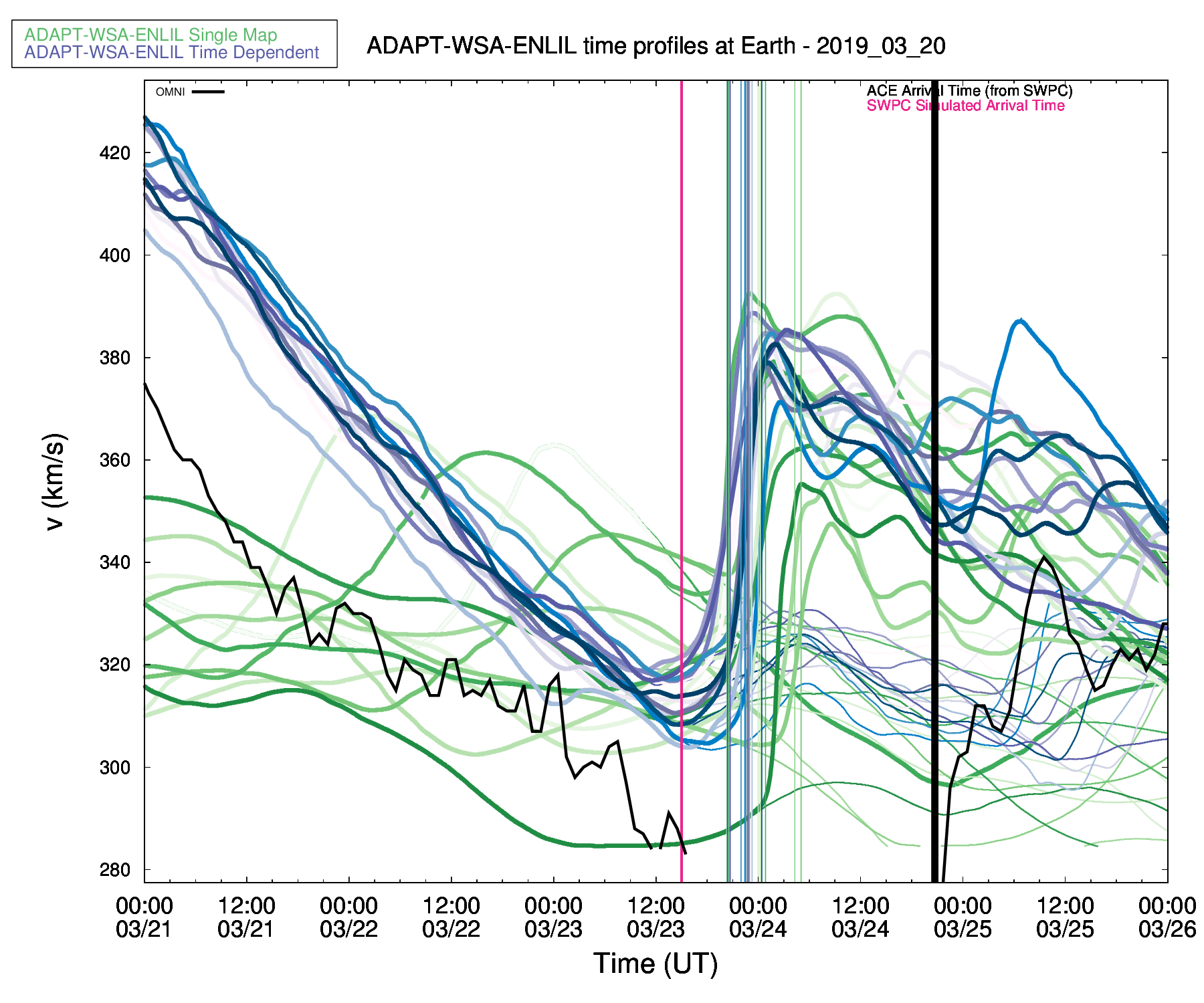}
\caption{Figure \ref{fig:adapt-time-series2} continued.\label{fig:adapt-time-series3}}
\end{figure*}

Error calculations for the median CME arrival for the ADAPT driven simulations are listed in Table \ref{tbl:results} as rows (d-e). More detailed errors for each event for simulations (d-e) are listed in Table \ref{tbl:errors4}. Figures \ref{fig:adapt-time-series}-\ref{fig:adapt-time-series3} show the simulated speed at Earth for the benchmark single map-driven runs (d) and time-dependent map-driven runs (e) for all 38 events. All 12 realizations of the single map driven simulations are plotted in shades of green and time-dependent runs in shades of purple with the thin lines (of the same color) showing the ambient simulation (with no CME).  Similar plots with all of the plasma parameters (speed, density, magnetic field, and temperature) time series and plots zoomed in near the CME arrival time (for better clarity) are available to download from the project webpage.  The realizations corresponding to the median arrival from the ADAPT simulations (d) and (e) are also plotted in dark and light orange in Figures \ref{fig:time-series}-\ref{fig:time-series3}, together with all of the other simulation variatons discussed in the previous sections.

On average the median spread/range of the ensemble CME arrival times for each event is 1.3 hours for (e) (with a minimum range of 0.3 hours, and maximum of 16.4 hours). The ADAPT arrival time errors are also plotted separately for each event in Figure \ref{fig:byevent} (together with all of the previous simulations) as dark orange squares  (single map) and light orange triangles (time-dependent) with error bars showing the range of arrival time errors of each ADAPT ensemble.  To get an idea of the accuracy of the ensemble spread, we can check if the observed CME arrival falls within the range of predicted arrivals from the ensemble for each event.   We found relatively poor performance here, with the observed arrival falling within the predicted range for only 3/42 (7\%) single-map driven ADAPT simulations (d) and 6/42 (14\%) of the time-dependent simulations (e).  If we increase each ensemble predicted range by $\pm$3 hours, this improves to 15/42 (36\%) and 14/42 (33\%) respectively.

For all events, the single map driven ADAPT simulations (d) were found to have ME=-2.4 hours, MAE=8.3 hours, and RMSE=10.1 hours, an increase compared to the MAE=6.6 hours of the benchmark (a).  The time dependent map driven ADAPT simulations (e) were found to have ME=-0.9 hours, MAE=7.4 hours, and RMSE=9.4 hours, an increase compared to the MAE=6.8 hours of the time-dependent benchmark (b). However, for the 5 event subset for 2017--2019 the errors decrease for both (d)  MAE=14.4 hours, and (e) MAE=11.8 hours (MAE$_{\rm TD - Single}$=$-2.6^{+4.2}_{-6.6}$), compared to (a) MAE=17.6 hours, and (b) MAE=16.2 hours.  For all events, the CME arrival time error of the time-dependent ADAPT simulations are about 1 hour less than the single map driven ADAPT simulations (MAE$_{\rm TD - Single}$=$-0.9{\pm 1.4}$).   
\subsection{Validation Discussion}\label{valid}
\begin{figure*}[ht]
\includegraphics[width=0.5\textwidth]{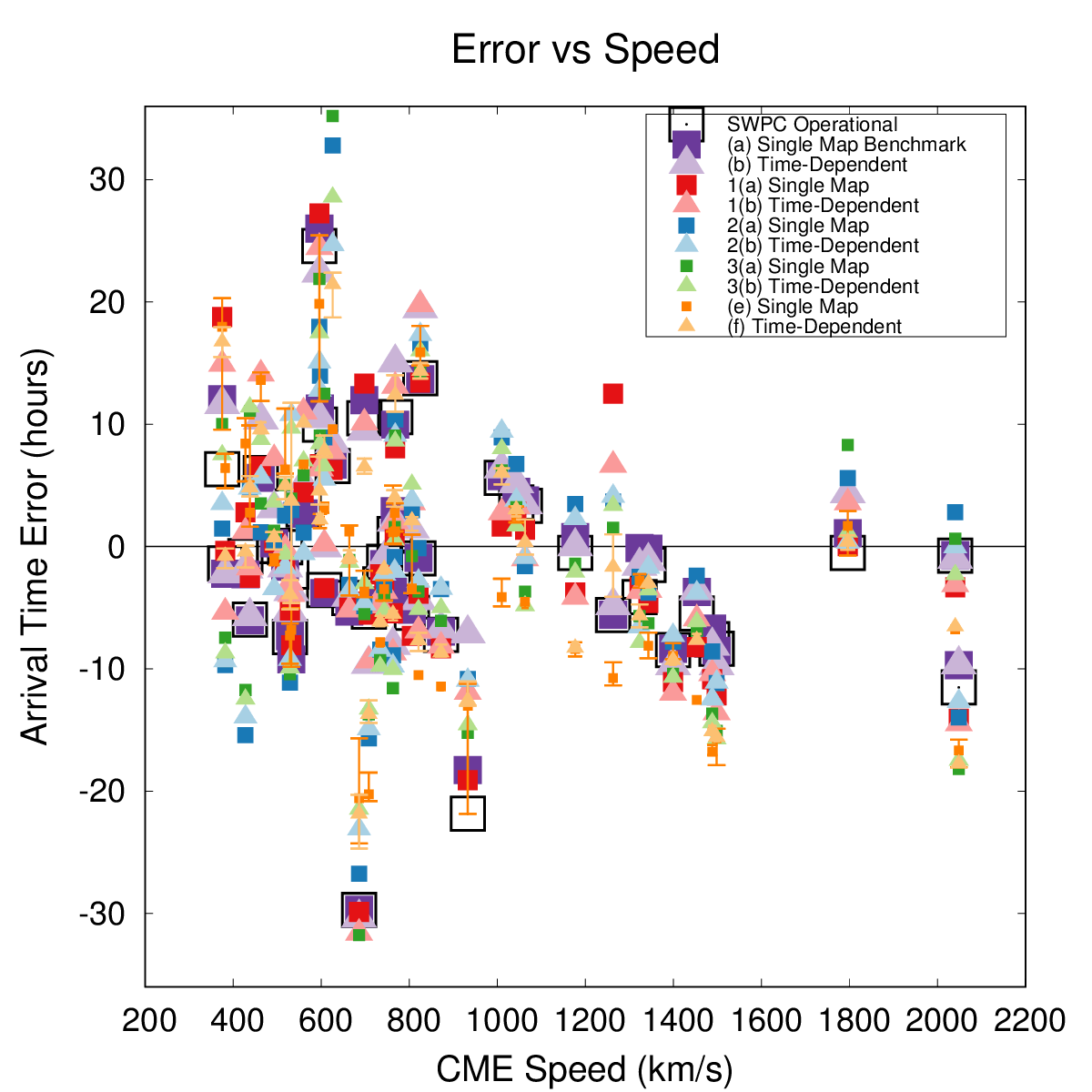}
\includegraphics[width=0.5\textwidth]{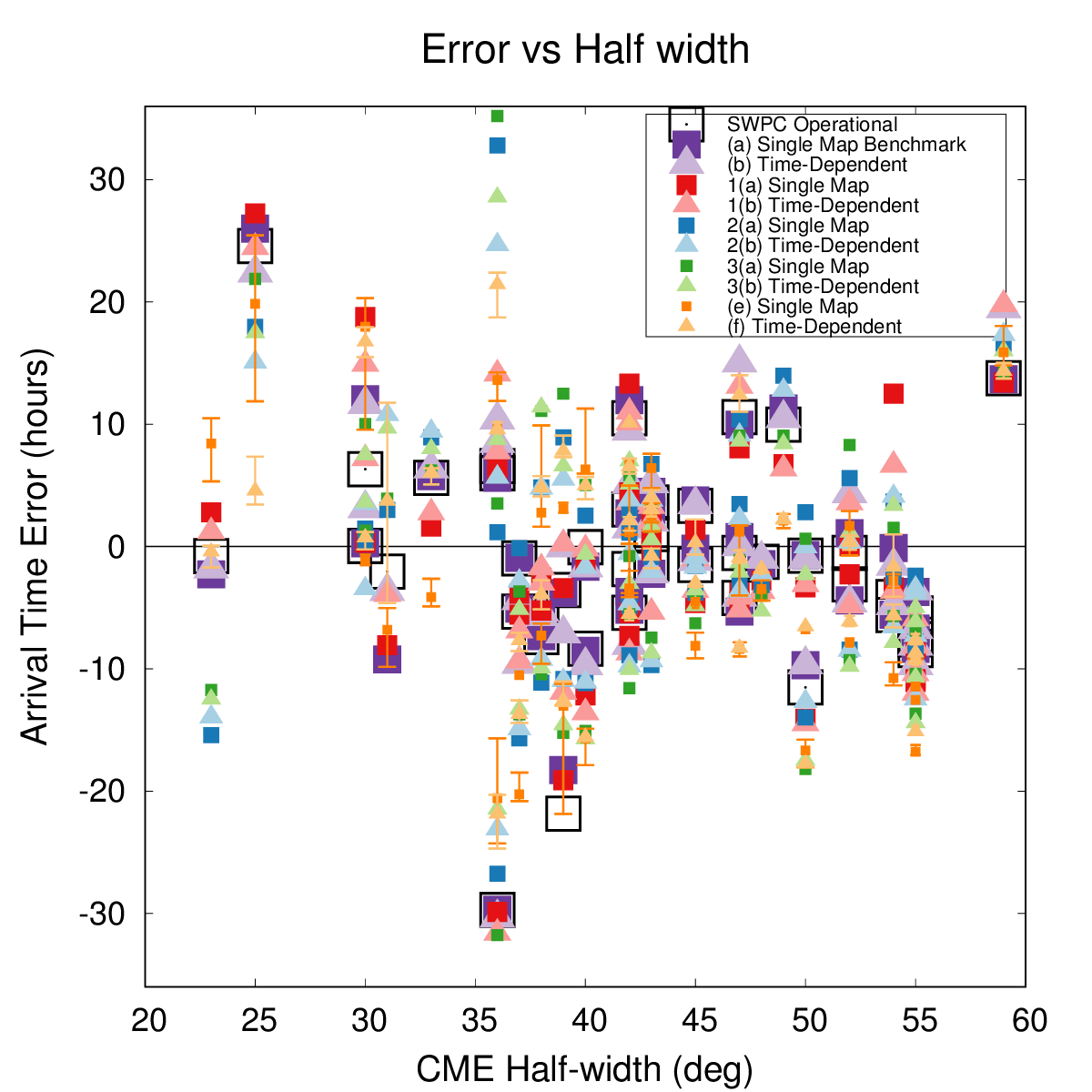}\\
\includegraphics[width=0.5\textwidth]{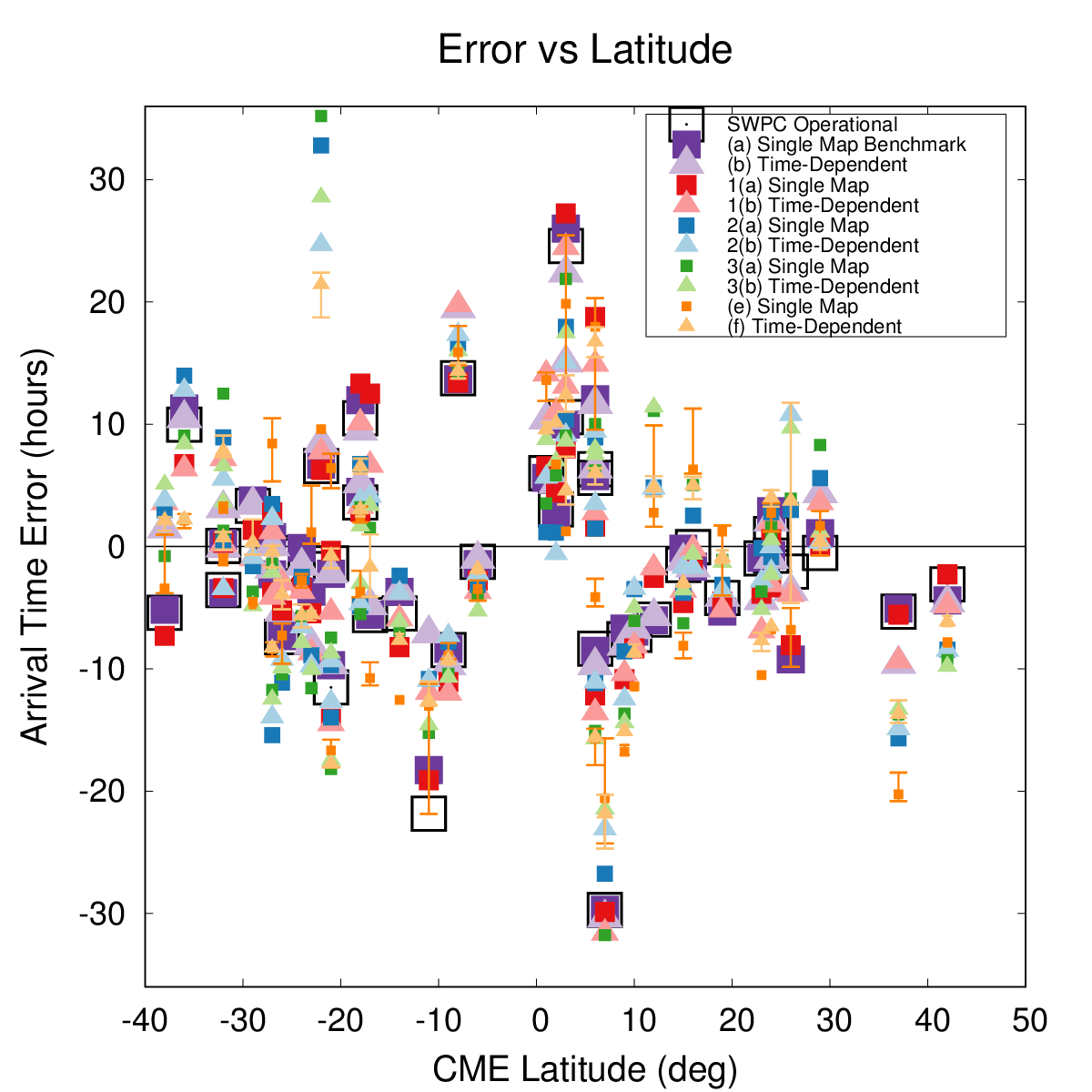}
\includegraphics[width=0.5\textwidth]{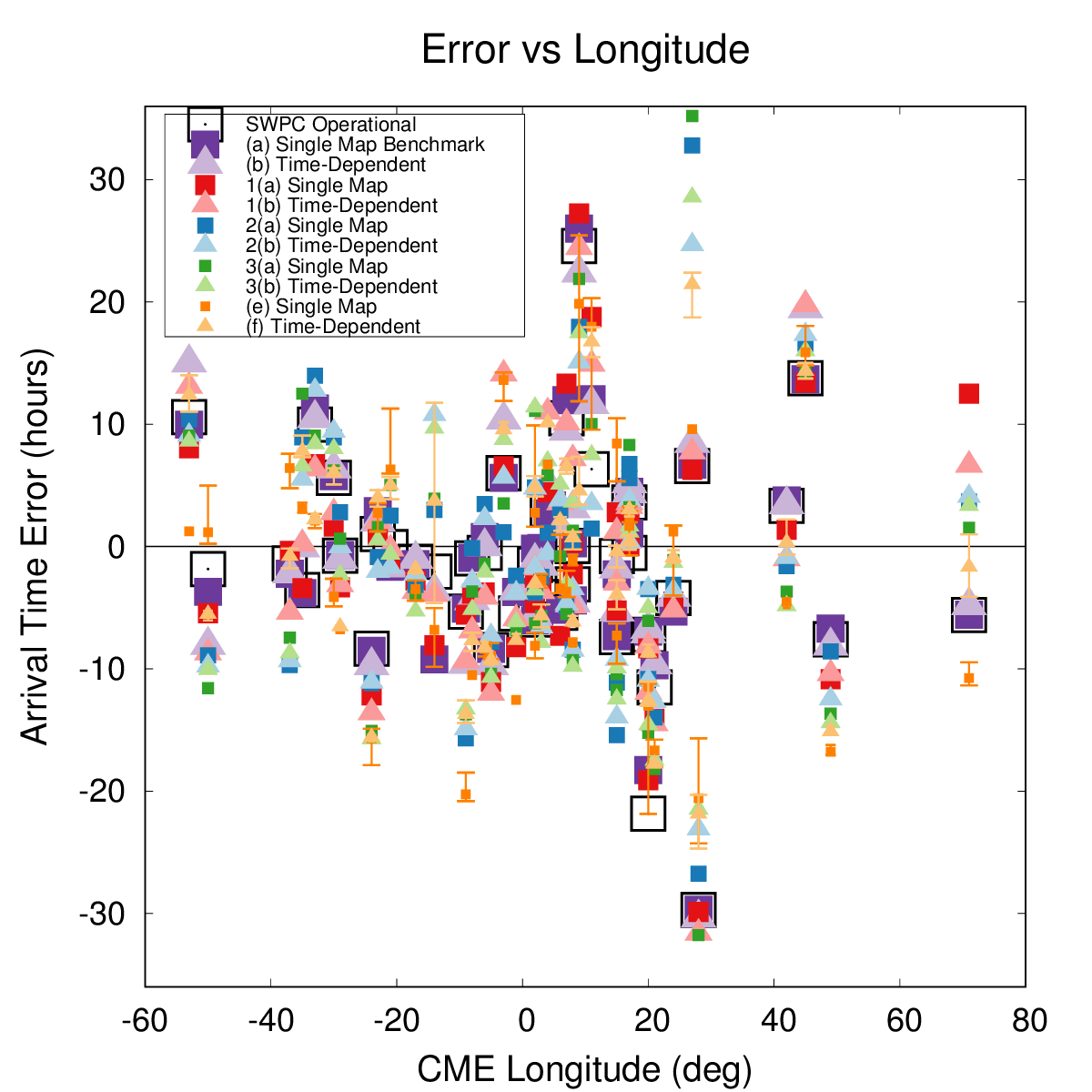}\\
\caption{Dependence of the arrival time error on CME input parameters.\label{fig:error_input}}
\end{figure*}

The validation results of the benchmark (a) and time-dependent simulations (b),  their variations 1(a-b), 2(a-b), 3(a-b), and ADAPT simulations (d) and (e) are all summarized together in Table \ref{tbl:results} and divided by period 2012--2014 (33 events), 2017--2019 (5 events), and all years.  For each pair of runs (single map and time-dependent), the table also shows the difference between the MAE of the time dependent and single map driven runs. We have visualized these arrival time errors separately for each event and simulation type in Figure \ref{fig:byevent} (error are listed in supplemental Tables  \ref{tbl:errors}-\ref{tbl:errors4}). In the supplemental plots at the end of the report we show the arrival time error vs CME input parameters in Figure \ref{fig:error_input}.

Switching from single map driven simulations to time-dependent ones decreases the CME arrival time error for all simulation types except for (a-b) and 1(a-b). For all events on average, the MAE for all of the simulations increases compared to the benchmark (a) and time-dependent (b) errors, with variation 3 showing the largest error increase. However for the 5 event subset from 2017--2019, the arrival time error decreased for simulations: 2(a) by MAE$_{\rm Bench (a)- Single 2(a)}$=2.7${\pm 4.2}$ hours,  2(b) by MAE$_{\rm TD (b)- TD 2(b)}$=1.6${\pm 4.7}$ hours, 3(b) by MAE$_{\rm TD (b)- TD 3(b)}$=0.7${\pm 5.5}$ hours, (d) by MAE$_{\rm Bench (a)- Single (d)}$=3.1${\pm 4.0}$ hours, (e) by MAE$_{\rm TD (b)- TD (e)}$=$4.4{\pm 7.2}$ hours. As discussed in Section \ref{magnetograms}, the GONGz product only has reliable zero-point correction after 2017.  While it is not possible to draw statistically significant conclusions for a sample size of 5 events, there is a clear trend for the GONGz inputs decreasing the error for 5 of the 8 simulation sets. The greatest decrease in CME arrival time error of 4.4 hours was for (e) time-dependent GONGz ADAPT WSA 4.5 ENLIL 2.9e a8b1 MAE=11.8 hours compared to MAE=16.2 hours for (b). 

\section{Summary}\label{summary}
In summary, this report describes the results of a three-year project from the NASA/NOAA MOU Annex with SWPC and CCMC.  The project consisted of multiple simulation validation studies for different model inputs, versions and settings for 38 historical events over the period of five years from 2012--2014 (33 events) and 2017--2019 (5 events). For all 38 events, within each model version/settings combination, the CME arrival time error decreased by 0.2 to 0.9 hours when using a sequence of time-dependent zero-point corrected magnetograms compared to using a single magnetogram as input. For all events, when using the older uncorrected magnetograms, the CME arrival time error increased for all new model versions/settings combinations compared to the benchmark.  However, for the 5 event subset from 2017--2019, the arrival time error decreased by MAE$_{\rm Bench (a)- Single (d)}$=3.1${\pm 4.0}$ hours for single map driven ADAPT (d) compared to the single map benchmark (a), by MAE$_{\rm TD (b)- TD (e)}$=$4.4{\pm 7.2}$ hours for time-dependent driven ADAPT (e) compared to the time-dependent benchmark (b), and by MAE$_{\rm Bench (a)- TD (e)}$=5.8$^{+8.6}_{-7.6}$ hours for time-dependent driven ADAPT (e) compared to the single map benchmark (a). 

The expectation from SWPC at the outset of this project was that by upgrading the initial operational WSA-ENLIL, a substantially improved representation of heliospheric dynamics would be obtained.  The upgraded configuration:

\begin{itemize}
\item  Employs improved observational inputs (zero point corrected GONGz synoptic maps).
\item  Accounts for well-established photospheric flux transport processes, currently unobservable (from Earth) on the far-side of the sun and constrained by data assimilation of new observations through implementation of the ADAPT model.
\item Accounts for dynamic evolution of solar surface by implementing time-dependent driving at the inner boundary to generate a more physically structured heliosphere
\item Improves the coronal modeling by updating WSA model
\item Takes advantage of new recalibrations of both the WSA and ENLIL models
\end{itemize}

Acknowledging the caveats previously noted, particularly regarding statistical concerns due to the limited size of the dataset along with the relevant history of GONG processing, the subset of results for more recent events occurring between 2017--2019 produced by the anticipated upgrade to the time-dependent ADAPT configuration (e) generally support this expectation, although recognition of the respective uncertainties makes such conclusions less certain. 

It should also be noted that the ADAPT validation results presented in this paper only serve to give an approximate lower bound on potential improvements afforded by the full implementation of ADAPT (e). Because ADAPT is designed to produce an ensemble of solutions to the modeled physical processes of interest, constrained by data-assimilation, the results should be interpreted and validated in that same vein - as an ensemble of solutions constraining the associated uncertainties. Our analysis of these ADAPT results was, by necessity, constrained to a simple use of the ensemble member representing the median CME arrival time for each event. In future work, we plan to implement more advanced validation methods to better understand the performance of the full ensemble and how the ensemble results can be fully used to benefit operational forecasts. These lie beyond the scope of this project. We also anticipate that further improvements to the WSA-ENLIL models can be realized by a more systematic recalibration of the free-parameters within each of those models.

An additional result of this undertaking is the recognition of the surprisingly complicated realities of such an endeavor, highlighting the critical role of interagency communication, thorough documentation, and version controlled software management.

\acknowledgments
MLM thanks C. Verbeke, T. Jensen, B. Brown, P. Riley and the CME Arrival Time and Impact Working Team members for useful discussions.


\bibliographystyle{apj}
\bibliography{reportref}  

\pagebreak

\section{Supplemental Figures \& Tables}
\begin{table*}[!ht]
\tabletypesize{\scriptsize}
\caption{Summary of CME observed and simulated arrival times along with the arrival time prediction errors for GONG WSA 2.2 ENLIL 2.6.2 a3b2. \label{tbl:errors}}
\begin{center}
\begin{tabular}{rc|cr|cr|cr}
\tableline
\tableline
Event	&	Arrival Time	&	Arrival Time	&	$\Delta t$	&	Arrival Time	&	$\Delta t$	&	Arrival Time	&	$\Delta t$	\\								
\#	&	Observed	&	SWPC Operational	&		&	Single map	&		&	Time-dependent	&		\\								
	&	{\scriptsize[UT]} 	&	{\scriptsize[UT]} 	&	(hours)	&	{\scriptsize[UT]} 	&	(hours)	&	{\scriptsize[UT]} 	&	(hours)	\\								
\tableline																							
1	&	2012-01-24T14:31Z	&	2012-01-24T14:00Z	&	-0.52	&	2012-01-24T15:37:51Z	&	1.10	&	2012-01-24T18:48:26Z	&	4.28	\\								
2	&	2012-02-14T07:03Z	&	2012-02-14T07:00Z	&	-0.05	&	2012-02-14T05:23:12Z	&	-1.65	&	2012-02-14T05:15:02Z	&	-1.78	\\								
3	&	2012-02-26T20:58Z	&	2012-02-26T20:00Z	&	-0.97	&	2012-02-26T20:02:29Z	&	-0.92	&	2012-02-26T16:30:16Z	&	-4.45	\\								
4	&	2012-03-08T10:45Z	&	2012-03-08T10:00Z	&	-0.75	&	2012-03-08T10:02:28Z	&	-0.70	&	2012-03-08T09:39:50Z	&	-1.08	\\								
5	&	2012-04-05T20:03Z	&	2012-04-05T18:00Z	&	-2.05	&	2012-04-05T10:49:07Z	&	-9.22	&	2012-04-05T16:19:48Z	&	-3.72	\\								
6	&	2012-04-21T09:25Z	&	2012-04-21T16:00Z	&	6.58	&	2012-04-21T16:07:38Z	&	6.70	&	2012-04-21T17:49:03Z	&	8.40	\\								
6	&	2012-04-23T02:33Z	&	2012-04-22T23:00Z	&	-3.55	&	2012-04-22T22:44:27Z	&	-3.80	&	2012-04-23T02:23:40Z	&	-0.15	\\								
7	&	2012-05-20T01:36Z	&	2012-05-19T20:00Z	&	-5.60	&	2012-05-19T20:00:13Z	&	-5.58	&	2012-05-19T20:43:28Z	&	-4.87	\\								
8	&	2012-06-16T09:01Z	&	2012-06-16T19:00Z	&	9.98	&	2012-06-16T20:18:50Z	&	11.28	&	2012-06-16T19:27:11Z	&	10.43	\\								
8	&	2012-06-16T19:31Z	&	2012-06-16T19:00Z	&	-0.52	&	2012-06-16T20:18:50Z	&	0.78	&	2012-06-16T19:27:11Z	&	-0.05	\\								
9	&	2012-07-14T17:28Z	&	2012-07-14T12:00Z	&	-5.47	&	2012-07-14T13:44:20Z	&	-3.72	&	2012-07-14T13:49:00Z	&	-3.65	\\								
10	&	2012-08-02T09:22Z	&	2012-08-02T08:00Z	&	-1.37	&	2012-08-02T07:07:50Z	&	-2.23	&	2012-08-02T07:05:06Z	&	-2.27	\\								
11	&	2012-09-03T11:23Z	&	2012-09-03T17:00Z	&	5.62	&	2012-09-03T17:08:26Z	&	5.75	&	2012-09-03T17:42:28Z	&	6.32	\\								
12	&	2012-09-30T22:13Z	&	2012-09-30T15:00Z	&	-7.22	&	2012-09-30T15:15:09Z	&	-6.95	&	2012-09-30T15:23:30Z	&	-6.82	\\								
13	&	2012-10-08T04:31Z	&	2012-10-08T15:00Z	&	10.48	&	2012-10-08T16:29:55Z	&	11.97	&	2012-10-08T13:55:43Z	&	9.40	\\								
14	&	2012-10-31T14:40Z	&	2012-10-31T21:00Z	&	6.33	&	2012-11-01T02:43:39Z	&	12.05	&	2012-11-01T02:15:51Z	&	11.58	\\								
15	&	2012-11-23T21:12Z	&	2012-11-23T17:00Z	&	-4.20	&	2012-11-23T15:52:19Z	&	-5.32	&	2012-11-23T17:01:07Z	&	-4.17	\\								
16	&	2013-01-16T23:00Z	&	2013-01-17T05:00Z	&	6.00	&	2013-01-17T04:37:15Z	&	5.62	&	2013-01-17T09:19:43Z	&	10.32	\\								
17	&	2013-01-19T16:47Z	&	2013-01-19T16:00Z	&	-0.78	&	2013-01-19T14:30:34Z	&	-2.27	&	2013-01-19T14:54:34Z	&	-1.87	\\								
18	&	2013-03-15T05:05Z	&	2013-03-15T02:00Z	&	-3.08	&	2013-03-15T00:40:35Z	&	-4.40	&	2013-03-15T00:26:21Z	&	-4.63	\\								
19	&	2013-03-16T01:50Z	&	2013-03-16T00:00Z	&	-1.83	&	2013-03-15T22:07:19Z	&	-3.70	&	2013-03-15T17:42:39Z	&	-8.12	\\								
19	&	2013-03-17T05:28Z	&	2013-03-16T21:00Z	&	-8.47	&	2013-03-16T21:05:25Z	&	-8.37	&	2013-03-16T19:41:19Z	&	-9.77	\\								
20	&	2013-04-13T22:15Z	&	2013-04-13T21:00Z	&	-1.25	&	2013-04-13T20:48:51Z	&	-1.43	&	2013-04-13T21:09:16Z	&	-1.08	\\								
21	&	2013-04-30T08:57Z	&	2013-04-30T09:00Z	&	0.05	&	2013-04-30T09:13:05Z	&	0.27	&	2013-04-30T12:00:00Z	&	3.05	\\								
22	&	2013-05-19T22:21Z	&	2013-05-19T14:00Z	&	-8.35	&	2013-05-19T13:50:00Z	&	-8.52	&	2013-05-19T12:33:18Z	&	-9.78	\\								
23	&	2013-05-24T17:35Z	&	2013-05-24T10:00Z	&	-7.58	&	2013-05-24T10:52:05Z	&	-6.70	&	2013-05-24T09:23:10Z	&	-8.18	\\								
24	&	2013-06-30T10:40Z	&	2013-06-30T14:00Z	&	3.33	&	2013-06-30T14:27:34Z	&	3.78	&	2013-06-30T14:03:28Z	&	3.38	\\								
25	&	2013-07-09T19:58Z	&	2013-07-09T23:00Z	&	3.03	&	2013-07-09T22:40:45Z	&	2.70	&	2013-07-10T01:01:08Z	&	5.05	\\								
26	&	2014-01-07T14:25Z	&	2014-01-07T09:00Z	&	-5.42	&	2014-01-07T09:14:45Z	&	-5.17	&	2014-01-07T15:49:12Z	&	1.40	\\								
27	&	2014-01-09T19:31Z	&	2014-01-09T08:00Z	&	-11.52	&	2014-01-09T09:48:29Z	&	-9.70	&	2014-01-09T09:48:08Z	&	-9.70	\\								
28	&	2014-03-25T19:25Z	&	2014-03-26T06:00Z	&	10.58	&	2014-03-26T05:23:28Z	&	9.97	&	2014-03-26T10:24:05Z	&	14.98	\\								
29	&	2014-04-20T10:24Z	&	2014-04-20T03:00Z	&	-7.40	&	2014-04-20T03:04:33Z	&	-7.32	&	2014-04-20T05:00:16Z	&	-5.38	\\								
29	&	2014-04-20T10:24Z	&	2014-04-20T14:00Z	&	3.60	&	2014-04-20T14:52:40Z	&	4.47	&	2014-04-20T15:30:23Z	&	5.10	\\								
30	&	2014-08-19T05:58Z	&	2014-08-19T00:00Z	&	-5.97	&	2014-08-18T23:55:24Z	&	-6.03	&	2014-08-19T00:12:25Z	&	-5.75	\\								
31	&	2014-09-06T14:19Z	&	2014-09-06T09:00Z	&	-5.32	&	2014-09-06T09:13:39Z	&	-5.08	&	2014-09-06T04:38:16Z	&	-9.67	\\								
32	&	2014-09-11T22:58Z	&	2014-09-12T00:00Z	&	1.03	&	2014-09-12T01:52:57Z	&	2.90	&	2014-09-12T00:56:41Z	&	1.97	\\								
33	&	2014-09-12T15:30Z	&	2014-09-12T14:00Z	&	-1.50	&	2014-09-12T15:19:45Z	&	-0.17	&	2014-09-12T14:13:39Z	&	-1.27	\\								
34	&	2017-07-01T16:26Z	&	2017-07-02T17:00Z	&	24.57	&	2017-07-02T18:26:53Z	&	26.00	&	2017-07-02T14:46:30Z	&	22.33	\\								
35	&	2017-07-16T05:14Z	&	2017-07-16T19:00Z	&	13.77	&	2017-07-16T18:55:54Z	&	13.68	&	2017-07-17T00:38:30Z	&	19.40	\\								
36	&	2017-09-06T23:08Z	&	2017-09-06T19:00Z	&	-4.13	&	2017-09-06T23:00:49Z	&	-0.12	&	2017-09-06T21:28:39Z	&	-1.65	\\								
37	&	2018-02-15T07:50Z	&	2018-02-14T10:00Z	&	-21.83	&	2018-02-14T13:32:11Z	&	-18.28	&	2018-02-15T00:39:57Z	&	-7.17	\\								
38	&	2019-03-24T20:43Z	&	2019-03-23T15:00Z	&	-29.72	&	2019-03-23T15:00:36Z	&	-29.70	&	2019-03-23T14:17:48Z	&	-30.42	\\								
\tableline													
\end{tabular}
\end{center}
\end{table*}

\begin{table*}[!ht]
\tabletypesize{\scriptsize}
\caption{Summary of CME observed and simulated arrival times along with the arrival time prediction errors for GONG WSA 2.2 ENLIL 2.9e a8b1. \label{tbl:errors1}}
\begin{center}
\begin{tabular}{rc|cr|cr}
\tableline
\tableline
Event	&	Arrival Time	&	Arrival Time	&	$\Delta t$	&	Arrival Time	&	$\Delta t$	\\
\#	&	Observed	&	Single map	&		&	Time-dependent	&		\\
	&	{\scriptsize[UT]} 	&	{\scriptsize[UT]} 	&	(hours)	&	{\scriptsize[UT]} 	&	(hours)	\\
\tableline											
1	&	2012-01-24T14:31Z	&	2012-01-24T14:33:30Z	&	0.03	&	2012-01-24T18:04:46Z	&	3.55	\\
2	&	2012-02-14T07:03Z	&	2012-02-14T05:47:31Z	&	-1.25	&	2012-02-14T06:49:51Z	&	-0.22	\\
3	&	2012-02-26T20:58Z	&	2012-02-26T17:06:57Z	&	-3.85	&	2012-02-26T14:02:27Z	&	-6.92	\\
4	&	2012-03-08T10:45Z	&	2012-03-08T07:24:16Z	&	-3.33	&	2012-03-08T07:35:58Z	&	-3.15	\\
5	&	2012-04-05T20:03Z	&	2012-04-05T11:58:08Z	&	-8.07	&	2012-04-05T16:05:28Z	&	-3.95	\\
6	&	2012-04-21T09:25Z	&	2012-04-21T15:41:22Z	&	6.27	&	2012-04-21T17:07:20Z	&	7.70	\\
6	&	2012-04-23T02:33Z	&	2012-04-22T23:08:33Z	&	-3.40	&	2012-04-23T02:44:50Z	&	0.18	\\
7	&	2012-05-20T01:36Z	&	2012-05-20T14:06:53Z	&	12.50	&	2012-05-20T08:13:35Z	&	6.62	\\
8	&	2012-06-16T09:01Z	&	2012-06-16T15:45:36Z	&	6.73	&	2012-06-16T15:20:53Z	&	6.32	\\
8	&	2012-06-16T19:31Z	&	2012-06-16T15:45:36Z	&	-3.75	&	2012-06-16T15:20:53Z	&	-4.17	\\
9	&	2012-07-14T17:28Z	&	2012-07-14T09:14:59Z	&	-8.22	&	2012-07-14T11:30:30Z	&	-5.95	\\
10	&	2012-08-02T09:22Z	&	2012-08-02T09:00:03Z	&	-0.35	&	2012-08-02T03:57:14Z	&	-5.40	\\
11	&	2012-09-03T11:23Z	&	2012-09-03T13:01:55Z	&	1.63	&	2012-09-03T14:06:21Z	&	2.72	\\
12	&	2012-09-30T22:13Z	&	2012-09-30T13:53:50Z	&	-8.32	&	2012-09-30T14:01:52Z	&	-8.18	\\
13	&	2012-10-08T04:31Z	&	2012-10-08T17:51:03Z	&	13.33	&	2012-10-08T14:35:22Z	&	10.07	\\
14	&	2012-10-31T14:40Z	&	2012-11-01T09:26:05Z	&	18.77	&	2012-11-01T05:33:39Z	&	14.88	\\
15	&	2012-11-23T21:12Z	&	2012-11-23T16:41:06Z	&	-4.50	&	2012-11-23T15:58:52Z	&	-5.22	\\
16	&	2013-01-16T23:00Z	&	2013-01-17T05:35:19Z	&	6.58	&	2013-01-17T13:04:26Z	&	14.07	\\
17	&	2013-01-19T16:47Z	&	2013-01-19T19:35:18Z	&	2.80	&	2013-01-19T17:58:31Z	&	1.18	\\
18	&	2013-03-15T05:05Z	&	2013-03-15T02:48:23Z	&	-2.27	&	2013-03-15T00:12:34Z	&	-4.87	\\
19	&	2013-03-16T01:50Z	&	2013-03-15T20:22:53Z	&	-5.45	&	2013-03-15T17:06:00Z	&	-8.73	\\
19	&	2013-03-17T05:28Z	&	2013-03-16T18:23:25Z	&	-11.07	&	2013-03-16T17:26:34Z	&	-12.02	\\
20	&	2013-04-13T22:15Z	&	2013-04-13T19:27:37Z	&	-2.78	&	2013-04-13T18:32:11Z	&	-3.70	\\
21	&	2013-04-30T08:57Z	&	2013-04-30T09:16:03Z	&	0.32	&	2013-04-30T16:06:45Z	&	7.15	\\
22	&	2013-05-19T22:21Z	&	2013-05-19T10:10:54Z	&	-12.17	&	2013-05-19T08:43:15Z	&	-13.62	\\
23	&	2013-05-24T17:35Z	&	2013-05-24T06:43:51Z	&	-10.85	&	2013-05-24T07:08:43Z	&	-10.43	\\
24	&	2013-06-30T10:40Z	&	2013-06-30T12:03:30Z	&	1.38	&	2013-06-30T09:37:05Z	&	-1.03	\\
25	&	2013-07-09T19:58Z	&	2013-07-10T00:24:13Z	&	4.43	&	2013-07-10T06:56:42Z	&	10.97	\\
26	&	2014-01-07T14:25Z	&	2014-01-07T07:06:28Z	&	-7.30	&	2014-01-07T17:58:09Z	&	3.55	\\
27	&	2014-01-09T19:31Z	&	2014-01-09T05:25:06Z	&	-14.08	&	2014-01-09T04:59:30Z	&	-14.52	\\
28	&	2014-03-25T19:25Z	&	2014-03-26T03:27:02Z	&	8.03	&	2014-03-26T08:30:47Z	&	13.08	\\
29	&	2014-04-20T10:24Z	&	2014-04-20T05:10:11Z	&	-5.22	&	2014-04-20T07:22:30Z	&	-3.02	\\
29	&	2014-04-20T10:24Z	&	2014-04-20T13:15:19Z	&	2.85	&	2014-04-20T13:39:50Z	&	3.25	\\
30	&	2014-08-19T05:58Z	&	2014-08-19T03:24:33Z	&	-2.55	&	2014-08-19T04:10:21Z	&	-1.78	\\
31	&	2014-09-06T14:19Z	&	2014-09-06T08:43:58Z	&	-5.58	&	2014-09-06T04:55:56Z	&	-9.38	\\
32	&	2014-09-11T22:58Z	&	2014-09-11T23:52:22Z	&	0.90	&	2014-09-12T00:58:06Z	&	2.00	\\
33	&	2014-09-12T15:30Z	&	2014-09-12T10:52:50Z	&	-4.62	&	2014-09-12T11:52:13Z	&	-3.62	\\
34	&	2017-07-01T16:26Z	&	2017-07-02T19:41:32Z	&	27.25	&	2017-07-02T16:52:55Z	&	24.43	\\
35	&	2017-07-16T05:14Z	&	2017-07-16T18:38:48Z	&	13.40	&	2017-07-17T00:59:09Z	&	19.75	\\
36	&	2017-09-06T23:08Z	&	2017-09-06T19:48:36Z	&	-3.32	&	2017-09-06T19:24:57Z	&	-3.72	\\
37	&	2018-02-15T07:50Z	&	2018-02-14T12:43:11Z	&	-19.10	&	2018-02-14T19:52:33Z	&	-11.95	\\
38	&	2019-03-24T20:43Z	&	2019-03-23T14:50:52Z	&	-29.87	&	2019-03-23T13:03:40Z	&	-31.65	\\
\tableline													
\end{tabular}
\end{center}
\end{table*}

\begin{table*}[!ht]
\tabletypesize{\scriptsize}
\caption{Summary of CME observed and simulated arrival times along with the arrival time prediction errors for GONGz WSA 4.5 ENLIL 2.6.2 a3b2. \label{tbl:errors2}}
\begin{center}
\begin{tabular}{rc|cr|cr}
\tableline
\tableline
Event	&	Arrival Time	&	Arrival Time	&	$\Delta t$	&	Arrival Time	&	$\Delta t$	\\
\#	&	Observed	&	Single map	&		&	Time-dependent	&		\\
	&	{\scriptsize[UT]} 	&	{\scriptsize[UT]} 	&	(hours)	&	{\scriptsize[UT]} 	&	(hours)	\\
\tableline											
1	&	2012-01-24T14:31Z	&	2012-01-24T20:07:18Z	&	5.60	&	2012-01-24T14:50:49Z	&	0.32	\\
2	&	2012-02-14T07:03Z	&	2012-02-14T09:35:50Z	&	2.53	&	2012-02-14T05:18:01Z	&	-1.73	\\
3	&	2012-02-26T20:58Z	&	2012-02-26T20:49:51Z	&	-0.13	&	2012-02-26T18:05:55Z	&	-2.87	\\
4	&	2012-03-08T10:45Z	&	2012-03-08T13:34:37Z	&	2.82	&	2012-03-08T10:41:50Z	&	-0.05	\\
5	&	2012-04-05T20:03Z	&	2012-04-05T23:03:50Z	&	3.00	&	2012-04-06T06:48:33Z	&	10.75	\\
6	&	2012-04-21T09:25Z	&	2012-04-22T18:13:02Z	&	32.80	&	2012-04-22T10:04:15Z	&	24.65	\\
6	&	2012-04-23T02:33Z	&	2012-04-23T11:28:47Z	&	8.92	&	2012-04-23T08:01:19Z	&	5.47	\\
7	&	2012-05-20T01:36Z	&	2012-05-20T05:17:50Z	&	3.68	&	2012-05-20T05:41:54Z	&	4.08	\\
8	&	2012-06-16T09:01Z	&	2012-06-16T23:00:41Z	&	13.98	&	2012-06-16T21:46:44Z	&	12.75	\\
8	&	2012-06-16T19:31Z	&	2012-06-16T23:00:41Z	&	3.48	&	2012-06-16T21:46:44Z	&	2.25	\\
9	&	2012-07-14T17:28Z	&	2012-07-14T15:05:51Z	&	-2.37	&	2012-07-14T13:37:34Z	&	-3.83	\\
10	&	2012-08-02T09:22Z	&	2012-08-01T23:39:34Z	&	-9.70	&	2012-08-01T23:58:45Z	&	-9.38	\\
11	&	2012-09-03T11:23Z	&	2012-09-03T20:19:05Z	&	8.93	&	2012-09-03T20:47:11Z	&	9.40	\\
12	&	2012-09-30T22:13Z	&	2012-09-30T18:43:11Z	&	-3.48	&	2012-09-30T18:45:18Z	&	-3.45	\\
13	&	2012-10-08T04:31Z	&	2012-10-07T23:38:49Z	&	-4.87	&	2012-10-07T23:46:16Z	&	-4.73	\\
14	&	2012-10-31T14:40Z	&	2012-10-31T16:08:49Z	&	1.47	&	2012-10-31T18:09:43Z	&	3.48	\\
15	&	2012-11-23T21:12Z	&	2012-11-23T18:04:21Z	&	-3.12	&	2012-11-23T17:32:30Z	&	-3.65	\\
16	&	2013-01-16T23:00Z	&	2013-01-17T00:10:21Z	&	1.17	&	2013-01-17T04:39:15Z	&	5.65	\\
17	&	2013-01-19T16:47Z	&	2013-01-19T01:21:59Z	&	-15.42	&	2013-01-19T02:48:57Z	&	-13.97	\\
18	&	2013-03-15T05:05Z	&	2013-03-14T20:39:09Z	&	-8.42	&	2013-03-14T20:33:36Z	&	-8.52	\\
19	&	2013-03-16T01:50Z	&	2013-03-15T16:56:16Z	&	-8.88	&	2013-03-15T16:01:36Z	&	-9.80	\\
19	&	2013-03-17T05:28Z	&	2013-03-16T21:40:45Z	&	-7.78	&	2013-03-16T22:09:28Z	&	-7.30	\\
20	&	2013-04-13T22:15Z	&	2013-04-13T19:05:45Z	&	-3.15	&	2013-04-13T19:57:10Z	&	-2.28	\\
21	&	2013-04-30T08:57Z	&	2013-04-30T09:20:50Z	&	0.38	&	2013-04-30T05:27:05Z	&	-3.48	\\
22	&	2013-05-19T22:21Z	&	2013-05-19T11:10:21Z	&	-11.17	&	2013-05-19T11:12:20Z	&	-11.13	\\
23	&	2013-05-24T17:35Z	&	2013-05-24T09:00:05Z	&	-8.57	&	2013-05-24T05:04:35Z	&	-12.50	\\
24	&	2013-06-30T10:40Z	&	2013-06-30T09:03:29Z	&	-1.60	&	2013-06-30T09:44:10Z	&	-0.92	\\
25	&	2013-07-09T19:58Z	&	2013-07-09T21:06:00Z	&	1.13	&	2013-07-09T19:21:42Z	&	-0.60	\\
26	&	2014-01-07T14:25Z	&	2014-01-07T17:23:49Z	&	2.97	&	2014-01-07T18:12:11Z	&	3.78	\\
27	&	2014-01-09T19:31Z	&	2014-01-09T05:32:16Z	&	-13.97	&	2014-01-09T06:48:26Z	&	-12.70	\\
28	&	2014-03-25T19:25Z	&	2014-03-26T05:49:09Z	&	10.40	&	2014-03-26T04:29:04Z	&	9.07	\\
29	&	2014-04-20T10:24Z	&	2014-04-19T23:17:00Z	&	-11.12	&	2014-04-20T01:08:12Z	&	-9.25	\\
29	&	2014-04-20T10:24Z	&	2014-04-20T17:09:43Z	&	6.75	&	2014-04-20T14:10:17Z	&	3.77	\\
30	&	2014-08-19T05:58Z	&	2014-08-19T10:48:38Z	&	4.83	&	2014-08-19T10:43:15Z	&	4.75	\\
31	&	2014-09-06T14:19Z	&	2014-09-05T22:37:17Z	&	-15.68	&	2014-09-05T23:22:39Z	&	-14.93	\\
32	&	2014-09-11T22:58Z	&	2014-09-11T22:06:35Z	&	-0.85	&	2014-09-11T20:54:18Z	&	-2.05	\\
33	&	2014-09-12T15:30Z	&	2014-09-12T11:42:39Z	&	-3.78	&	2014-09-12T13:45:25Z	&	-1.73	\\
34	&	2017-07-01T16:26Z	&	2017-07-02T10:24:20Z	&	17.97	&	2017-07-02T07:29:43Z	&	15.05	\\
35	&	2017-07-16T05:14Z	&	2017-07-16T21:50:15Z	&	16.60	&	2017-07-16T22:32:02Z	&	17.30	\\
36	&	2017-09-06T23:08Z	&	2017-09-06T20:38:27Z	&	-2.48	&	2017-09-06T16:31:33Z	&	-6.60	\\
37	&	2018-02-15T07:50Z	&	2018-02-14T20:59:05Z	&	-10.83	&	2018-02-14T20:51:41Z	&	-10.97	\\
38	&	2019-03-24T20:43Z	&	2019-03-23T17:57:29Z	&	-26.75	&	2019-03-23T21:35:36Z	&	-23.12	\\
\tableline													
\end{tabular}
\end{center}
\end{table*}

\begin{table*}[!ht]
\tabletypesize{\scriptsize}
\caption{Summary of CME observed and simulated arrival times along with the arrival time prediction errors for GONGz WSA 4.5 ENLIL 2.9e a8b1. \label{tbl:errors3}}
\begin{center}
\begin{tabular}{rc|cr|cr}
\tableline
\tableline
Event	&	Arrival Time	&	Arrival Time	&	$\Delta t$	&	Arrival Time	&	$\Delta t$	\\
\#	&	Observed	&	Single map	&		&	Time-dependent	&		\\
	&	{\scriptsize[UT]} 	&	{\scriptsize[UT]} 	&	(hours)	&	{\scriptsize[UT]} 	&	(hours)	\\
\tableline											
1	&	2012-01-24T14:31Z	&	2012-01-24T22:49:47Z	&	8.30	&	2012-01-24T15:40:03Z	&	1.15	\\
2	&	2012-02-14T07:03Z	&	2012-02-14T12:05:49Z	&	5.03	&	2012-02-14T06:23:42Z	&	-0.65	\\
3	&	2012-02-26T20:58Z	&	2012-02-26T17:17:55Z	&	-3.67	&	2012-02-26T15:47:57Z	&	-5.17	\\
4	&	2012-03-08T10:45Z	&	2012-03-08T11:24:13Z	&	0.65	&	2012-03-08T08:25:20Z	&	-2.32	\\
5	&	2012-04-05T20:03Z	&	2012-04-05T23:58:05Z	&	3.92	&	2012-04-06T05:42:36Z	&	9.65	\\
6	&	2012-04-21T09:25Z	&	2012-04-22T20:37:18Z	&	35.20	&	2012-04-22T13:58:20Z	&	28.55	\\
6	&	2012-04-23T02:33Z	&	2012-04-23T15:03:27Z	&	12.50	&	2012-04-23T09:07:47Z	&	6.57	\\
7	&	2012-05-20T01:36Z	&	2012-05-20T03:09:37Z	&	1.55	&	2012-05-20T04:56:41Z	&	3.33	\\
8	&	2012-06-16T09:01Z	&	2012-06-16T18:06:23Z	&	9.08	&	2012-06-16T17:23:02Z	&	8.37	\\
8	&	2012-06-16T19:31Z	&	2012-06-16T18:06:23Z	&	-1.40	&	2012-06-16T17:23:02Z	&	-2.12	\\
9	&	2012-07-14T17:28Z	&	2012-07-14T10:34:02Z	&	-6.88	&	2012-07-14T11:17:42Z	&	-6.17	\\
10	&	2012-08-02T09:22Z	&	2012-08-02T01:54:46Z	&	-7.45	&	2012-08-02T00:37:17Z	&	-8.73	\\
11	&	2012-09-03T11:23Z	&	2012-09-03T17:37:15Z	&	6.23	&	2012-09-03T19:21:53Z	&	7.97	\\
12	&	2012-09-30T22:13Z	&	2012-09-30T16:08:41Z	&	-6.07	&	2012-09-30T17:10:03Z	&	-5.03	\\
13	&	2012-10-08T04:31Z	&	2012-10-07T22:59:30Z	&	-5.52	&	2012-10-08T01:27:31Z	&	-3.05	\\
14	&	2012-10-31T14:40Z	&	2012-11-01T00:41:58Z	&	10.02	&	2012-10-31T22:09:07Z	&	7.48	\\
15	&	2012-11-23T21:12Z	&	2012-11-23T20:05:02Z	&	-1.10	&	2012-11-23T19:51:14Z	&	-1.33	\\
16	&	2013-01-16T23:00Z	&	2013-01-17T02:32:57Z	&	3.53	&	2013-01-17T07:44:11Z	&	8.73	\\
17	&	2013-01-19T16:47Z	&	2013-01-19T05:03:51Z	&	-11.72	&	2013-01-19T04:16:07Z	&	-12.50	\\
18	&	2013-03-15T05:05Z	&	2013-03-14T19:46:12Z	&	-9.30	&	2013-03-14T19:15:49Z	&	-9.82	\\
19	&	2013-03-16T01:50Z	&	2013-03-15T14:14:50Z	&	-11.58	&	2013-03-15T15:44:35Z	&	-10.08	\\
19	&	2013-03-17T05:28Z	&	2013-03-16T19:03:13Z	&	-10.40	&	2013-03-16T18:44:48Z	&	-10.72	\\
20	&	2013-04-13T22:15Z	&	2013-04-13T18:13:49Z	&	-4.02	&	2013-04-13T16:56:26Z	&	-5.30	\\
21	&	2013-04-30T08:57Z	&	2013-04-30T10:15:25Z	&	1.30	&	2013-04-30T12:33:19Z	&	3.60	\\
22	&	2013-05-19T22:21Z	&	2013-05-19T07:16:52Z	&	-15.07	&	2013-05-19T06:34:29Z	&	-15.77	\\
23	&	2013-05-24T17:35Z	&	2013-05-24T03:54:22Z	&	-13.67	&	2013-05-24T03:08:14Z	&	-14.43	\\
24	&	2013-06-30T10:40Z	&	2013-06-30T06:58:04Z	&	-3.68	&	2013-06-30T05:47:45Z	&	-4.87	\\
25	&	2013-07-09T19:58Z	&	2013-07-10T01:48:43Z	&	5.83	&	2013-07-10T02:57:06Z	&	6.98	\\
26	&	2014-01-07T14:25Z	&	2014-01-07T13:37:18Z	&	-0.78	&	2014-01-07T19:29:40Z	&	5.07	\\
27	&	2014-01-09T19:31Z	&	2014-01-09T01:19:03Z	&	-18.18	&	2014-01-09T02:00:08Z	&	-17.50	\\
28	&	2014-03-25T19:25Z	&	2014-03-26T04:30:21Z	&	9.08	&	2014-03-26T04:00:44Z	&	8.58	\\
29	&	2014-04-20T10:24Z	&	2014-04-19T23:57:59Z	&	-10.43	&	2014-04-20T00:26:29Z	&	-9.95	\\
29	&	2014-04-20T10:24Z	&	2014-04-20T13:37:39Z	&	3.22	&	2014-04-20T12:03:10Z	&	1.65	\\
30	&	2014-08-19T05:58Z	&	2014-08-19T17:03:50Z	&	11.08	&	2014-08-19T17:21:44Z	&	11.38	\\
31	&	2014-09-06T14:19Z	&	2014-09-06T00:34:17Z	&	-13.73	&	2014-09-06T00:59:42Z	&	-13.32	\\
32	&	2014-09-11T22:58Z	&	2014-09-12T00:34:32Z	&	1.60	&	2014-09-11T23:29:22Z	&	0.52	\\
33	&	2014-09-12T15:30Z	&	2014-09-12T09:13:22Z	&	-6.27	&	2014-09-12T11:54:02Z	&	-3.58	\\
34	&	2017-07-01T16:26Z	&	2017-07-02T14:19:14Z	&	21.88	&	2017-07-02T09:53:30Z	&	17.45	\\
35	&	2017-07-16T05:14Z	&	2017-07-16T19:32:03Z	&	14.30	&	2017-07-16T21:13:40Z	&	15.98	\\
36	&	2017-09-06T23:08Z	&	2017-09-06T17:26:52Z	&	-5.68	&	2017-09-06T15:14:18Z	&	-7.88	\\
37	&	2018-02-15T07:50Z	&	2018-02-14T16:35:57Z	&	-15.23	&	2018-02-14T17:14:00Z	&	-14.60	\\
38	&	2019-03-24T20:43Z	&	2019-03-23T12:56:50Z	&	-31.77	&	2019-03-23T23:12:41Z	&	-21.50	\\
\tableline													
\end{tabular}
\end{center}
\end{table*}

\begin{table*}[!ht]
\tabletypesize{\scriptsize}
\caption{Summary of CME observed and simulated arrival times along with the arrival time prediction errors for GONGz ADAPT WSA 4.5 ENLIL 2.9e a8b1 (median). \label{tbl:errors4}}
\begin{center}
\begin{tabular}{rc|cr|cr}
\tableline
\tableline
Event	&	Arrival Time	&	Arrival Time	&	$\Delta t$	&	Arrival Time	&	$\Delta t$	\\
\#	&	Observed	&	Single map	&		&	Time-dependent	&		\\
	&	{\scriptsize[UT]} 	&	{\scriptsize[UT]} 	&	(hours)	&	{\scriptsize[UT]} 	&	(hours)	\\
\tableline											
1	&	2012-01-24T14:31Z	&	2012-01-24T16:13:00Z	&	1.70	&	2012-01-24T14:53:47Z	&	0.37	\\
2	&	2012-02-14T07:03Z	&	2012-02-14T13:21:18Z	&	6.30	&	2012-02-14T12:01:58Z	&	4.97	\\
3	&	2012-02-26T20:58Z	&	2012-02-26T10:27:22Z	&	-10.50	&	2012-02-26T13:15:10Z	&	-7.70	\\
4	&	2012-03-08T10:45Z	&	2012-03-08T03:58:03Z	&	-6.77	&	2012-03-08T04:10:57Z	&	-6.57	\\
5	&	2012-04-05T20:03Z	&	2012-04-05T13:13:19Z	&	-6.82	&	2012-04-05T23:46:30Z	&	3.72	\\
6	&	2012-04-21T09:25Z	&	2012-04-21T19:01:53Z	&	9.60	&	2012-04-22T06:50:29Z	&	21.42	\\
6	&	2012-04-23T02:33Z	&	2012-04-23T05:40:40Z	&	3.12	&	2012-04-23T10:21:24Z	&	7.80	\\
7	&	2012-05-20T01:36Z	&	2012-05-19T14:50:30Z	&	-10.75	&	2012-05-19T23:54:13Z	&	-1.68	\\
8	&	2012-06-16T09:01Z	&	2012-06-16T11:08:19Z	&	2.12	&	2012-06-16T11:12:25Z	&	2.18	\\
8	&	2012-06-16T19:31Z	&	2012-06-16T11:08:19Z	&	-8.37	&	2012-06-16T11:12:25Z	&	-8.30	\\
9	&	2012-07-14T17:28Z	&	2012-07-14T04:55:40Z	&	-12.53	&	2012-07-14T09:47:55Z	&	-7.67	\\
10	&	2012-08-02T09:22Z	&	2012-08-02T15:47:30Z	&	6.42	&	2012-08-02T08:33:59Z	&	-0.80	\\
11	&	2012-09-03T11:23Z	&	2012-09-03T07:14:58Z	&	-4.13	&	2012-09-03T17:24:53Z	&	6.02	\\
12	&	2012-09-30T22:13Z	&	2012-09-30T10:45:10Z	&	-11.45	&	2012-09-30T13:30:52Z	&	-8.70	\\
13	&	2012-10-08T04:31Z	&	2012-10-08T00:50:08Z	&	-3.67	&	2012-10-08T11:01:03Z	&	6.50	\\
14	&	2012-10-31T14:40Z	&	2012-11-01T08:39:49Z	&	17.98	&	2012-11-01T07:24:37Z	&	16.73	\\
15	&	2012-11-23T21:12Z	&	2012-11-23T22:26:32Z	&	1.23	&	2012-11-23T20:16:56Z	&	-0.92	\\
16	&	2013-01-16T23:00Z	&	2013-01-17T12:37:06Z	&	13.62	&	2013-01-17T08:33:47Z	&	9.55	\\
17	&	2013-01-19T16:47Z	&	2013-01-20T01:13:47Z	&	8.43	&	2013-01-19T16:19:22Z	&	-0.45	\\
18	&	2013-03-15T05:05Z	&	2013-03-14T21:15:39Z	&	-7.82	&	2013-03-14T22:54:24Z	&	-6.17	\\
19	&	2013-03-16T01:50Z	&	2013-03-16T03:00:29Z	&	1.17	&	2013-03-15T20:17:25Z	&	-5.53	\\
19	&	2013-03-17T05:28Z	&	2013-03-16T20:30:39Z	&	-8.95	&	2013-03-16T20:08:28Z	&	-9.32	\\
20	&	2013-04-13T22:15Z	&	2013-04-13T18:45:30Z	&	-3.48	&	2013-04-13T20:23:01Z	&	-1.85	\\
21	&	2013-04-30T08:57Z	&	2013-04-30T07:49:12Z	&	-1.12	&	2013-04-30T09:40:03Z	&	0.72	\\
22	&	2013-05-19T22:21Z	&	2013-05-19T06:40:37Z	&	-15.67	&	2013-05-19T06:44:23Z	&	-15.60	\\
23	&	2013-05-24T17:35Z	&	2013-05-24T00:47:01Z	&	-16.78	&	2013-05-24T02:26:43Z	&	-15.13	\\
24	&	2013-06-30T10:40Z	&	2013-06-30T06:08:06Z	&	-4.52	&	2013-06-30T10:59:56Z	&	0.32	\\
25	&	2013-07-09T19:58Z	&	2013-07-10T02:41:37Z	&	6.72	&	2013-07-10T06:04:04Z	&	10.10	\\
26	&	2014-01-07T14:25Z	&	2014-01-07T10:59:49Z	&	-3.42	&	2014-01-07T16:31:15Z	&	2.10	\\
27	&	2014-01-09T19:31Z	&	2014-01-09T02:50:36Z	&	-16.67	&	2014-01-09T01:43:22Z	&	-17.78	\\
28	&	2014-03-25T19:25Z	&	2014-03-25T20:40:57Z	&	1.25	&	2014-03-26T07:48:05Z	&	12.38	\\
29	&	2014-04-20T10:24Z	&	2014-04-20T03:06:53Z	&	-7.28	&	2014-04-20T06:27:41Z	&	-3.93	\\
29	&	2014-04-20T10:24Z	&	2014-04-20T12:54:25Z	&	2.50	&	2014-04-20T13:24:23Z	&	3.00	\\
30	&	2014-08-19T05:58Z	&	2014-08-19T08:44:16Z	&	2.77	&	2014-08-19T10:46:18Z	&	4.80	\\
31	&	2014-09-06T14:19Z	&	2014-09-05T18:03:43Z	&	-20.25	&	2014-09-06T00:35:31Z	&	-13.72	\\
32	&	2014-09-11T22:58Z	&	2014-09-12T01:43:54Z	&	2.75	&	2014-09-12T02:57:39Z	&	3.98	\\
33	&	2014-09-12T15:30Z	&	2014-09-12T07:22:21Z	&	-8.12	&	2014-09-12T12:29:45Z	&	-3.00	\\
34	&	2017-07-01T16:26Z	&	2017-07-02T12:17:49Z	&	19.85	&	2017-07-01T20:59:56Z	&	4.55	\\
35	&	2017-07-16T05:14Z	&	2017-07-16T21:07:54Z	&	15.88	&	2017-07-16T19:34:16Z	&	14.33	\\
36	&	2017-09-06T23:08Z	&	2017-09-06T20:18:47Z	&	-2.82	&	2017-09-06T17:27:36Z	&	-5.67	\\
37	&	2018-02-15T07:50Z	&	2018-02-14T18:48:26Z	&	-13.02	&	2018-02-14T19:13:08Z	&	-12.60	\\
38	&	2019-03-24T20:43Z	&	2019-03-24T00:07:08Z	&	-20.58	&	2019-03-23T22:50:34Z	&	-21.87	\\
\tableline													
\end{tabular}
\end{center}
\end{table*}

\end{document}